\documentclass[a4paper,fleqn,usenatbib]{mnras}
\usepackage{newtxtext,newtxmath}
\usepackage[T1]{fontenc}
\usepackage{ae,aecompl}
\usepackage{graphicx}	
\usepackage{amsmath}	
\usepackage{amssymb}	
\usepackage{pdflscape}

\newcommand{\IUE}{{\it IUE}}
\newcommand{\HST}{{\it HST}}
\newcommand{\lbolint}{\ifmmode L(IR;X-rays) \else $L${\it (IR; X-rays)}\fi}
\newcommand{\ledd}{\ifmmode L_{Edd}\else $L_{Edd}$\fi}
\newcommand{\lbol}{\ifmmode L_{BOL}\else $L_{BOL}$\fi}
\newcommand{\Mdot}{\ifmmode \dot{M} \else $\dot{M}$\fi}
\newcommand{\oiii}{O\,{\sc iii}}
\newcommand{\feii}{Fe\,{\sc ii}}

\title[]{Spectral energy distribution variations of nearby Seyfert galaxies during AGN watch 
monitoring programs}

\author[E. Kilerci Eser and M. Vestergaard]{
Ece Kilerci Eser,$^{1,2}$\thanks{E-mail: ecekilerci@phys.nthu.edu.tw (EKE)}
M. Vestergaard,$^{2,3}$
\\
$^{1}$Institute of Astronomy , National Tsing Hua University, No. 101, Section 2, Kuang-Fu Road, Hsinchu, 30013, Taiwan\\
$^{2}$Dark Cosmology Centre, Niels Bohr Institute, University of Copenhagen,
Juliane Maries Vej 30, DK-2100 Copenhagen \O, Denmark\\
$^{3}$Steward Observatory, University of Arizona, 933 North Cherry Avenue, Tucson, AZ 85721.
}

\date{Accepted 2017 October 11. Received 2017 September 14; in original form 2017 March 20}

\pubyear{2017}

\begin{document}
\label{firstpage}
\pagerange{\pageref{firstpage}--\pageref{lastpage}}
\maketitle

\begin{abstract}

We present and analyse quasi-simultaneous multi-epoch spectral energy distributions 
(SEDs) of seven reverberation-mapped active galactic nuclei (AGNs) for which accurate black hole mass 
measurements and suitable archival data are available from the `AGN Watch' 
monitoring programs.   
We explore the potential of optical-UV and X-ray data, obtained within 2 d, 
to provide more accurate SED-based measurements of individual AGN and quantify 
the impact of source variability on key measurements typically used to characterize the 
black hole accretion process plus on bolometric correction factors at 5100\,\AA, 
1350\,\AA{} and for the 2--10 keV X-ray band, respectively.  
The largest SED changes occur on long time-scales ($\gtrsim$1 year). 
For our small sample, the 1$\micron$ to 10\,keV integrated accretion luminosity typically 
changes by 10 per cent on short time-scales (over 20 d), by $\sim$30 per cent over a year, but 
can change by 100 per cent or more for individual AGN. 
The extreme ultraviolet (EUV) gap is the most uncertain part of the intrinsic SED, introducing a 
$\sim$25 per cent uncertainty in the accretion-induced luminosity, relative to 
the model independent interpolation method that we adopt. 
That aside, our analysis shows that the uncertainty in the accretion-induced 
luminosity, the Eddington luminosity ratio and the bolometric correction factors 
can be reduced (by a factor of two or more) by use of the SEDs built from data 
obtained within 20 d. 
However, \Mdot{} and $\eta$ are mostly limited by the unknown EUV emission and the 
unknown details of the central engine and our aspect angle. 
\end{abstract}




\section{INTRODUCTION}\label{introduction}

Active galactic nuclei (AGNs) are powered by a central supermassive black hole accreting matter from the accretion disc surrounding it. 
The various physical processes connected with this mass accretion result in the AGN emitting radiation across the entire electromagnetic spectrum from radio through $\gamma$-rays.  The spectral energy distributions (SEDs) therefore carry important information about these processes and how black holes grow and affect their immediate and more distant environment. The SED typically consists of 4 $-$ 5 main components \citep[e.g.][]{Petersonbook}: (i) radio emission from a jet in $\sim$10 per cent of AGN; (ii) infrared emission from cool dust heated by stars in the host galaxy and hot dust heated by the AGN; (iii) optical-UV emission from the accretion disc, fast-moving dense gas in the broad-line region and slower-moving, more diffuse gas in the narrow-line region; (iv) X-ray emission from a region in close proximity of the black hole thought to be mostly in the form of a corona, and (v)  $\gamma$-ray emission, believed to be inverse Compton scattered synchrotron photons in the relativistic plasma in radio jets \citep[see also][]{McConville2014,Dermer2016}. Of these components, the radio component contributes at best only a minor portion of the total energy released and the $\gamma$-ray component is only observed (and observable) for a smaller fraction of AGN. Since the infrared (IR) component is reprocessed photons emitted at higher energies, the SED components that are most relevant for studying the accretion properties of AGN are the optical-UV and X-ray components \citep[e.g.][]{Sanders1989,Marconi2004,Suganuma2006}. The $\alpha$ accretion disc model of \citet{Shakura73} is considered the standard model to explain the optical-UV component of the SED by means of thermal emission from a geometrically thin, optically thick disc. 
 The peak of the SED and its tail obtained from the maximum effective temperature in this model do not match the observed Far-UV emission \citep{Stevans2014} and the soft X-ray excess component, 
the excess flux at soft X-rays below 2\,keV with respect to the extrapolated power law continuum in the 2$-$10 keV regime \citep[e.g.][]{Laor1997}. 
The up-scattered UV photons from the disc by a Comptonizing corona is believed to produce: 
(i) the soft X-ray excess component by the Compton up-scattering from a low-temperature, optically thick corona; and 
(ii) the high-energy X-ray power law component by Compton up-scattering from a high-temperature, optically thin corona \citep[e.g.][]{Czerny1987,Zheng1997,Telfer2002,Mehdipour2011,Mehdipour2015,Done2012,Jin2012a}. 
Although AGN accretion discs are not fully understood and the observed SEDs are not entirely consistent with the standard accretion disc model \citep[e.g.][]{KoratkarBlaes1999,Hubeny2001,Stevans2014}, it is still the commonly adopted model, mostly due to the lack of better alternative models. However, this may soon change as new promising models have appeared recently \citep[e.g.][]{GardnerDone2014, GardnerDone2016} 
and new high-quality {\it HST}/COS spectra of the far-UV and extreme-UV region of AGNs are obtained \citep[e.g.][and references therein]{Tilton2016}.

Several studies have constructed mean (or composite) SEDs to obtain the typical (accretion) properties of large AGN samples based on multiwavelength observations from radio to X-rays \citep[][]{Elvis94,Elvis2012,Richards2006,Shang2011}. 
Although the total power (i.e., energy per second) emitted by the AGN is represented by the bolometric luminosity, important information is also contained in the total integrated luminosity across the IR to X-ray regime, \lbolint, as well as broad-band luminosities, integrated over smaller energy ranges thereof. An often computed parameter is the mean bolometric correction (BC) factor, the scaling factor that brings a measured monochromatic luminosity, $\lambda L_{\lambda}$, to the average bolometric (or \lbolint) luminosity of the AGN population \citep{Elvis94,Richards2006,Runnoe2012}. The mean BC is a very practical way to infer the total integrated luminosity when only a single monochromatic luminosity measurement is available. However, there are also limitations to this practise.  Owing to the large range in SED properties \citep[e.g.][]{Elvis94,Richards2006}, the mean BCs provide only a crude estimate of the bolometric luminosity (or of \lbolint). 
Also, the value of the BC factor depends on the AGN sample from which it was derived  and on the assumed SED shape (\S \ref{S:bol}). 
While the mean BC factors provide a general sense of the intrinsic brightness of an AGN, they suffer from object to object differences and usually have large (up to 50 per cent) uncertainties \citep{Elvis94,Richards2006}. Therefore, these mean parameters should be used with caution. 
As emphasized by \citet{Richards2006}, measuring \lbolint\ directly is essential to obtain accurate measurements for individual sources. 
This is important because the total integrated luminosity, the Eddington luminosity ratio, $\lbol/\ledd$, the mass accretion rate and the accretion efficiency are key physical properties directly related to the accretion process: 
the total energy produced by the AGN depends directly on the black hole mass accretion rate, $\Mdot$, and the accretion efficiency, $\eta$.  And the Eddington luminosity ratio, related to $\Mdot$, affects the broad emission lines \citep{Collin2006} and the SED \citep[e.g.][]{Vasudevan2007}.

Obtaining a full SED for an individual AGN is challenging because the emission at all wavelengths tends to vary in time. This will provide 
an inaccurate view of the emission processes unless  simultaneous ground- and space-based observations are obtained. 
To avoid the effects of intrinsic variability on the SEDs of individual AGN, simultaneous SEDs have been 
investigated using space-based telescopes in a few studies \citep[e.g.][]{Brocksopp2006,Vasudevan2009a,Vasudevan2009b,Grupe2010}. 
Especially, the {\it{XMM$-$Newton}} and {\it{Swift}} telescopes \citep{Gehrels2004}  are capable of multiwavelength observations including optical-UV photometry and X-ray spectroscopy. 
\citet{Brocksopp2006} gather contemporaneous optical-UV and X-ray data to study the SEDs of 23 Palomar$-$Green quasars.
\citet{Vasudevan2007} constructed the SEDs of 14 quasars from the \citet{Brocksopp2006} sample with non-simultaneous data and found that the BCs are significantly different from the ones obtained from simultaneous SEDs.
Later, \citet{Vasudevan2009a} study the SEDs of 29 reverberation-mapped (RM) AGNs from the \citet{Peterson04} sample with simultaneous optical-UV and X-ray observations with 
{\it{XMM$-$Newton}}.
The authors fit the multicolour accretion disc model of \citet{Gierlinski1999} to their SEDs in order to measure  \lbolint\ and  calculate the Eddington luminosity ratio and X-ray BC factors for the 2--10\,keV band.
These authors also compare multiple observations for three sources and find that long-term optical-UV variability (over a few years) changes the SED shape and the integrated luminosity by up to a factor of two.
\citet{Vasudevan2009b} and \citet{Grupe2010} use simultaneous optical-UV and X-ray observations from {\it{Swift}} and present optical to X-ray SEDs of 26 and 92 AGNs, respectively.
 The individual single-epoch SEDs studied by those authors are not affected by intrinsic variations, but they have very few photometric data points in the optical-UV region. 
 Therefore, they use an accretion disc model or a power law with an exponential cut-off for the optical-UV part of the SED and derive the SED parameters based thereon. 
These studies provide ample evidence for the need of simultaneous optical-UV and X-ray observations when one aims to measure \lbolint\ directly from the SED.

Simultaneous multiwavelength SEDs are observationally expensive and therefore not always available. 
For this reason, limited studies of the intrinsic variability of AGN SEDs exist. 
We therefore set out to examine the intrinsic variations of the SEDs and their effects on the measured  \lbolint, Eddington luminosity ratio, mass accretion rate, accretion efficiency and BC for individual AGN. 
We make use of so-called single-epoch SEDs, snapshots of the AGN emission observed across the optical, UV and X-ray regime at a given observing epoch [i.e., observed (semi-)contemporaneously]. 
By generating such single-epoch SEDs at a range of epochs, we can study how the SED changes as the AGN accretion luminosity varies with time. 
We chose to investigate the SEDs of nearby RM AGNs \citep{Peterson04} for two reasons: 
(1) robust black hole mass measurements are vital to obtain reliable Eddington luminosity ratios and mass accretion rates; and (2) AGNs subjected
to repeated observations during monitoring campaigns are much more likely to have multiple epochs of simultaneously obtained data at a range of energies.

The RM technique \citep{Blandford82,Peterson93} utilises the variable nature of AGN to measure the distance and kinematics of the gas in the broad emission line region (BLR) of AGNs. 
This technique is mainly used to directly measure the masses of the central black holes (BHs) in local ($z\le0.3$) Type 1 
(broad-line) AGNs \citep[e.g.][]{Peterson04} but recent efforts by the Sloan Digital Sky Survey and DES/OzDES collaborations \citep{OzDES2014,OzDES2015,Shen2015a} investigate the interesting possibility of using this technique to measure cosmic distances \citep[][, see also]{Watson2011, KilerciEser2015}. 
Since the continuum photons emitted by the accretion disc photoionizes the broad-line gas, the broad emission-line flux varies in response to changes in the continuum flux with a delay corresponding to the light-travel time, $\tau$, of said photons.
RM measures the responsivity weighted distance to the gas as this travel time $\tau$, thereby defining a 
characteristic radius of the BLR, $R_{BLR}=c\tau$ ($c$ is the speed of light). 
The width of the broad emission line profiles and this radius of the line emitting region are consistent with a virialized BLR in the potential of the central BH \citep{PetersonWandel99,Peterson2000,Kollatschny2003,Peterson04,Bentz2010a}.   
This virial relationship allows us to measure the central BH mass by means of measurements of $R_{BLR}$ and the emission-line width \citep[e.g.][]{Peterson04}. The RM technique and the results of its application has further implications for our understanding of AGN and BHs. For example, the observed $R_{BLR} - L$ relationship \citep[e.g.][]{Kaspi2000,Kaspi2005,Bentz2006,Bentz09a,Bentz2013} for local RM AGNs allows us to estimate the BH mass at all redshifts indirectly via mass scaling relationships  \citep[e.g.][]{McLure2002,Vestergaard2002,Shields2003,GrupeMathur2004,GreeneHo2005,MathurGrupe2005, Kollmeier2006,Vestergaard06,McGill2008,Park2013}. Although, these mass values are afterall {\it estimates} given the approximative nature of the method, they are, in fact, very important for studies of the BH-galaxy connection in the local and more distant Universe \citep[e.g.][]{Kurk2007,Vestergaard2004,Vestergaard08,Vestergaard09, Jiang2010, Wang2016}, in part because these are the best estimates that we currently have. 
  
RM studies require intensive, time-consuming spectroscopic monitoring. 
Several monitoring programs have been designed for RM studies of nearby AGNs during the past three decades \citep[e.g.][and references therein]{Clavel91,Peterson91,Peterson92,Dietrich93,Peterson94,Stirpe94,Korista95,Kaspi96,Santos97,Collier98,Dietrich98,Peterson2000,Shemmer2001,Peterson2002,Peterson04,Denney2006, Bentz09b,Bentz2010a, Grier2012, Bentz2014, Derosa2015, Fausnaugh2016}. 
In order to get a reliable RM measurement, these observing campaigns have mainly targeted $\sim$60 low-luminosity, local, highly-variable AGNs \citep[hereafter referred as the RM sample; e.g.][]{BentzKatz2015}. 
Most of these RM programs, designed to obtain optical spectra nightly from ground-based telescopes, typically span a few months. 
However, as it turns out, even for the RM AGN, which are amongst the most observed objects, there are a limited
number of completely simultaneous multiwavelength observing programs. 
Space-based monitoring programs have been undertaken in the UV \citep[e.g.][]{Clavel89,Clavel90,Clavel91,Reichert94,Korista95,Crenshaw96,Ulrich96,Rodriguez-Pascual97,Wanders97,O'Brien98,Collier2001,Peterson2005,Derosa2015} and X-rays \citep[e.g.][]{Warwick96,Leighly1997,Turner2001,Edelson2015} only for a couple of AGN. 
These RM programs mainly aimed at measuring the broad emission-line response times to the continuum variations, driving the line emission, and at determining the structure and kinematics of the BLR. 
The monitored AGN show both long-term and short-term continuum variability in the optical and UV  \citep[e.g.][]{Clavel90,Clavel91,Korista95,Cackett2015,Derosa2015,Fausnaugh2016,McHardy2016,Edelson2017}. 
Another important conclusion reached from these monitoring campaigns is that the optical and UV continua vary together without a phase difference \citep[e.g.][]{Clavel90,Clavel91,Peterson93,Korista95}. 
Optical continuum has a lower variability amplitude compared to the UV continuum  \citep[e.g.][]{Clavel90,Clavel91,Peterson93,Reichert94,Korista95,KilerciEser2015,Fausnaugh2016} and in general the amplitude of variation increases with decreasing wavelength \citep[e.g.][]{Clavel91,Krolik1991,Korista95}. 
The X-ray continuum also shows rapid variability, on time-scales as short as hours \citep[e.g.][]{Nandra91}.
Although some observations show that the X-ray and optical-UV fluxes are correlated \citep[e.g.][]{Clavel92}, others do not \citep[e.g.][]{Done1990}. 
 
The `International AGN WATCH' \citep[][]{Peterson2002} is a network of more than 100 astronomers that aimed to obtain multiwavelength monitoring of a number of AGNs 
by using the {\it International Ultraviolet Explorer (IUE), Hubble Space Telescope (HST)} and ground-based telescopes.
The `AGN Watch' RM monitoring campaigns revealed a clear view of the continuum variations in the optical, UV and the X-rays, however propagation of multiwavelength continuum variations into the broad-band SED properties has, to our knowledge, not been investigated during these campaigns.  
The multiwavelength simultaneous monitoring data sets have a great potential to reveal such SED variations. 
Accretion rate variations can be determined by studying the SED variations and important physical insight into the continuum variability can be gained. 
Our goal here is to investigate \lbolint , Eddington luminosity ratio, mass accretion rate and accretion efficiency variations
based on the simultaneous optical to X-rays SEDs obtained in the `AGN Watch' program. 
Quantifying these variations is a first step towards understanding their possible effects on the broad-emission line properties which will be investigated in a future study (Kilerci Eser \& Vestergaard, in preparation).

In this work, we quantify \lbolint , Eddington ratio, mass accretion rate and accretion efficiency variations
of seven RM AGN monitored in the `AGN Watch' program
and study the SEDs of individual objects at multiple epochs. 
We obtain  \lbolint\ measurements directly from the SEDs of simultaneous optical-UV and X-ray data. 
In Sections 2 and 3  we present the sample and the data, respectively. 
In \S 4 we describe the corrections applied to the data before we generate the SEDs (\S5). We present and discuss our results from analysing the SEDs and their temporal variations in \S 6 and summarize our conclusions in \S 7.
We use a cosmology with $H_0=72$\,km\,s$^{-1}$\,Mpc$^{-1}$, $\Omega_\Lambda = 0.7$ and $\Omega_{\rm m}=0.3$ throughout. 
Also, we use the base 10 logarithm throughout.

\begin{table*}
           \centering
           \caption{
           Local Seyfert sample. Columns: (1) Object name. (2) Redshifts are adopted from NASA/IPAC Extragalactic data base (NED). (3) E(B$-$V) values are based on \citet{Schlafly2011} dust maps and adopted from NED.  
           (4) Distances in Mpc are computed from the redshifts \citep{Wright2006}; uncertainties of the distances are calculated assuming a 500 km s$^{-1}$ uncertainty in the recession velocities \citep[similar to][]{Bentz2013}. The distances of NGC\,4151 and NGC\,3783 are adopted from \citet{Bentz2013}. (5) Host galaxy flux densities in $10^{-15}$ erg s$^{-1}$ cm$^{-2}$ \AA$^{-1}$ for the relevant aperture sizes (column 6) and orientations (column 7) are adopted from \citet{Bentz09a,Bentz2013,KilerciEser2015}. (6) Aperture size in square arcsec (arcsec $\times$ arcsec). (7)  Position angle in degrees ($\degr$). (8) The black hole mass, listed in units of $10^{6}$ $M_{\sun}$, are based on RM analysis and adopted from \citet{BentzKatz2015}. These masses are computed using the scaling factor of $<f> = 4.3 \pm 1.1$ from \citet{Grier2013b}. (9) Radio loudness information, based on the $R=f(5GHz)/f(B)$ values ($\log(R) > 1 = $ radio loud) calculated from the fluxes adopted from NED.
	}
           \label{tab:table1}
            \begin{tabular}{lcrccccrc}
           \hline
           Object & $z$ & E(B$-$V) & Distance & Host flux density & Aperture                                                     & PA          & BH mass               & Radio loud\\
                      &        & [mag]   & [Mpc]      &[$10^{-15}$ erg s$^{-1}$ cm$^{-2}$ \AA$^{-1}$]& [arcsec $\times$ arcsec] & [$\degr$] & [10$^6 M_{\sun}$] & \\
            (1) & (2) & (3) & (4) & (5) & (6) & (7) & (8) & (9) \\
           \hline
Fairall\,9     &0.04702  & 0.023 &202.8$\pm$7.2&2.99 $\pm$ 0.15& $4.0 \times9.0$ &0.0 &$199.1^{+39.2}_{-60.9}$ &No\\
NGC\,3783& 0.00973 &0.105  &25.1$\pm$5.0&4.72 $\pm$ 0.47 &$5.0 \times10.0$&0.0&$ 23.5^{+4.6}_{-4.5}$&No\\ 
NGC\,4151& 0.00332 & 0.024 &16.6$\pm$3.3&15.00 $\pm$1.40&$5.0 \times7.5$  &90.0&$ 35.9^{+4.5}_{-4.1}$&No\\
NGC\,5548& 0.01717 & 0.018 &72.5$\pm$7.0&3.75 $\pm$ 0.38 &$5.0 \times7.5$  &90.0 &$ 52.2^{+2.0}_{-2.0}$ &No\\  
3C\,390.3   & 0.05610 & 0.063 &243.5$\pm$7.2&0.83 $\pm$ 0.04&$5.0 \times7.5$  &90.0&$434.5^{+41.9 }_{-48.5}$ &Yes\\
Mrk\,509    &0.03440 &0.051&147.1$\pm$7.3 &2.43 $\pm$ 0.12&$5.0 \times7.6$  &90.0&$111.9^{+9.4 }_{-9.4}$ &No\\ 
NGC\,7469& 0.01632 & 0.061 &68.8$\pm$7.0&8.43 $\pm$ 0.78 &$5.0 \times7.5$  &90.0&$9.0^{+1.1 }_{-1.1}$&No\\ 
		\hline
	\end{tabular}
\end{table*}

\section{The Sample }\label{S:Sample}

To investigate the temporal variations of the SEDs we focus our study on the 
$\sim$60 local RM AGNs with robust black hole mass measurements  \citep[][and references therein]{Peterson04,Bentz09b,Denney10,Bentz2013,Grier2013,BentzKatz2015}.  
The  multiwavelength spectroscopic monitoring data publicly available for this sample provide a particularly suited data base to probe intrinsic variations in the \lbolint , Eddington ratio, mass accretion rate and accretion efficiency
as gleaned from quasi-simultaneous SEDs. 

Our sample size is defined by the availability of multiple sets of epochs of optical-UV and X-ray data, 
typically observed within a few d of each other, from the International AGN Watch\footnote[1]{http://www.astronomy.ohio-state.edu/$\sim$agnwatch/} 
data base \citep{Peterson2002} the {\it IUE}, {\it HST}, and {\it FUSE} archives, and the literature in general. 
Our final sample consists of seven RM AGN (NGC\,5548, NGC\,7469, NGC\,3783, Mrk\,509, NGC\,4151, 3C\,390.3, Fairall\,9).  
We list several source properties of these AGN in Table \ref{tab:table1}, namely: object name (column 1); redshift (column 2); 
Galactic colour excess E(B$-$V) value based on the dust maps from \citet{Schlafly2011} (column 3); 
distance (column 4); adopted host galaxy flux measurement from \citet{Bentz09b,Bentz2013} (column 5); 
the aperture size and position angle used for the host galaxy flux measurements (columns 6 and 7, respectively); 
 the adopted central black holes mass ($M_{BH}$) estimate from  \citet[][]{BentzKatz2015} (column 8). 

\section{Data}\label{S:data}

The International AGN Watch data base \citep{Peterson2002} provides optical and UV 
light curves and spectra from several monitoring campaigns that ran between 1988 and 2001.  
To generate SEDs, we select AGNs from this data base where simultaneous optical-UV and X-ray data are available. 
We match the Julian Dates (JD) of the optical observation with the closest JD of the UV observation to within two d. 
We extract additional UV data from the {\it IUE}, {\it HST} and {\it FUSE} archives which contain UV observations that are not officially 
part of the AGN Watch data base \citep{Peterson2002} but obtained for other reasons during the optical monitoring campaigns. 
Once the simultaneous optical-UV data epochs are selected, we then located archival X-ray observations obtained during these epochs from the literature.  
While we aim to obtain the contemporaneous data sets within a few d, for a couple of epochs 
(see the footnotes to Table \ref{tab:table4}) 
we take the closest X-ray data to within three weeks in order to increase the number of available SED epochs which would otherwise be rather limited. 

\subsection{Optical-UV Data}\label{S:opdata} 

We extract the ground-based optical and the space-based (\IUE\ ) UV spectra from the International 
AGN Watch data base. 
Two UV spectra of Mrk\,509, obtained by \IUE\ during the optical monitoring campaign in 1990, are not provided in the AGN Watch data base \citep{Peterson2002}, but are available 
in the Barbara A. Mikulski Archive for Space Telescopes (MAST\footnote[2]{Multimission Archive at STScI}) \IUE\  archive\footnote[3]{http://archive.stsci.edu/iue/}. 
We extract the processed \IUE\ spectra observed during JD=2448187.7 and  JD=2448187.8 and create an average spectrum for further analysis. 
Details of the observations, data processing and calibration are provided by the reference study listed in Table \ref{tab:table2}. 
We place all the optical spectra on an absolute flux scale using the [\oiii]$\lambda 5007$ line flux provided by the reference works listed in Table \ref{tab:table2}. 
For NGC\,5548, we use the updated fluxes presented by \citet{Peterson2013}. 

\begin{table}
           \centering
           \caption{ References for the Optical, UV and X-ray Data. Columns: (1) Object name; For NGC\,4151 see \citet{KilerciEser2015} for data selection and measurements. (2) Optical Data References are 1: \citet{Santos97}; 2: \citet{Stirpe94}; 3: \citet{Kaspi96}; 4: \citet{KilerciEser2015}; 5: \citet{Peterson91}; 6: \citet{Peterson92}; 7: \citet{Peterson2013}; 8: \citet{Dietrich98}; 9: \citet{Carone96}; 10: \citet{Collier98}. 
           (3) UV Data References are: 11: \citet{Rodriguez-Pascual97}; 12: \citet{Reichert94}; 13: \citet{Crenshaw96}; 14: \citet{Clavel91}; 15: \citet{Clavel92}; 16: \citet{O'Brien98}; 17: \citet{ Wanders97}; Names of the used UV spectra used in IUE archive are ads/Sa.IUE\#swp39925 and \#swp39926. 
           (4) X-ray Data References are: 15: \citet{Clavel92}; 18: \citet{Alloin95}; 19: \citet{Warwick96}; 20: \citet{Nandra91}; 21: \citet{Magdziarz98}; 22: \citet{Pounds94}; and 23: \citet{Nandra2000}; Tartarus \citep{Turner2001}: http://tartarus.gsfc.nasa.gov/.
	}
           \label{tab:table2}
           \begin{tabular}{lccc}
           \hline
           Object & Optical data & UV data & X-ray data\\
           (1) & (2) & (3) & (4) \\
           \hline
Fairall\,9                                  &1            &11                                             & Tartarus \\
NGC\,3783                             &2             &12                                             &18                                           \\
NGC\,4151			     &3, 4         &13                                            &19                                            \\
NGC\,5548                             &5, 6, 7     &14, 15                                      &15, 20, 21                                \\
3C\,390.3                               &8             &16                                             & Tartarus  \\
Mrk\,509                                 &9            & IUE archive                              &22                                          \\  
NGC\,7469                             &10, 4       &17                                             & 23                                          \\
		\hline
	\end{tabular}
\end{table}          
           
\subsubsection{Continuum emission measurements}\label{S:opuv_cont} 

We model each of the optical and UV spectra with a power law continuum in the observed wavelength ranges listed in columns (2) and (3) of Table \ref{tab:table3}.  
We parametrize the continuum as $F(\lambda)= N\times\ (\lambda /\lambda({\rm ref}))^{\alpha}$, where $\alpha$ is the slope and $N$ is the normalization at the reference wavelength, $\lambda({\rm ref})$. For the optical spectra, $\lambda({\rm ref})$ = 5100\,\AA, while $\lambda({\rm ref})$ = 1350\,\AA{} for the UV spectra. 
The best-fitting values of these parameters are determined by the Levenberg-Marquardt least-squares fit IDL routine \citep[MPFIT][]{Markwardt2009}. 
Since error spectra are not available, in order to constrain the uncertainty of the best-fitting continuum we compute the root mean square (rms) of the continuum flux density, $\sigma_{C}$ , 
within the continuum windows relative to the best-fitting continuum. 
We use the best-fitting continuum and $\sigma_{C}$ to construct extreme continua levels: 
(a) best-fitting continuum$+\sigma_{C}$; (b) best-fitting continuum$-\sigma_{C}$; (c) the bluest continuum possible;   
and (d) the reddest slope possible within the measurement error, $\sigma_{C}$.   
We adopt the largest difference between the best-fitting continuum and these four extreme continuum levels as our conservative uncertainty on the continuum level setting. 
Similarly, we take the difference between the best-fitting slope and the slopes of the bluest and the reddest continuum levels and adopt the largest difference as the slope uncertainty. 
We measure the monochromatic continuum flux density at rest frame $\lambda({\rm ref})$ (i.e., 5100\,\AA\ or 1350\,\AA) as the average flux density in the second (optical) or first (UV) continuum window listed in columns (2) and (3) of Table \ref{tab:table3}, respectively. 
The $1\sigma$ rms flux density in this continuum window measured relative to the mean value is adopted as the monochromatic continuum flux density uncertainty. 

\begin{table*}
           \centering
           \caption{Wavelength windows for optical, UV continuum fit. All wavelengths are in the observed frame. The second optical continuum window contains $\lambda_{rest}=5100$\,\AA.
	}
           \label{tab:table3}
           \begin{tabular}{lcc}
           \hline
           Object & Optical continuum windows & UV continuum windows \\
            & (\AA)& (\AA) \\
            (1) & (2) & (3) \\
             \hline
Fairall\,9                                  &4950-4970; 5328-5352 & 1380-1390; 1500-1520; 1780-1810; 1860-1900\\ 
NGC\,3783                             &4800-4820; 5130-5170 & 1350-1370; 1450-1470; 1710-1725; 1800-1820\\ 
NGC\,4151                             &4590-4610; 5100-5125 & 1260-1290; 1420-1460; 1805-1835 \\ 
NGC\,5548                             &4790-4800; 5182-5192 &  1350-1385; 1445-1475; 1710-1730; 1845-1870\\
3C\,390.3                               &4750-4790; 5460-5470 &  1345-1385; 1500-1520; 1800-1830; 1890-1910\\
Mrk\,509                                 &4905-4915; 5270-5280 &  1380-1390; 1500-1520; 1780-1810; 1860-1900\\   
NGC\,7469                             &4840-4890; 5176-5200 &  1306-1327; 1473-1495; 1730-1750; 1805-1835\\ 
\hline
	\end{tabular}
\end{table*}

\subsection{X-ray data}\label{S:xraydata} 
We adopt X-ray measurements based on observations by the {\it GINGA}, {\it ASCA}, {\it RXTE}, {\it EXOSAT} and {\it ROSAT} satellites 
from the works listed in column (3) of Table \ref{tab:table2} which also present the details of the X-ray data processing and analysis. 
For two objects, we extract X-ray measurements from the Tartarus data base \citep{Turner2001}.   

\subsubsection{X-ray power law continuum measurements}\label{S:Xray_cont} 

The `primary' continuum (i.e., that is not reprocessed) in AGN X-ray spectra is typically well described by a power law function \citep{Haardt1991}.  
More complex models are available to account for other spectral features in the spectra such as the hard reflection component \citep{Ross1993} and the fluorescent Fe\,K$\alpha$ line \citep[e.g.][]{Fabian2000}. 
Although the AGN X-ray emission can extend up to 100\,keV and beyond \citep[e.g.][]{Molina2013}, we restrict our analysis to the available X-ray data between 2 and 10 keV.  
 The lack of 0.3$-$2\,keV soft X-ray data motivate our method for computing the integrated luminosity described later in section \ref{S:bomeasurements}.
 
The reference works listed in Table \ref{tab:table2} provide the results of the X-ray spectral analysis. 
This includes a power law fit to the observed fluxes between 2 and 10 keV. 
To construct the SEDs we adopt the quoted photon index $\Gamma$ and the broad-band flux between 2 and 10 keV, $F$(2$-$10\,keV). 
Since we are interested in the continuum emitted by the AGN, we adopt the unabsorbed X-ray power law component from the model of the X-ray data. 
We express the broad-band flux between 2 and 10 keV as given in the XSPEC \citep{Arnaud1996} manual\footnote[4]{https://heasarc.gsfc.nasa.gov/xanadu/xspec/manual/XSmodelPowerlaw.html}
\begin{equation}\label{Eq:x1}
F(2 - 10\, \rm keV)=\int_{2 keV}^{10 keV}f_{E}\,dE=\int_{2 keV}^{10 keV}N E^{-\alpha} dE\,
 \end{equation} 
\noindent where  $f_{E}=N\,E^{-\alpha}$ is the monochromatic flux at energy $E$, $N$ is the normalization, $\alpha$ is the power law index: $\alpha=\Gamma-1$, 
where $\Gamma$ is the photon index; $F$(2$-$10\,keV) is in units of (\rm erg\, \rm cm$^{-2}$\, \rm s$^{-1}$\, \rm keV$^{-1}$).
We follow this formalism to calculate the monochromatic fluxes between 2 and 10 keV. 
 
For NGC\,3783 we adopt the broad-band flux between 1 and 2\,keV, $F$(1$-$2\,keV), observed with {\it ROSAT} \citep{Alloin95} 
because there are no X-ray observations at energies between 2\,keV and 10\,keV. 
We estimate the 2--10\,keV flux for this AGN by extrapolating the 1-2\,keV fluxes, assuming a $\Gamma$ of 1.9 presented by \citet{Alloin95}. 
We note that the adopted X-ray measurements for NGC\,3783 are not robust because the X-ray exposure was very short ($\sim$402 s) 
due to operational problems of {\it{ROSAT}} \citep[see][for details]{Alloin95}. 

\begin{figure}
\begin{center}$
\begin{array}{c}
\includegraphics[scale=0.2]{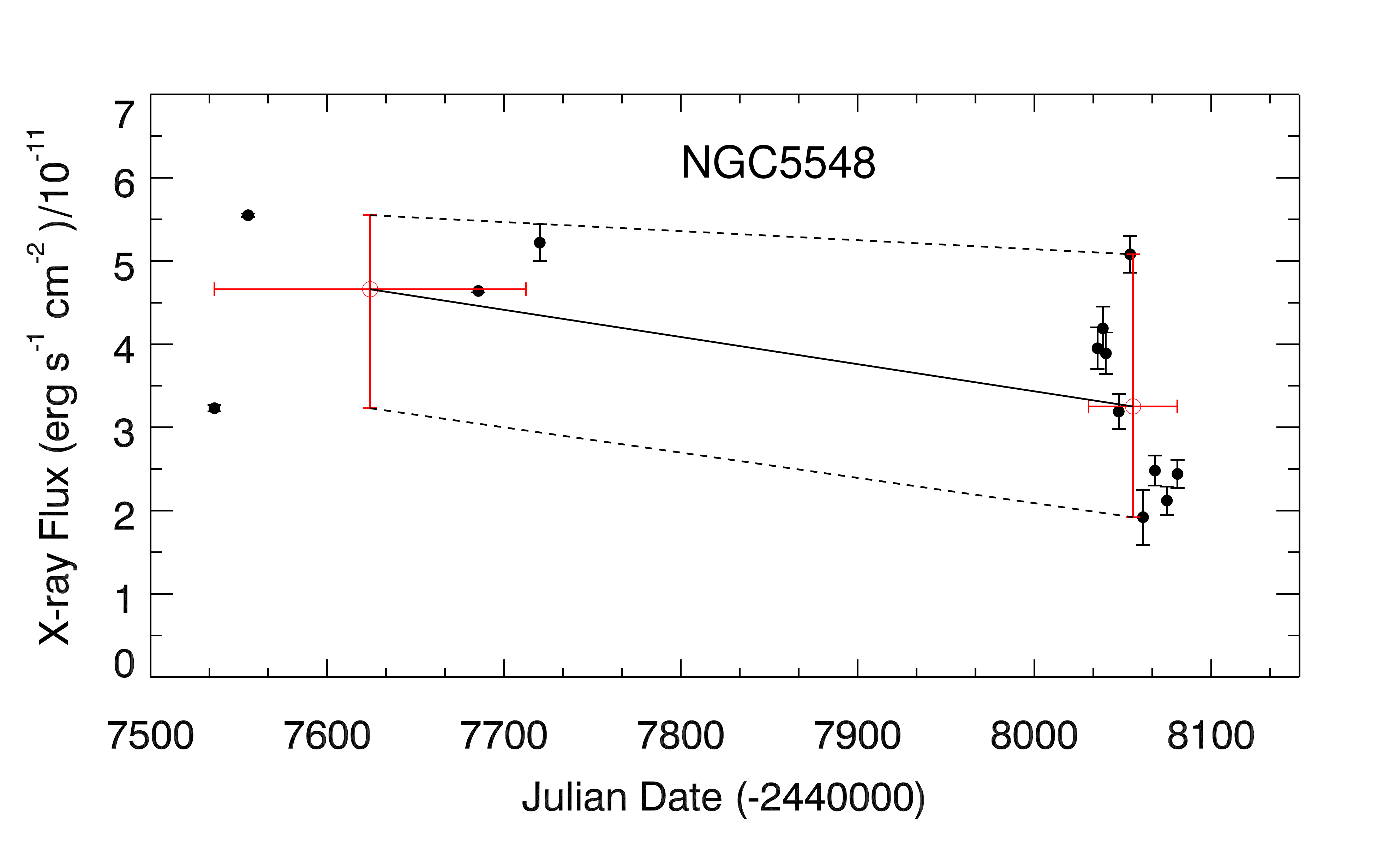}\\ 
\includegraphics[scale=0.2]{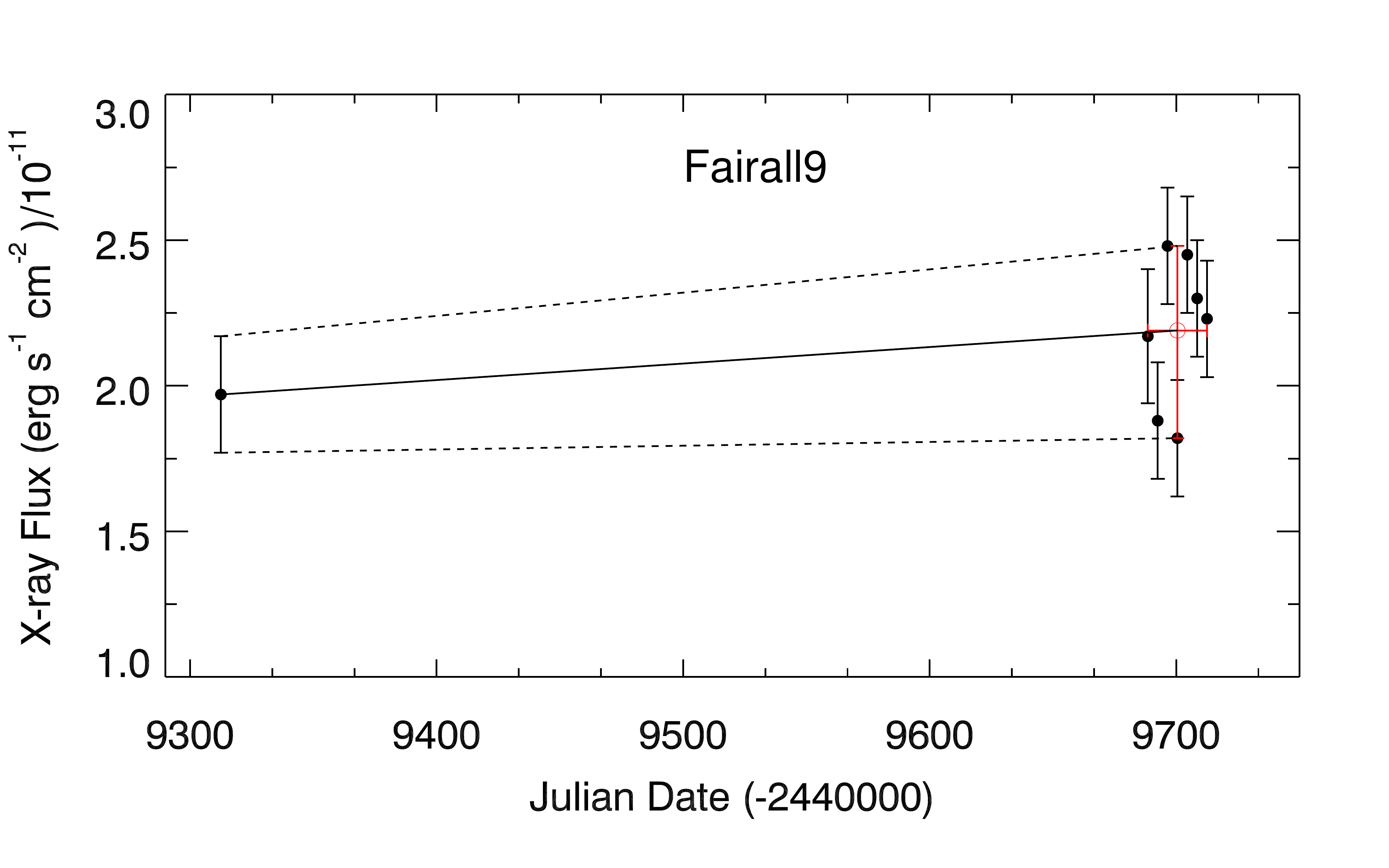}\\ 
\end{array}$
\end{center}
\caption{X-ray flux, $F_{\rm 2-10 keV}$, interpolation for NGC\,5548 and Fairall\,9. The black filled circles are the observed X-ray fluxes. 
For NGC\,5548 the open red circles represent the mean $F_{\rm 2-10 keV}$ values over year 1(JDs:7500$-$7750) and year 2 (JDs:8000$-$8100). 
For Fairall\,9  the open red circle represents the mean $F_{\rm 2-10 keV}$ values between JDs  9688$-$9712.
The solid black lines show the interpolated X-ray fluxes. The dashed grey lines indicate the $\pm 1\sigma$ uncertainties of the interpolated fluxes based on the $\pm 1\sigma$ of the mean fluxes.} 
\label{fig:xrayinterp}
\end{figure}

\subsubsection{Estimated X-ray fluxes}\label{S:extraX-rays} 

For NGC\,5548, there are several epochs for which there are simultaneous optical-UV data but no X-ray observations  within two d. 
In order to optimise the number of epochs with SEDs available, we opt to estimate the observed $F$(2$-$10\,keV) flux during those epochs 
that are straddled by actual X-ray measurements. This way we obtain a guideline measure of the likely X-ray flux level.
We have X-ray data within year 1 (JDs = 2447534 $-$ 2447722) and within year 2 (JDs = 2448035 $-$ 2448078) of the AGN Watch monitoring campaigns.  
The available data are shown in Fig. \ref{fig:xrayinterp} as the black filled points. 
We estimate the X-ray flux level between the two epochs by linearly interpolating between the average 
flux levels (red open circles in Fig. \ref{fig:xrayinterp}, top panel) measured for each of these two epochs. 
The interpolation is shown by a black solid line in the top panel of Fig. \ref{fig:xrayinterp}.
The uncertainties in the interpolated X-ray fluxes (dashed grey lines) are based on the interpolated $\pm 1\sigma$ uncertainties of the mean X-ray fluxes shown by the red circles.
For the epochs during year 1 without direct X-ray measurements we adopt a $\Gamma$ value estimated as the mean of the year 1 photon index values, $\Gamma=1.57\pm0.12$. 
 For the epochs between year 1 and 2, we estimate the photon index as the mean of all observed $\Gamma$ values, $\Gamma=1.65\pm0.11$. 
Then, we use equation (\ref{Eq:x1}) to calculate $F$(2$-$10\,keV). 
Table \ref{tab:table4} lists the SEDs that are based on these estimated X-ray fluxes (epoch numbered from 12 to 45). 

For Fairall\,9, one X-ray observation at JD=24429312 and seven X-ray observations between JDs =2449688$-$2449712 (for references see Table \ref{tab:table2})  are available.
There are many simultaneous optical and UV data between JDs=2449470$-$2449688 without X-ray measurements. 
Again, we choose to estimate the $F$(2$-$10\,keV) flux for the epochs with missing X-ray data by linearly interpolating between 
the X-ray flux on JD=24429312  
and the mean $F(2 - 10 keV)$ observed between JDs =2449688$-$2449712. 
Fig. \ref{fig:xrayinterp} (bottom) shows the available X-ray data and the interpolation. 
Again, we adopt the mean of the observed photon index values, $\Gamma=1.9\pm0.11$, for the (interpolated) epochs between JDs=2449470$-$2449688.
The SEDs based on the estimated X-ray fluxes are listed in Table \ref{tab:table4} from epochs 1 to 26.

\section{Analysis}\label{S:analysis}

\subsection{ Corrections applied to the data}\label{S:Dataprep}
 
\subsubsection{Galactic extinction correction}
The intrinsic emission produced by the AGN is contaminated by dust and gas extinction in the AGN host galaxy, the intergalactic medium (IGM) and our Galaxy. 
Since the extinctions caused by the AGN host galaxy and IGM, respectively, are not very well constrained, we only apply a Galactic reddening correction.
In this case, we use the 
reddening curve of \citet{Cardelli89} with the colour excess values adopted from NASA/IPAC Extragalactic data base based on the dust maps of \citet{Schlafly2011} to correct the observed optical-UV flux densities, $F_{\lambda_{obs}}$ ($\lambda_{obs}$ is the observed wavelength).
The Galactic colour excess E(B$-$V) for each object is listed in column (3) of Table \ref{tab:table1}. 
Galactic absorption correction is included when modelling the X-ray power law.  

\subsubsection{Correction for host galaxy starlight contamination}
The AGN host galaxy emission contributes to the nuclear continuum emission in the observed optical and UV spectra. 
Therefore it is necessary to remove the host galaxy contamination to assure that the observed spectra represents only the intrinsic AGN emission.
\citet{Bentz09a,Bentz2013} measured the observed host galaxy flux densities at rest frame 5100\,\AA\ for the RM sample using \HST\ optical images. 
These authors use the exact apertures that were used for the spectroscopic observations we are analysing. 
We thus base our correction for the stellar flux contribution on these measurements.  
The stellar fluxes are corrected for Galactic extinction as described above.
\citet{Kinney96} provide template spectra for different morphological galaxy types between 1200\AA\ $-$1$\micron$. 
We scale each of the elliptical, Sa, and S0 galaxy templates presented by \citet{Kinney96} to match the mean 5090\AA\ $-$ 5115\AA\ flux density to 
that reported by \citet{Bentz2013} and listed in Table \ref{tab:table1}. 
 For each AGN we use the appropriate template for the galaxy type quoted in that work 
and subtract the scaled host galaxy template from the observed optical-UV spectra of each of our objects. 

We note that the aperture sizes of the optical and UV data are different. 
While the optical spectra are obtained with 4 arcsec $\times$ 9 arcsec, 5 arcsec $\times$ 10 arcsec, 5 arcsec $\times$ 7.5 arcsec, 
5 arcsec $\times$ 7.6 arcsec apertures, all of the UV spectra are obtained with the large \IUE\ aperture (10 arcsec $\times$ 20 arcsec). 
The galaxy template spectra of \citet{Kinney96} are also obtained through apertures of 10 arcsec $\times$ 20 arcsec; this holds for both the UV and optical host galaxy spectra. 
Since we do not have the original UV images of the host galaxies on which the \citet{Kinney96} templates are based, it is not possible to make an accurate correction for the different aperture in the UV. 
As an estimate of how much of a difference the larger UV apertures can make in the host galaxy starlight corrected UV fluxes, we use the ratio of the aperture areas in the UV and optical. 
This ratio gives us an upper limit in the scaling factor difference in the host galaxy contribution since the surface brightness decreases with distance from the centre. 
We list the aperture area ratios for each AGN in our sample in Table \ref{tab:tablenew4}. 
The upper limit in the flux differences from the fractional areas is between a factor of 4.0 and 5.6. 
So, if the \IUE\ aperture was aligned with the optical aperture, then we would expect that the UV flux would be no more than a factor of 5.6 higher 
than predicted from the scaled template based on the optical measurements of \citet{Bentz09a,Bentz2013}. 
When we subtract from our observed UV continuum a 5.6 times higher UV ($1000\AA < \lambda < 2000\AA$) host galaxy flux density we measure a difference in the nuclear UV flux density of only 1$-$11 per cent 
with four AGN showing as little as 1 per cent difference.
Since this is a small difference we consider the applied host galaxy starlight contamination correction in both the optical and UV as reasonable.

\begin{table}
           \centering
           \caption{UV and optical aperture area ratios and the percentage differences in the host galaxy starlight corrected UV fluxes.
	}
           \label{tab:tablenew4}
           \begin{tabular}{llcr}
           \hline
           Object & Host & Aperture  & Nuclear UV  \\
           & type & area ratio & flux density \\
            & &(\IUE /ground-based) &  difference (per cent) \\
            (1) & (2) & (3) & (4) \\
             \hline
Fairall\,9                                  &Sa & 5.6 &4\\ 
NGC\,3783                             &Sa & 4 &1 \\    
NGC\,4151                            &Sa & 5.3 & 1\\
NGC\,5548                           &S0 & 5.3 & 1\\ 
3C\,390.3                                &Sa & 5.3 & 11\\ 
Mrk\,509                                 &elliptical & 5.3 & 1\\                                               
NGC\,7469                                &Sa & 5.3 & 7\\                         
\hline
	\end{tabular}
\end{table}  

\subsubsection{Optical-UV luminosity}

We compute the optical and UV nuclear power law continuum and monochromatic luminosities as 
$L_{\nu_{rest}}= 4\pi\,D_{L}^{2}\,F_{\nu_{o}}/(1+z)$ \citep[e.g.][]{Petersonbook}, where $z$ is the redshift, $\nu_{rest}$ is the rest frame frequency, $D_{L}$ is the 
luminosity distance and $F_{\nu_{o}}$ is the observed flux density at observed frequency ${\nu_{o}}$. We list $z$ and $D_{L}$ in Table \ref{tab:table1}. 

\subsubsection{X-ray luminosity}
We apply the following power law $K$-correction \citep[][]{Petersonbook} to calculate the X-ray luminosities at energy $E$
\begin{equation}\label{Eq:xk}
L_{\nu}(E)=4 \pi D_{L}^{2}\times\ F_{obs}\times\ (1+z)^{\Gamma-2}=\frac{F_{\nu}(E) 4 \pi D_{L}^{2}}{(1+z)^{(2-\Gamma)}}\,.
 \end{equation}

\section{ GENERATING THE OPTICAL-UV-X-RAY SPECTRAL ENERGY DISTRIBUTIONS} \label{S:sedssection}

\begin{figure*}
\begin{center}$
\begin{array}{cc}
\includegraphics[scale=0.4]{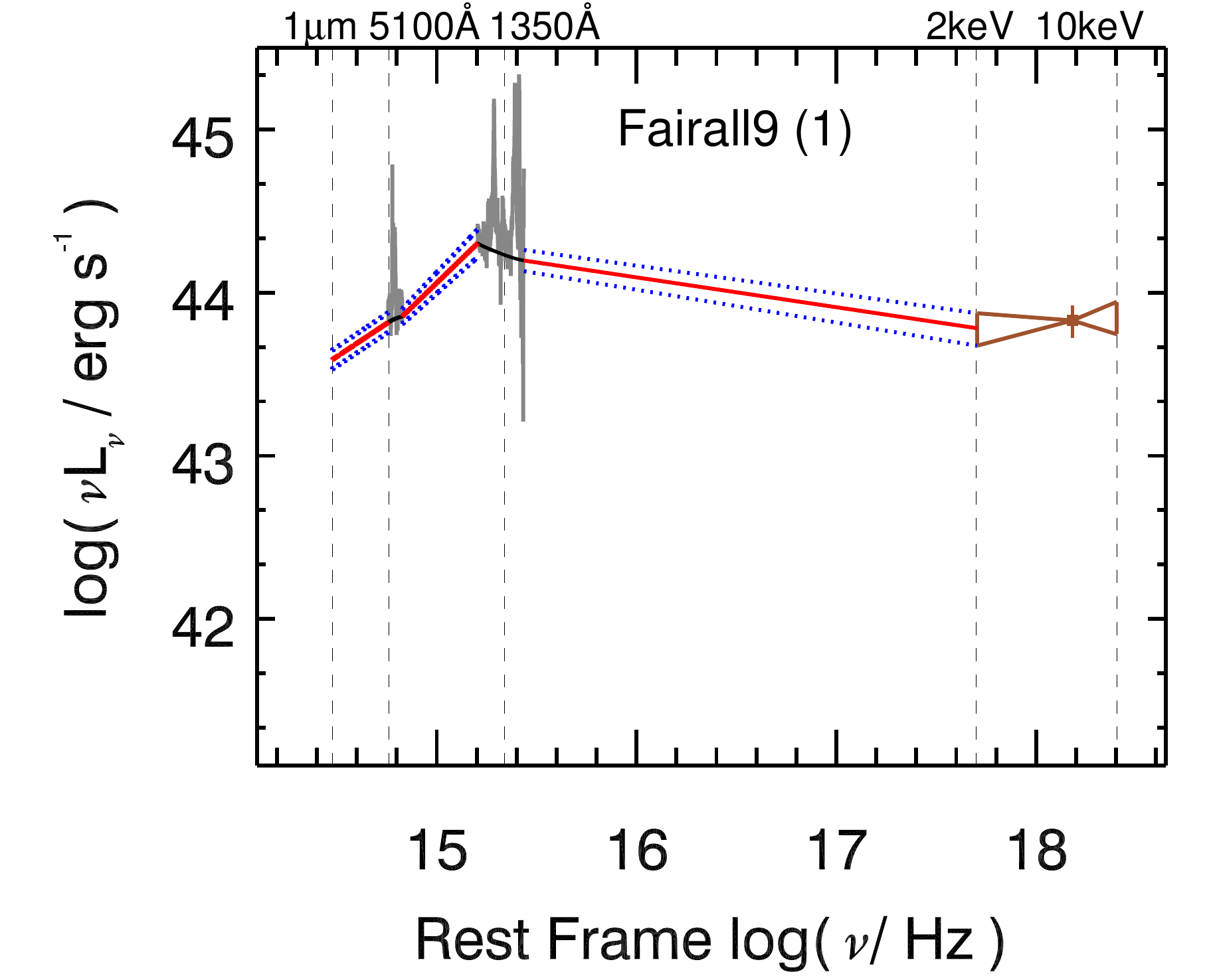} &
\includegraphics[scale=0.4]{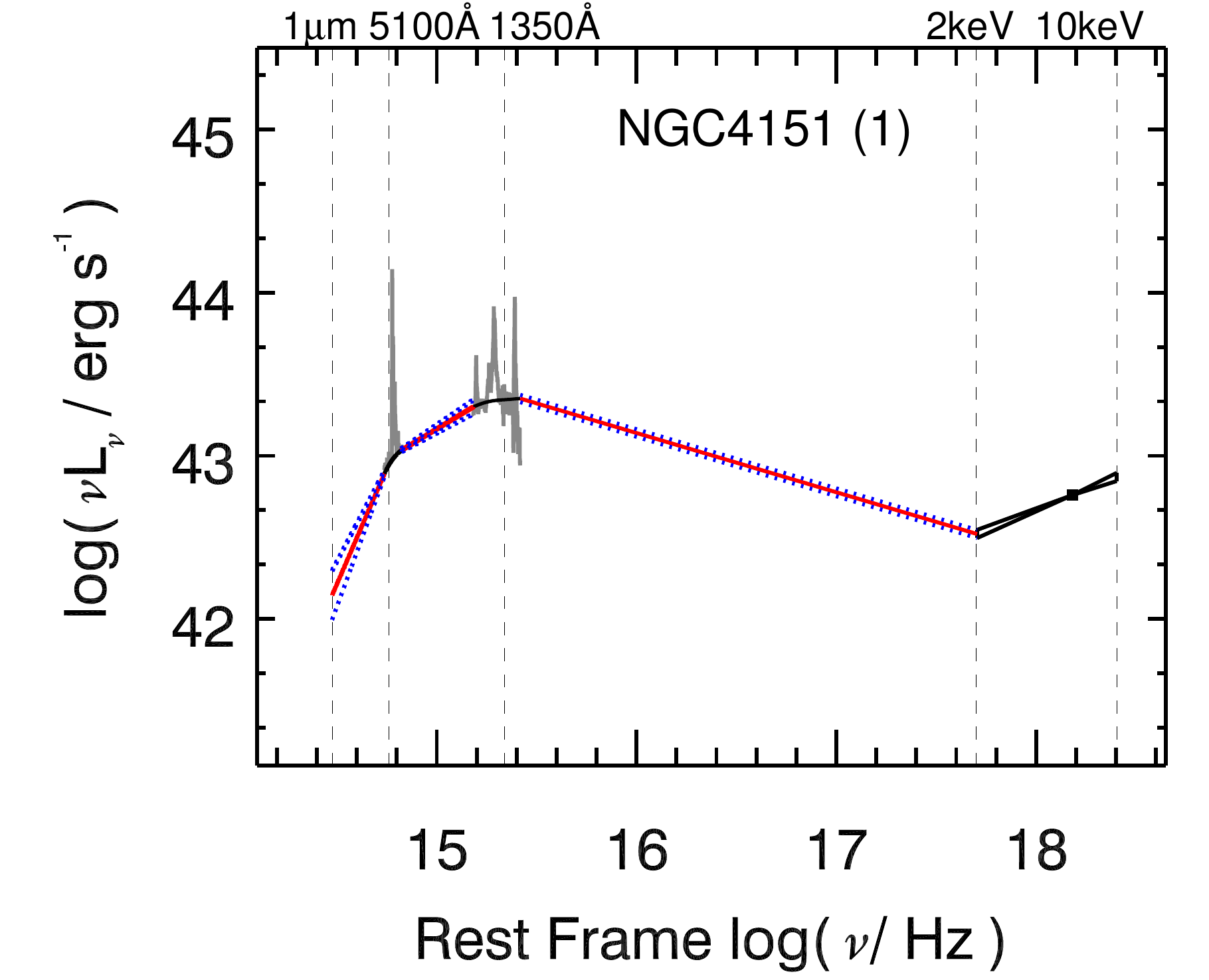}\\
\includegraphics[scale=0.4]{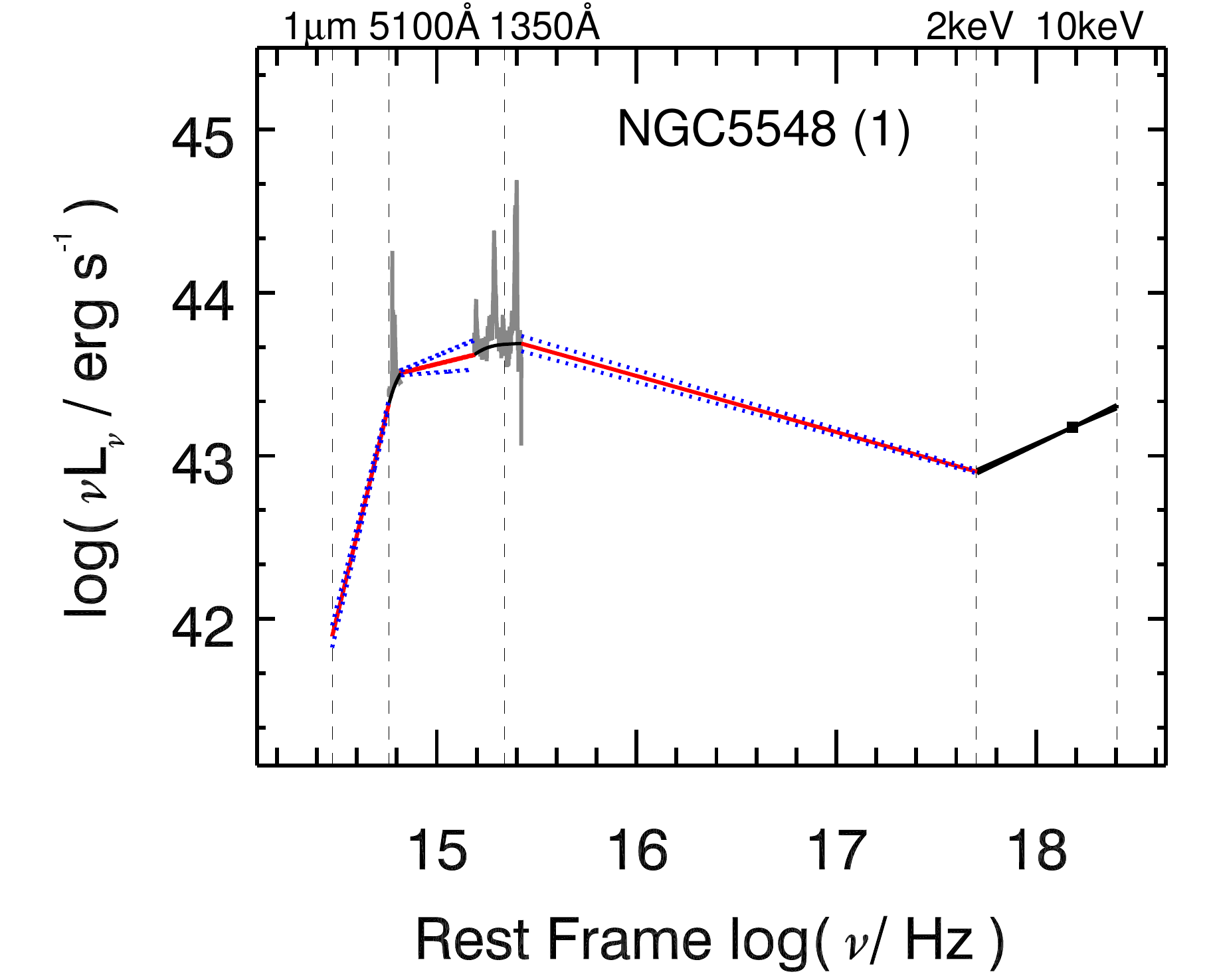} &
\includegraphics[scale=0.4]{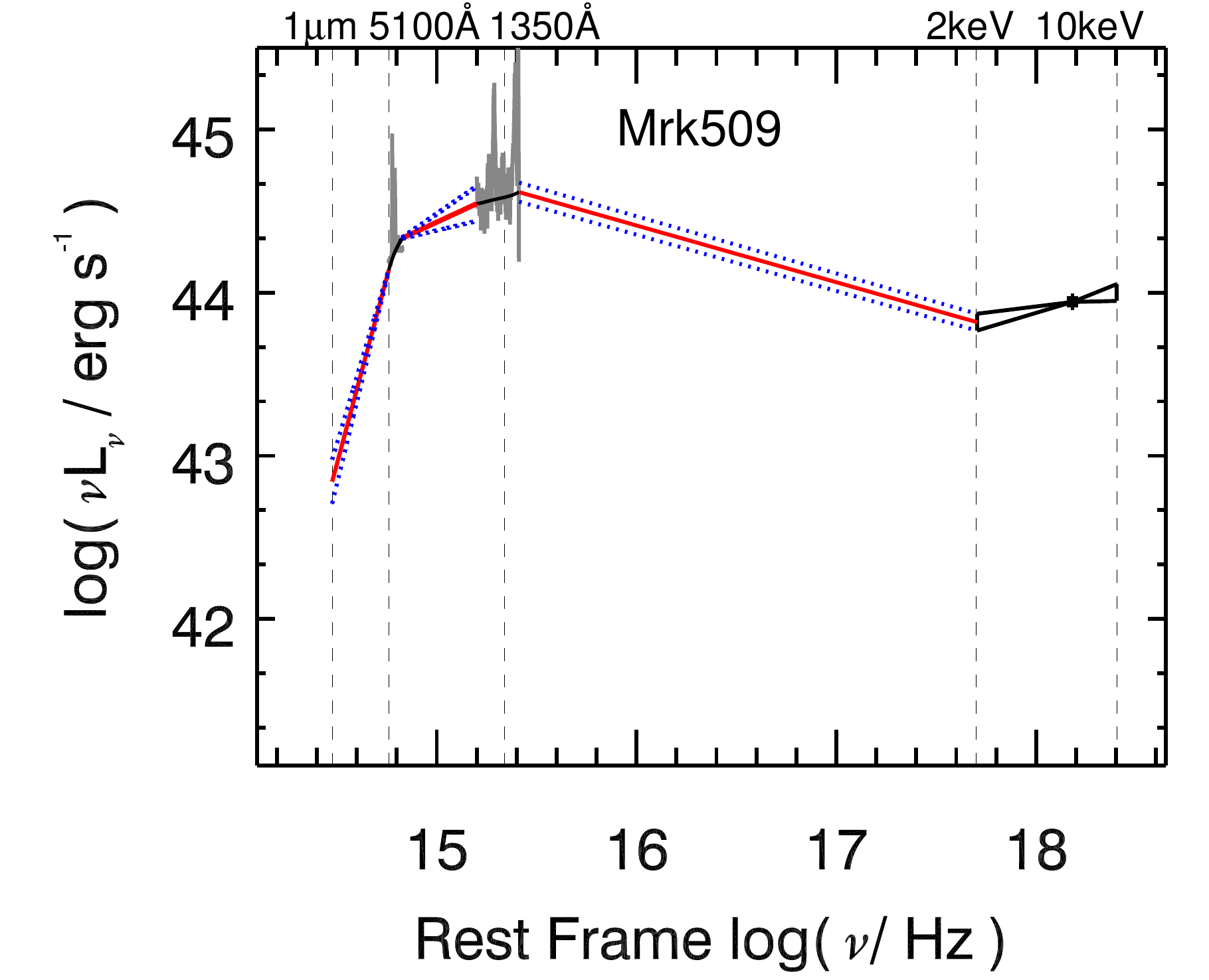}\\
\end{array}$
\end{center}
\caption{Four representative quasi-simultaneous SEDs of our sample. 
We present the rest of the SEDs in Appendix \ref{S:Appendixseds}. 
If more than one epoch is available for a given object, the epoch number is listed in the parenthesis; Table \ref{tab:table4} lists the relevant Julian Dates, JDs, of each epoch number.  
Optical and UV spectra are shown in grey and the solid black lines on the spectra show the best-fitting continuum. 
The optical continuum is extrapolated to 1$\micron$ (\S \ref{S:sedssection}) and the level of the 1$\micron$ luminosity is set by the extrapolation (see \S \ref{S:sedssectionextrapol} for details).
The solid red line represents the linear interpolation between available data sets from 1$\micron$ to 10 keV of the optical, UV and X-ray  continua. 
The `butterfly-shape' symbol at high energies represents the best power law fit slope and its $1 \sigma$ confidence levels of the 2--10\,keV X-ray continuum. 
A brown `butterfly-shape' symbol represents the X-ray continuum based on the \textit{estimated} X-ray fluxes (\S \ref{S:extraX-rays}) for NGC\,5548 and Fairall\,9. 
The frequency and luminosity ranges are the same in each panel to allow direct comparison.  
 }
\label{fig:fig2}
\end{figure*}

We combine the multiwavelength data described in \S \ref{S:data} to build single-epoch SEDs for each of the seven sources in our sample. 
We show the SEDs in log$(\nu L_{\nu})$ versus log$(\nu)$ representation in Fig. \ref{fig:fig2}; 
source names and SED epochs (in parenthesis) are listed in each panel (Table \ref{tab:table4} lists the specific JDs of the SED epochs). 
The optical and UV spectra are shown in grey and the underlying AGN continuum fit is shown as the black solid thin lines. 
The X-ray data are represented by a `butterfly shape' that indicates the 2\,keV to 10\,keV X-ray continuum (\S\ \ref{S:Xray_cont}) represented by the best-fitting power law slope 
and its $1 \sigma$ confidence limit.
We mark a couple of wavelength regions of interest by vertical dashed lines for reference. 

\subsection{ Estimating the continuum level in the SED gaps} \label{S:sedssectiongaps} 

Due to Galactic H\,{\sc i} gas absorption in the extreme ultraviolet (EUV) region, SEDs have gaps between the optical-UV and X-ray regions. 
We have two options for estimating the intensity levels in these gaps: (a) we can adopt specific accretion disc models, e.g., a thin accretion disc \citep{Shakura73} or a thin disc plus an X-ray corona \citep{Jin2012a};  
or (b) we can adopt the most simple approach of linearly interpolating the continuum emission between the observed data sets \citep[e.g.][]{Elvis94}.
We chose linear interpolation in order to avoid any model dependence to enter our analysis  (\S~\ref{S:bolmodel}). 
However, we briefly examine option (a) in \S~\ref{S:bolmodel} and Appendix~\ref{S:optxangf}. 
We interpolate between the continuum gaps from the optical to the X-rays in log$-$log space.  

\subsection{Extrapolation to 1$\mu m$} \label{S:sedssectionextrapol} 

For completeness and comparison with other studies, we present the SEDs starting from 1$\micron$ and measure the 1$\micron$ $-$ 10\,keV integrated luminosity, $L(1\mu m-10\,keV)$.  
Since our reddest data point is at $\sim$5400\AA\ we linearly extrapolate the best-fitting optical continuum fit to 1$\micron$. 
The applied linear extrapolations (or interpolations) are shown as the red solid lines in Fig. \ref{fig:fig2}. 
The dotted blue lines represent the $1\sigma$ uncertainty  in the interpolation based on the observed $\pm 1\sigma$ continuum uncertainties (described in \S \ref{S:Xray_cont} and \S \ref{S:Dataprep}). 
The extrapolation of the optical continuum to 1 $\micron$ is reasonable because adopting e.g., the $\nu^{1/3}$ dependance as expected for a pure, 
thin accretion disc provides an insignificant difference to $L(1\mu m-10\,keV)$ 
(\S \ref{S:bomeasurements} presents the analysis), 
which is the parameter of main interest here.

 \section{RESULTS AND DISCUSSION} \label{S:results} 

 \subsection{The multi-epoch SEDs} \label{S:theseds}

\begin{figure*} 
\begin{center}$
\begin{array}{c}
\includegraphics[scale=1.0]{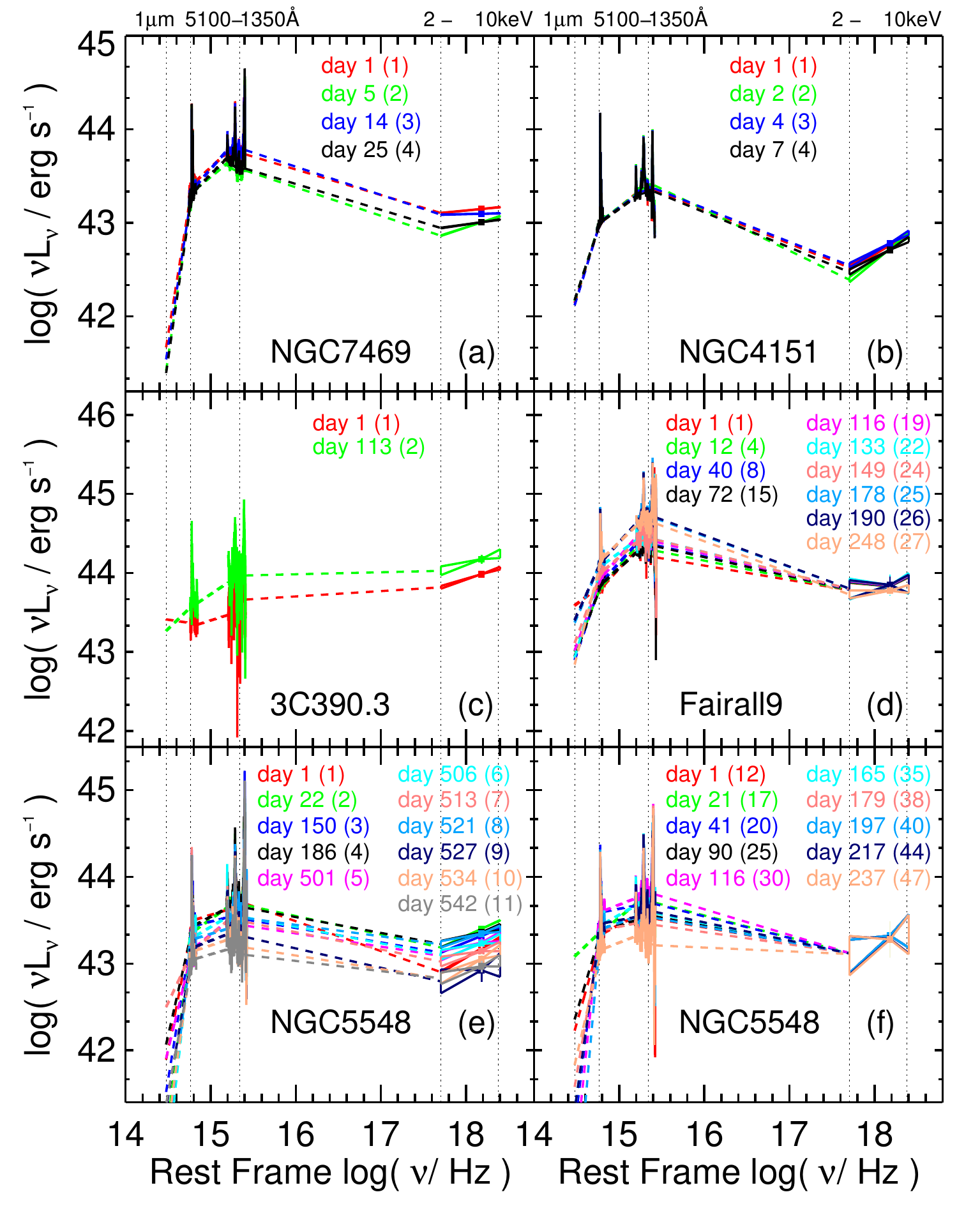}\\ 
\end{array}$
\end{center}
\caption{Multi-epoch SEDs of our sample AGN. SEDs at different epochs are shown in different colours. The day numbers and the SED epochs are shown by the same colour. 
The day number is the difference between the Julian Date of the epoch in question and the Julian Date of the first SED epoch. 
Dashed lines show the linearly extrapolated/interpolated regions and the butterfly shape represents the X-ray continuum (\S\ \ref{S:sedssection}). 
For panels `d' (Fairall\,9) and `f' (NGC\,5548) the X-ray fluxes are \textit{estimated} which explains the low X-ray variance there. }  
\label{fig:fig3}
\end{figure*}

With multi-epoch quasi-simultaneous SEDs, it is possible to compare the changing SED shape over time for individual AGN as its continuum emission varies.  
In Fig. \ref{fig:fig3} we show the observed single-epoch  SEDs for the five AGNs where we have SEDs at multiple epochs. 
To distinguish the SEDs, each epoch is colour coded. 
The dashed lines show the extrapolated and interpolated regions and the `butterfly' shape represents the X-ray continuum and its $1\sigma$ uncertainty. 
For NGC\,4151 and NGC\,7469 for which we only have SEDs spanning very short periods of seven  and 25 d, respectively, we do not see a dramatic change in the overall SED shape.
This is also true for 3C\,390.3 for which we only have two epochs in a time span of 113 d. 
For Fairall\,9, we have 27 epochs spanning 248 d. 
Here, we see the most dramatic changes in the optical-UV region and the change in the X-ray flux is relatively small as expected, since 
we {\it{estimate}} the average X-ray level by interpolation (\S \ref{S:extraX-rays}). 
For NGC\,5548, we have a total of 47 SED epochs, spanning 566 d. 
Amongst those, 11 SEDs are based on simultaneous optical-UV and X-ray data to within two d. 
We show these SEDs in panel (e) of Fig.~\ref{fig:fig3}. 
These simultaneous SEDs span a time period of  1.5 years (542 d) and here we see dramatic changes (e.g larger than 100 per cent; \S~\ref{S:SEDvar}) across the entire SED.  
In panel (f) we show ten representative SEDs of NGC\,5548 with simultaneous optical-UV data (within two d) and estimated X-ray continuum (\S ~\ref{S:extraX-rays}). 
As the X-ray fluxes are interpolated, they clearly contain no variability information. 
We quantify these SED changes in \S~\ref{S:SEDvar}.  

Fig.~\ref{fig:fig3} indicates that the amplitude of the SED variations depend on the time span over which these variations are measured:
long term variations appear to result in more dramatic changes in the SEDs (\S~\ref{S:SEDvar}). 
The UV spectral region typically exhibits higher amplitude variations compared to the optical region. 
This is consistent with the general variability characteristics of AGNs that higher frequency emission exhibits larger variation 
amplitudes \citep[e.g.][]{Clavel91,Korista95,Vandenberk2004} than lower energy emission. 

\subsection{The total accretion luminosity} \label{S:bol}

The bolometric luminosity of an AGN is the total luminosity of the source integrated across the electromagnetic spectrum.
However, the bulk luminosity produced by the central supermassive black hole and accretion disc system is the total luminosity 
from $\sim$1$\micron$ to $\sim$10\,kev$-$100\,keV \citep[e.g.][]{Vasudevan2009a,Jin2012a}. 
The total  luminosity from 13.6\,eV to 13.6\,keV is the ionizing luminosity 
that produces the broad emission lines in the BLR by photoionization of the BLR gas.  
Since the data in 10$-$100\,keV range are unavailable, we adopt an upper energy limit of 10\,keV for the SEDs. 
Because we are interested in the total luminosity produced by the accretion alone, 
we define the accretion luminosity, $L_{BOL}(acc)$, as the integrated luminosity from 1$\micron$ to 10\,keV. 
We thereby exclude the radio through IR regime since the IR emission is reprocessed higher energy emission and the radio flux, if present, does not carry much energy.

\subsubsection{Accretion luminosity measurements: the interpolation method} \label{S:bomeasurements}

\begin{figure} 
\begin{center}$
\begin{array}{c}
\includegraphics[scale=0.5]{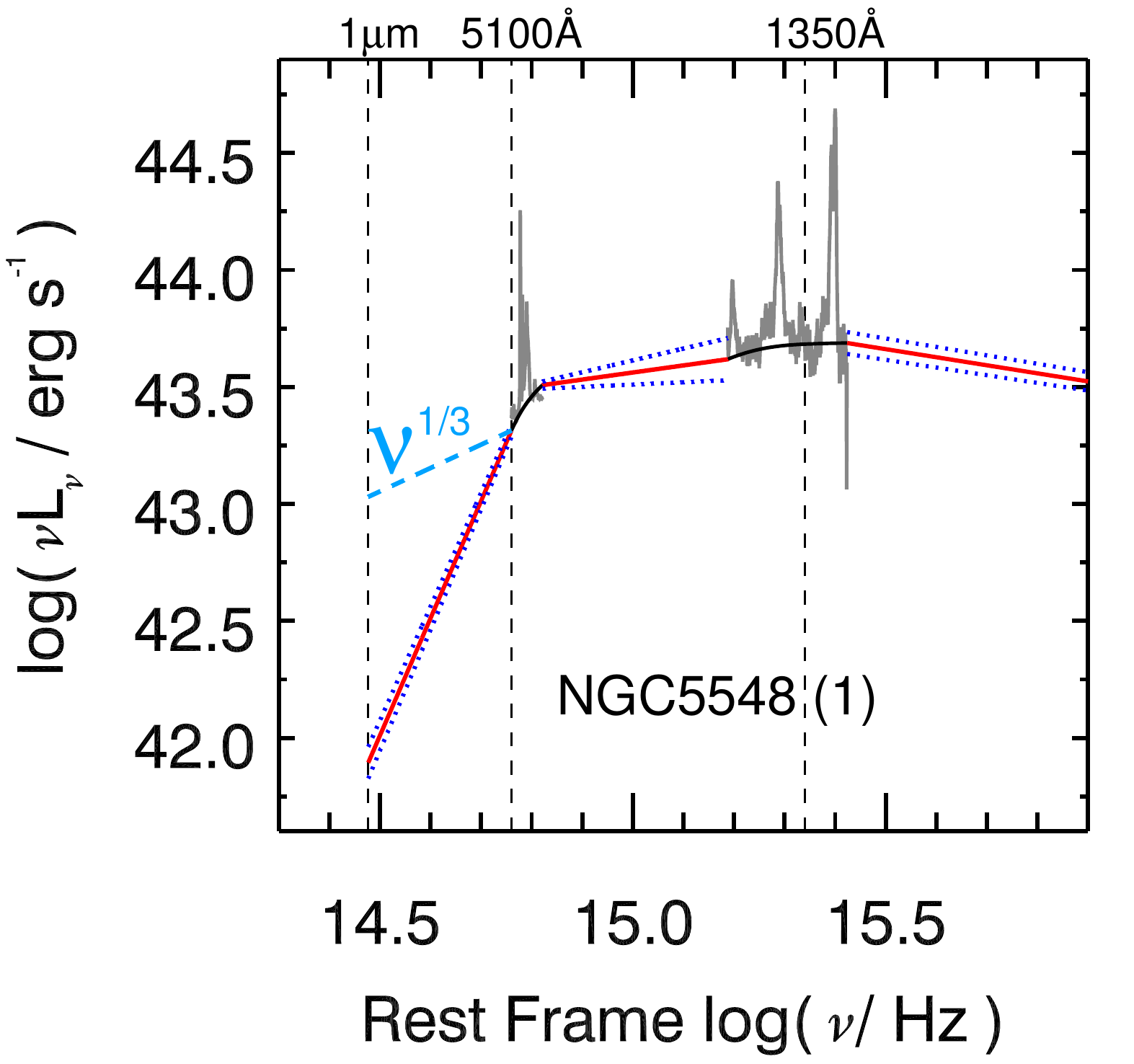}\\ 
\end{array}$
\end{center}
\caption{One representative SED of NGC5548 (see Fig. \ref{fig:fig2} for symbols and colour code).
The $\nu^{1/3}$ disc model is shown as the light blue, dashed line. The linear extrapolation from optical continuum to 1$\micron$ is shown by the red solid line. 
The $\nu^{1/3}$ model is not consistent with the linear extrapolation of the optical continuum and may overestimate the luminosity. However, its effect on the integrated luminosity is negligible.} 
\label{fig:fig4}
\end{figure}

 We do not have simultaneous near-IR spectra for our sample. Although we expect variability time-scales in the near-IR to be lower in amplitude and longer than in the optical or UV, the 1 micron continuum luminosity can vary by $\sim$0.4dex in a few years \citep{Landt2011}. In order to avoid additional uncertainties from non-simultaneous near-IR data, to 
estimate the accretion luminosity between 1$\micron$ and the optical continuum, we extrapolate the optical continuum 
from the lowest energy data point of each SED at $\sim$5400\AA\ to 1$\micron$. 
We test two extrapolation methods: (A) we linearly extrapolate the best-fitting optical continuum power law from $\sim$5400\AA\ to 1$\micron$  using the best-fitting continuum slope (\S\ \ref{S:opuv_cont}); and
(B) we use an accretion disc model of the form $F_{\nu} \propto \nu^{1/3}$ \citep{Shakura73} between 1$\micron$ and the longest optical wavelength. 
We show the linear extrapolation and the accretion disc model for the near-IR region for one representative SED in Fig. \ref{fig:fig4}. 
To quantify the difference in the accretion luminosities caused by choosing 
one of the two approaches, we calculate $L_{BOL}(acc)$ by integrating the SED from 1$\micron$ to 10\,keV for both cases for all sources. 
For each object the difference in $L_{BOL}(acc)$ between methods A and B is very small. 
For the sample as a whole we obtain an average difference of 1.6$\pm$0.7 per cent , or 0.007$\pm$0.003\,dex 
(uncertainties quoted are the standard deviations of the mean differences of all measurements in all epochs or for all objects). 
 As this is an insignificant change we do not use method B for our analysis later. 
 Hereafter $L_{BOL}(acc)$ refers to the measurement based on the interpolation outlined by method (A). 
 We also note that, since the available X-ray data do not cover the soft X-ray range between 0.3 and 2\,keV, we opt not to apply any model that is more complicated than our simplistic interpolation.

We tabulate $L_{BOL}(acc)$ in column (4) of Table \ref{tab:table4}. 
In this table, object names, SED epoch numbers and  SED Julian Date intervals are listed in columns (1), (2) and (3), respectively. 
The errors on the $L_{BOL}(acc)$ values are obtained by propagating the measurement uncertainties on the continuum  emission. 
Namely, we integrate the SEDs over the $\pm 1\sigma\ $ luminosity ranges 
(the blue dotted lines in Fig. \ref{fig:fig2}) of each data point and obtain the upper and lower accretion luminosity, i.e., $L_{BOL}(acc)$ $\pm1\sigma\ $. 
These uncertainties do not include the uncertainty due to our lack of knowledge of the intrinsic SED shape in the EUV region. 
By adding the emission line fluxes to the continuum measurements we find  
the emission line flux contribution to $L_{BOL}(acc)$  is an additional 5 per cent. 
Although a small fraction since the emission lines are not part of the accretion luminosity, but is reprocessed emission, 
we opt to omit the emission line contribution.

\begin{figure*}
\begin{center}$
\begin{array}{c}
\includegraphics[scale=0.4]{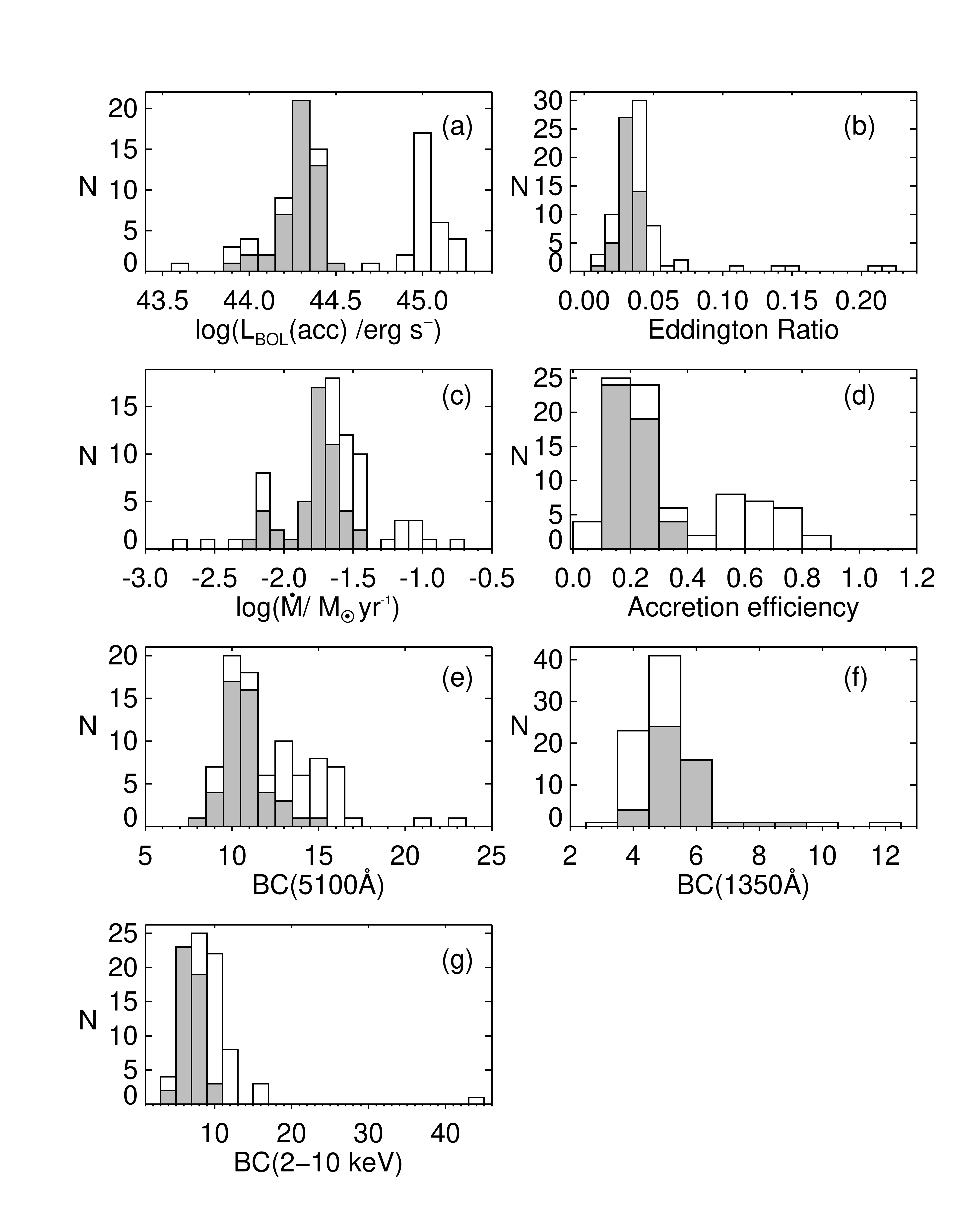} \\
\end{array}$
\end{center}
\caption{Distributions of the accretion luminosities (panel a), Eddington luminosity ratios (panel b), mass accretion rates (panel c), accretion efficiencies (panel d) and BCs (panels e, f, g) of our AGN sample. 
All the SED epochs listed in Table \ref{tab:table4} are included in the histograms, except in panel (d) the two outlier values of 3C\,390.3 are not shown. Grey shaded histograms represent NGC\,5548.}
\label{fig:fig5}
\end{figure*}

The distribution of $L_{BOL}(acc)$ for our sample of seven objects is shown in Fig. \ref{fig:fig5}; the grey histograms represent NGC\,5548.
The accretion luminosities cover the range between 43.56 and 45.25 dex which is in the mid-to-upper luminosity range observed for Seyferts \citep{Elvis94,Vasudevan2009a}. 

\begin{landscape}
\begin{table}
\caption{Results from the SEDs. Columns: (1) Object name. (2) SED epoch number in. (3) Julian date interval of the SED as $-2400000$. (4) Log of accretion luminosity in erg s$^{-1}$. Accretion Luminosity measured from linearly interpolated SED between 1$\micron$ and 10\,keV. (5) Eddington luminosity ratio calculated as $\lambda_{Edd}=L_{BOL}(acc) /L_{Edd}$ where $L_{Edd}=M_{BH} \times 1.3 \times 10^{38}$.
(6) Log of optical luminosity at $\lambda$ = 5100\,\AA\ in  erg s$^{-1}$. (7) Optical BC at $\lambda$ = 5100\,\AA. BC for the stated monochromatic luminosity is calculated as $L_{BOL}(acc)=$BC($\lambda$)$\times \lambda L_{\lambda}$. (8)  Log of optical luminosity at $\lambda$ = 1350\,\AA\ in  erg s$^{-1}$. 
(9) UV BC at $\lambda$ = 1350\,\AA. (10) Log of 2$-$10\,keV luminosity in erg s$^{-1}$. (11) BC for the 2$-$10 keV X-ray band. (12) (a) Only the optical and UV data are simultaneous at this epoch, the X-ray flux is estimated based on the available observations, see \S \ref{S:extraX-rays} for details; (b) Only the X-ray and UV data are simultaneous, the optical data are from JD 2449724; (c) The X-ray flux is extrapolated from 0.1\,keV\,$-$\,2\,keV band to 2\,keV\,$-$\,10 keV by assuming the same X-ray spectral slope; and (d) The optical and UV data are simultaneous at JD 2449861 whereas the X-ray data are from JD 2449842.}
\label{tab:table4}
\begin{tabular}{cccccccccccc} 
\hline
Object & Epoch & SED JD & $\log [L_{BOL}(acc)$ & $\lambda_{Edd}$ & $\log [\lambda L_{\lambda}$ (5100 \AA) & BC(5100 \AA) &$\log [\lambda L_{\lambda}$ (1350 \AA) &BC(1350 \AA) & $\log [L_{\rm 2-10\,keV}$ & BC(2--10\,keV)  & Notes\\
 & & ($-2400000$) & /erg s$^{-1}$] &  & /erg s$^{-1}$] & & /erg s$^{-1}$] & & /erg s$^{-1}$] & & \\
(1) & (2) & (3) & (4) & (5) & (6) & (7) & (8) & (9) & (10) & (11) & (12)  \\
 \hline
Fairall\,9& 1 &49473.0$ - $49475.5&  $44.94^{+0.06}_{-0.06}$  &$  0.035^{+0.007}_{-0.013}$&$  43.83^{+ 0.06}_{- 0.06}$&$  13.14^{+ 1.86}_{- 3.38}$&$44.22^{+ 0.07}_{- 0.07}$&$ 5.25^{+ 0.93}_{- 1.46}$&$44.02^{+ 0.05}_{- 0.05}$&$ 8.33^{+ 0.95}_{- 1.99}$ & a \\
	& 2&49477.0$ - $49479.5&  $44.96^{+0.06}_{-0.06}$  &$  0.036^{+0.007}_{-0.014}$&$  43.83^{+ 0.06}_{- 0.06}$&$  13.48^{+ 2.08}_{- 3.58}$&$44.24^{+ 0.07}_{- 0.07}$&$ 5.15^{+ 0.93}_{- 1.45}$&$44.02^{+ 0.05}_{- 0.05}$&$ 8.54^{+ 0.98}_{- 2.04}$ & a \\ 
	& 3&49482.6$ - $49484.5&  $44.97^{+0.06}_{-0.06}$  &$  0.037^{+0.007}_{-0.014}$&$  43.79^{+ 0.05}_{- 0.05}$&$  15.10^{+ 1.71}_{- 3.78}$&$44.28^{+ 0.05}_{- 0.05}$&$ 4.89^{+ 0.64}_{- 1.27}$&$44.02^{+ 0.08}_{- 0.08}$&$ 8.84^{+ 1.78}_{- 2.47}$ & a \\
	& 4&49486.6$ - $49487.5&  $44.98^{+0.05}_{-0.05}$  &$  0.038^{+0.007}_{-0.014}$&$  43.78^{+ 0.04}_{- 0.04}$&$  15.71^{+ 1.42}_{- 3.57}$&$44.30^{+ 0.06}_{- 0.06}$&$ 4.81^{+ 0.67}_{- 1.21}$&$44.03^{+ 0.05}_{- 0.05}$&$ 8.95^{+ 1.04}_{- 2.09}$ & a \\
	& 5 &49490.0$ - $49491.5&  $44.96^{+0.05}_{-0.05}$  &$  0.037^{+0.007}_{-0.014}$&$  43.82^{+ 0.04}_{- 0.04}$&$  13.76^{+ 1.32}_{- 3.23}$&$44.26^{+ 0.06}_{- 0.06}$&$ 5.03^{+ 0.74}_{- 1.31}$&$44.03^{+ 0.05}_{- 0.05}$&$ 8.64^{+ 1.00}_{- 2.06}$ & a \\
	& 6 &49497.0$ - $49499.5&  $45.01^{+0.06}_{-0.06}$  &$  0.041^{+0.008}_{-0.016}$&$  43.82^{+ 0.12}_{- 0.12}$&$  15.46^{+ 5.12}_{- 6.25}$&$44.34^{+ 0.06}_{- 0.06}$&$ 4.64^{+ 0.66}_{- 1.26}$&$44.03^{+ 0.05}_{- 0.05}$&$ 9.54^{+ 1.10}_{- 2.42}$ & a \\ 
	& 7 &49510.0$ - $49511.5&  $45.07^{+0.05}_{-0.05}$  &$  0.047^{+0.009}_{-0.017}$&$  43.85^{+ 0.04}_{- 0.04}$&$  16.44^{+ 1.52}_{- 3.85}$&$44.44^{+ 0.05}_{- 0.05}$&$ 4.23^{+ 0.55}_{- 1.07}$&$44.03^{+ 0.05}_{- 0.05}$&$10.95^{+ 1.33}_{- 2.64}$ & a \\ 
	& 8 &49514.0$ - $49515.5&  $45.02^{+0.05}_{-0.05}$  &$  0.042^{+0.008}_{-0.016}$&$  43.82^{+ 0.03}_{- 0.03}$&$  15.87^{+ 1.27}_{- 3.64}$&$44.36^{+ 0.06}_{- 0.06}$&$ 4.55^{+ 0.66}_{- 1.18}$&$44.03^{+ 0.05}_{- 0.05}$&$ 9.76^{+ 1.18}_{- 2.34}$ & a \\
	& 9 &49518.0$ - $49519.5&  $45.05^{+0.08}_{-0.07}$  &$  0.045^{+0.009}_{-0.018}$&$  43.85^{+ 0.03}_{- 0.03}$&$  15.67^{+ 1.41}_{- 4.21}$&$44.41^{+ 0.10}_{- 0.10}$&$ 4.38^{+ 1.13}_{- 1.58}$&$44.03^{+ 0.05}_{- 0.05}$&$10.42^{+ 1.30}_{- 2.89}$ & a \\
	&10 &49523.0$ - $49524.5&  $45.09^{+0.04}_{-0.04}$  &$  0.050^{+0.010}_{-0.018}$&$  43.87^{+ 0.04}_{- 0.04}$&$  16.66^{+ 1.86}_{- 3.89}$&$44.48^{+ 0.04}_{- 0.04}$&$ 4.12^{+ 0.45}_{- 0.96}$&$44.03^{+ 0.05}_{- 0.05}$&$11.66^{+ 1.40}_{- 2.71}$ & a \\	
	&11 &49530.0$ - $49531.5&  $45.04^{+0.05}_{-0.05}$  &$  0.043^{+0.009}_{-0.016}$&$  43.86^{+ 0.03}_{- 0.03}$&$  14.98^{+ 0.93}_{- 3.34}$&$44.39^{+ 0.06}_{- 0.06}$&$ 4.47^{+ 0.64}_{- 1.15}$&$44.03^{+ 0.05}_{- 0.05}$&$10.15^{+ 1.22}_{- 2.43}$ & a \\ 
	&12 &49533.0$ - $49535.5&  $45.04^{+0.04}_{-0.04}$  &$  0.044^{+0.009}_{-0.016}$&$  43.84^{+ 0.03}_{- 0.03}$&$  15.84^{+ 1.06}_{- 3.37}$&$44.39^{+ 0.05}_{- 0.05}$&$ 4.43^{+ 0.52}_{- 1.03}$&$44.03^{+ 0.05}_{- 0.05}$&$10.24^{+ 1.25}_{- 2.35}$ & a \\	
	&13 &49537.0$ - $49539.5&  $44.99^{+0.05}_{-0.05}$  &$  0.039^{+0.008}_{-0.014}$&$  43.81^{+ 0.04}_{- 0.04}$&$  15.27^{+ 1.37}_{- 3.50}$&$44.31^{+ 0.06}_{- 0.06}$&$ 4.80^{+ 0.74}_{- 1.25}$&$44.03^{+ 0.05}_{- 0.05}$&$ 9.05^{+ 1.08}_{- 2.14}$ & a \\ 		
	&14 &49542.0$ - $49543.5&  $45.00^{+0.10}_{-0.09}$  &$  0.040^{+0.008}_{-0.017}$&$  43.84^{+ 0.03}_{- 0.03}$&$  14.49^{+ 1.23}_{- 4.22}$&$44.32^{+ 0.14}_{- 0.14}$&$ 4.77^{+ 1.87}_{- 2.30}$&$44.03^{+ 0.05}_{- 0.05}$&$ 9.36^{+ 1.25}_{- 2.85}$ & a \\	
	&15 &49546.0$ - $49547.5&  $45.00^{+0.04}_{-0.04}$  &$  0.040^{+0.008}_{-0.014}$&$  43.79^{+ 0.03}_{- 0.03}$&$  16.04^{+ 1.16}_{- 3.34}$&$44.32^{+ 0.05}_{- 0.05}$&$ 4.71^{+ 0.53}_{- 1.06}$&$44.03^{+ 0.05}_{- 0.05}$&$ 9.24^{+ 1.11}_{- 2.06}$ & a \\	
	&16 &49559.0$ - $49561.5&  $44.96^{+0.06}_{-0.06}$  &$  0.037^{+0.007}_{-0.014}$&$  43.78^{+ 0.03}_{- 0.03}$&$  15.12^{+ 1.28}_{- 3.62}$&$44.26^{+ 0.08}_{- 0.08}$&$ 4.99^{+ 1.05}_{- 1.53}$&$44.03^{+ 0.05}_{- 0.05}$&$ 8.53^{+ 1.05}_{- 2.13}$ & a \\	
	&17 &49565.0$ - $49567.5&  $45.01^{+0.06}_{-0.06}$  &$  0.041^{+0.008}_{-0.016}$&$  43.83^{+ 0.05}_{- 0.05}$&$  15.08^{+ 1.67}_{- 3.81}$&$44.34^{+ 0.07}_{- 0.07}$&$ 4.65^{+ 0.84}_{- 1.35}$&$44.03^{+ 0.05}_{- 0.05}$&$ 9.48^{+ 1.13}_{- 2.38}$ & a \\		
	&18 &49577.0$ - $49579.5&  $45.01^{+0.06}_{-0.06}$  &$  0.041^{+0.008}_{-0.015}$&$  43.89^{+ 0.02}_{- 0.02}$&$  13.12^{+ 0.75}_{- 2.98}$&$44.33^{+ 0.07}_{- 0.07}$&$ 4.78^{+ 0.87}_{- 1.37}$&$44.04^{+ 0.05}_{- 0.05}$&$ 9.40^{+ 1.16}_{- 2.31}$ & a \\ 		
	&19 &49589.0$ - $49591.5&  $45.05^{+0.05}_{-0.05}$  &$  0.045^{+0.009}_{-0.017}$&$  43.88^{+ 0.03}_{- 0.03}$&$  14.80^{+ 0.95}_{- 3.32}$&$44.41^{+ 0.05}_{- 0.05}$&$ 4.41^{+ 0.60}_{- 1.12}$&$44.04^{+ 0.05}_{- 0.05}$&$10.43^{+ 1.29}_{- 2.53}$ & a \\  	
	&20 &49593.0$ - $49595.5&  $45.08^{+0.05}_{-0.05}$  &$  0.048^{+0.010}_{-0.018}$&$  43.95^{+ 0.03}_{- 0.03}$&$  13.47^{+ 1.14}_{- 3.02}$&$44.44^{+ 0.05}_{- 0.05}$&$ 4.36^{+ 0.51}_{- 1.04}$&$44.04^{+ 0.05}_{- 0.05}$&$11.05^{+ 1.37}_{- 2.60}$ & a \\
	&21 &49597.0$ - $49599.5&  $45.09^{+0.04}_{-0.04}$  &$  0.049^{+0.010}_{-0.018}$&$  43.88^{+ 0.04}_{- 0.04}$&$  16.16^{+ 1.65}_{- 3.62}$&$44.47^{+ 0.04}_{- 0.04}$&$ 4.17^{+ 0.40}_{- 0.92}$&$44.04^{+ 0.05}_{- 0.05}$&$11.26^{+ 1.45}_{- 2.60}$ & a \\
        &22 &49607.5$ - $49609.3&  $45.07^{+0.06}_{-0.06}$  &$  0.047^{+0.009}_{-0.019}$&$  44.00^{+ 0.02}_{- 0.02}$&$  11.90^{+ 0.49}_{- 3.00}$&$44.43^{+ 0.07}_{- 0.07}$&$ 4.45^{+ 0.78}_{- 1.36}$&$44.04^{+ 0.07}_{- 0.07}$&$10.84^{+ 1.78}_{- 3.10}$ & a \\
	&23 &49617.0$ - $49619.5&  $45.04^{+0.05}_{-0.05}$  &$  0.044^{+0.009}_{-0.016}$&$  43.94^{+ 0.03}_{- 0.03}$&$  12.55^{+ 0.91}_{- 2.72}$&$44.37^{+ 0.05}_{- 0.05}$&$ 4.66^{+ 0.55}_{- 1.10}$&$44.04^{+ 0.05}_{- 0.05}$&$10.00^{+ 1.27}_{- 2.34}$ & a \\ 
	&24 &49623.0$ - $49624.5&  $45.07^{+0.05}_{-0.05}$  &$  0.046^{+0.009}_{-0.017}$&$  43.94^{+ 0.03}_{- 0.03}$&$  13.30^{+ 0.84}_{- 2.92}$&$44.42^{+ 0.05}_{- 0.05}$&$ 4.45^{+ 0.58}_{- 1.10}$&$44.04^{+ 0.05}_{- 0.05}$&$10.60^{+ 1.35}_{- 2.53}$ & a \\
	&25 &49653.0$ - $49653.5&  $45.24^{+0.04}_{-0.04}$  &$  0.069^{+0.014}_{-0.026}$&$  44.10^{+ 0.03}_{- 0.03}$&$  13.70^{+ 1.25}_{- 3.26}$&$44.68^{+ 0.04}_{- 0.04}$&$ 3.63^{+ 0.43}_{- 0.91}$&$44.04^{+ 0.05}_{- 0.05}$&$15.64^{+ 2.12}_{- 3.94}$ & a \\
	&26 &49664.0$ - $49665.5&  $45.25^{+0.04}_{-0.04}$  &$  0.070^{+0.015}_{-0.026}$&$  44.10^{+ 0.02}_{- 0.02}$&$  14.03^{+ 1.13}_{- 3.03}$&$44.69^{+ 0.03}_{- 0.03}$&$ 3.60^{+ 0.38}_{- 0.81}$&$44.04^{+ 0.05}_{- 0.05}$&$15.95^{+ 2.36}_{- 3.86}$ & a \\ 
	&27 &49712.5$ - $49724.2&  $45.18^{+0.04}_{-0.04}$  &$  0.060^{+0.013}_{-0.023}$&$  44.07^{+ 0.02}_{- 0.02}$&$  12.86^{+ 1.37}_{- 3.01}$&$44.62^{+ 0.05}_{- 0.05}$&$ 3.61^{+ 0.53}_{- 0.92}$&$43.97^{+ 0.03}_{- 0.03}$&$16.22^{+ 1.86}_{- 3.84}$ & b \\
  Mrk\,509& 1&48187.0$ - $48188.3&  $45.20^{+0.05}_{-0.05}$  &$  0.114^{+0.010}_{-0.024}$&$  44.18^{+ 0.01}_{- 0.01}$&$  10.66^{+ 0.17}_{- 2.04}$&$44.59^{+ 0.07}_{- 0.07}$&$ 4.13^{+ 0.68}_{- 1.04}$&$44.12^{+ 0.03}_{- 0.03}$&$12.16^{+ 0.75}_{- 2.42}$ & \\
NGC\,3783& 1 &48801.0$ - $48802.5&  $43.56^{+0.05}_{-0.05}$  &$  0.012^{+0.003}_{-0.004}$&$  42.62^{+ 0.06}_{- 0.06}$&$   8.71^{+ 1.44}_{- 2.53}$&$43.06^{+ 0.03}_{- 0.03}$&$ 3.17^{+ 0.34}_{- 0.83}$&$41.91^{+ 0.07}_{- 0.07}$&$43.92^{+ 8.79}_{- 12.96}$ & c \\
NGC\,4151& 1&49326.0$ - $49327.0&  $43.95^{+0.02}_{-0.02}$  &$  0.020^{+0.003}_{-0.004}$&$  42.96^{+ 0.01}_{- 0.01}$&$   9.74^{+ 0.89}_{- 1.70}$&$43.34^{+ 0.03}_{- 0.03}$&$ 4.02^{+ 0.42}_{- 0.73}$&$42.91^{+ 0.01}_{- 0.01}$&$10.84^{+ 0.98}_{- 1.88}$ & \\
	& 2&49327.4$ - $49328.0&  $43.94^{+0.02}_{-0.02}$  &$  0.019^{+0.003}_{-0.004}$&$  42.97^{+ 0.01}_{- 0.01}$&$   9.37^{+ 0.97}_{- 1.74}$&$43.39^{+ 0.02}_{- 0.02}$&$ 3.60^{+ 0.41}_{- 0.69}$&$42.86^{+ 0.01}_{- 0.01}$&$12.04^{+ 1.25}_{- 2.23}$ & \\
	& 3&49328.5$ - $49329.5&  $43.96^{+0.02}_{-0.02}$  &$  0.020^{+0.003}_{-0.004}$&$  42.96^{+ 0.01}_{- 0.01}$&$   9.93^{+ 0.92}_{- 1.67}$&$43.35^{+ 0.02}_{- 0.02}$&$ 4.04^{+ 0.41}_{- 0.70}$&$42.94^{+ 0.01}_{- 0.01}$&$10.55^{+ 0.97}_{- 1.77}$ & \\
	& 4&49331.0$ - $49332.0&  $43.92^{+0.02}_{-0.02}$  &$  0.018^{+0.003}_{-0.004}$&$  42.96^{+ 0.01}_{- 0.01}$&$   9.10^{+ 0.85}_{- 1.58}$&$43.32^{+ 0.02}_{- 0.02}$&$ 3.97^{+ 0.41}_{- 0.71}$&$42.87^{+ 0.01}_{- 0.01}$&$11.39^{+ 1.08}_{- 1.98}$ & \\
NGC\,5548& 1&47534.0$ - $47536.0&  $44.31^{+0.04}_{-0.04}$  &$  0.031^{+0.002}_{-0.006}$&$  43.36^{+ 0.02}_{- 0.02}$&$   8.91^{+ 0.63}_{- 1.66}$&$43.68^{+ 0.05}_{- 0.05}$&$ 4.19^{+ 0.58}_{- 0.92}$&$43.32^{+ 0.01}_{- 0.01}$&$ 9.61^{+ 0.42}_{- 1.71}$ & \\
	& 2&47555.0$ - $47557.2&  $44.38^{+0.03}_{-0.03}$  &$  0.037^{+0.003}_{-0.006}$&$  43.23^{+ 0.03}_{- 0.03}$&$  14.09^{+ 1.33}_{- 2.65}$&$43.67^{+ 0.05}_{- 0.05}$&$ 5.12^{+ 0.70}_{- 1.09}$&$43.56^{+ 0.00}_{- 0.00}$&$ 6.67^{+ 0.39}_{- 1.15}$ & \\ 
	& 3&47684.0$ - $47685.5&  $44.28^{+0.04}_{-0.04}$  &$  0.029^{+0.001}_{-0.005}$&$  43.28^{+ 0.05}_{- 0.05}$&$   9.97^{+ 1.22}_{- 2.06}$&$43.52^{+ 0.06}_{- 0.06}$&$ 5.82^{+ 0.91}_{- 1.34}$&$43.48^{+ 0.00}_{- 0.00}$&$ 6.31^{+ 0.20}_{- 1.07}$ & \\
	& 4&47719.0$ - $47721.3&  $44.39^{+0.04}_{-0.04}$  &$  0.037^{+0.001}_{-0.006}$&$  43.33^{+ 0.03}_{- 0.03}$&$  11.44^{+ 0.72}_{- 1.96}$&$43.65^{+ 0.06}_{- 0.06}$&$ 5.47^{+ 0.77}_{- 1.16}$&$43.53^{+ 0.02}_{- 0.02}$&$ 7.11^{+ 0.31}_{- 1.17}$ & \\
	& 5&48035.7$ - $48036.7&  $44.21^{+0.06}_{-0.06}$  &$  0.025^{+0.001}_{-0.005}$&$  43.08^{+ 0.08}_{- 0.08}$&$  13.43^{+ 2.64}_{- 3.78}$&$43.45^{+ 0.08}_{- 0.08}$&$ 5.78^{+ 1.26}_{- 1.72}$&$43.41^{+ 0.03}_{- 0.03}$&$ 6.31^{+ 0.48}_{- 1.35}$ & \\
\hline
	\end{tabular}
\end{table}
\end{landscape}

\begin{landscape}
\begin{table}
\contcaption{}
\label{tab:table4continued}
\begin{tabular}{cccccccccccc} 
\hline
Object & Epoch & SED JD & $\log L_{BOL}(acc)$ & $\lambda_{Edd}$ & $\log \lambda L_{\lambda}$ (5100 \AA) & BC(5100\,\AA) &$\log \lambda L_{\lambda}$ (1350\,\AA) &BC(1350\,\AA) & $\log L_{\rm 2-10 keV}$ & BC(2--10\, keV)  & Notes\\
 & & ($-2400000$) & /erg s$^{-1}$] &  & /erg s$^{-1}$] & & /erg s$^{-1}$] & & /erg s$^{-1}$] & & \\
(1) & (2) & (3) & (4) & (5) & (6) & (7) & (8) & (9) & (10) & (11) & (12)  \\
 \hline
NGC\,5548 & 6&48040.0$ - $48041.0&  $44.26^{+0.05}_{-0.05}$  &$  0.028^{+0.001}_{-0.005}$&$  43.13^{+ 0.11}_{- 0.11}$&$  13.41^{+ 3.95}_{- 4.70}$&$43.54^{+ 0.06}_{- 0.06}$&$ 5.29^{+ 0.72}_{- 1.24}$&$43.41^{+ 0.03}_{- 0.03}$&$ 7.17^{+ 0.46}_{- 1.43}$ & \\
	& 7&48047.0$ - $48048.5&  $44.20^{+0.05}_{-0.05}$  &$  0.024^{+0.001}_{-0.005}$&$  43.11^{+ 0.05}_{- 0.05}$&$  12.48^{+ 1.69}_{- 3.12}$&$43.49^{+ 0.07}_{- 0.07}$&$ 5.23^{+ 0.92}_{- 1.43}$&$43.32^{+ 0.03}_{- 0.03}$&$ 7.66^{+ 0.51}_{- 1.68}$ & \\ 
	& 8&48054.3$ - $48056.0&  $44.30^{+0.06}_{-0.05}$  &$  0.030^{+0.001}_{-0.006}$&$  43.11^{+ 0.05}_{- 0.05}$&$  15.31^{+ 1.75}_{- 3.40}$&$43.51^{+ 0.09}_{- 0.09}$&$ 6.08^{+ 1.35}_{- 1.78}$&$43.52^{+ 0.02}_{- 0.02}$&$ 5.93^{+ 0.28}_{- 1.16}$ & \\ 
	& 9&48061.4$ - $48062.4&  $44.02^{+0.09}_{-0.08}$  &$  0.016^{+0.001}_{-0.004}$&$  43.04^{+ 0.05}_{- 0.05}$&$   9.65^{+ 1.31}_{- 2.85}$&$43.32^{+ 0.09}_{- 0.09}$&$ 5.06^{+ 1.26}_{- 1.83}$&$43.10^{+ 0.07}_{- 0.07}$&$ 8.33^{+ 1.53}_{- 2.56}$ & \\
	&10&48068.0$ - $48069.0&  $43.98^{+0.07}_{-0.06}$  &$  0.015^{+0.001}_{-0.003}$&$  42.99^{+ 0.05}_{- 0.05}$&$   9.93^{+ 1.35}_{- 2.54}$&$43.21^{+ 0.09}_{- 0.09}$&$ 5.98^{+ 1.38}_{- 1.89}$&$43.21^{+ 0.03}_{- 0.03}$&$ 5.93^{+ 0.47}_{- 1.36}$ & \\ 
        &11&48075.0$ - $48077.0&  $43.92^{+0.07}_{-0.06}$  &$  0.013^{+0.001}_{-0.003}$&$  42.92^{+ 0.06}_{- 0.06}$&$  10.13^{+ 1.66}_{- 2.70}$&$43.12^{+ 0.09}_{- 0.09}$&$ 6.40^{+ 1.46}_{- 1.99}$&$43.14^{+ 0.03}_{- 0.03}$&$ 6.02^{+ 0.58}_{- 1.37}$ & \\
	&12&47509.0$ - $47510.0&  $44.29^{+0.07}_{-0.07}$  &$  0.029^{+0.002}_{-0.006}$&$  43.24^{+ 0.04}_{- 0.04}$&$  11.06^{+ 1.29}_{- 2.62}$&$43.53^{+ 0.06}_{- 0.06}$&$ 5.71^{+ 0.94}_{- 1.51}$&$43.48^{+ 0.08}_{- 0.08}$&$ 6.34^{+ 1.27}_{- 1.70}$ & a \\ 
	&13 &47512.0$ - $47514.0&  $44.31^{+0.06}_{-0.06}$  &$  0.031^{+0.002}_{-0.007}$&$  43.29^{+ 0.04}_{- 0.04}$&$  10.46^{+ 1.10}_{- 2.62}$&$43.57^{+ 0.06}_{- 0.06}$&$ 5.56^{+ 0.80}_{- 1.50}$&$43.48^{+ 0.08}_{- 0.08}$&$ 6.76^{+ 1.31}_{- 1.89}$ & a \\ 
	&14 &47517.0$ - $47518.0&  $44.33^{+0.06}_{-0.06}$  &$  0.032^{+0.002}_{-0.007}$&$  43.30^{+ 0.03}_{- 0.03}$&$  10.62^{+ 0.86}_{- 2.28}$&$43.60^{+ 0.05}_{- 0.05}$&$ 5.31^{+ 0.62}_{- 1.22}$&$43.48^{+ 0.08}_{- 0.08}$&$ 6.94^{+ 1.35}_{- 1.79}$ & a \\
	&15 &47522.0$ - $47524.0&  $44.32^{+0.06}_{-0.06}$  &$  0.032^{+0.002}_{-0.007}$&$  43.31^{+ 0.03}_{- 0.03}$&$  10.46^{+ 0.88}_{- 2.30}$&$43.60^{+ 0.05}_{- 0.05}$&$ 5.36^{+ 0.71}_{- 1.30}$&$43.48^{+ 0.08}_{- 0.08}$&$ 6.93^{+ 1.36}_{- 1.82}$ & a \\
	&16 &47525.0$ - $47526.0&  $44.37^{+0.05}_{-0.05}$  &$  0.036^{+0.002}_{-0.007}$&$  43.33^{+ 0.03}_{- 0.03}$&$  11.04^{+ 0.73}_{- 2.31}$&$43.68^{+ 0.03}_{- 0.03}$&$ 4.87^{+ 0.41}_{- 1.05}$&$43.48^{+ 0.08}_{- 0.08}$&$ 7.70^{+ 1.48}_{- 1.98}$ & a \\
	&17&47528.0$ - $47530.0&  $44.39^{+0.06}_{-0.06}$  &$  0.038^{+0.002}_{-0.009}$&$  43.36^{+ 0.07}_{- 0.07}$&$  10.88^{+ 1.86}_{- 3.22}$&$43.71^{+ 0.05}_{- 0.05}$&$ 4.87^{+ 0.60}_{- 1.32}$&$43.48^{+ 0.08}_{- 0.08}$&$ 8.12^{+ 1.56}_{- 2.37}$ & a \\ 
	&18&47538.0$ - $47539.0&  $44.42^{+0.06}_{-0.06}$  &$  0.040^{+0.002}_{-0.009}$&$  43.36^{+ 0.04}_{- 0.04}$&$  11.48^{+ 1.15}_{- 2.93}$&$43.76^{+ 0.04}_{- 0.04}$&$ 4.52^{+ 0.49}_{- 1.17}$&$43.48^{+ 0.08}_{- 0.08}$&$ 8.56^{+ 1.64}_{- 2.44}$ & a \\ 
	&19&47543.0$ - $47544.0&  $44.41^{+0.06}_{-0.06}$  &$  0.039^{+0.002}_{-0.009}$&$  43.30^{+ 0.05}_{- 0.05}$&$  12.88^{+ 1.70}_{- 3.37}$&$43.76^{+ 0.05}_{- 0.05}$&$ 4.47^{+ 0.58}_{- 1.16}$&$43.48^{+ 0.08}_{- 0.08}$&$ 8.37^{+ 1.62}_{- 2.33}$ & a \\
	&20&47549.0$ - $47550.0&  $44.38^{+0.06}_{-0.06}$  &$  0.037^{+0.002}_{-0.008}$&$  43.41^{+ 0.02}_{- 0.02}$&$   9.45^{+ 0.61}_{- 2.09}$&$43.70^{+ 0.05}_{- 0.05}$&$ 4.86^{+ 0.60}_{- 1.19}$&$43.48^{+ 0.08}_{- 0.08}$&$ 7.91^{+ 1.53}_{- 2.12}$ & a \\
	&21 &47561.0$ - $47562.0&  $44.29^{+0.08}_{-0.07}$  &$  0.029^{+0.002}_{-0.007}$&$  43.29^{+ 0.11}_{- 0.11}$&$  10.00^{+ 2.86}_{- 3.72}$&$43.52^{+ 0.06}_{- 0.06}$&$ 5.77^{+ 0.96}_{- 1.68}$&$43.48^{+ 0.08}_{- 0.08}$&$ 6.33^{+ 1.27}_{- 1.86}$ & a \\ 
	&22 &47576.0$ - $47577.0&  $44.23^{+0.07}_{-0.07}$  &$  0.026^{+0.002}_{-0.006}$&$  43.14^{+ 0.04}_{- 0.04}$&$  12.47^{+ 1.43}_{- 3.21}$&$43.44^{+ 0.07}_{- 0.07}$&$ 6.19^{+ 1.16}_{- 1.84}$&$43.48^{+ 0.08}_{- 0.08}$&$ 5.59^{+ 1.12}_{- 1.60}$ & a \\  
	&23 &47586.0$ - $47587.0&  $44.25^{+0.07}_{-0.07}$  &$  0.027^{+0.002}_{-0.006}$&$  43.21^{+ 0.04}_{- 0.04}$&$  11.12^{+ 1.28}_{- 2.60}$&$43.48^{+ 0.07}_{- 0.07}$&$ 5.98^{+ 1.08}_{- 1.63}$&$43.48^{+ 0.08}_{- 0.08}$&$ 5.89^{+ 1.20}_{- 1.58}$ & a \\
	&24 &47594.0$ - $47595.0&  $44.23^{+0.07}_{-0.07}$  &$  0.026^{+0.002}_{-0.006}$&$  43.22^{+ 0.05}_{- 0.05}$&$  10.31^{+ 1.51}_{- 2.61}$&$43.44^{+ 0.06}_{- 0.06}$&$ 6.27^{+ 1.09}_{- 1.70}$&$43.48^{+ 0.08}_{- 0.08}$&$ 5.62^{+ 1.15}_{- 1.53}$ & a \\ 
	&25 &47598.0$ - $47599.0&  $44.30^{+0.07}_{-0.07}$  &$  0.030^{+0.005}_{-0.005}$&$  43.30^{+ 0.04}_{- 0.04}$&$   9.90^{+ 1.80}_{- 1.95}$&$43.54^{+ 0.06}_{- 0.06}$&$ 5.67^{+ 1.21}_{- 1.28}$&$43.48^{+ 0.08}_{- 0.08}$&$ 6.50^{+ 1.58}_{- 1.51}$ & a \\
	&26 &47601.0$ - $47602.0&  $44.32^{+0.06}_{-0.06}$  &$  0.032^{+0.002}_{-0.007}$&$  43.31^{+ 0.03}_{- 0.03}$&$  10.23^{+ 0.88}_{- 2.36}$&$43.59^{+ 0.05}_{- 0.05}$&$ 5.33^{+ 0.67}_{- 1.32}$&$43.48^{+ 0.08}_{- 0.08}$&$ 6.88^{+ 1.34}_{- 1.85}$ & a \\
	&27 &47613.0$ - $47614.0&  $44.38^{+0.06}_{-0.06}$  &$  0.036^{+0.002}_{-0.008}$&$  43.36^{+ 0.03}_{- 0.03}$&$  10.53^{+ 0.71}_{- 2.39}$&$43.69^{+ 0.05}_{- 0.05}$&$ 4.85^{+ 0.66}_{- 1.24}$&$43.48^{+ 0.08}_{- 0.08}$&$ 7.88^{+ 1.52}_{- 2.14}$ & a \\
	&28 &47617.0$ - $47618.0&  $44.39^{+0.06}_{-0.06}$  &$  0.037^{+0.002}_{-0.008}$&$  43.39^{+ 0.04}_{- 0.04}$&$  10.13^{+ 1.01}_{- 2.35}$&$43.71^{+ 0.04}_{- 0.04}$&$ 4.79^{+ 0.52}_{- 1.13}$&$43.48^{+ 0.08}_{- 0.08}$&$ 8.09^{+ 1.56}_{- 2.15}$ & a \\
	&29 &47621.0$ - $47622.0&  $44.46^{+0.05}_{-0.05}$  &$  0.044^{+0.002}_{-0.011}$&$  43.41^{+ 0.05}_{- 0.05}$&$  11.30^{+ 1.50}_{- 3.13}$&$43.84^{+ 0.04}_{- 0.04}$&$ 4.21^{+ 0.42}_{- 1.11}$&$43.48^{+ 0.08}_{- 0.08}$&$ 9.56^{+ 1.84}_{- 2.80}$ & a \\  
	&30 &47624.0$ - $47625.0&  $44.44^{+0.05}_{-0.05}$  &$  0.042^{+0.002}_{-0.009}$&$  43.41^{+ 0.04}_{- 0.04}$&$  10.68^{+ 0.97}_{- 2.48}$&$43.80^{+ 0.04}_{- 0.04}$&$ 4.40^{+ 0.43}_{- 1.03}$&$43.48^{+ 0.08}_{- 0.08}$&$ 9.06^{+ 1.73}_{- 2.42}$ & a \\
	&31 &47657.0$ - $47658.0&  $44.37^{+0.06}_{-0.06}$  &$  0.036^{+0.001}_{-0.009}$&$  43.38^{+ 0.03}_{- 0.03}$&$   9.98^{+ 0.62}_{- 2.42}$&$43.68^{+ 0.05}_{- 0.05}$&$ 4.93^{+ 0.61}_{- 1.31}$&$43.48^{+ 0.08}_{- 0.08}$&$ 7.77^{+ 1.49}_{- 2.21}$ & a \\ 
          &32 &47661.0$ - $47662.0&  $44.38^{+0.06}_{-0.06}$  &$  0.037^{+0.002}_{-0.008}$&$  43.39^{+ 0.04}_{- 0.04}$&$   9.77^{+ 1.11}_{- 2.36}$&$43.69^{+ 0.05}_{- 0.05}$&$ 4.97^{+ 0.69}_{- 1.26}$&$43.48^{+ 0.08}_{- 0.08}$&$ 7.93^{+ 1.55}_{- 2.14}$ & a \\  
	&33 &47665.0$ - $47666.0&  $44.38^{+0.06}_{-0.06}$  &$  0.037^{+0.002}_{-0.008}$&$  43.37^{+ 0.03}_{- 0.03}$&$  10.29^{+ 0.68}_{- 2.40}$&$43.70^{+ 0.05}_{- 0.05}$&$ 4.81^{+ 0.64}_{- 1.25}$&$43.48^{+ 0.08}_{- 0.08}$&$ 7.89^{+ 1.52}_{- 2.18}$ & a \\ 
  &34 &47668.0$ - $47669.0&  $44.33^{+0.07}_{-0.07}$  &$  0.033^{+0.002}_{-0.007}$&$  43.36^{+ 0.03}_{- 0.03}$&$   9.32^{+ 0.80}_{- 2.21}$&$43.60^{+ 0.06}_{- 0.06}$&$ 5.42^{+ 0.87}_{- 1.48}$&$43.48^{+ 0.08}_{- 0.08}$&$ 7.05^{+ 1.39}_{- 1.95}$ & a \\
	&35 &47673.0$ - $47674.0&  $44.33^{+0.06}_{-0.06}$  &$  0.032^{+0.001}_{-0.008}$&$  43.30^{+ 0.05}_{- 0.05}$&$  10.52^{+ 1.22}_{- 2.84}$&$43.60^{+ 0.06}_{- 0.06}$&$ 5.32^{+ 0.75}_{- 1.50}$&$43.48^{+ 0.08}_{- 0.08}$&$ 6.97^{+ 1.34}_{- 2.04}$ & a \\
	&36 &47677.0$ - $47678.0&  $44.30^{+0.07}_{-0.07}$  &$  0.031^{+0.002}_{-0.007}$&$  43.30^{+ 0.03}_{- 0.03}$&$  10.18^{+ 0.80}_{- 2.48}$&$43.56^{+ 0.06}_{- 0.06}$&$ 5.57^{+ 0.83}_{- 1.53}$&$43.48^{+ 0.08}_{- 0.08}$&$ 6.61^{+ 1.29}_{- 1.88}$ & a \\
	&37 &47679.0$ - $47680.0&  $44.31^{+0.06}_{-0.06}$  &$  0.031^{+0.001}_{-0.007}$&$  43.28^{+ 0.05}_{- 0.05}$&$  10.74^{+ 1.29}_{- 2.81}$&$43.58^{+ 0.05}_{- 0.05}$&$ 5.41^{+ 0.70}_{- 1.44}$&$43.48^{+ 0.08}_{- 0.08}$&$ 6.68^{+ 1.29}_{- 1.89}$ & a \\
	&38 &47686.0$ - $47688.0&  $44.25^{+0.07}_{-0.07}$  &$  0.027^{+0.002}_{-0.007}$&$  43.25^{+ 0.05}_{- 0.05}$&$   9.94^{+ 1.41}_{- 2.70}$&$43.46^{+ 0.07}_{- 0.07}$&$ 6.20^{+ 1.19}_{- 1.87}$&$43.48^{+ 0.08}_{- 0.08}$&$ 5.80^{+ 1.17}_{- 1.68}$ & a \\ 
	&39 &47699.0$ - $47700.0&  $44.27^{+0.07}_{-0.07}$  &$  0.028^{+0.002}_{-0.006}$&$  43.31^{+ 0.03}_{- 0.03}$&$   9.04^{+ 0.84}_{- 1.99}$&$43.48^{+ 0.07}_{- 0.07}$&$ 6.14^{+ 1.09}_{- 1.64}$&$43.48^{+ 0.08}_{- 0.08}$&$ 6.05^{+ 1.23}_{- 1.60}$ & a \\ 
	&40 &47705.0$ - $47706.0&  $44.28^{+0.07}_{-0.07}$  &$  0.029^{+0.002}_{-0.007}$&$  43.20^{+ 0.07}_{- 0.07}$&$  12.07^{+ 2.27}_{- 3.66}$&$43.54^{+ 0.06}_{- 0.06}$&$ 5.58^{+ 0.88}_{- 1.59}$&$43.48^{+ 0.08}_{- 0.08}$&$ 6.28^{+ 1.23}_{- 1.82}$ & a \\ 
	&41 &47709.0$ - $47710.0&  $44.33^{+0.06}_{-0.06}$  &$  0.033^{+0.001}_{-0.008}$&$  43.27^{+ 0.05}_{- 0.05}$&$  11.41^{+ 1.42}_{- 3.06}$&$43.62^{+ 0.05}_{- 0.05}$&$ 5.14^{+ 0.68}_{- 1.40}$&$43.48^{+ 0.08}_{- 0.08}$&$ 7.04^{+ 1.35}_{- 2.02}$ & a \\
	&42 &47713.0$ - $47714.0&  $44.34^{+0.06}_{-0.06}$  &$  0.033^{+0.002}_{-0.008}$&$  43.31^{+ 0.05}_{- 0.05}$&$  10.59^{+ 1.44}_{- 2.75}$&$43.62^{+ 0.06}_{- 0.06}$&$ 5.24^{+ 0.77}_{- 1.39}$&$43.48^{+ 0.08}_{- 0.08}$&$ 7.12^{+ 1.40}_{- 1.97}$ & a \\ 
	&43 &47716.0$ - $47717.0&  $44.34^{+0.06}_{-0.06}$  &$  0.033^{+0.002}_{-0.007}$&$  43.29^{+ 0.03}_{- 0.03}$&$  11.07^{+ 0.85}_{- 2.49}$&$43.62^{+ 0.06}_{- 0.06}$&$ 5.18^{+ 0.73}_{- 1.32}$&$43.48^{+ 0.08}_{- 0.08}$&$ 7.11^{+ 1.39}_{- 1.91}$ & a \\
\hline
	\end{tabular}
\end{table}
\end{landscape}  

\begin{landscape}
\begin{table}
\contcaption{}
\label{tab:table4continued2}
\begin{tabular}{cccccccccccc} 
\hline
Object & Epoch & SED JD & $\log L_{BOL}(acc)$ & $\lambda_{Edd}$ & $\log \lambda L_{\lambda}$ (5100 \AA) & BC(5100\,\AA) &$\log \lambda L_{\lambda}$ (1350\,\AA) &BC(1350\,\AA) & $\log L_{\rm 2-10 keV}$ & BC(2--10\, keV)  & Notes\\
 & & [$-2400000$] & /erg s$^{-1}$] &  & /erg s$^{-1}$] & & /erg s$^{-1}$] & & /erg s$^{-1}$] & & \\
(1) & (2) & (3) & (4) & (5) & (6) & (7) & (8) & (9) & (10) & (11) & (12)  \\
 \hline
NGC\,5548 &44 &47725.0$ - $47726.0&  $44.31^{+0.07}_{-0.07}$  &$  0.031^{+0.003}_{-0.007}$&$  43.30^{+ 0.03}_{- 0.03}$&$  10.33^{+ 1.01}_{- 2.46}$&$43.59^{+ 0.06}_{- 0.06}$&$ 5.25^{+ 0.81}_{- 1.40}$&$43.45^{+ 0.10}_{- 0.10}$&$ 7.28^{+ 1.94}_{- 2.23}$ & a \\ 
	&45 &47729.0$ - $47730.0&  $44.22^{+0.08}_{-0.08}$  &$  0.025^{+0.003}_{-0.006}$&$  43.31^{+ 0.03}_{- 0.03}$&$   8.06^{+ 1.10}_{- 1.90}$&$43.37^{+ 0.09}_{- 0.09}$&$ 7.09^{+ 1.87}_{- 2.32}$&$43.45^{+ 0.10}_{- 0.10}$&$ 5.87^{+ 1.68}_{- 1.80}$ & a \\
	&46 &47736.0$ - $47737.0&  $44.09^{+0.10}_{-0.10}$  &$  0.019^{+0.004}_{-0.009}$&$  43.05^{+ 0.05}_{- 0.05}$&$  10.90^{+ 2.40}_{- 5.20}$&$43.15^{+ 0.12}_{- 0.12}$&$ 8.67^{+ 3.20}_{- 4.87}$&$43.45^{+ 0.10}_{- 0.10}$&$ 4.35^{+ 1.41}_{- 2.21}$ & a \\
	&47 &47745.0$ - $47746.0&  $44.12^{+0.10}_{-0.09}$  &$  0.020^{+0.003}_{-0.005}$&$  43.06^{+ 0.05}_{- 0.05}$&$  11.64^{+ 2.36}_{- 3.25}$&$43.24^{+ 0.10}_{- 0.10}$&$ 7.70^{+ 2.31}_{- 2.74}$&$43.45^{+ 0.10}_{- 0.10}$&$ 4.76^{+ 1.50}_{- 1.56}$& a \\
3C\,390.3& 1&49729.1$ - $49732.8&  $44.69^{+0.08}_{-0.07}$  &$  0.009^{+0.001}_{-0.002}$&$  43.37^{+ 0.10}_{- 0.10}$&$  21.31^{+ 6.16}_{- 6.85}$&$43.61^{+ 0.19}_{- 0.19}$&$12.03^{+ 6.53}_{- 6.75}$&$44.15^{+ 0.01}_{- 0.01}$&$ 3.48^{+ 0.36}_{- 0.61}$ & \\
	& 2 &49842.5$ - $49861.0&  $44.93^{+0.09}_{-0.08}$  &$  0.016^{+0.003}_{-0.003}$&$  43.57^{+ 0.07}_{- 0.07}$&$  23.06^{+ 5.41}_{- 6.00}$&$43.93^{+ 0.17}_{- 0.17}$&$10.03^{+ 5.02}_{- 5.15}$&$44.35^{+ 0.03}_{- 0.03}$&$ 3.87^{+ 0.65}_{- 0.78}$& d \\
NGC\,7469& 1&50248.0$ - $50249.0&  $44.38^{+0.03}_{-0.03}$  &$  0.210^{+0.025}_{-0.040}$&$  43.30^{+ 0.06}_{- 0.06}$&$  12.05^{+ 1.90}_{- 2.57}$&$43.71^{+ 0.03}_{- 0.03}$&$ 4.63^{+ 0.38}_{- 0.77}$&$43.34^{+ 0.00}_{- 0.00}$&$10.83^{+ 0.17}_{- 1.56}$ & \\
	& 2&50253.0$ - $50254.0&  $44.20^{+0.05}_{-0.03}$  &$  0.141^{+0.016}_{-0.029}$&$  43.20^{+ 0.08}_{- 0.08}$&$  10.00^{+ 2.00}_{- 2.57}$&$43.55^{+ 0.04}_{- 0.04}$&$ 4.53^{+ 0.44}_{- 0.86}$&$43.18^{+ 0.00}_{- 0.00}$&$10.64^{+ 0.08}_{- 1.73}$ & \\ 
        & 3&50262.0$ - $50263.0&  $44.39^{+0.04}_{-0.03}$  &$  0.215^{+0.026}_{-0.042}$&$  43.24^{+ 0.07}_{- 0.07}$&$  14.08^{+ 2.57}_{- 3.31}$&$43.75^{+ 0.03}_{- 0.03}$&$ 4.31^{+ 0.35}_{- 0.73}$&$43.30^{+ 0.00}_{- 0.00}$&$12.27^{+ 0.25}_{- 1.84}$ & \\
	& 4&50273.0$ - $50274.0&  $44.24^{+0.04}_{-0.03}$  &$  0.153^{+0.018}_{-0.030}$&$  43.20^{+ 0.08}_{- 0.08}$&$  11.10^{+ 2.22}_{- 2.79}$&$43.57^{+ 0.03}_{- 0.03}$&$ 4.64^{+ 0.36}_{- 0.79}$&$43.20^{+ 0.00}_{- 0.00}$&$11.07^{+ 0.08}_{- 1.69}$ & \\ 
\hline
	\end{tabular}
\end{table}
\end{landscape}  

\subsubsection{Accretion luminosity based on the SED model of \citet{Korista97}} \label{S:bolmodel}

\begin{figure}
\begin{center}$
\begin{array}{c}
\includegraphics[scale=0.7]{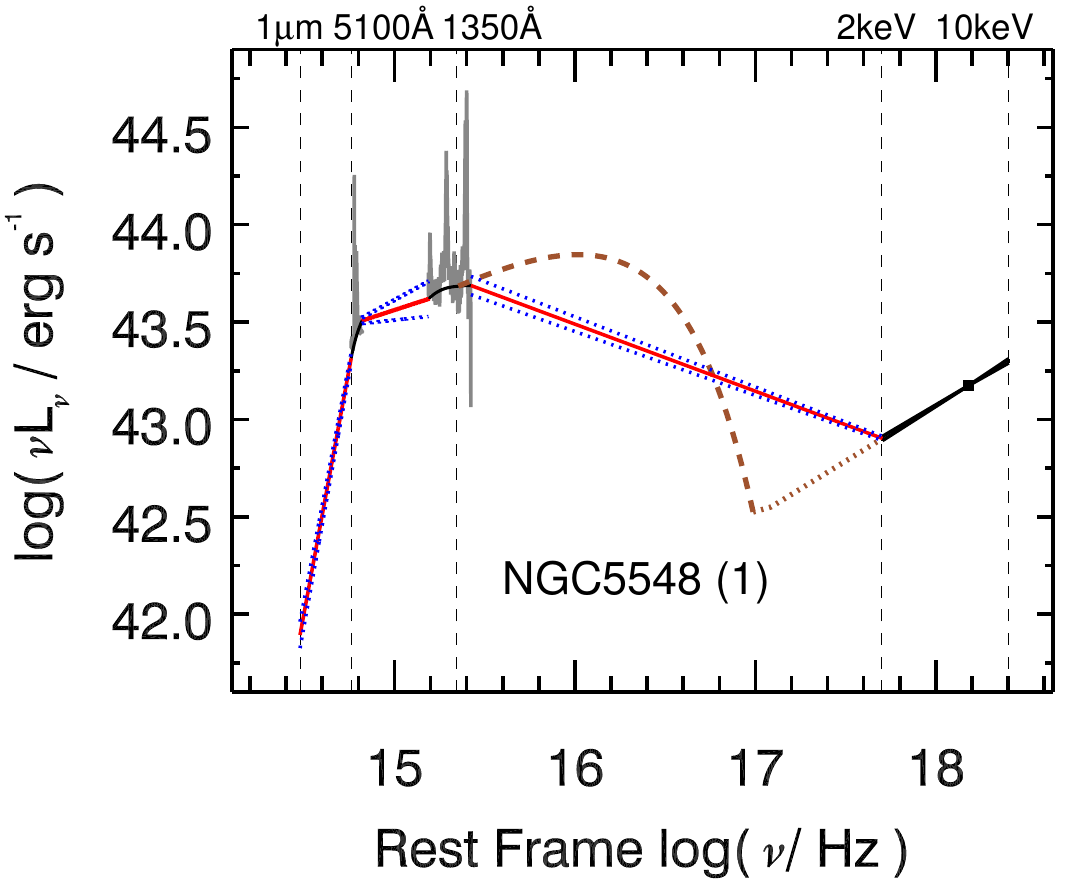}\\ 
\end{array}$
\end{center}
\caption{Comparison of the thermal accretion disc model of \citet{Korista97} to the linear interpolation adopted here for a sample SED of Seyfert galaxy NGC\,5548 (see Fig. \ref{fig:fig2} for symbols and colour code).
In the UV to X-ray region the SED model of \citet{Korista97} is shown as the brown dashed line. The dotted brown line shows the X-ray extrapolation.} 
\label{fig:fig6}
\end{figure}

\begin{figure}
\begin{center}$
\begin{array}{c}
\includegraphics[scale=0.5]{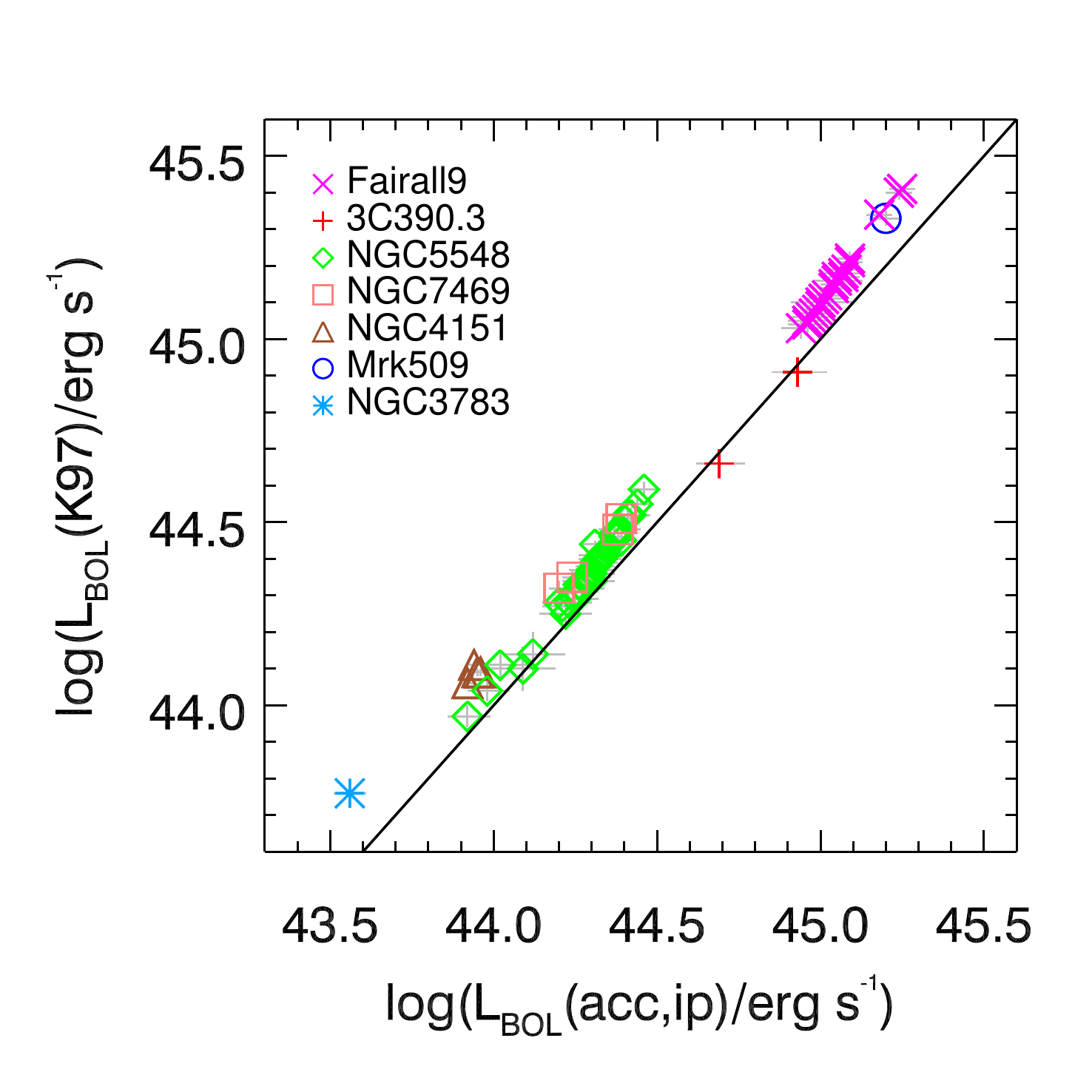}\\ 
\end{array}$
\end{center}
\caption{Comparison of the accretion luminosities based on \citet{Korista97} model and linear interpolation. The solid line shows a one-to-one relationship. 
Fig. shows that the \citet{Korista97} SED model gives higher luminosities (on average 23.3 per cent) compared to the linear interpolation. }
\label{fig:fig7}
\end{figure}

The largest gap in the SED is in the EUV region, from $\sim$1100\AA\ to 2\,keV. 
To bridge this gap it is common practice to linearly interpolate between the UV and X-ray data points \citep[e.g.][]{Elvis94,Laor1997} 
because it is simple and does not impose any model assumptions. 
\citet{Runnoe2012} show that linear interpolation in the EUV region underestimates the accretion luminosity relative to the SED model of \citet{Korista97}. 
We also compute the accretion luminosity by this model in order to quantify the differences.
The continuum SED adopted from the work of \citet{Korista97} is an empirical SED of the ionizing continuum generated from the observed average SEDs of radio-quiet AGN \citep{Elvis94}. 
The photoionization calculations by the authors show that such an ionizing continuum is successful at producing the observed broad emission line properties. 
Their SED is a combination of a thermal UV bump of the form $f_{\nu}\propto \nu^{-0.5} \exp(-h\nu/kT_{cut})$ and a power law component for the X-ray emission.  
At energies of 13.6 eV and above a power law with a slope of $-1.0$ ($f_{\nu}\propto \nu^{-1}$) is used. 
We follow \citet{Korista97} and adopt their typical values for the SED parameters including the 
cutoff temperature $T_{cut}$ at $10^{6} K$ and the peak energy of the UV bump $E_{peak}$ at 44\,eV. 
Fig. \ref{fig:fig6} compares the \citet{Korista97} model (brown dashed line) to the linear interpolation in the EUV region for a sample SED. 
We normalize the \citet{Korista97} model to the UV continuum luminosity at 1350\,\AA\ and extrapolate the observed X-ray power law component 
towards the UV region until it intersects the model (brown dotted line). We integrate the SED between 1$\micron$ $-$10 keV and obtain the accretion luminosity, $L_{BOL}(K97)$.  
In Fig. \ref{fig:fig7} we compare the accretion luminosities obtained from the two methods (interpolation, $L_{BOL}(acc,ip)$ and the \citet{Korista97} model). 
We confirm the results of \citet{Runnoe2012} that the \citet{Korista97} SED model yields slightly higher luminosities: 
$L_{BOL}(K97)$ is on average 23.3 per cent\ (equivalent to 0.09\,dex) higher than $L_{BOL}(acc,ip)$ and the $1 \sigma$ standard deviation is 10.0 per cent (equivalent to 0.03\,dex). 
This comparison indicates that the uncertainty in the $L_{BOL}(acc,ip)$  
values, due to an assumed different EUV spectral shape, is at least  $\sim$23 per cent on average.  
And, obviously, if the \citet{Korista97} SED model represents the intrinsic AGN SED, then we systematically underestimate 
the accretion luminosities by $\sim$23 per cent by use of the simple interpolated SED adopted here.  

It is worth noting that the mean percent difference between the two luminosities is large and comparable to the observed $L_{BOL}(acc,ip)$ variability 
amplitudes (\S \ref{S:bolvar}) of individual sources. 
Even with the thermal bump of the \citet{Korista97} SED model included we may not entirely capture the intrinsic SED and, furthermore, 
the BLR clouds may  actually see a different continuum than we do \citep[e.g.][]{Korista97b}. 
Because it is unclear which of the two methods gives the best estimate of $L_{BOL}(acc)$, 
we opt for the simpler method that provides the minimum accretion luminosity by measuring directly the data 
and which has the advantage of being model independent. 
We also applied an accretion disc model \citep[the \textit{optxagnf} model of][]{Done2012} to one SED to infer $L_{BOL}(acc)$.  
The details of this analysis is given in Appendix \ref{S:optxangf}. 
The accretion disc model is not robust enough for $L_{BOL}(acc)$ measurements but it provides at best a somewhat conservative 
lower limit that is \textit{23 per cent} lower compared to $L_{BOL}(acc,ip)$. 
As it happens, $L_{BOL}(acc,ip)$ values lie between the two extremes. 
Hereafter, we use $L_{BOL}(acc)$ as the total luminosity generated by the accretion process as estimated by interpolating the SEDs.

\subsection{The BC factor} \label{S:bolometriccor}

It is not straightforward to obtain and coordinate observing time on a large range of telescopes covering the entire electromagnetic spectrum over which AGN emit.
Therefore, it is not always possible to obtain simultaneous multiwavelength data 
from which to construct the SED that allow us to directly measure the accretion luminosity. 
A practical way to overcome this problem is to apply mean 
BCs to monochromatic luminosities and estimate the accretion luminosity. 
However, this may not be a good approach for all applications, especially if the AGN is strongly varying. 
Also, the mean BC values obtained for large samples have large sample spreads \citep[e.g.][]{Elvis94,Richards2006}. 
In order to check the difference between those BCs and the ones obtained from quasi-simultaneous SEDs of individual sources, 
in the following, we determine single-epoch BCs at wavelengths 5100\,\AA{} and 1350\,\AA\ and for the 2$-$10 keV X-ray band, widely used to estimate AGN accretion luminosities. 

\subsubsection{BC measurements} \label{S:bcmeasurements}

We calculate the bolometric corrections by assuming: 
\begin{equation}\label{E:BC} 
BC(\lambda)= L_{BOL}(acc)/\lambda L_{\lambda}.
\end{equation}
\noindent Here 
BC($\lambda$) is the bolometric correction at rest wavelength $\lambda$ 
and $L_{\lambda}$ is the monochromatic luminosity. 
The monochromatic optical luminosity at 5100\,\AA, $\lambda L_{\lambda}$(5100\,\AA), 
the monochromatic UV luminosity at 1350\,\AA, $\lambda L_{\lambda}$(1350\,\AA)
and the total X-ray band luminosity, $L_{\rm 2-10\,keV}$, measured from our data are listed in Table \ref{tab:table4}. 
The BC factors at these energies, BC(5100\,\AA), BC(1350\,\AA) and BC(2--10\,keV), 
 are listed in columns (7), (9) and (11) of Table 4, respectively, for our objects.
The uncertainties on the BC values are estimated based on Equation \ref{E:BC} and standard error propagation rules \citep[][Chapter 4.4]{Taylor}. 

In Fig. \ref{fig:fig5} we present the distributions of the optical (top), UV (middle) and X-ray (bottom) BCs for our sample. 
It is noteworthy that there is a considerable range of BC factors even for our small sample, 
namely: BC(5100\,\AA)=8.0 -- 23.1, BC(1350\,\AA)=3.2 -- 12.0 and BC(2--10 keV) = 3.5 -- 44.0.  
However, they are fully consistent with the results of larger surveys such as \citet{Elvis94,Elvis2012,Richards2006}.

\subsection{The Eddington luminosity ratio} \label{S:Eddratio}
The Eddington luminosity ratio, $\lambda_{Edd}$, is the ratio of the accretion luminosity to the Eddington luminosity, $L_{Edd}$, 
which scales with the mass of the central black hole ($M_{BH}$) and is defined as \citep[e.g.][]{Petersonbook}
\begin{equation}\label{Eq:eddratio}
\lambda_{Edd}=\frac{L_{BOL}(acc)}{L_{Edd}}=\frac{L_{BOL}(acc)}{1.26\times 10^{38}(M_{BH}/M_{\odot})}
 \end{equation}
\noindent  
In the accretion process the accreting matter is under the influence of gravitational and radiation forces. 
Note that, we here use the accretion luminosity, $L_{BOL}(acc)$ as this is the luminosity that 
balances the radiation pressure.  
$L_{Edd}$ is the maximum luminosity satisfying the balance between the two forces in the case of spherical accretion and 
is therefore often used as a measure of the maximum luminosity that is possible for the central engine.  

To compute $\lambda_{Edd}$, we adopt the 
up-to-date RM-based $M_{\rm BH}$ measurements from the AGN Black Hole Mass data base \citep{BentzKatz2015}, 
the $L_{BOL}(acc)$ values of column (4) of Table \ref{tab:table4}; the $\lambda_{Edd}$values are tabulated in column (5) of Table  \ref{tab:table4}. 
We propagate the uncertainties in $L_{BOL}(acc)$ and $M_{BH}$ through equation (\ref{Eq:eddratio}) to estimate the uncertainties in the $\lambda_{Edd}$ values. 
The Eddington ratio distribution of our sample is shown in Fig. \ref{fig:fig5} covering a range from 0.01 to 0.22. 
For the majority of our sample, the Eddington ratios are rather small ($\lambda_{Edd} \le 0.1$). 
Only for NGC\,7469, is the Eddington ratio moderate ($0.14\le \lambda_{Edd} \le0.22$). 
We note that, our $\lambda_{Edd}$ values agree with those of \citet{Vasudevan2009a} to within the measurement uncertainties for those objects 
where $L_{BOL}(acc)$ did not change significantly. 

\subsection{Mass accretion rate and accretion efficiency measurements} \label{S:mdotandeta}
 The intrinsic continuum luminosity of AGN is expected to be produced by the accretion of material on to the central supermassive BH. 
The mass  accretion rate, \Mdot, is directly proportional to the emitted accretion luminosity: $L=\eta \dot{M} c^{2}$, where $\eta$ is the efficiency of the energy conversion. 
Both \Mdot\ and $\eta$ can be determined from accretion disc models \citep[e.g.][]{DavisLaor2011,Raimundo2012}.
In particular, \Mdot{} can be inferred when the monochromatic optical continuum luminosity at 4392\,\AA , $L(4392\,\AA)$, the black hole mass and the disc inclination are known \citep{Raimundo2012}.

We apply equation (3) and equation (4) of \citet{Raimundo2012} to our sample to estimate \Mdot{} and $\eta$, respectively. 
For \Mdot{} we use the RM-based $M_{\rm BH}$ measurements (listed in Table \ref{tab:table1}) of \citet{BentzKatz2015} and assume a common inclination of $i$=36 $\degr$ ($cos(i)=0.8$; this is a reasonable assumption since our sample has only face-on, low inclination AGN, otherwise known as broad-lined (or type 1) AGN).  
We measure $L(4392\,\AA)$ from the optical spectra of 3C390.3 since  $\lambda$4392 is covered by the spectra but for the remaining AGN, we use the interpolated continuum in the SED between the optical and UV spectra. 
To infer the accretion efficiency $\eta$ we use the directly measured $L_{BOL}(acc)$ values from the SEDs and note that they may likely be $\sim$23 per cent too low, thereby providing $\eta$ values $\sim$20 per cent too high (\S \ref{S:bomeasurements}). 
We tabulate the \Mdot{} and $\eta$ values in columns (4) and (5) of Table  \ref{tab:table5}, respectively. The errors listed are those propagated from the measurement uncertainties on the parameters that  \Mdot{} and $\eta$ depend on, namely ($\lambda L_{\lambda}$, $M_{\rm BH}$) and ($L_{\rm BOL}$, \Mdot), respectively.
The distributions of \Mdot{} and $\eta$ for our sample of seven AGNs is shown in Fig. \ref{fig:fig5} in panels (c) and (d), respectively. 
Our sample consists of moderately accreting BHs ($ 0.002 < \dot{M}/M_{\odot} yr^{-1} <  0.194 $) with a wide range of accretion efficiencies ($ 0.03 < \eta < 0.87 $, with a single exception discussed below. However, given the likely underestimate of $L_{BOL}(acc)$, the values are more likely to be: $ 0.02 < \eta < 0.7 $).

\citet{Raimundo2012} determine the mass accretion rate of 22 local AGNs with simultaneous optical-UV and 
X-ray imaging observations from the {\it Swift}/BAT catalogue of \citet{Vasudevan2009b}. 
We have four AGNs common with their sample, namely 3C\,390.3, NGC\,5548, Mrk\,509, NGC\,7469. 
Our $ \dot{M}$ and $\eta$ measurements agree with their results to within our measurement uncertainties, except for 3C\,390.3  
for which our $L(4392\,\AA)$ values are significantly lower (by 1 dex).  Two issues may explain this difference.  
For one, our two studies adopt different approaches to correct for the contribution of star light. 
While we use stellar light measurements based on high-resolution (0.0575 arcsec) {\it HST} imaging \citep[][ \S~\ref{S:analysis}]{Bentz09a}, 
\citet{Vasudevan2009b} model the 2-dimensional light distribution of the stars and the nuclear source as observed with the {\it Swift}/UV/Optical Telescope (UVOT) instrument. The authors estimate that with the larger UVOT point-spread function of $\sim$2\,arcsec, their nuclear fluxes may be too large by a factor of a few. While this cannot explain the full factor 10 difference, long-term variability is a likely strong contributor thereto since the data used by \citet{Raimundo2012} were obtained in 2008 May, 13 years later than those analysed here. This is supported by the \citet{Dietrich2012} study, which measures an average optical luminosity that is a factor $\sim$4 higher than observed in 1995.

Based on our $L(4392\,\AA)$ measurements we obtain extreme accretion efficiency values for  3C\,390.3, much higher than the efficiency limit of 0.4 for a maximally rotating black hole \citep{Novikov1973}. 
These unrealistic $\eta$ values for 3C\,390.3 cannot be explained by the likely too low $L_{BOL}(acc)$ values  
(\S~\ref{S:bomeasurements}) as it is only a 20 per cent effect. Thus, this suggests that either the optical-UV emission of this AGN is not modelled well by a thin accretion disc or at least one of the parameters that affect $\eta$ is inaccurate, including the source inclination angle $i$ and the black hole mass.  We assumed $i=36\degr (cos(i)=0.8)$, but it can, in fact, range from $i=0\degr$ to 60$\degr$. 
A higher inclination of 60$\degr$ will, in fact, reduce the inferred accretion efficiency  significantly from 5.246 to 2.592 (SED epoch 1) and from  3.664 to 1.810 (SED epoch 2). 
In addition, the black hole mass has a measurement uncertainty that is mainly due to the unknown geometry and dynamics of the broad line region, as expressed by the dimensionless scaling factor $f$.
However, even combining the high inclination with the 35 per cent lower value of $f$ of 2.8 \citep{Graham2011} instead of 4.3 \citep{Grier2013b} cannot reduce the inferred accretion efficiency below a value of 1.0, let alone a value of 0.4 for a maximally rotating black hole.  Only for $f \sim$1 will the inferred $\eta$ in this case attain a value of order 0.4 or less. While this may be a coincidence, it is interesting to note that the recent study of \citet{Shankar2017} advocates that $f$=1 due to a systematic bias in the stellar velocity dispersions measured for quiescent galaxies in the $M - \sigma$ relation, from which $f$ is determined. 

While for 3C390.3 a range of poorly known parameters may explain this discrepant $\eta$ value, this case does serve as a reminder that the $\eta$ estimate is merely a crude estimate when details of the central engine are not well known.

\begin{table}
           \centering
           \caption{Optical 4392\,\AA{} continuum luminosities, mass accretion rates and accretion efficiencies.  Columns: (3) Base 10 logarithm of the monochromatic luminosity at a wavelength of $\lambda$= 4392\,\AA{} in units of erg $s^{-1}$. (4) Base 10 logarithm of the mass accretion rate in units of M$_{\sun} yr^{-1}$. (5) Accretion efficiency calculated as explained in \S~\ref{S:mdotandeta}. 
	}
           \label{tab:table5}
           \begin{tabular}{lcccc}
           \hline
           Object & SED  & $\log [\lambda L(4392\,$\AA) & $\log[ \dot{M}$ & $\eta$\\
                      & epoch &  /erg $s^{-1}$]                     & /M$_{\sun} yr^{-1}$] &  \\                     
            (1) & (2)  & (3) & (4) & (5) \\
             \hline
Fairall\,9&1&$43.862^{+ 0.047 }_{- 0.047 }$ & $-1.676^{+ 0.089 }_{- 0.122 }$ & $0.737^{+0.168 }_{-0.286  }$\\
Fairall\,9&2&$43.861^{+ 0.051 }_{- 0.051 }$ & $-1.677^{+ 0.091 }_{- 0.122 }$ & $0.758^{+0.178 }_{-0.297  }$\\
Fairall\,9&3&$43.854^{+ 0.035 }_{- 0.035 }$ & $-1.689^{+ 0.084 }_{- 0.119 }$ & $0.804^{+0.173 }_{-0.312  }$\\
Fairall\,9&4&$43.883^{+ 0.026 }_{- 0.026 }$ & $-1.644^{+ 0.081 }_{- 0.118 }$ & $0.739^{+0.153 }_{-0.277  }$\\
Fairall\,9&5&$43.929^{+ 0.028 }_{- 0.028 }$ & $-1.576^{+ 0.082 }_{- 0.118 }$ & $0.610^{+0.127 }_{-0.231  }$\\
Fairall\,9&6&$43.925^{+ 0.090 }_{- 0.090 }$ & $-1.581^{+ 0.115 }_{- 0.133 }$ & $0.685^{+0.207 }_{-0.292  }$\\
Fairall\,9&7&$43.996^{+ 0.024 }_{- 0.024 }$ & $-1.475^{+ 0.081 }_{- 0.118 }$ & $0.615^{+0.127 }_{-0.233  }$\\
Fairall\,9&8&$43.932^{+ 0.023 }_{- 0.023 }$ & $-1.571^{+ 0.081 }_{- 0.117 }$ & $0.685^{+0.140 }_{-0.259  }$\\
Fairall\,9&9&$43.961^{+ 0.025 }_{- 0.025 }$ & $-1.527^{+ 0.081 }_{- 0.118 }$ & $0.663^{+0.138 }_{-0.267  }$\\
Fairall\,9&10&$43.912^{+ 0.036 }_{- 0.036 }$ & $-1.601^{+ 0.085 }_{- 0.119 }$ & $0.877^{+0.191 }_{-0.332  }$\\
Fairall\,9&11&$43.995^{+ 0.017 }_{- 0.017 }$ & $-1.477^{+ 0.079 }_{- 0.117 }$ & $0.576^{+0.116 }_{-0.216  }$\\
Fairall\,9&12&$43.937^{+ 0.018 }_{- 0.018 }$ & $-1.564^{+ 0.080 }_{- 0.117 }$ & $0.709^{+0.144 }_{-0.262  }$\\
Fairall\,9&13&$43.931^{+ 0.025 }_{- 0.025 }$ & $-1.572^{+ 0.081 }_{- 0.118 }$ & $0.642^{+0.132 }_{-0.242  }$\\
Fairall\,9&14&$43.896^{+ 0.021 }_{- 0.021 }$ & $-1.625^{+ 0.080 }_{- 0.117 }$ & $0.750^{+0.159 }_{-0.316  }$\\
Fairall\,9&15&$43.899^{+ 0.020 }_{- 0.020 }$ & $-1.620^{+ 0.080 }_{- 0.117 }$ & $0.732^{+0.149 }_{-0.268  }$\\
Fairall\,9&16&$43.918^{+ 0.022 }_{- 0.022 }$ & $-1.592^{+ 0.080 }_{- 0.117 }$ & $0.633^{+0.130 }_{-0.243  }$\\
Fairall\,9&17&$43.976^{+ 0.030 }_{- 0.030 }$ & $-1.506^{+ 0.083 }_{- 0.118 }$ & $0.580^{+0.122 }_{-0.224  }$\\
Fairall\,9&18&$44.012^{+ 0.016 }_{- 0.016 }$ & $-1.451^{+ 0.079 }_{- 0.117 }$ & $0.510^{+0.102 }_{-0.193  }$\\
Fairall\,9&19&$43.996^{+ 0.017 }_{- 0.017 }$ & $-1.475^{+ 0.080 }_{- 0.117 }$ & $0.598^{+0.121 }_{-0.225  }$\\
Fairall\,9&20&$44.016^{+ 0.026 }_{- 0.026 }$ & $-1.445^{+ 0.082 }_{- 0.118 }$ & $0.594^{+0.123 }_{-0.223  }$\\
Fairall\,9&21&$44.042^{+ 0.024 }_{- 0.024 }$ & $-1.405^{+ 0.081 }_{- 0.118 }$ & $0.552^{+0.115 }_{-0.205  }$\\
Fairall\,9&22&$44.125^{+ 0.012 }_{- 0.012 }$ & $-1.281^{+ 0.079 }_{- 0.116 }$ & $0.399^{+0.079 }_{-0.158  }$\\
Fairall\,9&23&$44.025^{+ 0.022 }_{- 0.022 }$ & $-1.431^{+ 0.080 }_{- 0.117 }$ & $0.522^{+0.107 }_{-0.194  }$\\
Fairall\,9&24&$44.046^{+ 0.018 }_{- 0.018 }$ & $-1.401^{+ 0.080 }_{- 0.117 }$ & $0.516^{+0.104 }_{-0.193  }$\\
Fairall\,9&25&$44.197^{+ 0.023 }_{- 0.023 }$ & $-1.174^{+ 0.081 }_{- 0.117 }$ & $0.456^{+0.096 }_{-0.175  }$\\
Fairall\,9&26&$44.192^{+ 0.011 }_{- 0.011 }$ & $-1.181^{+ 0.079 }_{- 0.116 }$ & $0.473^{+0.100 }_{-0.177  }$\\
Fairall\,9&27&$44.215^{+ 0.012 }_{- 0.012 }$ & $-1.146^{+ 0.079 }_{- 0.116 }$ & $0.374^{+0.083 }_{-0.144  }$\\
NGC\,3783&1&$42.676^{+ 0.047 }_{- 0.047 }$ & $-2.527^{+ 0.088 }_{- 0.085 }$ & $0.214^{+0.050 }_{-0.071  }$\\
NGC\,4151&1&$43.042^{+ 0.010 }_{- 0.010 }$ & $-2.162^{+ 0.052 }_{- 0.048 }$ & $0.228^{+0.035 }_{-0.047  }$\\
NGC\,4151&2&$43.050^{+ 0.011 }_{- 0.011 }$ & $-2.150^{+ 0.052 }_{- 0.048 }$ & $0.218^{+0.035 }_{-0.047  }$\\
NGC\,4151&3&$43.046^{+ 0.010 }_{- 0.010 }$ & $-2.156^{+ 0.052 }_{- 0.048 }$ & $0.230^{+0.036 }_{-0.047  }$\\
NGC\,4151&4&$43.042^{+ 0.010 }_{- 0.010 }$ & $-2.162^{+ 0.052 }_{- 0.048 }$ & $0.214^{+0.033 }_{-0.044  }$\\
NGC\,5548&1&$43.512^{+ 0.018 }_{- 0.018 }$ & $-1.620^{+ 0.024 }_{- 0.023 }$ & $0.149^{+0.010 }_{-0.028  }$\\
NGC\,5548&2&$43.396^{+ 0.024 }_{- 0.024 }$ & $-1.794^{+ 0.028 }_{- 0.027 }$ & $0.266^{+0.024 }_{-0.049  }$\\
NGC\,5548&3&$43.443^{+ 0.034 }_{- 0.034 }$ & $-1.724^{+ 0.038 }_{- 0.035 }$ & $0.179^{+0.017 }_{-0.034  }$\\
NGC\,5548&4&$43.455^{+ 0.021 }_{- 0.021 }$ & $-1.705^{+ 0.027 }_{- 0.026 }$ & $0.218^{+0.014 }_{-0.037  }$\\
NGC\,5548&5&$43.202^{+ 0.053 }_{- 0.053 }$ & $-2.085^{+ 0.055 }_{- 0.050 }$ & $0.350^{+0.050 }_{-0.084  }$\\
NGC\,5548&6&$43.327^{+ 0.067 }_{- 0.067 }$ & $-1.898^{+ 0.069 }_{- 0.060 }$ & $0.255^{+0.044 }_{-0.062  }$\\
NGC\,5548&7&$43.183^{+ 0.044 }_{- 0.044 }$ & $-2.114^{+ 0.047 }_{- 0.043 }$ & $0.367^{+0.042 }_{-0.086  }$\\
NGC\,5548&8&$43.294^{+ 0.026 }_{- 0.026 }$ & $-1.948^{+ 0.031 }_{- 0.029 }$ & $0.309^{+0.023 }_{-0.063  }$\\
NGC\,5548&9&$43.202^{+ 0.028 }_{- 0.028 }$ & $-2.085^{+ 0.032 }_{- 0.030 }$ & $0.225^{+0.022 }_{-0.063  }$\\
NGC\,5548&10&$43.155^{+ 0.030 }_{- 0.030 }$ & $-2.155^{+ 0.034 }_{- 0.032 }$ & $0.242^{+0.021 }_{-0.056  }$\\
NGC\,5548&11&$43.060^{+ 0.037 }_{- 0.037 }$ & $-2.298^{+ 0.040 }_{- 0.037 }$ & $0.293^{+0.032 }_{-0.069  }$\\
NGC\,5548&12&$43.362^{+ 0.031 }_{- 0.031 }$ & $-1.845^{+ 0.034 }_{- 0.032 }$ & $0.239^{+0.024 }_{-0.055  }$\\
NGC\,5548&13&$43.354^{+ 0.038 }_{- 0.038 }$ & $-1.857^{+ 0.041 }_{- 0.038 }$ & $0.261^{+0.027 }_{-0.064  }$\\
NGC\,5548&14&$43.430^{+ 0.022 }_{- 0.022 }$ & $-1.743^{+ 0.027 }_{- 0.026 }$ & $0.206^{+0.015 }_{-0.044  }$\\
NGC\,5548&15&$43.405^{+ 0.024 }_{- 0.024 }$ & $-1.780^{+ 0.028 }_{- 0.027 }$ & $0.225^{+0.018 }_{-0.049  }$\\
NGC\,5548&16&$43.462^{+ 0.019 }_{- 0.019 }$ & $-1.695^{+ 0.024 }_{- 0.024 }$ & $0.205^{+0.013 }_{-0.043  }$\\
NGC\,5548&17&$43.397^{+ 0.058 }_{- 0.058 }$ & $-1.792^{+ 0.060 }_{- 0.053 }$ & $0.271^{+0.040 }_{-0.075  }$\\
NGC\,5548&18&$43.503^{+ 0.027 }_{- 0.027 }$ & $-1.633^{+ 0.031 }_{- 0.030 }$ & $0.198^{+0.015 }_{-0.049  }$\\
NGC\,5548&19&$43.511^{+ 0.031 }_{- 0.031 }$ & $-1.622^{+ 0.035 }_{- 0.033 }$ & $0.188^{+0.017 }_{-0.045  }$\\
NGC\,5548&20&$43.601^{+ 0.014 }_{- 0.014 }$ & $-1.486^{+ 0.021 }_{- 0.021 }$ & $0.130^{+0.008 }_{-0.029  }$\\
NGC\,5548&21&$43.435^{+ 0.069 }_{- 0.069 }$ & $-1.735^{+ 0.070 }_{- 0.061 }$ & $0.185^{+0.034 }_{-0.054  }$\\
NGC\,5548&22&$43.352^{+ 0.023 }_{- 0.023 }$ & $-1.860^{+ 0.028 }_{- 0.027 }$ & $0.217^{+0.020 }_{-0.054  }$\\
NGC\,5548&23&$43.424^{+ 0.019 }_{- 0.019 }$ & $-1.752^{+ 0.025 }_{- 0.024 }$ & $0.179^{+0.017 }_{-0.040  }$\\
\hline
	\end{tabular}
\end{table}    

\begin{table}
           \centering
           \contcaption{}
           \label{tab:table5continued}
           \begin{tabular}{lllll}
           \hline
           Object & SED  & $\log [\lambda L(4392\,$\AA) & $\log[ \dot{M}$ & $\eta$\\
                      & epoch &  /erg $s^{-1}$]                     & /M$_{\sun} yr^{-1}$] &  \\                     
            (1) & (2)  & (3) & (4) & (5) \\
             \hline
 NGC\,5548&24&$43.426^{+ 0.032 }_{- 0.032 }$ & $-1.748^{+ 0.035 }_{- 0.034 }$ & $0.169^{+0.019 }_{-0.040  }$\\
NGC\,5548&25&$43.398^{+ 0.030 }_{- 0.030 }$ & $-1.791^{+ 0.034 }_{- 0.032 }$ & $0.216^{+0.037 }_{-0.040  }$\\
NGC\,5548&26&$43.481^{+ 0.022 }_{- 0.022 }$ & $-1.666^{+ 0.027 }_{- 0.026 }$ & $0.172^{+0.013 }_{-0.039  }$\\
NGC\,5548&27&$43.561^{+ 0.013 }_{- 0.013 }$ & $-1.546^{+ 0.021 }_{- 0.020 }$ & $0.149^{+0.009 }_{-0.033  }$\\
NGC\,5548&28&$43.501^{+ 0.027 }_{- 0.027 }$ & $-1.636^{+ 0.031 }_{- 0.030 }$ & $0.188^{+0.015 }_{-0.042  }$\\
NGC\,5548&29&$43.574^{+ 0.033 }_{- 0.033 }$ & $-1.526^{+ 0.037 }_{- 0.035 }$ & $0.173^{+0.016 }_{-0.045  }$\\
NGC\,5548&30&$43.616^{+ 0.021 }_{- 0.021 }$ & $-1.465^{+ 0.026 }_{- 0.025 }$ & $0.142^{+0.009 }_{-0.032  }$\\
NGC\,5548&31&$43.571^{+ 0.014 }_{- 0.014 }$ & $-1.532^{+ 0.021 }_{- 0.021 }$ & $0.142^{+0.007 }_{-0.034  }$\\
NGC\,5548&32&$43.462^{+ 0.036 }_{- 0.036 }$ & $-1.695^{+ 0.039 }_{- 0.037 }$ & $0.211^{+0.022 }_{-0.049  }$\\
NGC\,5548&33&$43.526^{+ 0.021 }_{- 0.021 }$ & $-1.598^{+ 0.026 }_{- 0.026 }$ & $0.168^{+0.011 }_{-0.039  }$\\
NGC\,5548&34&$43.460^{+ 0.027 }_{- 0.027 }$ & $-1.697^{+ 0.031 }_{- 0.030 }$ & $0.189^{+0.017 }_{-0.045  }$\\
NGC\,5548&35&$43.470^{+ 0.029 }_{- 0.029 }$ & $-1.683^{+ 0.033 }_{- 0.031 }$ & $0.181^{+0.015 }_{-0.046  }$\\
NGC\,5548&36&$43.467^{+ 0.015 }_{- 0.015 }$ & $-1.688^{+ 0.022 }_{- 0.022 }$ & $0.173^{+0.012 }_{-0.042  }$\\
NGC\,5548&37&$43.429^{+ 0.031 }_{- 0.031 }$ & $-1.744^{+ 0.034 }_{- 0.033 }$ & $0.199^{+0.017 }_{-0.049  }$\\
NGC\,5548&38&$43.394^{+ 0.034 }_{- 0.034 }$ & $-1.798^{+ 0.037 }_{- 0.035 }$ & $0.196^{+0.021 }_{-0.050  }$\\
NGC\,5548&39&$43.440^{+ 0.016 }_{- 0.016 }$ & $-1.728^{+ 0.023 }_{- 0.022 }$ & $0.174^{+0.015 }_{-0.038  }$\\
NGC\,5548&40&$43.415^{+ 0.041 }_{- 0.041 }$ & $-1.765^{+ 0.044 }_{- 0.041 }$ & $0.197^{+0.022 }_{-0.051  }$\\
NGC\,5548&41&$43.441^{+ 0.031 }_{- 0.031 }$ & $-1.727^{+ 0.035 }_{- 0.033 }$ & $0.202^{+0.017 }_{-0.051  }$\\
NGC\,5548&42&$43.434^{+ 0.039 }_{- 0.039 }$ & $-1.736^{+ 0.042 }_{- 0.039 }$ & $0.209^{+0.023 }_{-0.051  }$\\
NGC\,5548&43&$43.451^{+ 0.016 }_{- 0.016 }$ & $-1.712^{+ 0.022 }_{- 0.022 }$ & $0.197^{+0.013 }_{-0.044  }$\\
NGC\,5548&44&$43.471^{+ 0.015 }_{- 0.015 }$ & $-1.682^{+ 0.022 }_{- 0.022 }$ & $0.175^{+0.016 }_{-0.041  }$\\
NGC\,5548&45&$43.337^{+ 0.021 }_{- 0.021 }$ & $-1.883^{+ 0.027 }_{- 0.026 }$ & $0.223^{+0.030 }_{-0.053  }$\\
NGC\,5548&46&$43.183^{+ 0.033 }_{- 0.033 }$ & $-2.113^{+ 0.036 }_{- 0.034 }$ & $0.280^{+0.058 }_{-0.131  }$\\
NGC\,5548&47&$43.178^{+ 0.032 }_{- 0.032 }$ & $-2.121^{+ 0.036 }_{- 0.034 }$ & $0.309^{+0.056 }_{-0.081  }$\\
3C\,390.3&1&$43.352^{+ 0.041 }_{- 0.041 }$ & $-2.779^{+ 0.057 }_{- 0.061 }$ & $5.246^{+0.902 }_{-1.204  }$\\
3C\,390.3&2&$43.616^{+ 0.041 }_{- 0.041 }$ & $-2.383^{+ 0.057 }_{- 0.061 }$ & $3.664^{+0.761 }_{-0.890  }$\\
Mrk\,509&1&$44.338^{+ 0.005 }_{- 0.005 }$ & $-0.713^{+ 0.035 }_{- 0.035 }$ & $0.146^{+0.012 }_{-0.030  }$\\
NGC\,7469&1&$43.452^{+ 0.037 }_{- 0.037 }$ & $-0.948^{+ 0.060 }_{- 0.060 }$ & $0.037^{+0.006 }_{-0.008  }$\\
NGC\,7469&2&$43.371^{+ 0.046 }_{- 0.046 }$ & $-1.070^{+ 0.065 }_{- 0.064 }$ & $0.033^{+0.005 }_{-0.008  }$\\
NGC\,7469&3&$43.403^{+ 0.043 }_{- 0.043 }$ & $-1.021^{+ 0.063 }_{- 0.062 }$ & $0.045^{+0.007 }_{-0.010  }$\\
NGC\,7469&4&$43.364^{+ 0.046 }_{- 0.046 }$ & $-1.081^{+ 0.065 }_{- 0.064 }$ & $0.037^{+0.006 }_{-0.008  }$\\            
\hline
	\end{tabular}
\end{table}

\subsection{Time variations in the SEDs}\label{S:SEDvar}

We investigate in the following the effects of continuum variations on the SED parameters by quantifying the temporal variations of the 
accretion luminosity, $\lambda_{Edd}$, $ \dot{M}$, $\eta$ and BCs.

\subsubsection{Time variations in accretion luminosity and Eddington ratio} \label{S:bolvar}

\begin{figure}
\begin{center}$
\begin{array}{c}
\includegraphics[scale=0.45]{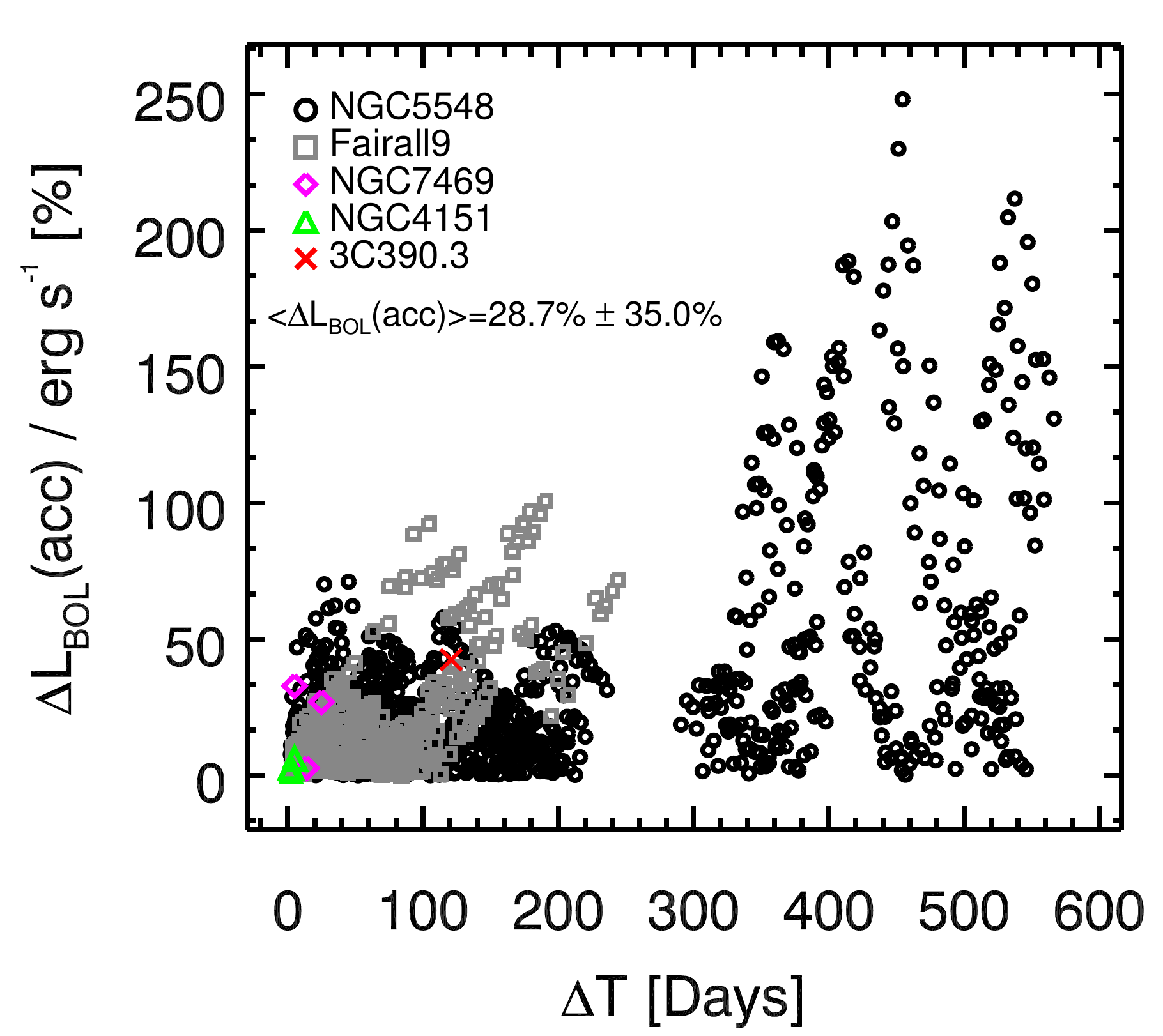}\\
\includegraphics[scale=0.45]{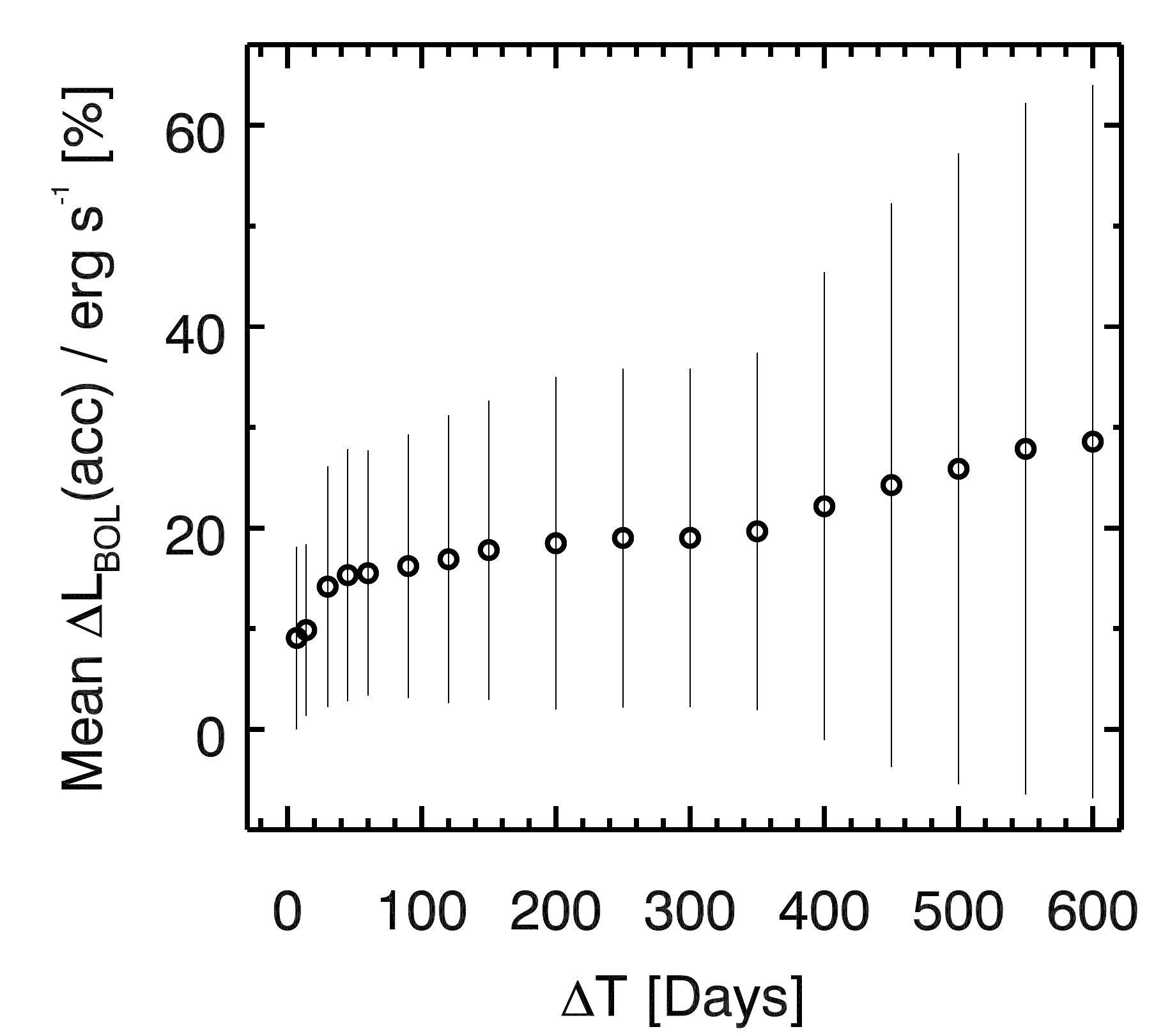}\\
\end{array}$
\end{center}
\caption{Temporal variations of $L_{BOL}(acc)$. Top panel: accretion luminosity percent differences between SED epochs versus the time differences. 
All of the available SED epochs are used for individual sources. Large $\Delta L_{BOL}(acc)$ differences ($>$ 100 per cent) is seen as $\Delta T$ exceeds a year. 
Bottom panel: average percent differences versus the binned time intervals. Vertical bars are the standard deviation (1$\sigma$) of the distribution around the mean values. 
The average $\Delta L_{BOL}(acc)$ and its 1$\sigma$ rms tend to increase with $\Delta T$ to $\sim$30 per cent across 600 d.}
\label{fig:fig8}
\end{figure}

The mean and rms values for the accretion luminosity $L_{BOL}(acc)$ are tabulated in Table \ref{tab:table6}. 
The rms (i.e., 1$\sigma$) variations  in $L_{BOL}(acc)$ (column 6), listed as the percent deviation relative to the mean $L_{BOL}(acc)$, is between $\sim$4 per cent and $\sim$38 per cent. 
This rms value depends strongly on the probed time span (\S \ref{S:theseds}): SED variations over a week appear to have a negligible effect on $L_{BOL}(acc)$, but SED changes over a year may cause significant luminosity variations. 
The temporal variations in $L_{BOL}(acc)$ of our sample are illustrated in Fig. \ref{fig:fig8}. 
The top panel shows $\Delta L_{BOL}(acc)$ as a function of time span for the six AGN (colour-coded) observed for multiple epochs; 
$\Delta L_{BOL}(acc)$ is the percent difference, {\it relative to the lower luminosity}, between any two $L_{BOL}(acc)$ values, while 
$\Delta T$ is the difference in the mean JDs of the SEDs. 
For each source, we use all of the available SED epochs when computing the statistics.  
Note that the long term variations (of a year or more) are available only for NGC\,5548.  
For NGC\,5548 the $\Delta L_{BOL}(acc)$ range expands and becomes larger than 100 per cent as $\Delta T$ exceeds $\sim$ 300 d. 
For our sample of seven objects, of which five have multi-epoch SEDs, the mean $\Delta L_{BOL}(acc)$ is about 30 per cent over $\sim$ 600 d (see also the lower panel). 
In the bottom panel, we show the mean $\Delta L_{BOL}(acc)$ within a set of time intervals; here error bars show the rms deviations. 
This panel shows the average $\Delta L_{BOL}(acc)$ and the rms around it to increase with $\Delta T$.  
Based on the NGC\,5548 data, we conclude that the changes in the optical-UV and X-ray continuum over the time span of a year may alter $L_{BOL}(acc)$ significantly (by up to $\sim$200 per cent). 
This will have importance when one is generating the SED from non-simultaneously observed data (\S \ref{S:lbolunc4}). 

The Eddington ratio, $\lambda_{Edd}$, of individual AGN changes with the same amplitude as the integrated $L_{BOL}(acc)$ (Equation \ref{Eq:eddratio} and Table \ref{tab:table6}).  
For the vast majority of our sample we probe 20$-$39 per cent rms variations. 
The highest $\lambda_{Edd}$ rms variations are moderate, around 32 per cent (NGC\,5548) and 39 per cent (3C\,390.3). 
For individual AGN, the probed $\lambda_{Edd}$ variations are not as dramatic as $L_{BOL}(acc)$.

\begin{landscape}
\begin{table}
           \centering
           \caption{Basic statistics of $L_{BOL}(acc)$ (shown as `$L_{BOL}$' in the table-headings for short), Eddington ratios, mass accretion rates, accretion efficiencies and BCs. Mean measurements are listed with the standard deviations of the parent measurement distribution. Columns: (1) Object  name. (2) Mean UV luminosity at  $\lambda$= 1350\,\AA{} in $10^{43}$ erg s$^{-1}$. (3) Mean optical luminosity at  $\lambda$= 5100\,\AA{} in $10^{43}$ erg s$^{-1}$. (4) Mean of the log of accretion luminosity in erg s$^{-1}$. (5), (7), (9), (11) The rms variation, $\sigma$, from the previous column is expressed as the percent deviation relative to the mean. (6) Mean Eddington ratio. (8) Mean of the log accretion rate in $M_{\sun}$ yr$^{-1}$. (10) Mean accretion efficiency. (12) Mean optical BC. (13) Mean UV BC. (14) Mean X-ray BC. NGC\,5548 ($\star$) statistics are based only on the SEDs with simultaneous optical, UV and  X-ray data. NGC\,5548 (all) statistics includes all available SEDs presented in this work. Sample mean represents the mean values for the sample objects and the 1$\sigma$ scatter around the sample mean.
	}
           \label{tab:table6}
           \begin{tabular}{lccccccccccrrr} 
           \hline
           Object & Mean  & Mean & Mean & $L_{BOL}$ & Mean & $\lambda_{Edd}$ &  Mean & $\dot{M}$ & Mean & $\eta$ &  Mean  & Mean & Mean \\
                 & $\lambda L$(1350\,\AA) & $\lambda L$(5100\,\AA) & $ \log [L_{BOL}$& rms &  $\lambda_{Edd}$ & rms & $\log [\dot{M}/$ & rms &  $\eta$ & rms & BC & BC & BC\\
                 &[$10^{43}$                       &[$10^{43}$                       &  /erg s$^{-1}$]            & varia- &                     & varia- & $M_{\sun}$]  & varia- & & varia- & (5100\,\AA) & (1350\,\AA) &  (2--10\,keV)\\
                 &erg s$^{-1}$]                          & erg s$^{-1}$]                      &              & tion &                     & tion & yr$^{-1}$]  & tion & & tion &   &   &   \\
                 &                   &                    &   & ( per cent) & & ( per cent) &  & ( per cent) & & ( per cent) & & \\                 
            (1) & (2)  & (3) & (4) & (5) & (6) & (7) & (8) & (9) & (10) & (11) & (12) & (13) & (14) \\
             \hline
Fairall\,9                                  &25.56$\pm$8.35&7.77$\pm$1.89 &45.04$\pm$0.08&20.1&0.045$\pm$0.009& 19 & $-$1.49$\pm$0.15 &41& 0.62$\pm$0.12&19 &14.59$\pm$1.31 & 4.52$\pm$0.43 &10.48$\pm$2.16\\ 
NGC\,4151                               &2.24$\pm$0.15  &0.92$\pm$0.01&43.94$\pm$0.02& 3.7&0.019$\pm$0.001 & 4& $-$2.16$\pm$0.01 &0 &0.22$\pm$0.01  &5& 9.53$\pm$0.37 & 3.91$\pm$0.21 &11.20$\pm$0.66\\ 
NGC\,5548 ($\star$)&3.17$\pm$1.18&1.46$\pm$0.48&44.20$\pm$0.16&32.4&0.026$\pm$0.008 & 32 & $-$1.95$\pm$0.22 &49 & 0.26$\pm$0.07  & 27&11.70$\pm$2.14 & 5.49$\pm$0.61 & 7.00$\pm$1.16\\ 
NGC\,5548 (all) &3.83$\pm$1.24&1.88$\pm$0.45 &44.29$\pm$0.11&22.1&0.031$\pm$0.007 & 22 & $-$1.78$\pm$0.19 &36 & 0.21$\pm$0.05 &24 & 10.84$\pm$1.37 & 5.49$\pm$0.84 & 6.99$\pm$1.12\\ 
3C\,390.3                                  &6.29$\pm$3.14&3.03$\pm$0.97 &44.81$\pm$0.17&38.1&0.013$\pm$0.005 & 39 & $-$2.58$\pm$0.28 &47 &  4.45$\pm$1.12  & 25&22.18$\pm$1.24 &11.03$\pm$1.41& 3.67$\pm$0.28\\ 
NGC\,7469                                &4.50$\pm$1.03 &1.73$\pm$0.19&44.30$\pm$0.10&21.3 &0.180$\pm$0.038&21 & $-$1.03$\pm$0.06 &14 &  0.04$\pm$0.01 &25&11.81$\pm$1.73 & 4.53$\pm$0.15&11.20$\pm$0.73\\ 
Sample Mean  &8.48$\pm$9.66&3.07$\pm$2.74 &44.48$\pm$0.44&0.99&0.042$\pm$0.034 & 82 & $-$1.69$\pm$0.33 &97 & 0.43$\pm$0.67& 156&14.29$\pm$5.68 & 6.24$\pm$3.26& 8.09$\pm$3.47\\ 
\hline
	\end{tabular}
\end{table}  
\end{landscape}

\subsubsection{Time variations of mass accretion rate \Mdot{}  and accretion efficiency, $\eta$} \label{S:mdotetavar}
\begin{figure*}
\begin{center}$
\begin{array}{ll}
\includegraphics[scale=0.45]{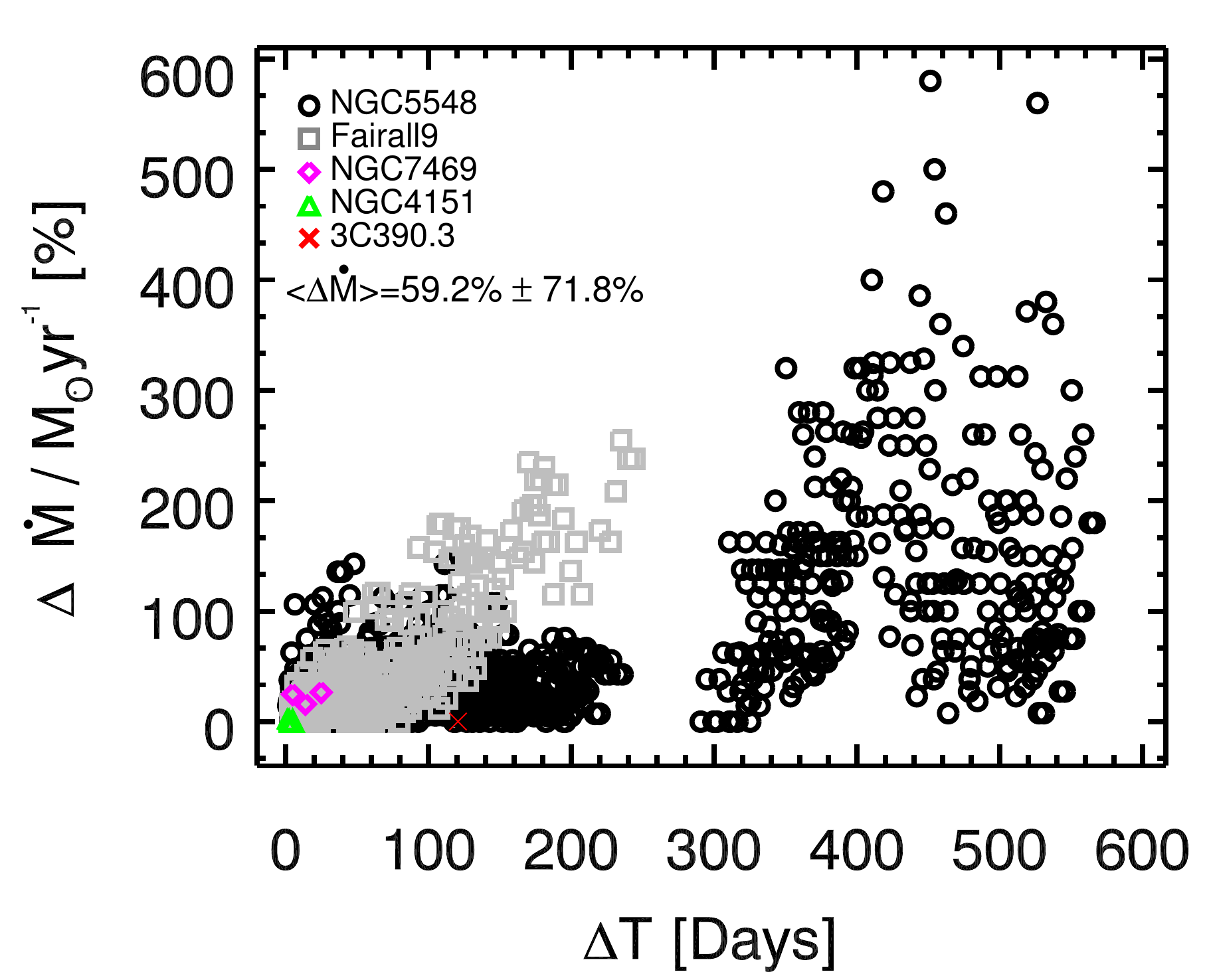}&
\includegraphics[scale=0.45]{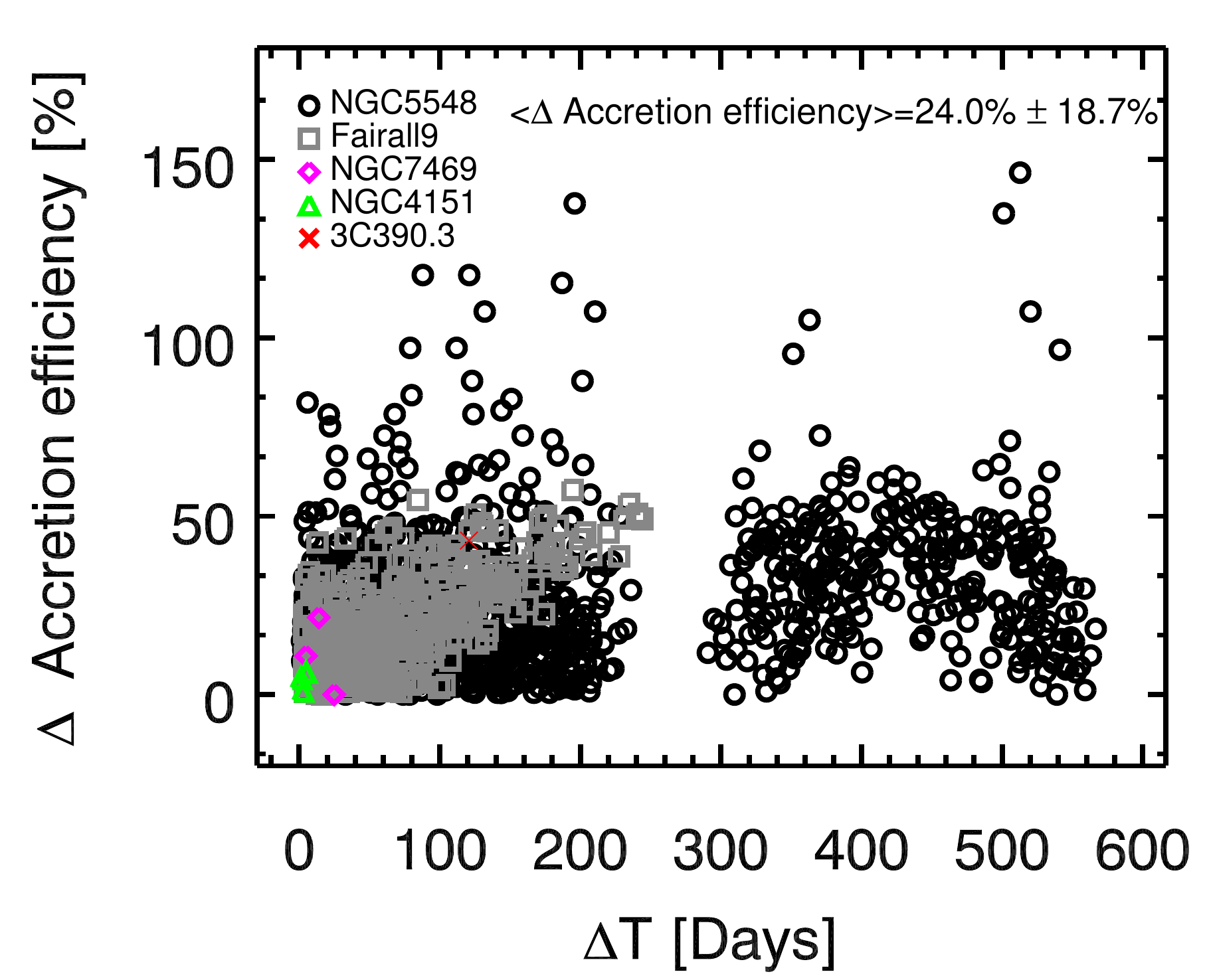}\\
\includegraphics[scale=0.45]{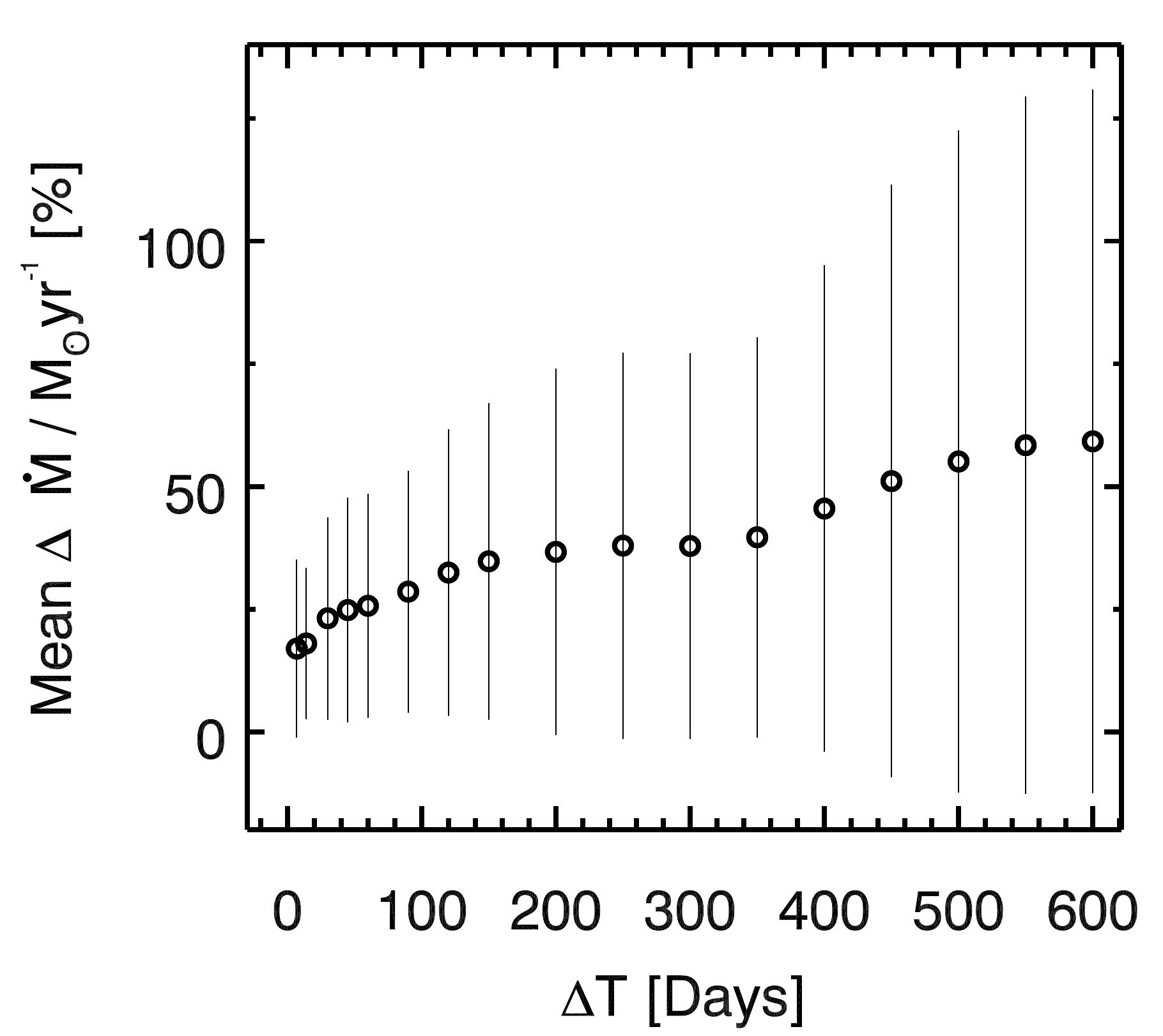}&
\includegraphics[scale=0.45]{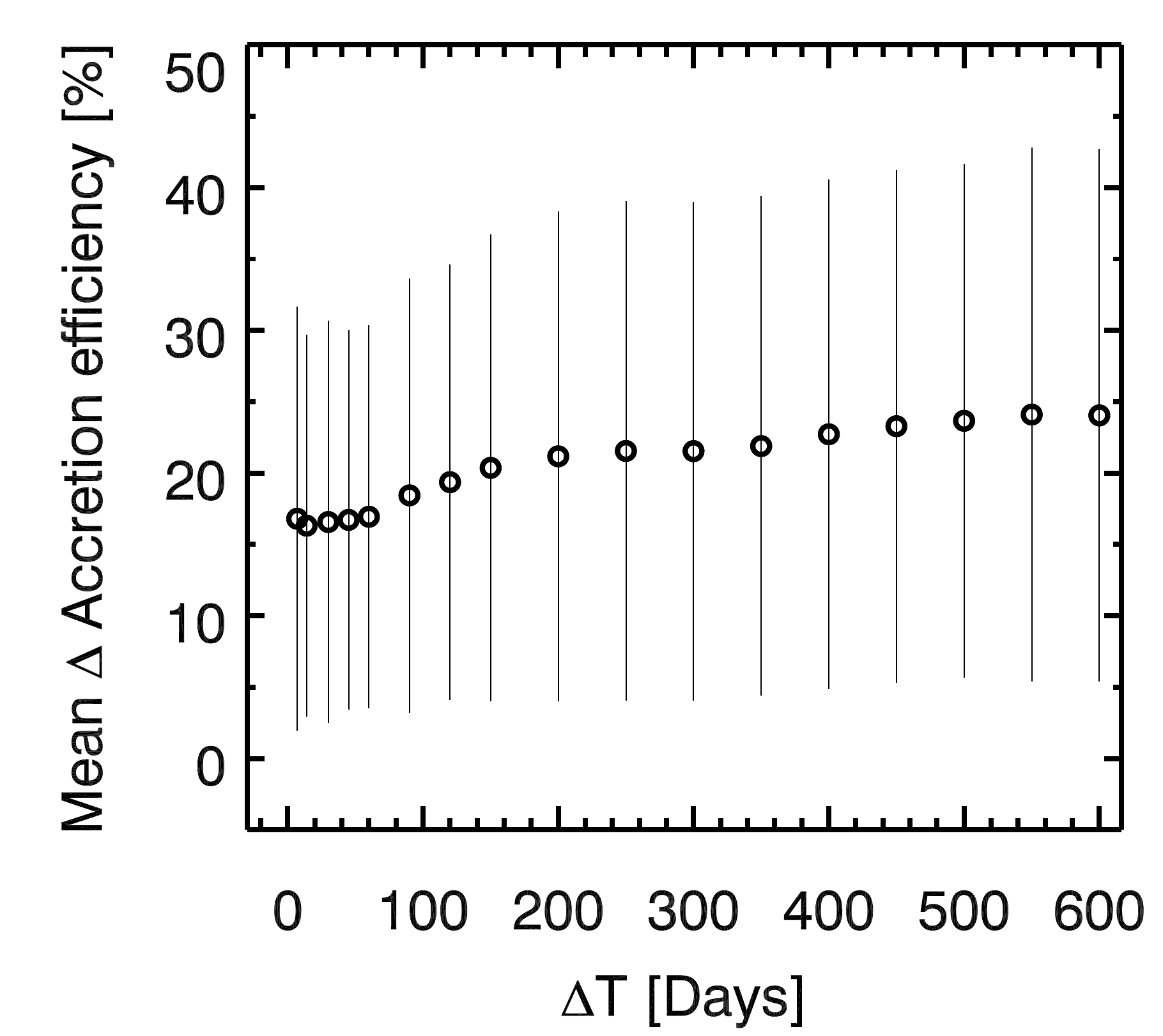}\\
\end{array}$
\end{center}
\caption{Temporal variations of mass accretion rate (left) and accretion efficiency (right). 
Top panel: mass accretion rate (left) and accretion efficiency (right) percent differences between all available SED epochs as function of $\Delta T$. 
Variations in accretion efficiencies do not increase with $\Delta T$. 
Bottom panel: the average $\Delta \dot{M}$ and its 1$\sigma$ deviation increases with $\Delta T$. 
The average $\Delta$ accretion efficiency  is almost constant over time, while the 1$\sigma$ spread (vertical bars) around this mean increases slightly. }
\label{fig:fig9}
\end{figure*}

 The mass accretion rate and accretion efficiency of individual AGN also change with the luminosity. 
 We list the mean and 1$\sigma$ variations of $ \dot{M}$ and $\eta$ in Table \ref{tab:table6}. 
The rms of the \Mdot{} variations 
is significant (between 36 per cent  and 49 per cent) for Fairall\,9, NGC\,5548 and 3C\,390.3, while it is low (10 per cent) for NGC\,7469 and negligible for NGC\,4151. 
Fig. \ref{fig:fig9} illustrates that  the temporal variations in $ \dot{M}$ depend on the time range probed.
 For NGC\,5548, the $\Delta \dot{M}$ range expands to extreme values (larger than 300 per cent) between  $\Delta T \sim$ 360$-$570 d. 
 The average $\Delta \dot{M}$ and its 1$\sigma$ deviation (Fig. \ref{fig:fig9} bottom panel) increases slightly with time. 
 
For most AGN in our sample the 1$\sigma$ variation in $\eta$ is moderate between 19 per cent  and 27 per cent, but negligible for NGC\,4151. 
The sample average percent differences in $\eta$ (Fig. \ref{fig:fig9} top right)  do not tend to increase with $\Delta T$; it is essentially constant in time, also considering the large distribution in values indicated by the vertical bars in 
Fig. \ref{fig:fig9}, bottom right.

\begin{figure}
\includegraphics[width=\columnwidth]{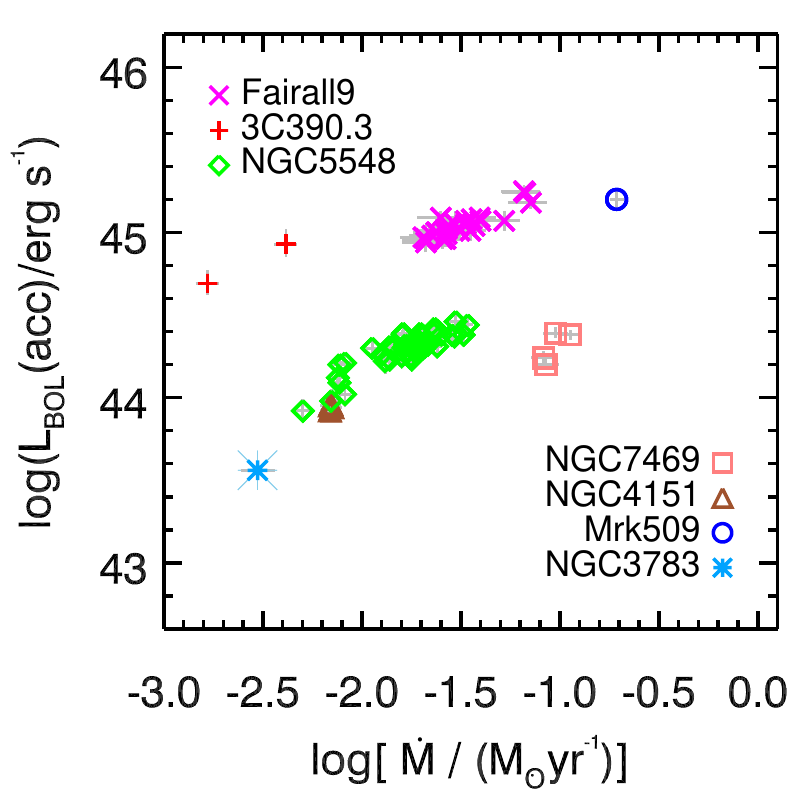}
\caption{Accretion luminosity versus accretion rate for the seven AGN in our sample; the $L_{BOL}(acc)$ and $\dot{M}$ measurements are simultaneous measurements for each SED epoch. 
The $L_{BOL}(acc)$ measurements are based on interpolation of quasi-simultaneous multiwavelength data. The $\dot{M}$ measurements are based on  
a simple accretion disc model, the optical luminosity at 4392\,\AA\ and an estimate of disc inclination and $M_{BH}$ \citep{Raimundo2012}. 
The accretion luminosity tends to increase with increasing $\dot{M}$, showing a correlation as expected. The difference in $L_{BOL}(acc)$ for a given \Mdot{} for the AGNs shown is due mainly to the efficiency of the mass to energy conversion. In addition, the fraction of the energy emitted at IR and radio energies is not accounted for by $L_{BOL}(acc)$, shown here; these fractions will vary from AGN to AGN, changing the $L_{BOL}(acc)$ measured here.
}
\label{fig:fig10}
\end{figure}

The temporal variations show that the average variability amplitude of $ \dot{M}$ is larger than that of $L_{BOL}(acc)$. 
These two parameters are highly correlated as we expect mass accretion rate to drive the $L_{BOL}(acc)$ variations (\S\,\ref{S:mdotandeta}). 
Fig. \ref{fig:fig10}, showing $L_{BOL}(acc)$ versus $ \dot{M}$ for our sample, also illustrates this. 
For NGC\,5548, for which we have the most epochs, we find\footnote{We use the Bayesian regression method of \citet{Kelly2007}, implemented in IDL as `LINMIX ERR.pro'.}  that the accretion luminosity can be expressed as $ \log[L_{BOL}(acc)] = 45.21+ 0.51 \times \log[\dot{M}]$ (with a scatter of 0.02).
Individual AGN exhibits a relationship that is offset from this due in part to the different value of $\eta$ and a slightly different distribution of luminosity drawn from \Mdot{} on the accretion luminosity, shown here, and the infrared and radio luminosities, which are not included in $L_{BOL}(acc)$. 

\subsubsection{BC variations in time} \label{S:bcvar}

The continuum variations that alter $L_{BOL}(acc)$, affect the BCs similarly. 
In order to evaluate the robustness of BCs with respect to SED variations, we quantify the BC variations and examine its behaviour over time. 
The rms variations in the mean BCs range between $\sim$4 per cent and $\sim$20 per cent. 
This shows that the BCs for individual objects are not constant but change with the accretion state of the AGN. 
In the case of NGC\,5548, for which we have the most reliable measurements based on 11 SED epochs with simultaneous optical/UV and X-ray data,  
the rms variations in the mean BCs is moderate (between 11 per cent and 18 per cent) over $\sim$540 d.  

\begin{figure*}
\begin{center}$
\begin{array}{ll}
\includegraphics[scale=0.75]{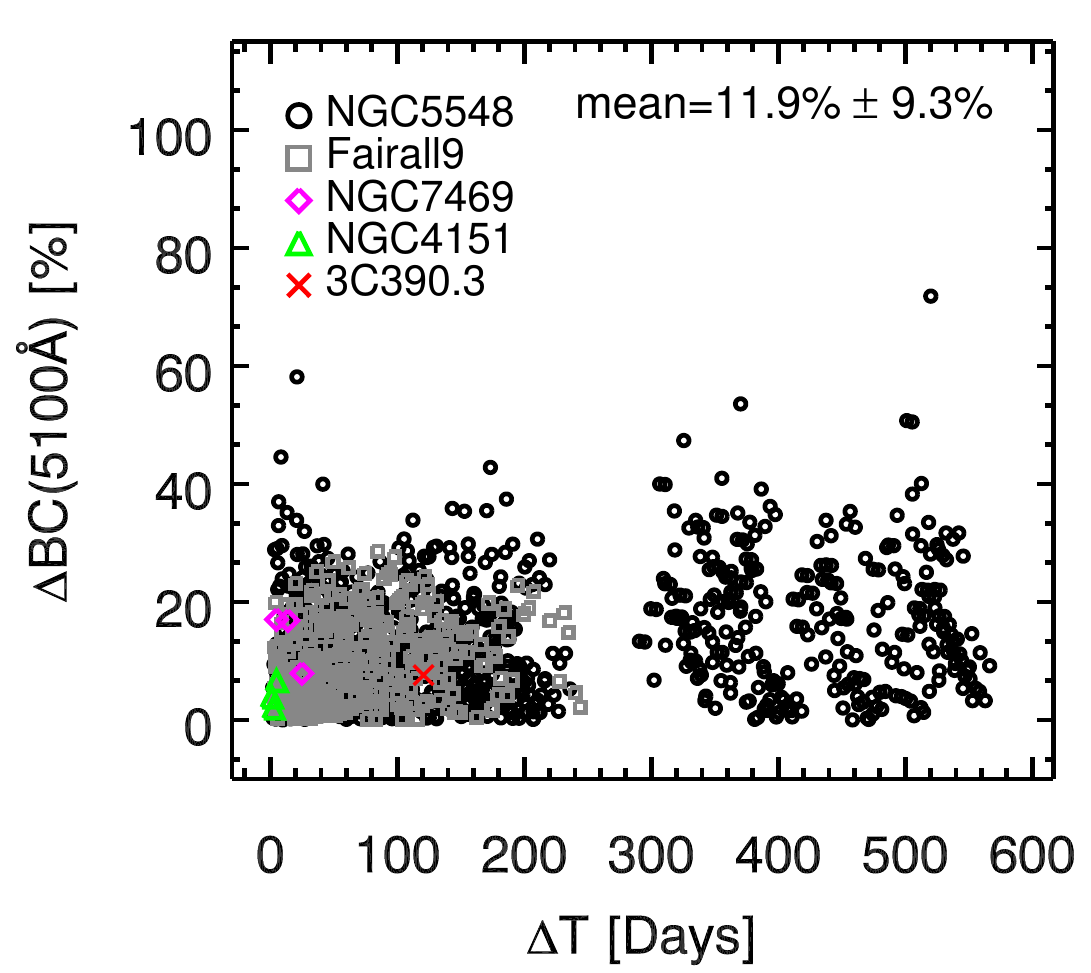}&
\includegraphics[scale=0.75]{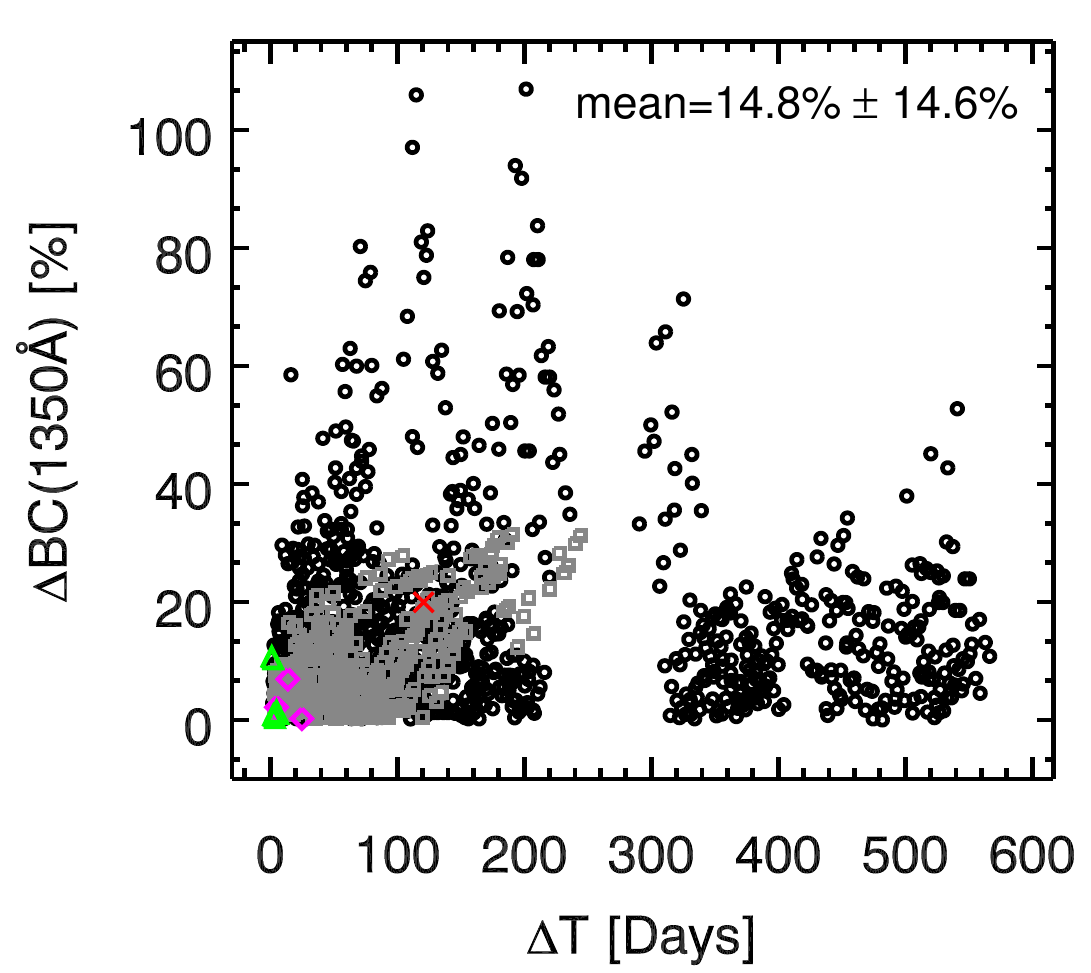}\\
\includegraphics[scale=0.75]{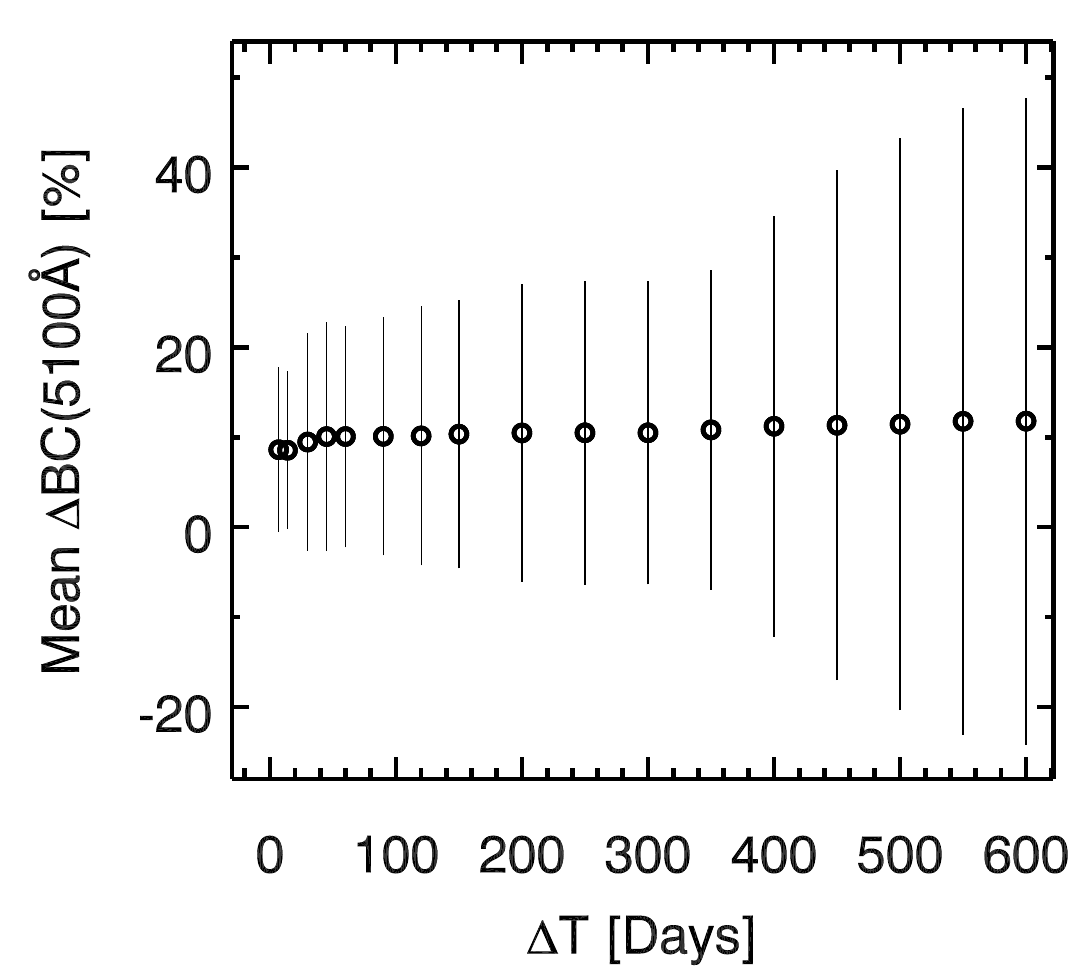}&
\includegraphics[scale=0.75]{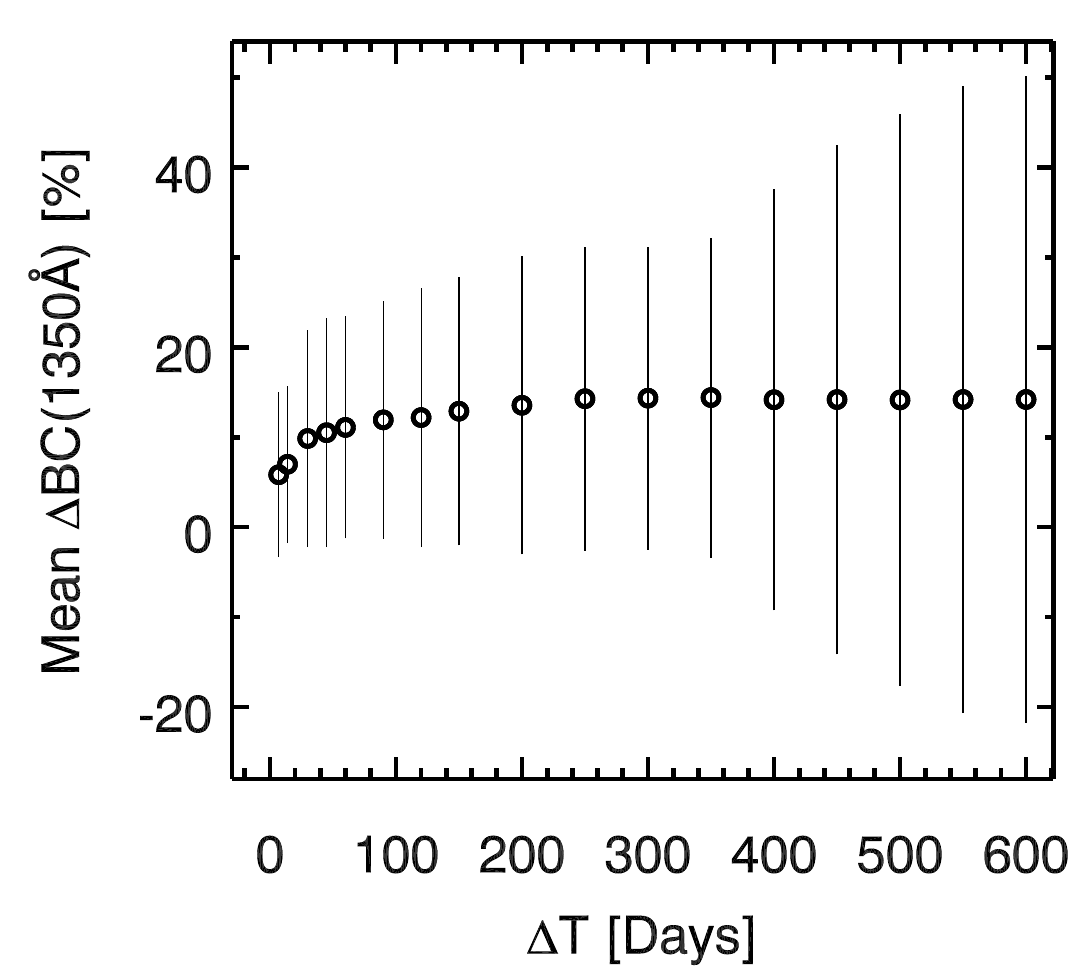}\\
\end{array}$
\end{center}
\caption{Temporal variations of BC(5100\,\AA) (left) and BC(1350\,\AA) (right).  
Top panel: BC(5100\,\AA) (left) and BC(1350\,\AA) (right) percent differences between all available SED epochs versus $\Delta T$. 
Variations in BCs do not tend to increase with increasing $\Delta T$. 
Bottom panel: average $\Delta BC(5100\,\AA)$ and $\Delta BC(1350\,\AA)$ tend to be constant over time, while the 1$\sigma$ spread (vertical bars) around this mean increases. }
\label{fig:fig11}
\end{figure*}

The temporal variations of BC(5100\,\AA) and BC(1350\,\AA) of our sample, based on all available SED epochs, are shown in Fig.~\ref{fig:fig11}. 
Here, $\Delta BC(5100\,\AA)$ and $\Delta BC(1350\,\AA)$ are the percent differences relative to the lower BC value between any two BCs. 
The top panels show that the mean percent differences of BC(5100\,\AA) and BC(1350\,\AA)  are 11.9 per cent and 14.8 per cent, respectively. 
While BC(1350\,\AA) vary more and reach higher differences (up to 100 per cent), BC(5100\,\AA) vary less. 
The bottom panels illustrate the mean and 1$\sigma$ distributions of $\Delta BC(5100\,\AA)$ and $\Delta BC(1350\,\AA)$ for a set of time intervals. 
They show that the mean variations in the optical and UV regime do not increase with $\Delta T$, as seen for $\Delta L_{BOL}$, but remain almost constant at $\sim$10 per cent--15 per cent, yet the 1$\sigma$ spread around the mean does increase especially for time spans larger than a year. 
We note that the amplitude of the BC variations are typically lower than those of $L_{BOL}(acc)$ (cf. Fig.s~\ref{fig:fig8} and~\ref{fig:fig11}).

We refrain from quantifying the temporal variations of BC(2--10\,keV) since most of the X-ray fluxes of Fairall\,9 and NGC\,5548 are estimated, carrying no variability information. 

\subsubsection{Long term SED variations based on literature SEDs} \label{S:longterm}

 To investigate the long term SED variations of our sample we use literature SEDs obtained from simultaneous {\it XMM$-$Newton}, {\it NuSTAR}, {\it Swift}, {\it Chandra}, {\it INTEGRAL}, {\it HST},  
{\it Far Ultraviolet Spectroscopic Explorer}  ({\it FUSE}) data. 
Here we present the comparison of our AGN Watch SEDs with the SEDs presented by \citet{Kaspi2001}, \citet{Scott2005}, \citet{Mehdipour2011} and \citet{Mehdipour2015}, for NGC\,3783, 
NGC\,7469, Mrk\,509 and NGC\,5548, respectively. 

\begin{figure*}
$
\begin{array}{ll}
\includegraphics[width=\columnwidth]{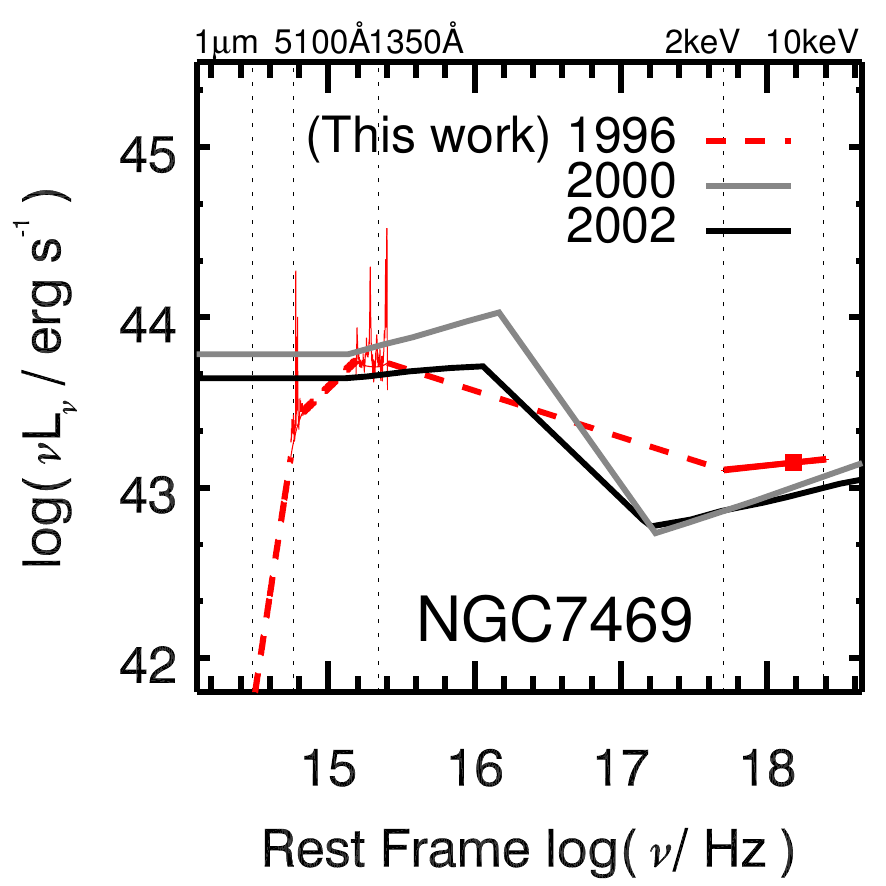}&
\includegraphics[width=\columnwidth]{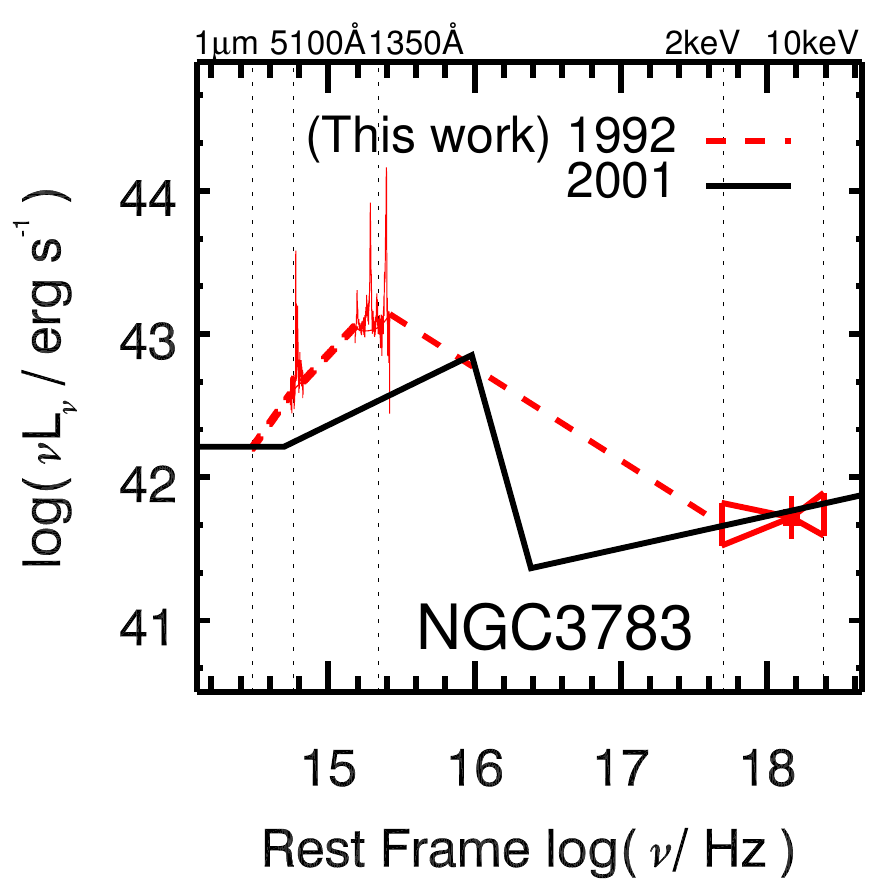}\\
\end{array}$
\caption{  Left-hand panel: comparison between the SEDs of NGC\,7469 over six years. The grey solid line is the SED of NGC\,7469 from 2000 \citep{Kriss2003}. The black solid line is the 
SED from 2002 \citep{Scott2005}. The red lines show our AGN Watch SED of NGC\,7469 from 1996 June 13. 
Right-hand panel: comparison between the SEDs of NGC\,3783 obtained in 1992 (our AGN watch SED, red lines) and 
the broken power law continuum model SED (black solid line) of \citet{Kaspi2001} based on year 2000 {\it Chandra}/HETGS X-ray observations \citep{Kaspi2001}. 
}
\label{fig:fig12}
\end{figure*}

\citet{Scott2005} use {\it Chandra}, {\it FUSE}, {\it HST} data and build a simultaneous SED of NGC\,7469 from 2002. 
They also present one SED from  \citet{Kriss2003} that is based on {\it XMM$-$Newton} and {\it FUSE} observations separated by one year. 
\citet{Scott2005} fit the hard X-ay continuum with a power law. They correct for Galactic reddening and fit the UV continuum with a power law. 
They use the UV and X-ray spectral continuum best-fitted power laws and assume $f_{\nu} \propto \nu^{-1}$ at long wavelengths. 
In Fig. \ref{fig:fig12} we compare our NGC\,7469 (1) SED from 1996 with the SEDs presented by \citet{Scott2005}  
All of these SEDs are based on best-fitted continua and therefore obtained by a similar approach. 
The SEDs given by \citet{Scott2005} display a 35 per cent $L_{BOL}(acc)$ decrease from 2000 to 2002.
In contrast, the $L_{BOL}(acc)$ based on our SEDs increased 72 per cent and 12 per cent  in  a time span of 4 and 6 years, respectively.

\begin{figure*}
$
\begin{array}{ll}
\includegraphics[width=\columnwidth]{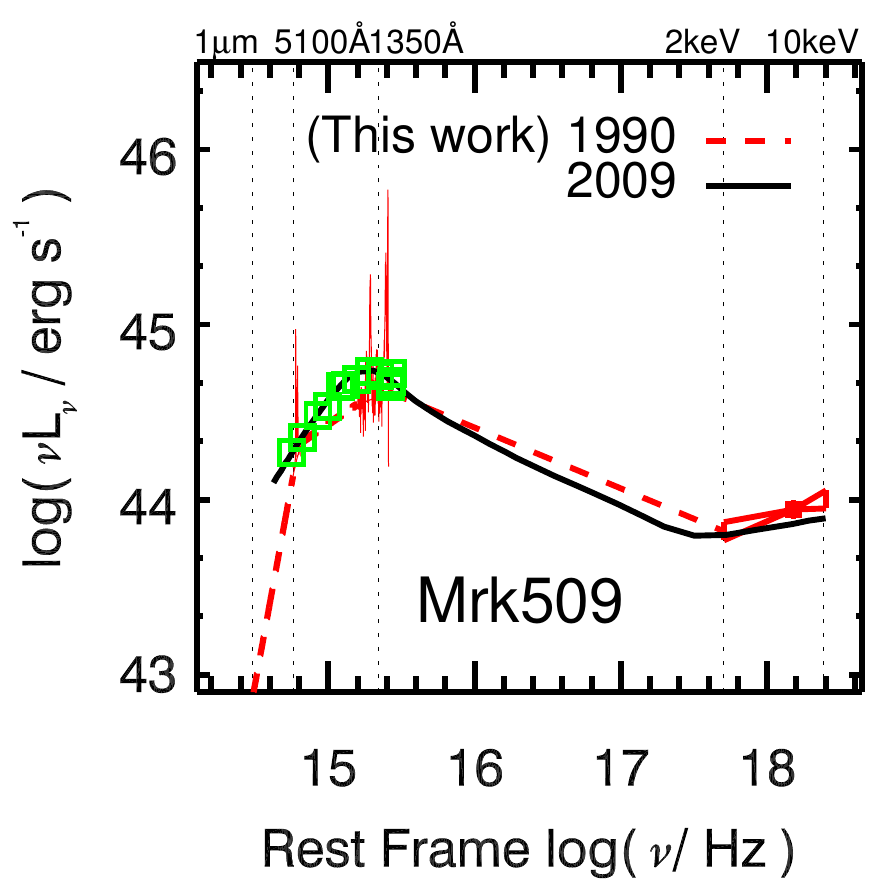}&
\includegraphics[width=\columnwidth]{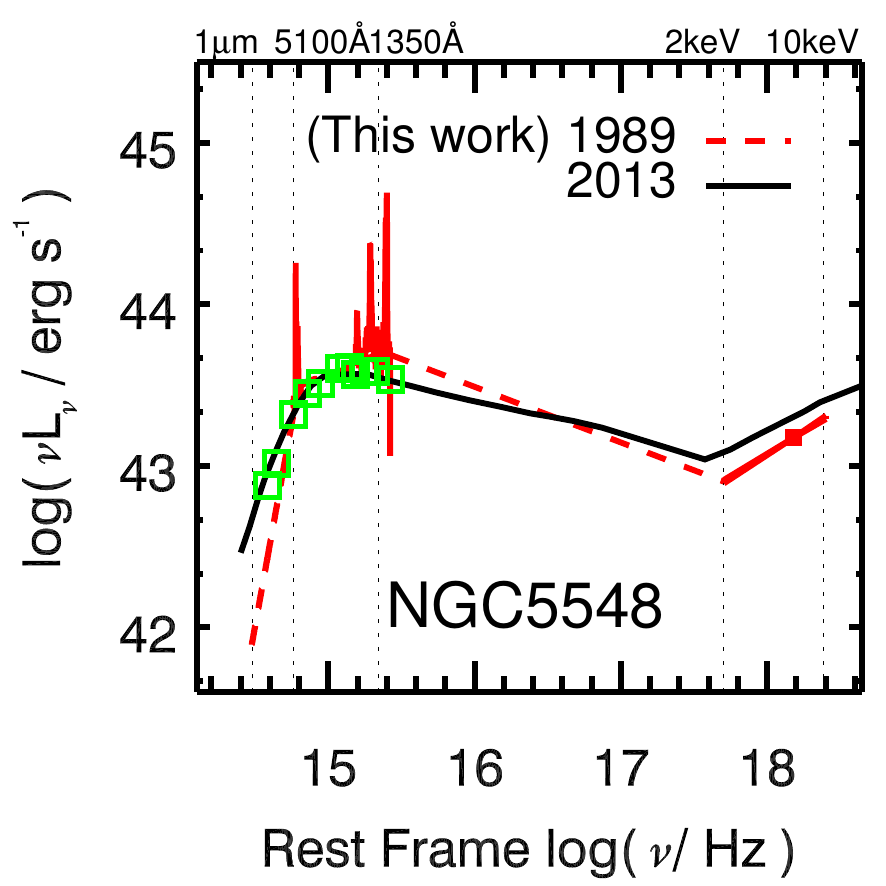}\\
\end{array}$
\caption{  Left-hand panel: two SEDs of Mrk\,509, observed 19 years apart. 
The black solid line is the best-fitting model for the 2009-10-19 data  \citep[from fig. 12 of][]{Mehdipour2011}; the green squares are their optical/UV measurements. 
The red lines represent our SED from 1990. The two SEDs display little variation from 1990 to 2009. 
Right-hand panel: comparison between the SEDs of NGC\,5548 obtained in this work (SED (1) is from 1989 January 07) and the one observed in 2013 by \citet{Mehdipour2015}. 
The black solid line is the best-fitting continuum model for the stacked data from the Summer of 2013 of NGC\,5548 \citep[from fig. 10 of][]{Mehdipour2015}. 
The green squares are contemporaneous {\it XMM$-$Newton}/OM, {\it Swift}/UVOT and {\it HST}/COS data \citep[][]{Mehdipour2015}. We do not see a dramatic SED change in 24 years. 
However, note that our SED does not have a soft X-ray excess component and it is only based on interpolated X-ray data. } 
\label{fig:fig13}
\end{figure*}

\citet{Mehdipour2011} present simultaneous SEDs of Mrk\,509 from {\it XMM$-$Newton}, {\it Swift}, {\it HST} Cosmic Origin Spectrograph (COS) taken in 2009 and {\it FUSE} observations  
from the year 2000. 
The optical/UV data in their SED include {\it V}, {\it B}, {\it U}, UVW1, UVM2 , UVW2  photometry from {\it XMM$-$Newton}'s OM and Swift's UVOT, plus {\it HST}/COS and FUSE spectroscopy. 
To establish the intrinsic continuum emission the authors correct for Galactic reddening, X-ray absorption, stellar flux from the host galaxy and nuclear broad and narrow line emission. 
They model the optical/UV continuum and the soft X-rays with the Comptonization model of \citet{Titarchuk1994} and use a power law for the hard X-rays.  
Their fig. 13 shows a 33 per cent flux variability over 100 d, assuming pure disc blackbody emission.  
In Fig. \ref{fig:fig13} (left-hand panel) we compare our SED of MRK\,509 from 1990 with the best-fitting broad-band model for {\it XMM$-$Newton} Obs. 2 (from 2009 October 19) of \citet{Mehdipour2011}. 
Clearly, the SED has changed very little between 1990 and 2009. The $L_{BOL}(acc)$ from 1990 is only 9 per cent lower than that based on the 2009 data.

 \citet{Mehdipour2015} determine the SED of NGC\,5548 from near-infrared to hard X-rays using stacked simultaneous 
{\it XMM$-$Newton}, {\it NuSTAR}, {\it Swift}, {\it INTEGRAL}, {\it HST}/COS observations obtained in the summer of 2013. 
Their SED is based on {\it I}, {\it R}, {\it V}, {\it B}, {\it U}, UVW1, UVM2 , UVW2 and {\it HST}/COS photometry, {\it Swift}/UVOT UV, {\it XMM$-$Newton}/OM optical grism spectroscopy and soft to 
hard X-rays (0.3$-$2\, keV, 2$-$10\,keV, 10$-$30\,keV, 30$-$80\,keV) spectroscopy. 
In order to uncover the intrinsic continuum they correct for the Galactic reddening, host galaxy stellar contamination, BLR and NLR emission lines, blended \feii\ and 
Balmer continuum, Galactic H\,{\sc i} absorption, X-ray emission lines, the warm X-ray absorption by the obscurer (when necessary) in their broad-band SED modelling. 
The authors also determine the SED of NGC\,5548 from archival {\it XMM$-$Newton} observations obtained in 2000 and 2001. 
They model the near-IR/optical/UV and soft X-ray excess part of the SED with the warm Comptonization model of \citet{Titarchuk1994} and use a power law model for hard X-ray continuum component. 
They report the integrated $L_{\rm0.001-10\,keV}$ luminosities for four different epochs. 
Their measurements correspond to a 14 per cent difference in $L_{\rm0.001-10\,keV}$ over 1 year (between 2000 December 24 and 2001 July 09 ), and 3 per cent difference over 13 years (between 2000 and 2013). 
In Fig. \ref{fig:fig13} (right panel) we compare our NGC\,5548 (1) SED from 1989 with the SED of \citet{Mehdipour2015} obtained in 2013.  
The two SEDs do not show a dramatic difference and the $L_{BOL}(acc)$ difference is 12 per cent. 
For NGC\,5548 we see larger (typically 30 per cent and can be up to 100 per cent) $L_{BOL}(acc)$ variations in the first few years of the AGN Watch monitoring. 
A possible explanation for the low SED and $L_{BOL}(acc)$ changes is that NGC\,5548  has been in a low luminosity state since 2005 \citep{Bentz2007,Bentz09b}. 

Both for NGC\,5548 and Mrk\,509, the data used for the two SEDs are not the same, since our SED does not have soft X-ray data. 
The $L_{BOL}(acc)$ changes depend on what is assumed for the EUV (for our sample the uncertainty in the $L_{BOL}(acc)$ is 23 per cent \S \ref{S:bolmodel}). 
Since we cannot produce the SED models of \citet{Mehdipour2011} and \citet{Mehdipour2015} without soft X-ray data, it is not possible to make a robust comparison. 
Therefore, the measured long-term $L_{BOL}(acc)$ variations quoted here are not typical. 

 \citet{Kaspi2001} analyse a {\it Chandra}/HETGS spectrum of NGC\,3783 from 2000 July to determine the origin of the X-ray and UV absorbers. 
They present one SED (their fig. 7) based on the {\it Chandra} data. Their SED is a broken power law model obtained from X-ray observations and model assumptions, 
but their SED does not include a flux calibrated optical spectrum. 
Given the lack of UV-optical data and the fact that this is a model-dominated SED,  a direct comparison has little value. 
However, it appears that the Xray spectra are consistent both in X-ray slope and flux level.  Hence, it is possible that the SED has not changed much.

 The long-term SED comparisons of Mrk\,509, NGC\,5548 and NGC\,3783 show that longer time span does not necessarily mean higher SED variations: 
the $L_{BOL}(acc)$ percent differences are lower compared to typical 30 per cent difference for short-term (\S~\ref{S:bolvar}). 
On the other hand, for NGC\,7469 the difference changes from 12 per cent to 72 per cent. 
\citet{Vasudevan2009a} also show that the bolometric luminosity of Mrk\,335 and NGC\,3227 vary by 99 per cent and 68 per cent over 7 and 6 years, respectively. 
But they report no variation for NGC\,4051 over 1 year. These results indicate that the long-term SED variations of the RM sample depends on individual source properties and are nontrivial to predict. 

 \subsection{Uncertainties in the accretion luminosities and the implications} \label{S:lbolunc} 
 
The uncertainties in the AGN $L_{BOL}(acc)$ measurements may originate from the uncertainty in the SED or from applying a mean BC to a monochromatic luminosity.  
There are two potential effects related to the SEDs: the effects of source variability when generating the SEDs and the selected method to bridge the gaps in the SED. 

\subsubsection{Interpolation of the SEDs versus applying mean BCs} \label{S:lbolunc3} 

\begin{figure*} 
\begin{center}$
\begin{array}{c}
\includegraphics[scale=0.6]{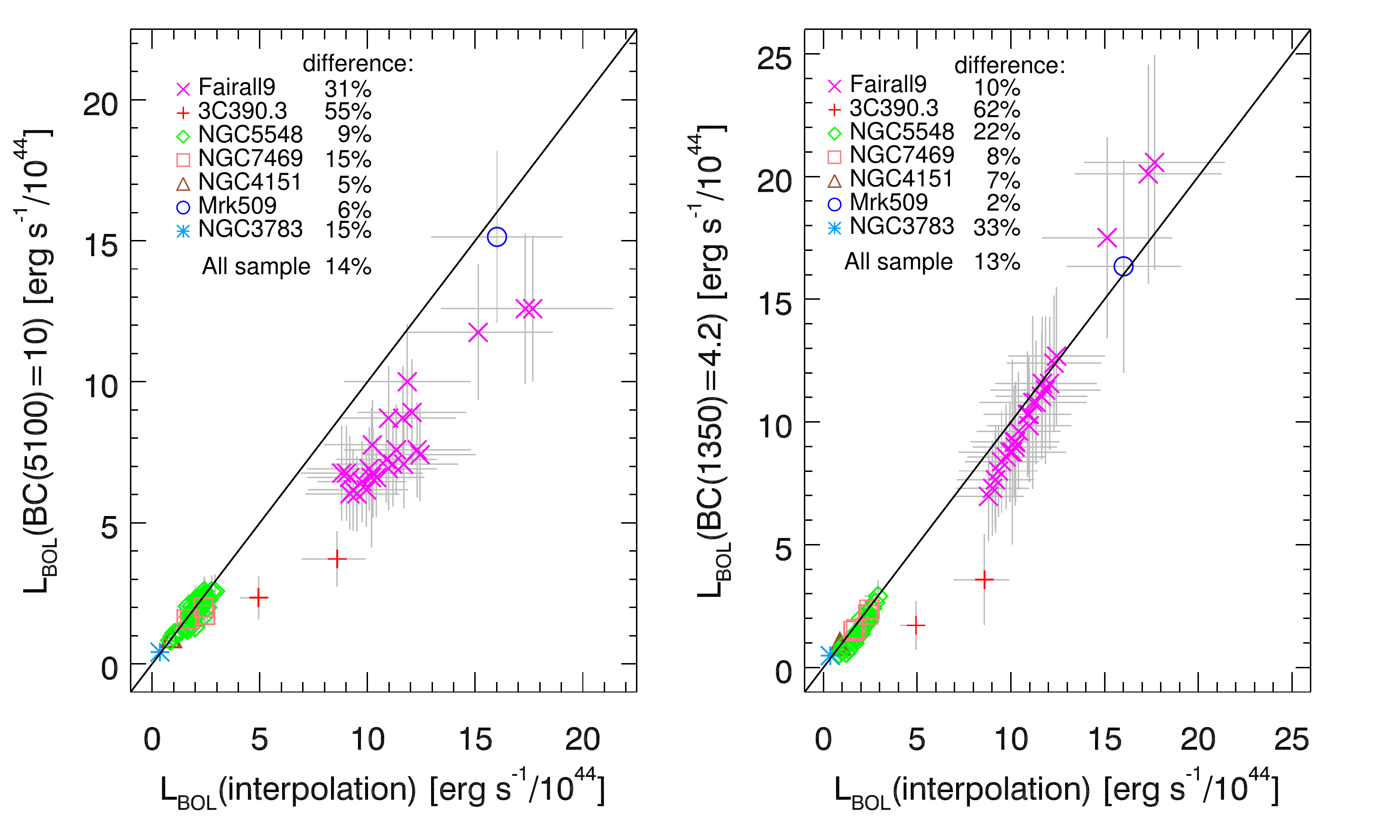}\\
\end{array}$
\end{center}
\caption{Comparison between $L_{BOL}(acc)$ measured by integrating the SEDs (x-axes) and estimated (y-axes) by applying 
a mean BC(5100\,\AA) of 10 (left-hand panel) and a mean BC(1350\,\AA) of 4.2 (right-hand panel). 
The solid lines show where the luminosities are equal. 
The luminosity difference measured for each source is quantified as the mean of the residuals between the two bolometric luminosities, and is listed in each panel.} 
\label{fig:lbolsedlbolbcdif}
\end{figure*}

As shown in Fig. \ref{fig:lbolsedlbolbcdif}, if we were to use a constant mean BC to calculate  $L_{BOL}(acc)$ instead of 
integrating over the SEDs, we would {\it on average} underestimate (or overestimate) $L_{BOL}(acc)$ by  $\sim$13$-$14 per cent, depending on the AGN. Yet, the difference can range between 2 per cent and 62 per cent.
The typical uncertainty of $L_{BOL}(acc)$, when measured by integrating the SED and including the continuum measurement uncertainties only, is $\sim$22 per cent. When using monochromatic luminosities and a mean BC the uncertainty can be significantly higher, or lower, depending on the AGN. This shows that, especially for nearby and/or low-luminosity AGN that tend to vary with larger luminosity amplitudes (than the more luminous quasars, say), the use of a BC to estimate $L_{BOL}(acc)$ is particularly problematic. There is no robust indicator of whether or not one happens to obtain a reliable measure of $L_{BOL}(acc)$ this way. 
For such moderate to highly variable sources, one will clearly need to observe the optical-UV-X-ray SED to get a reasonable handle on $L_{BOL}(acc)$ to better than a factor of 2.
 
\subsubsection{Effects of source variability in the SEDs} \label{S:lbolunc4}  
Temporal variations in $L_{BOL}(acc)$  (Section \ref{S:bolvar}) provide an estimate of the expected effect of the intrinsic source variability on the accretion luminosity.  
The mean of the accretion luminosity variations over 20 d is about 10 per cent (Fig. \ref{fig:fig8} bottom panel) and even for one year it does not exceed 20 per cent for the few objects studied here. 
This indicates that if we were to combine multiwavelength data obtained within 20 d or one year, this would result in a $\sim$10 per cent or $\sim$20 per cent uncertainty in the $L_{BOL}(acc)$ measurements, respectively. 
This is helpful, for selecting the time span over which the data that define the single-epoch SEDs are to be obtained. 
This also shows that using semi-simultaneous data can eliminate at least 10 per cent uncertainty in the $L_{BOL}(acc)$ measurements. 
We also note that the typical uncertainty in the $L_{BOL}(acc)$ measurements themselves are about 22 per cent for our sample objects (this is mainly due to the uncertainties of the continuum measurements), and thus using semi-simultaneous data as we did for a couple of cases does not reduce such measurement uncertainties.  One caveat to this conclusion is that the uncertainty may change for another sample of objects.

\subsubsection{Interpolation versus modelling of the SEDs} \label{S:lbolunc2} 
Our lack of knowledge of the intrinsic SED at the EUV range alone brings 
a $\sim$23 per cent (\S \ref{S:bolmodel}) uncertainty to our $L_{BOL}(acc)$ estimates. 
This uncertainty is slightly larger than the $L_{BOL}(acc)$ uncertainty originating from the optical, UV and X-ray continuum {\it measurement} uncertainties of $\sim$22 per cent. 
This unknown EUV emission will always limit $L_{BOL}(acc)$ measurements and it cannot be eliminated by using simultaneous multiwavelength data; the non-simultaneity of the data is a minor limitation in this particular context. 

 \subsubsection{Uncertainties in the mass accretion rates and accretion efficiencies} \label{S:lbolunc5} 
 The estimated $ \dot{M}$ and $\eta$ values highly depend on the assumed inclination angle, $L_{BOL}(acc)$ estimates and the adopted $M_{\rm BH}$. 
 For broad-lined AGN with unknown aspect angles to the disc normal,
the possible $i$-range between zero and 60$\degr${} alone 
brings a significant  uncertainty  of a factor of 2.8 to both to $ \dot{M}$ and $\eta$. 
While \Mdot{} can indeed vary by a factor of $\sim$3 (\S~\ref{S:mdotetavar}), typically, the use of simultaneous SEDs alone does not reduce the uncertainty in the $ \dot{M}$ and $\eta$ estimates significantly.
More accurate accretion disc and broad-line region structure details are needed.

 \subsubsection{Uncertainties in the Eddington luminosity ratios} \label{S:lbolunc1} 
 Applying a mean BC is the most practical way to estimate the bolometric luminosity and the Eddington luminosity ratio 
 since this does not involve generating the SED for the AGN in question. 
 In case of RM AGN, if we ignore the statistical accuracy of the RM BH masses themselves \citep[of a factor of 2.9;][]{onken2004}, the largest contribution to the $\lambda_{Edd}$ uncertainty is from the applied mean BC(5100\,\AA). 
In the literature, the mean BC(5100\,\AA) values obtained from the mean SEDs of large AGN samples are between $\sim$8.0 and 13.0 \citep{Elvis94,Richards2006,Runnoe2012}. 
The standard deviation of these mean BCs is between 0.4 and 5.0, so the bolometric luminosity values estimated from these are uncertain by up to $\sim$50 per cent. 
This can in principle be reduced to $\sim$22 per cent (\S \ref{S:lbolunc2}) by using simultaneous multiwavelength SEDs; yet that will only provide a snapshot BC value for a highly variable AGN. 
Even applying the mean BC values reported by the key studies of \citet{Elvis94,Richards2006} and \citet{Runnoe2012} to monochromatic luminosity measurements of {\it individual} AGNs can result in an error by a factor larger than that indicated by the statistics of these studies owing to source variability.

\section{Conclusions}\label{S:conc}
In our aim to study the SEDs and their time variations for the entire local sample of RM AGNs of \citet{Peterson04}, 
we collected quasi-simultaneous optical, UV and X-ray data to within two d from the AGN Watch data base,  
the Mikulski Archive for Space Telescopes, the Tartarus archives and from the literature. 
We present multi-epoch quasi-simultaneous optical-to-X-ray SEDs of seven AGN from the RM sample, namely NGC\,4151, NGC\,7469, 3C390.3, Fairall 9 and NGC\,5548 for which the available data base span a range of 7 d, 25 d, 113 d, 248 d and 566 d, respectively; for NGC\,5548, the range of strict simultaneous SEDs within 2 d is 542 d.
We use these SEDs to measure the accretion luminosity directly (as oppose to scaling a monochromatic luminosity)  
and infer the Eddington ratio, mass accretion rate and accretion efficiency by adopting the robustly measured RM-based black hole masses \citep{Peterson04,BentzKatz2015}.
The main conclusions of this work are given below.
\begin{enumerate}

\item
Our analysis shows that the amplitude of SED variations of individual sources depends on the time span. 
On the longest time span of 542 -- 566 d (NGC\,5548) that we probe in this work, we see dramatic SED changes. 

\item
We use quasi-simultaneous optical, UV and X-ray data and linearly interpolate in log$-$log space to generate the SEDs from 1$\micron$ to 10\,keV to measure the accretion luminosity, $L_{BOL}(acc)$.
The applied interpolation provides a model-independent first order approach for measuring the accretion luminosity. 
Based on the range of possible and reasonable methods by which we can estimate the SED shape in the EUV range we calculate the uncertainty in our integrated luminosity $L_{BOL}(acc)$ to be of order $\sim$23 per cent, based on commonly used accretion disc models of the optical-UV-X-ray emission. Since this is a significant fraction, the highest uncertainty in the absolute value of the measured $L_{BOL}(acc)$ is our inability to extract the intrinsic SED in the EUV regime and all studies suffer this effect.

\item
Our analysis of the multi-epoch temporal SED variations of individual objects shows that typical 
$L_{BOL}(acc)$ variations over 20 d are about 10 per cent and about 20 per cent over a year. 
The mean $\Delta L_{BOL}(acc)$ slightly increases 
with longer time differences; it is about 30  per cent for 600 d.  
However, for individual objects, like NGC\,5548, $L_{BOL}(acc)$ can vary in excess of 100 per cent over 300 d or more.
Therefore, combining data obtained at different times may add at least an additional 10 per cent uncertainty to $L_{BOL}(acc)$ measurements. 
Thus, to limit the typical  (relative) uncertainty to 10 per cent, obtaining optical, UV and X-ray measurements to within 20 d is needed.

\item
The mass accretion rate and accretion efficiency of individual AGN also vary in time. 
NGC\,5548 shows significant changes (larger than 300 per cent) over a year. 
The accretion efficiency is less variable in comparison. 
In fact, the mass accretion rate has the highest variability amplitude amongst the parameters investigated in this work.
With the caveat that our analysis is limited to a small sample of AGN and only span 542 d, it does indicate that the unknown emission in the EUV gap plus the unknown structure and inclination of the central engine, not source variability,  is the dominant limitation for accurate measurements of the mass accretion rate and the accretion efficiency for all AGN.

\item
We find that the BC factors at a given wavelength (1350\,\AA{} or 5100\,\AA) for individual AGN are not constant in time but change with the AGN  continuum level. 
The sample standard deviations of the mean BCs are between $\sim$4 per cent\ and $\sim$20 per cent.
The two BC factors examined changed on average by 10 per cent--15 per cent but changes of order 100 per cent are also recorded for our AGN sample.
While the sample average temporal variations of BC(5100\,\AA) and BC(1350\,\AA{}) of our sample are similar,  BC(1350\,\AA{}) tends to vary more compared to BC(5100\,\AA) for individual AGN.

\item
We demonstrate that $L_{BOL}(acc)$ estimated based on a bolometric scaling from L(1350\,\AA) and L(5100\,\AA)  
give similar results as the $L_{BOL}(acc)$ measurement based on integration over quasi-simultaneous optical, UV and X-ray data for our handful of nearby Seyfert 1 galaxies. 
We find that using a constant mean BC instead of integrating over the SEDs, gives on average a $\sim$14 per cent difference in $L_{BOL}(acc)$ for these AGN. The caveat is that only for one AGN (NGC\,5548) do we have SEDs spanning more than $\sim$200 d and the amplitude variations depend on the time span.

\item
The use of simultaneous multiwavelength SEDs can reduce the uncertainty in the Eddington luminosity ratio, $\lambda_{Edd}$, from $\sim$50 per cent 
(based on the use of mean BC factors) to $\sim$22 per cent (the typical measurement uncertainty).

\item
Our study emphasises the usefulness of simultaneous multiwavelength (at least optical, UV and X-rays) data in measuring the accretion luminosity from the SEDs.
A larger sample of AGN with multiple multiwavelength observations would be required for exploring further the accretion emission variations and their effects on the observed SED since our data base does not provide sufficient statistics to study long-term variations over $\sim$200 d and is only based on five AGNs. 

\end{enumerate}

\subsection*{Acknowledgments}  
The authors acknowledge fruitful discussions with Sandra Raimundo.
MV gratefully acknowledges support from a
FREJA Fellowship granted by the Dean of the Faculty of Natural Sciences at the
University of Copenhagen, a Marie Curie International Incoming Fellowship 
(REA FP7--PEOPLE--2009--IIF; Proposal No 255190; Acronym: `BHmass') 
and support from the Danish Council for Independent
Research via grant no. DFF 4002--00275. EKE gratefully acknowledges support from and hospitality of the Dark Cosmology Centre during her stay at the University of Copenhagen where the majority of this work was performed. 
During the preparation of this work the Dark Cosmology Centre was funded by the Danish National Research Foundation. 
This research has made use of the NASA/IPAC Extragalactic data base (NED), which 
is operated by the Jet Propulsion Laboratory, California Institute of Technology, 
under contract with the National Aeronautics and Space Administration. 
Some of the data presented in this paper were obtained from the Mikulski Archive for Space Telescopes (MAST). STScI is operated by the Association of Universities for Research in Astronomy, Inc., under NASA contract NAS5-26555. Support for MAST for non-HST data is provided by the NASA Office of Space Science via grant NNX09AF08G and by other grants and contracts.



\bibliographystyle{mnras}
\bibliography{sedpaper}




\appendix

\section{SEDs OF OUR SAMPLE} \label{S:Appendixseds}

\begin{figure*}
\begin{center}$
\begin{array}{cc}
\includegraphics[width=0.5\linewidth,scale=1.5]{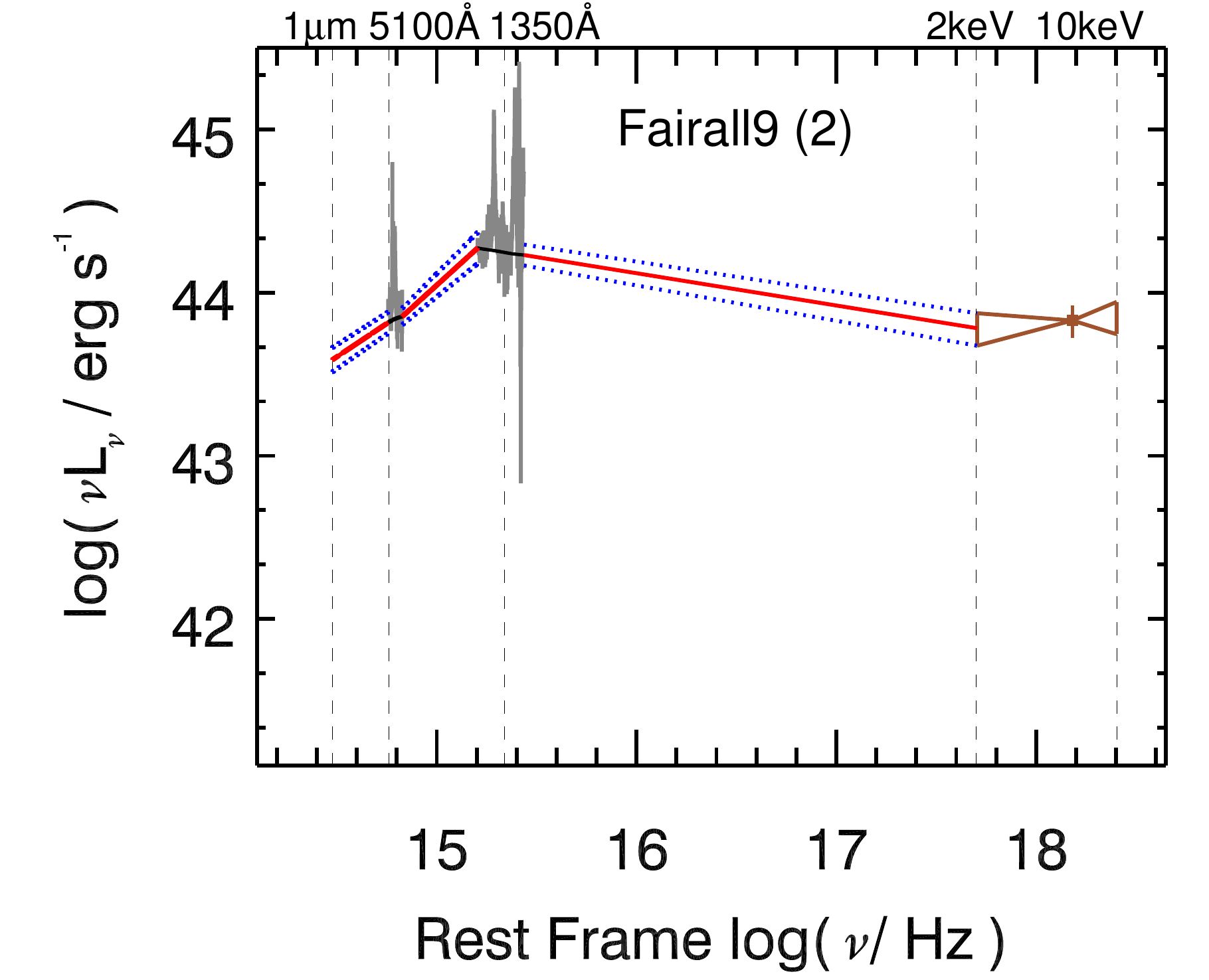} &
\includegraphics[width=0.5\linewidth,scale=1.5]{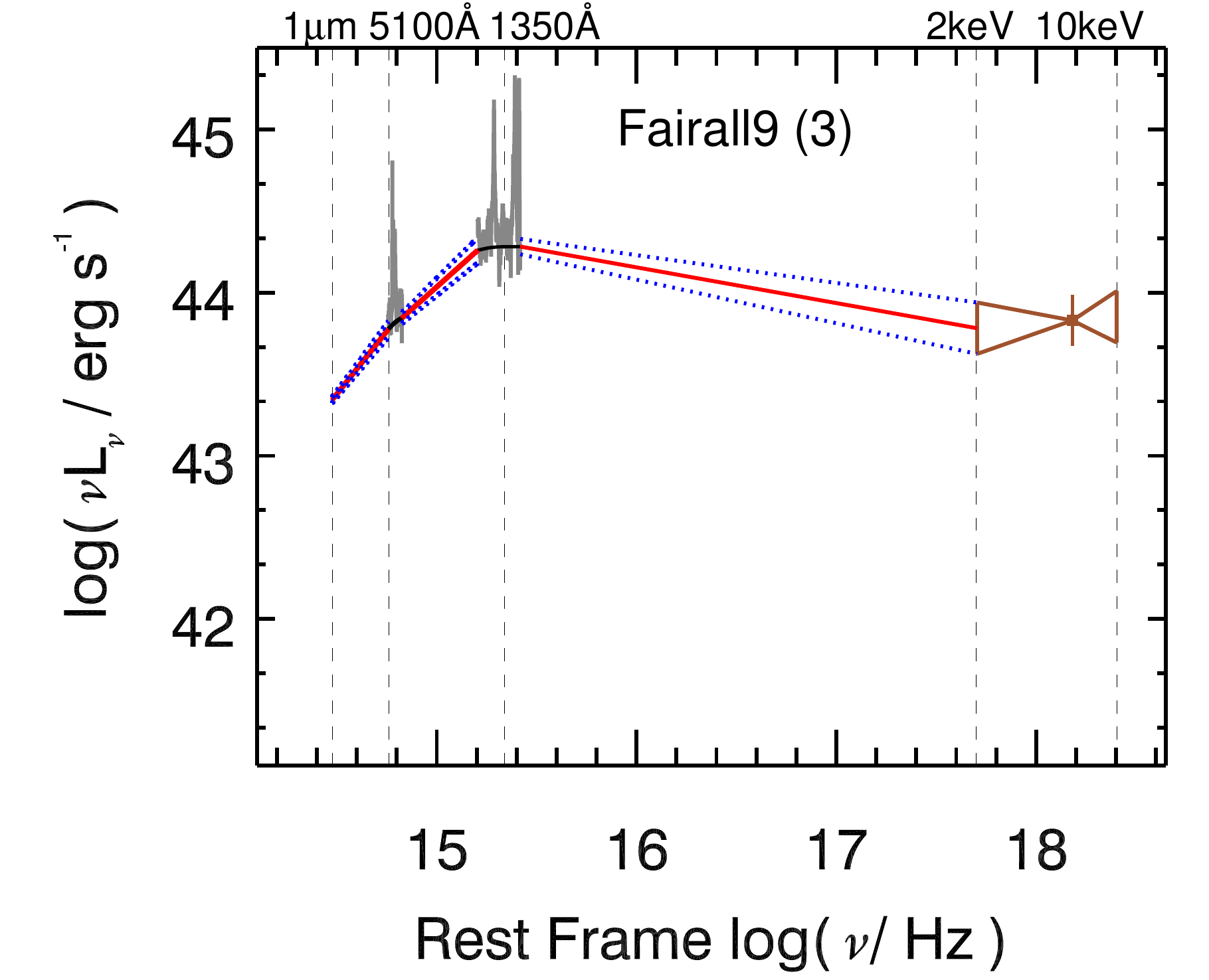}\\
\includegraphics[width=0.5\linewidth,scale=1.5]{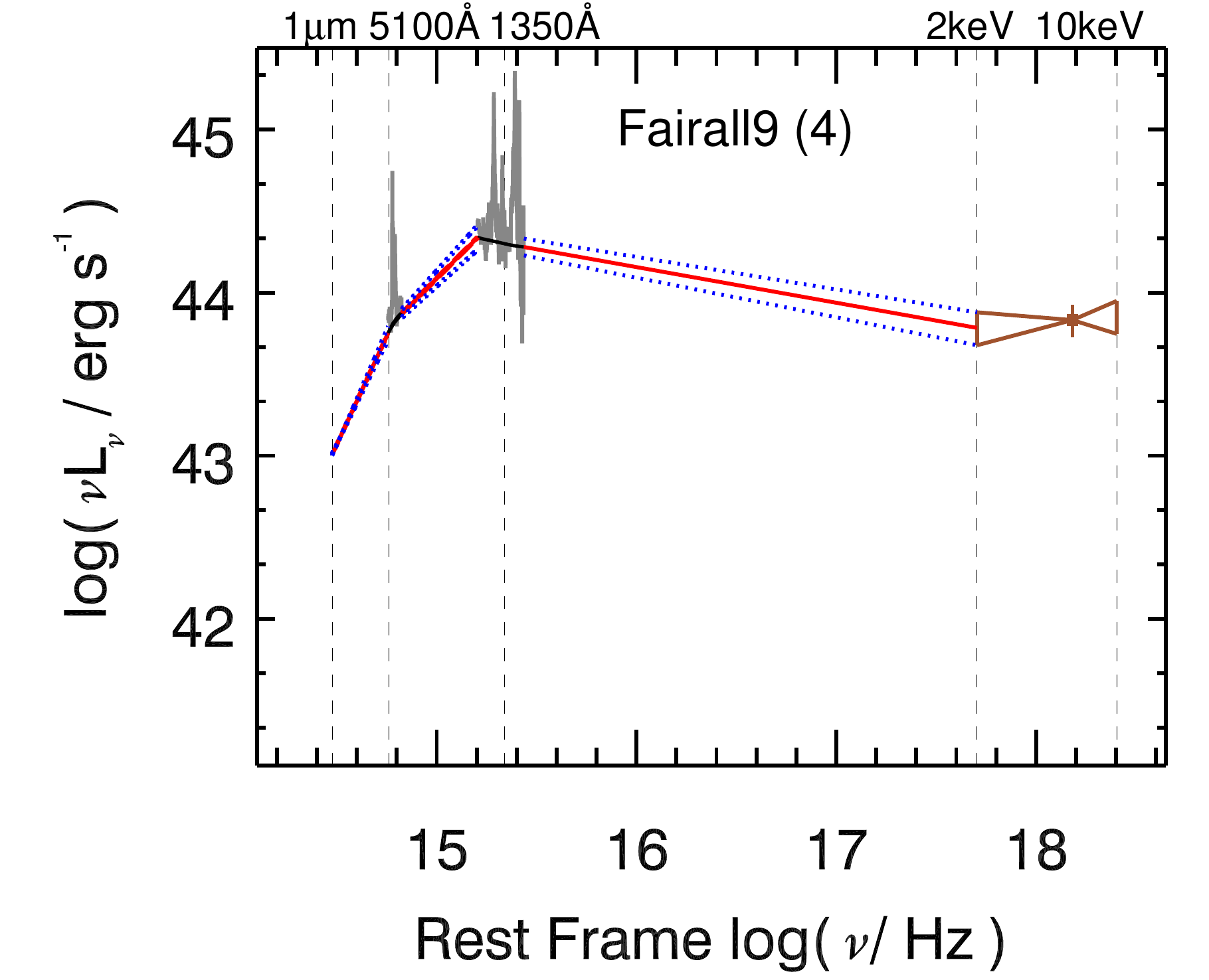} &
\includegraphics[width=0.5\linewidth,scale=1.5]{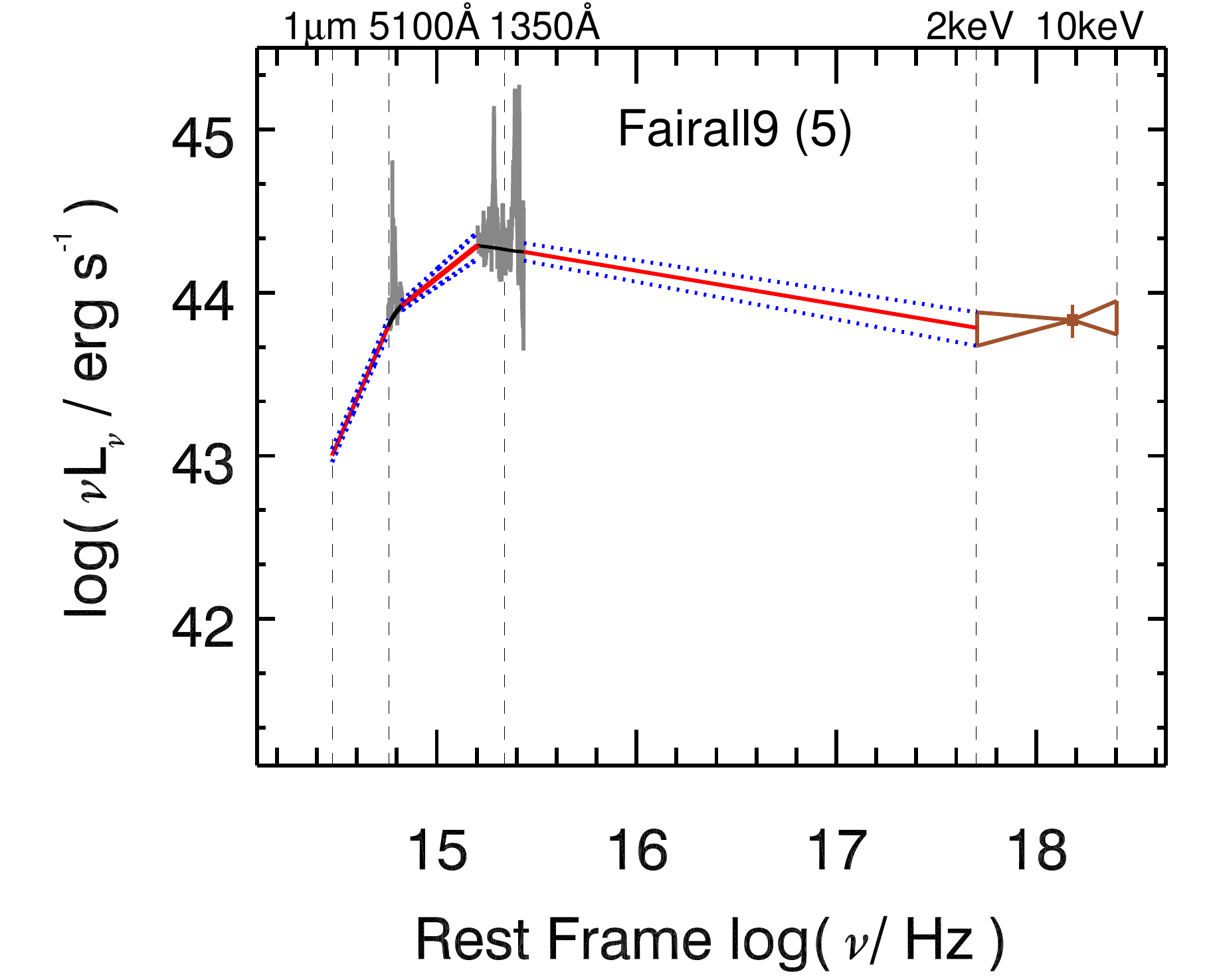}\\
\includegraphics[width=0.5\linewidth,scale=1.5]{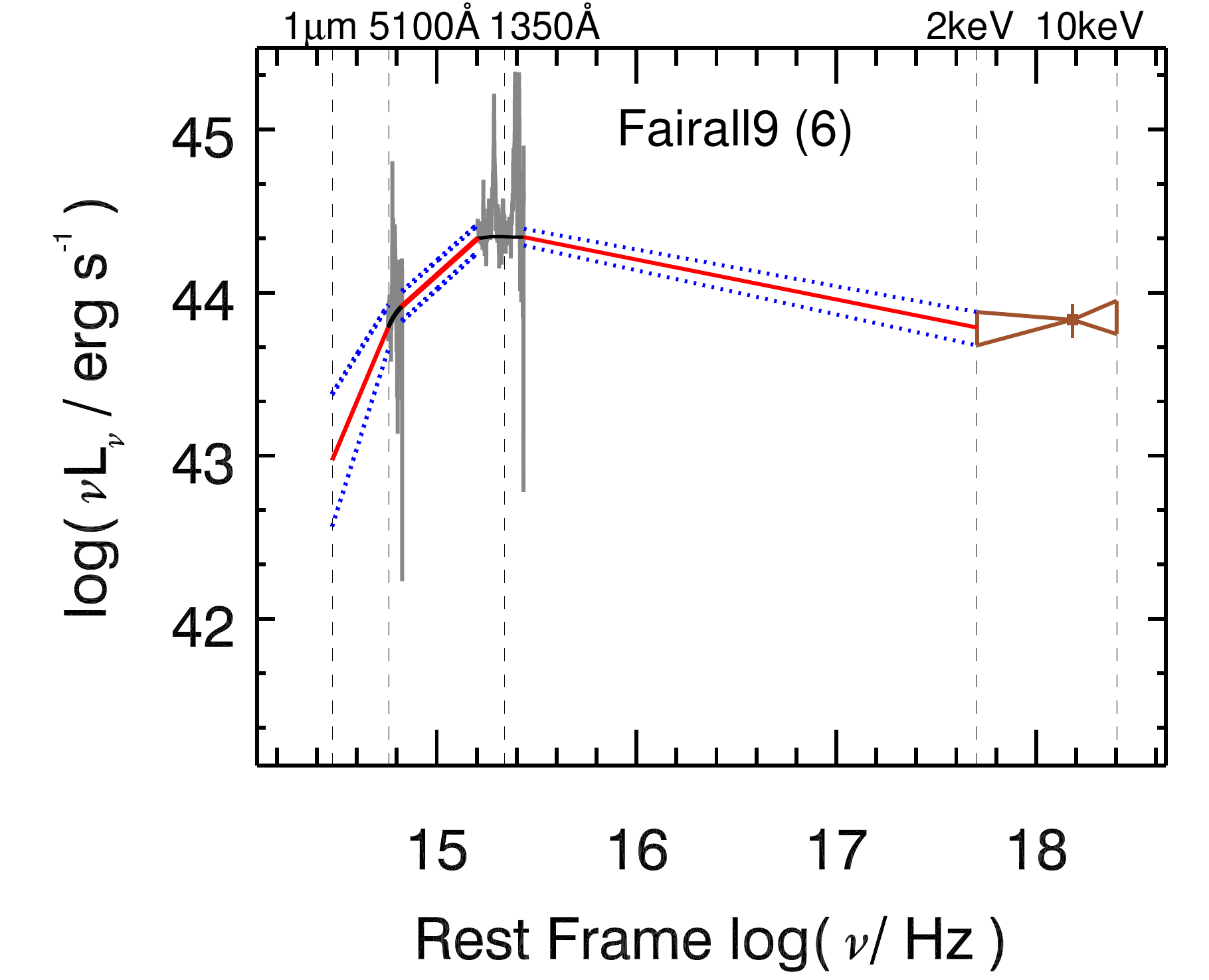} &
\includegraphics[width=0.5\linewidth,scale=1.5]{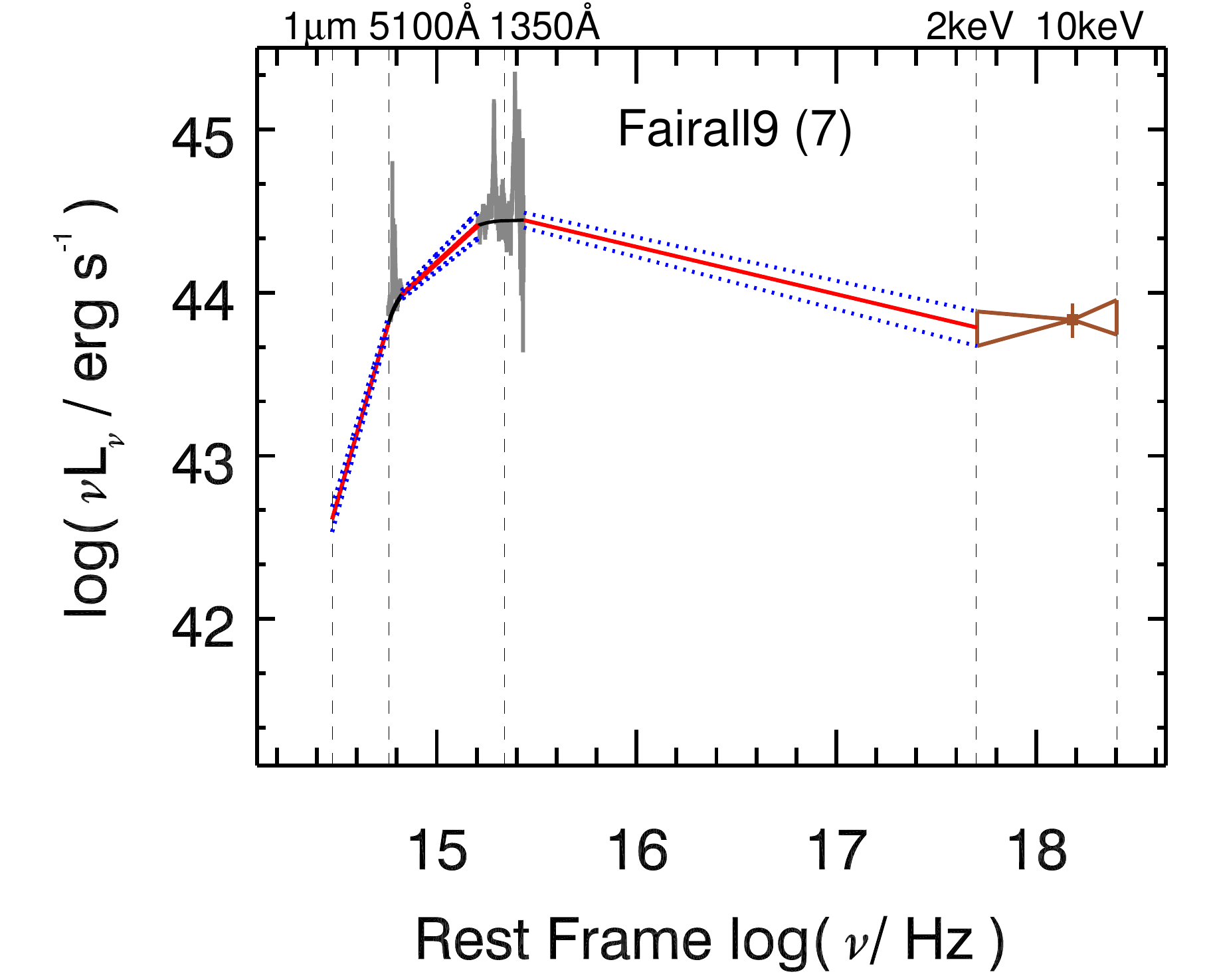}\\
\end{array}$
\end{center}
\caption{SEDs of our sample AGN for a given epoch (listed in parenthesis after the AGN name). The Julian Dates of these epochs are listed in Table \ref{tab:table4}. The best-fitting continuum to the observed spectrum is shown as the solid black lines. 
The linearly extrapolated or interpolated regions are shown as the solid red lines and the butterfly shape shows the X-ray continuum. For Fairall\,9 (27), only the X-ray and UV data are simultaneous, the optical data are obtained 12 d later.  
The X-ray data of NGC\,3783 are extrapolated from the 0.1--2\,keV observations. For 3C390.3 (2) the optical and UV data are simultaneous, but the X-ray data are obtained 20 d earlier. }
\label{fig:allsedsap}
\end{figure*}

\begin{figure*}
\begin{center}$
\begin{array}{cc}
\includegraphics[width=0.5\linewidth,scale=1.5]{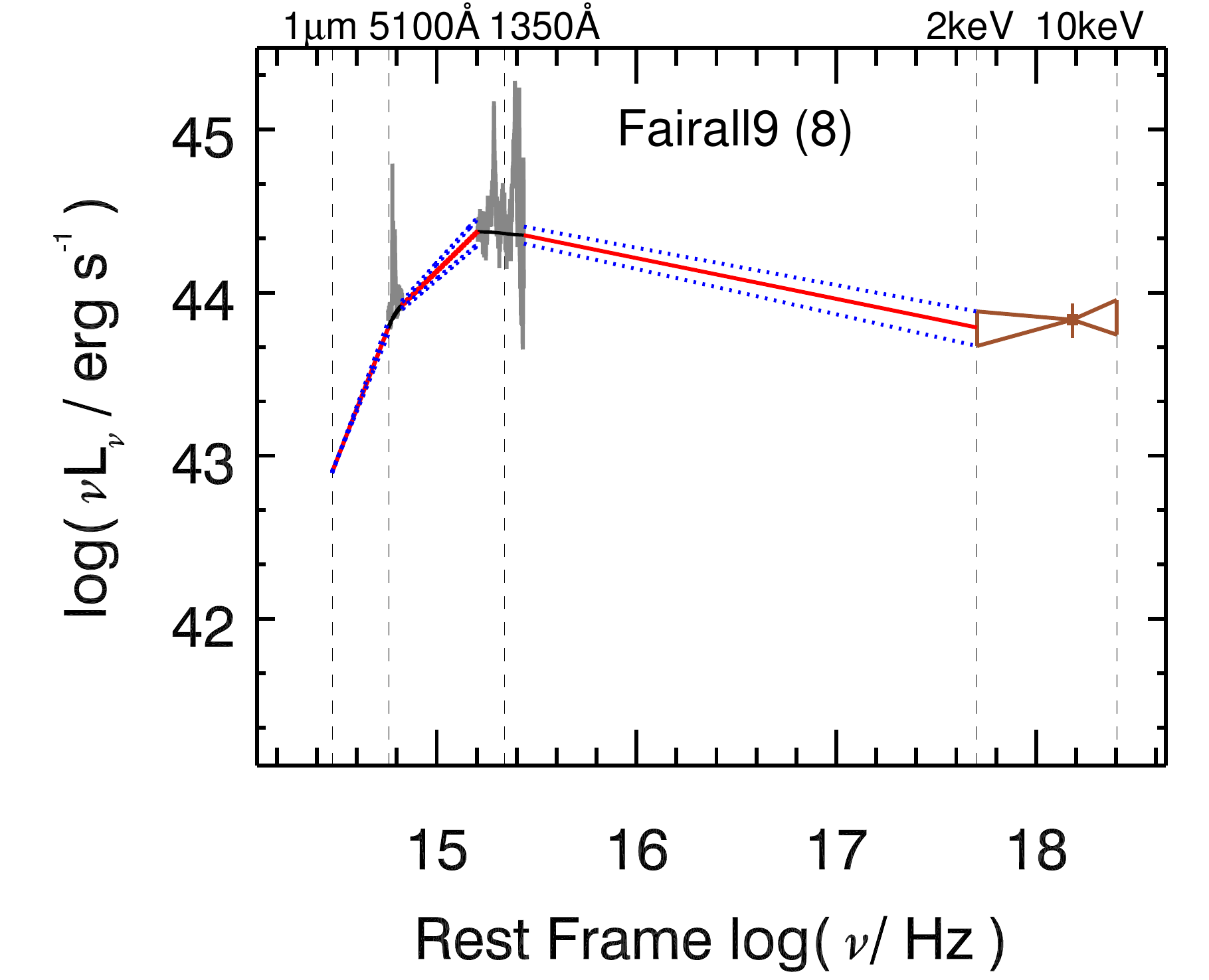} &
\includegraphics[width=0.5\linewidth,scale=1.5]{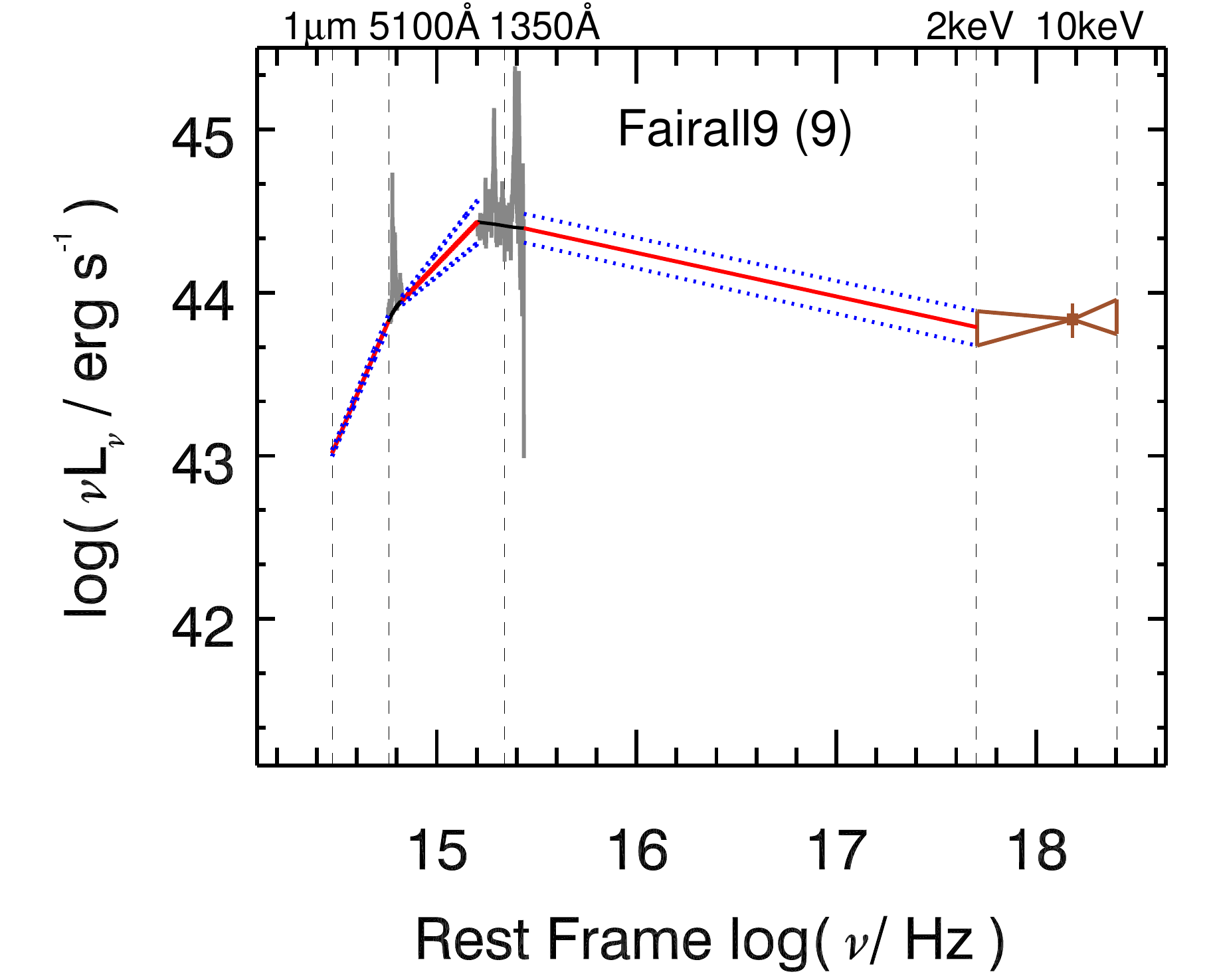}\\
\includegraphics[width=0.5\linewidth,scale=1.5]{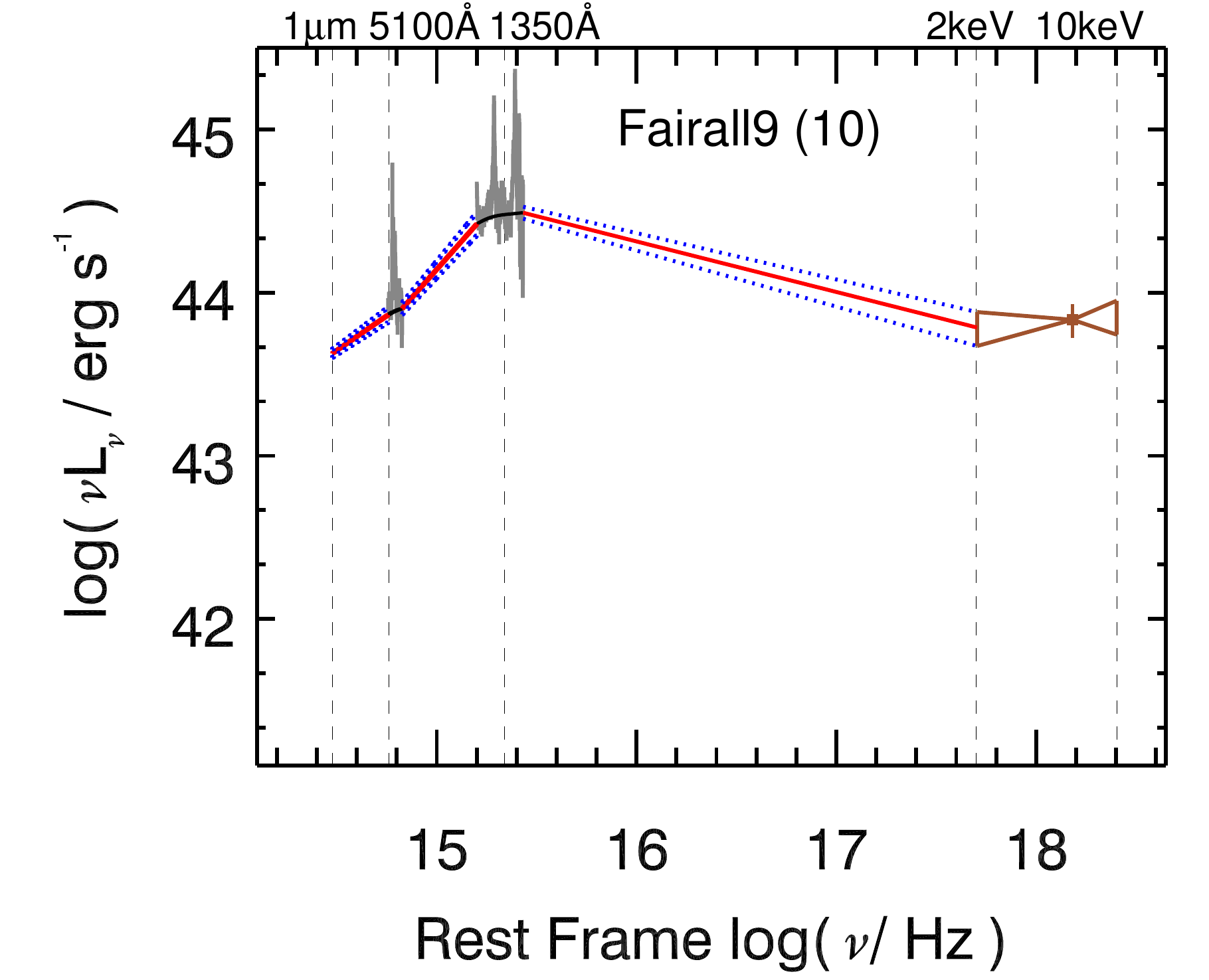} &
\includegraphics[width=0.5\linewidth,scale=1.5]{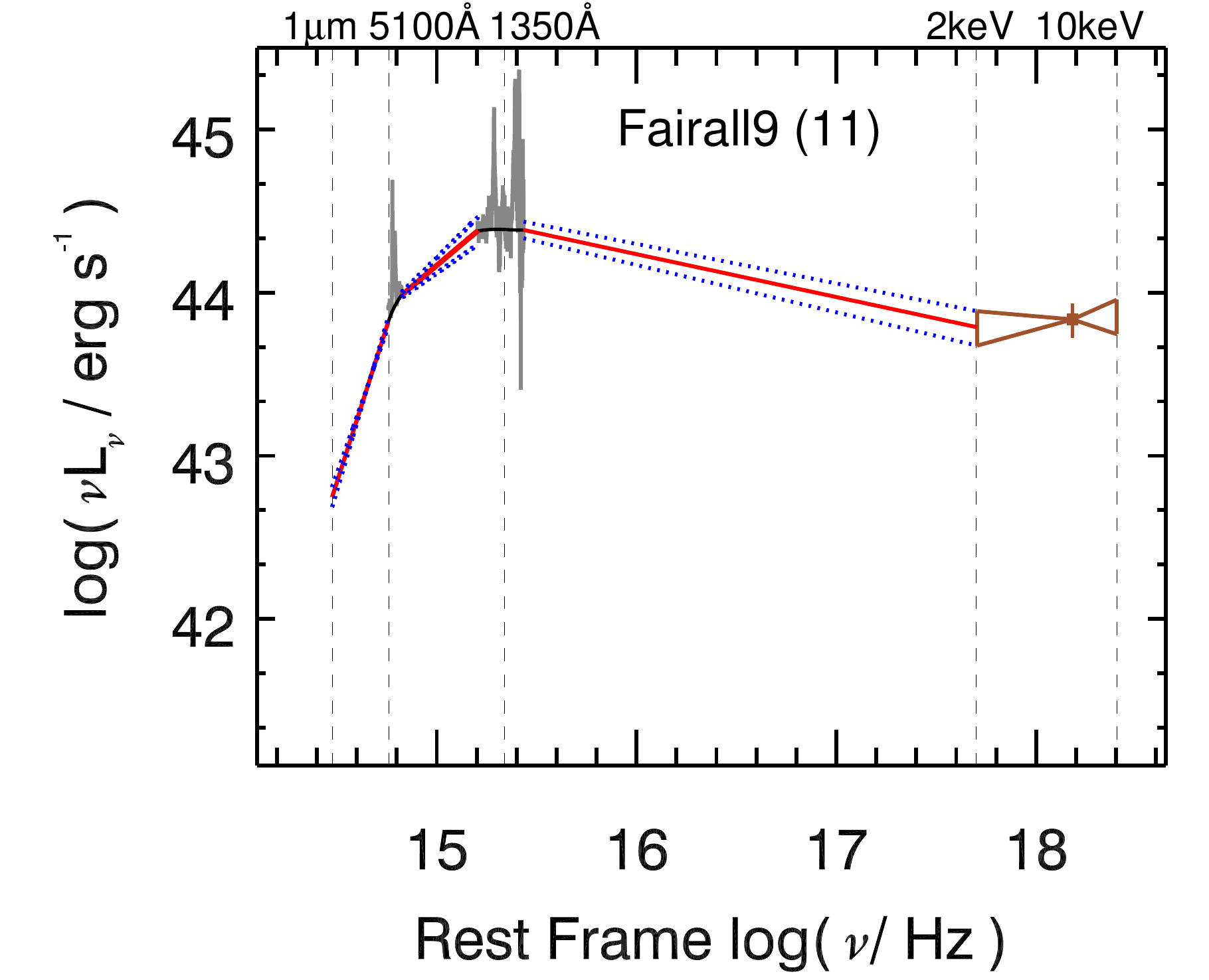}\\
\includegraphics[width=0.5\linewidth,scale=1.5]{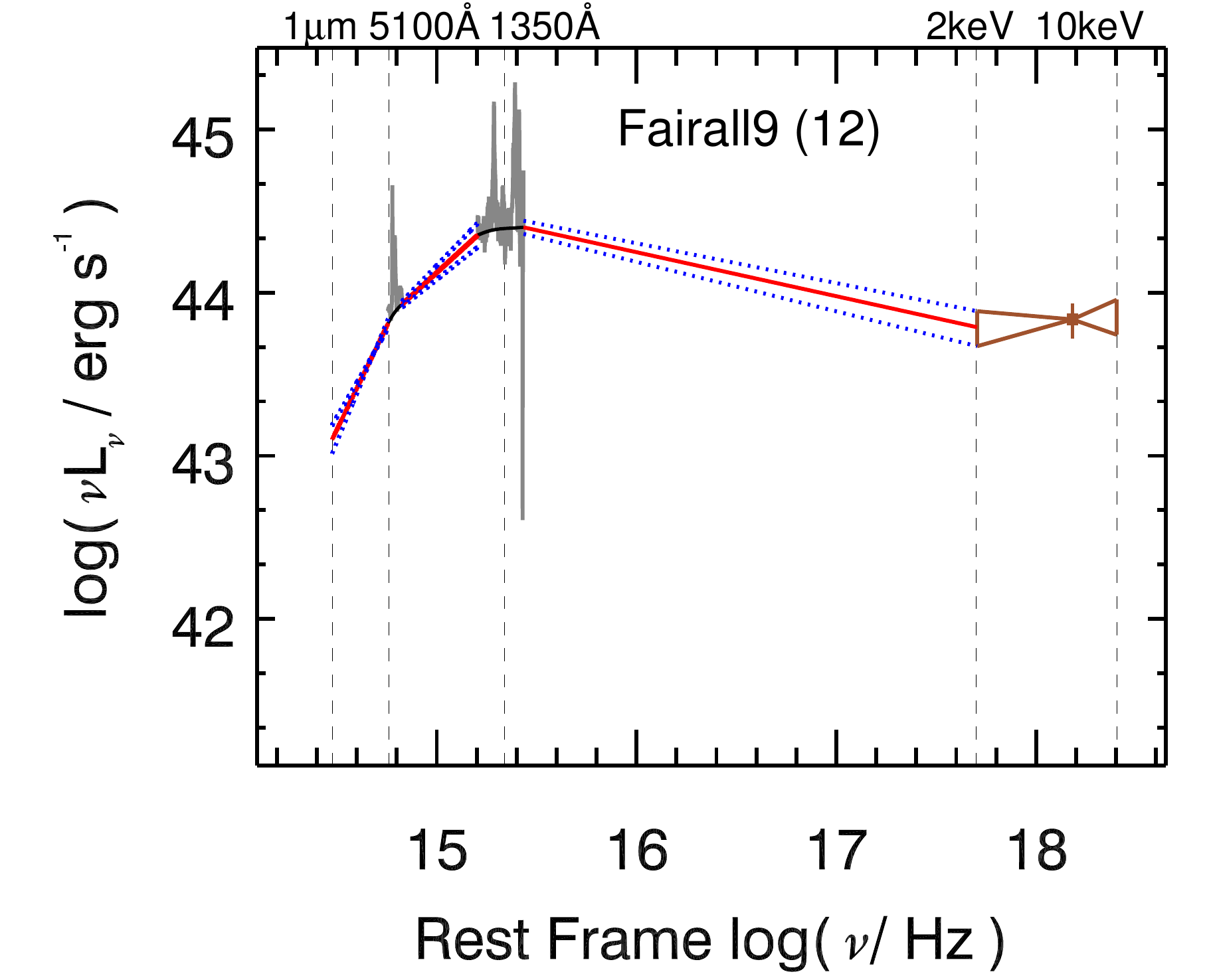} &
\includegraphics[width=0.5\linewidth,scale=1.5]{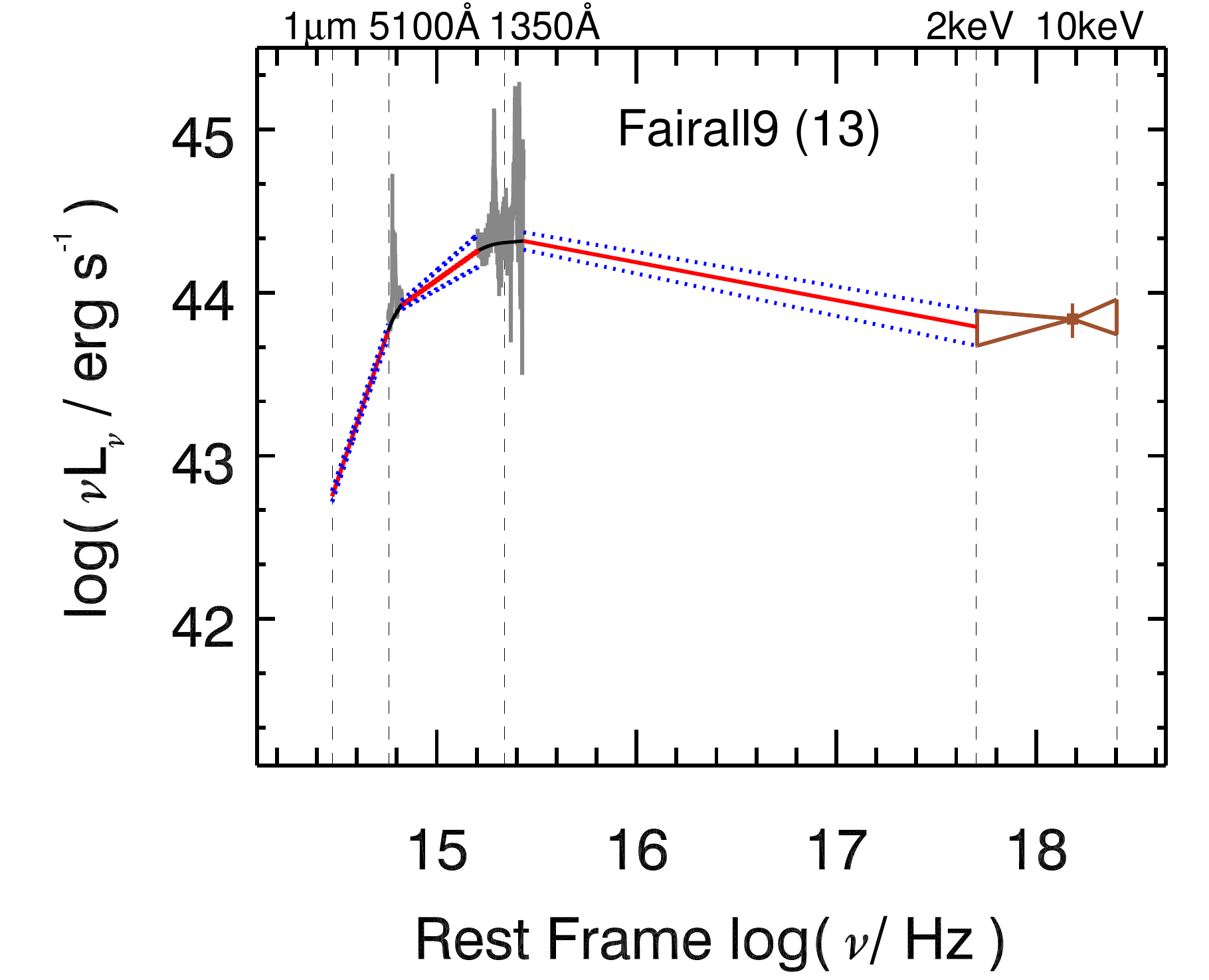}\\
\end{array}$
\end{center}
\contcaption{}
\end{figure*}

\begin{figure*}
\begin{center}$
\begin{array}{cc}
\includegraphics[width=0.5\linewidth,scale=1.5]{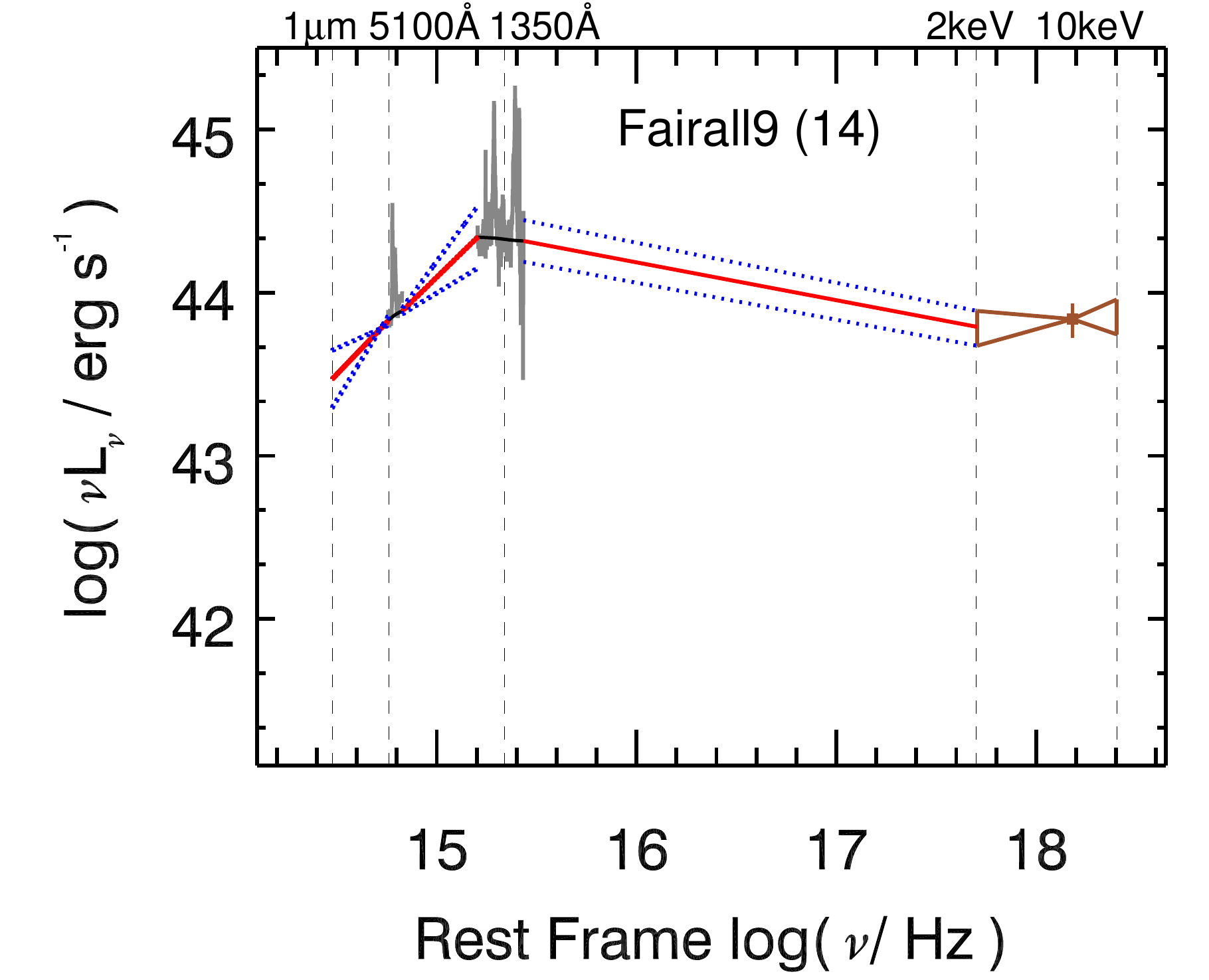} &
\includegraphics[width=0.5\linewidth,scale=1.5]{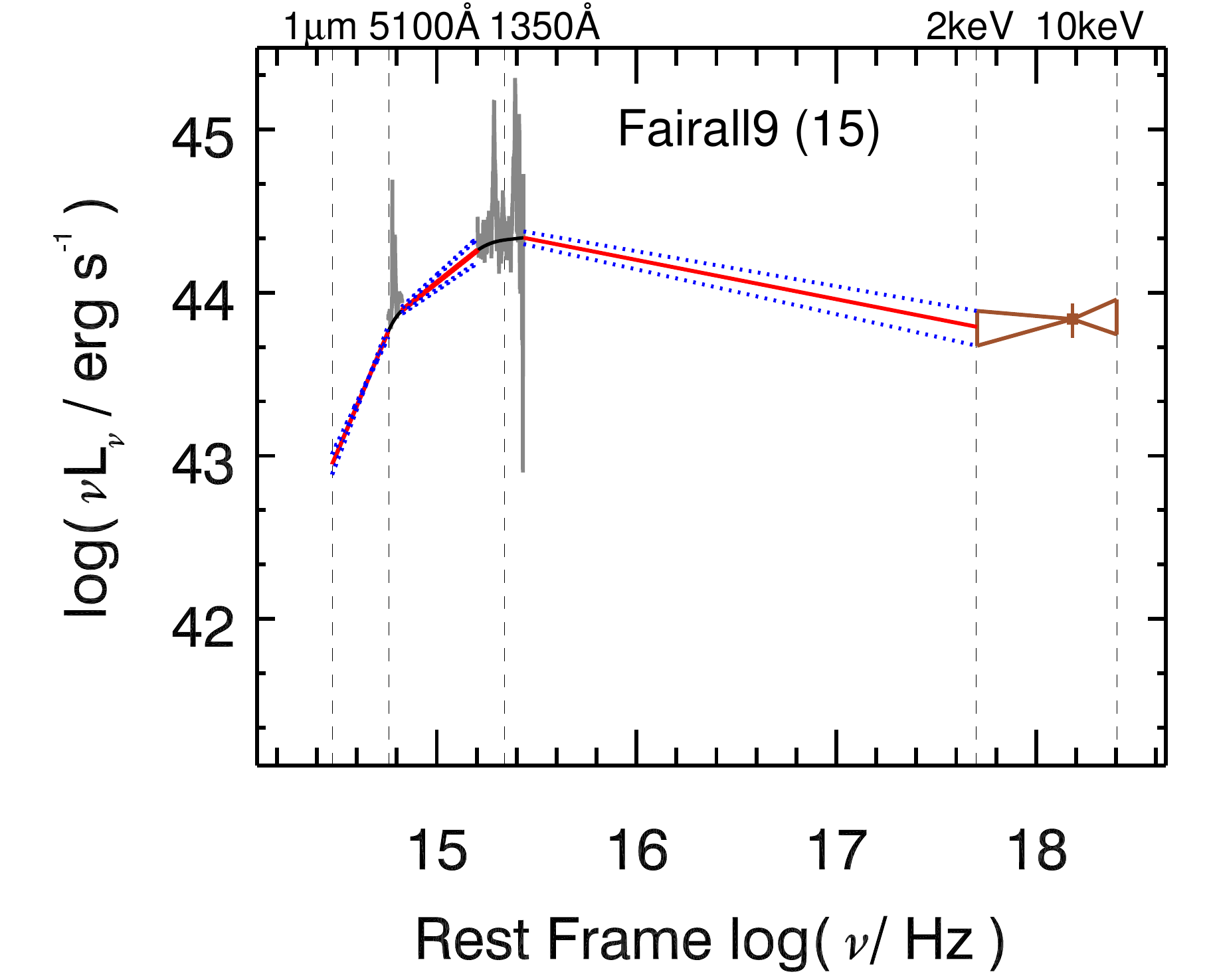}\\
\includegraphics[width=0.5\linewidth,scale=1.5]{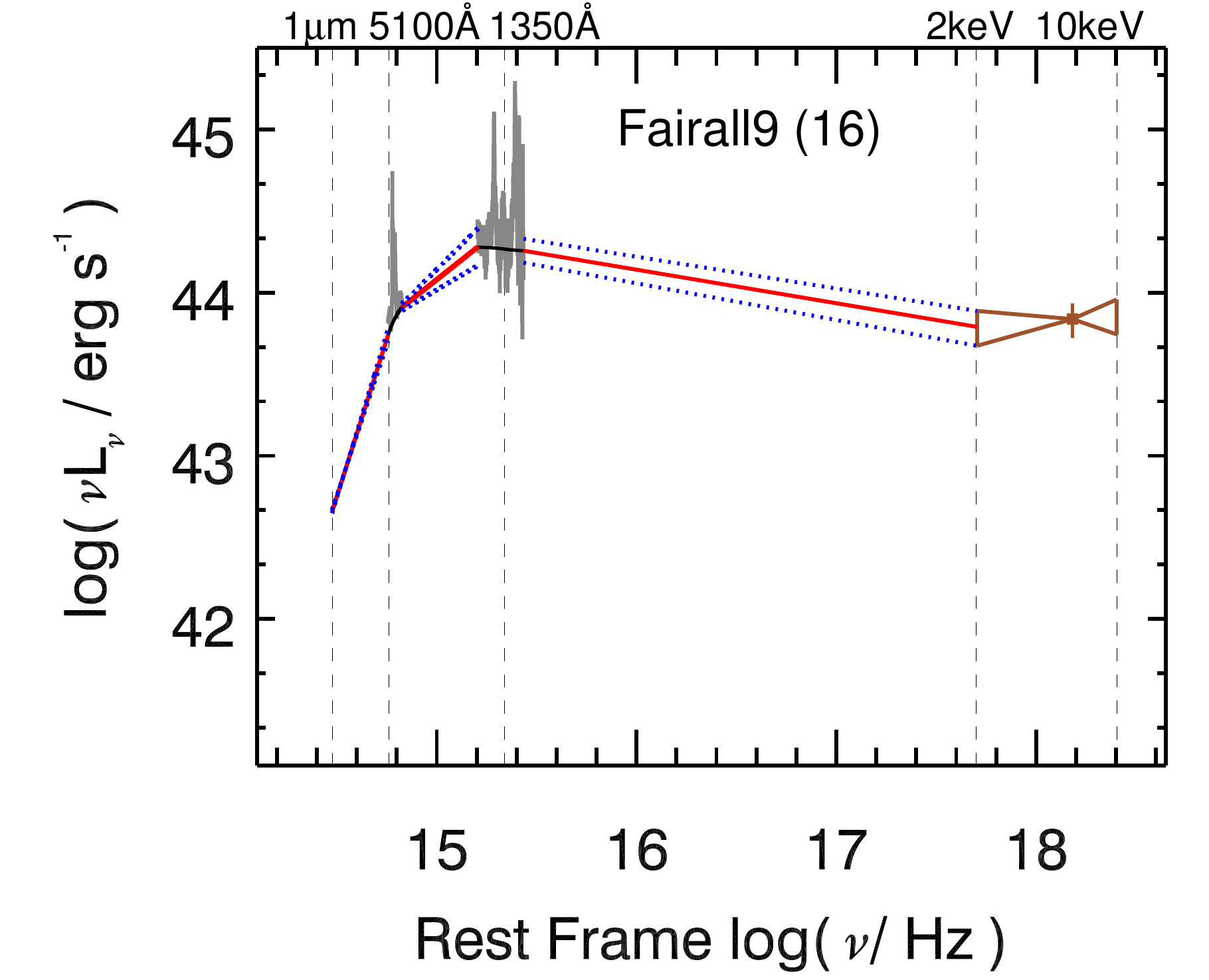} &
\includegraphics[width=0.5\linewidth,scale=1.5]{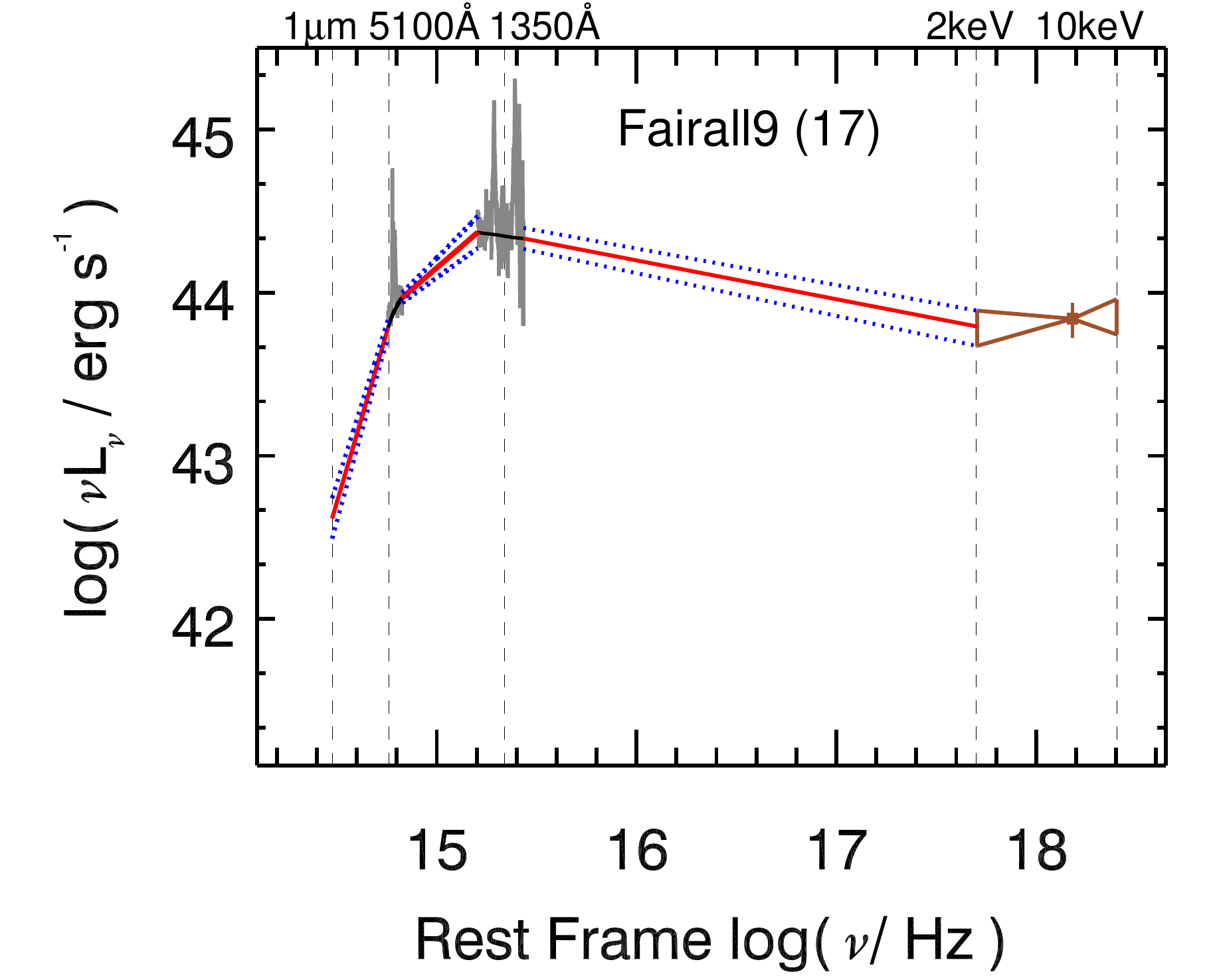}\\
\includegraphics[width=0.5\linewidth,scale=1.5]{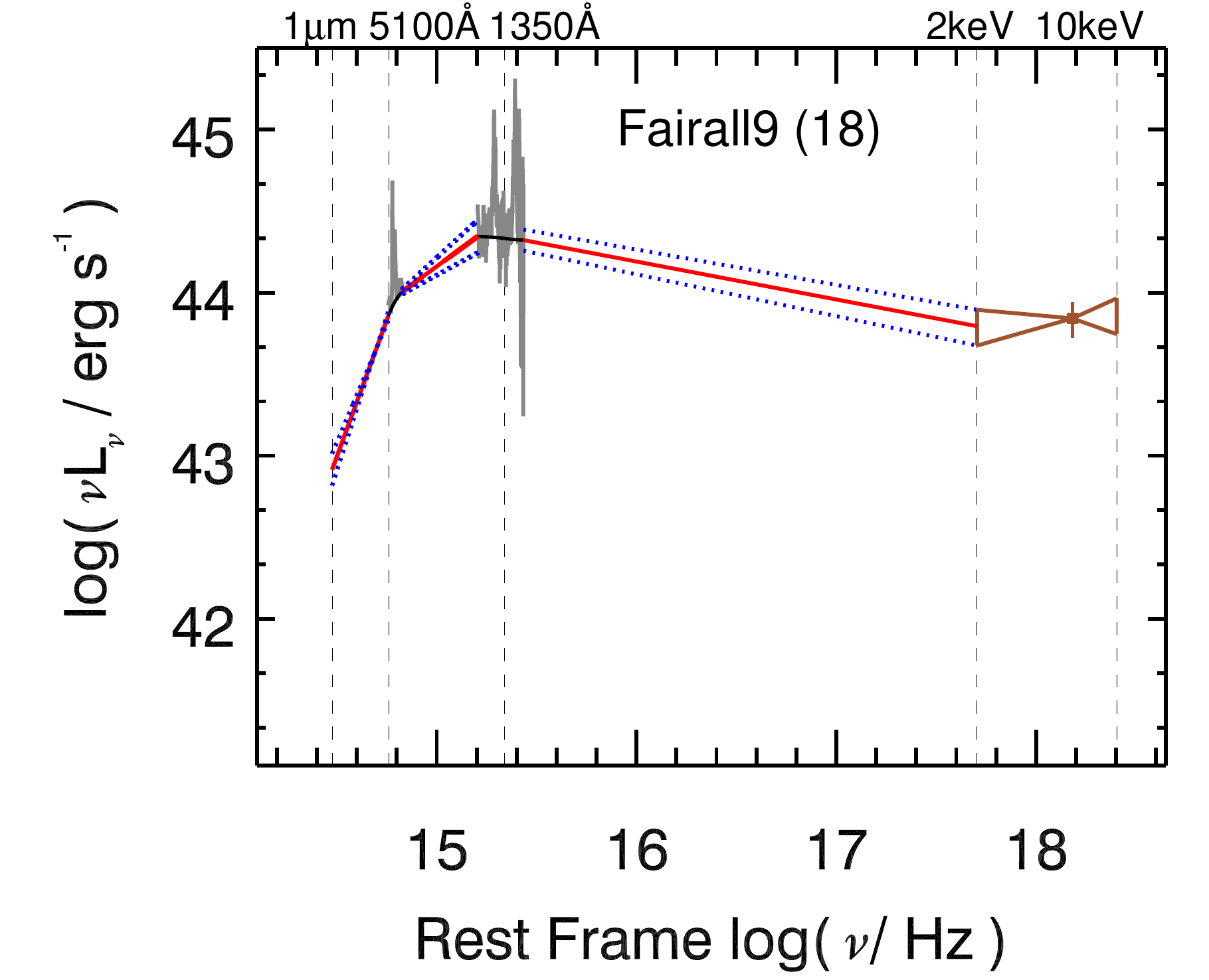} &
\includegraphics[width=0.5\linewidth,scale=1.5]{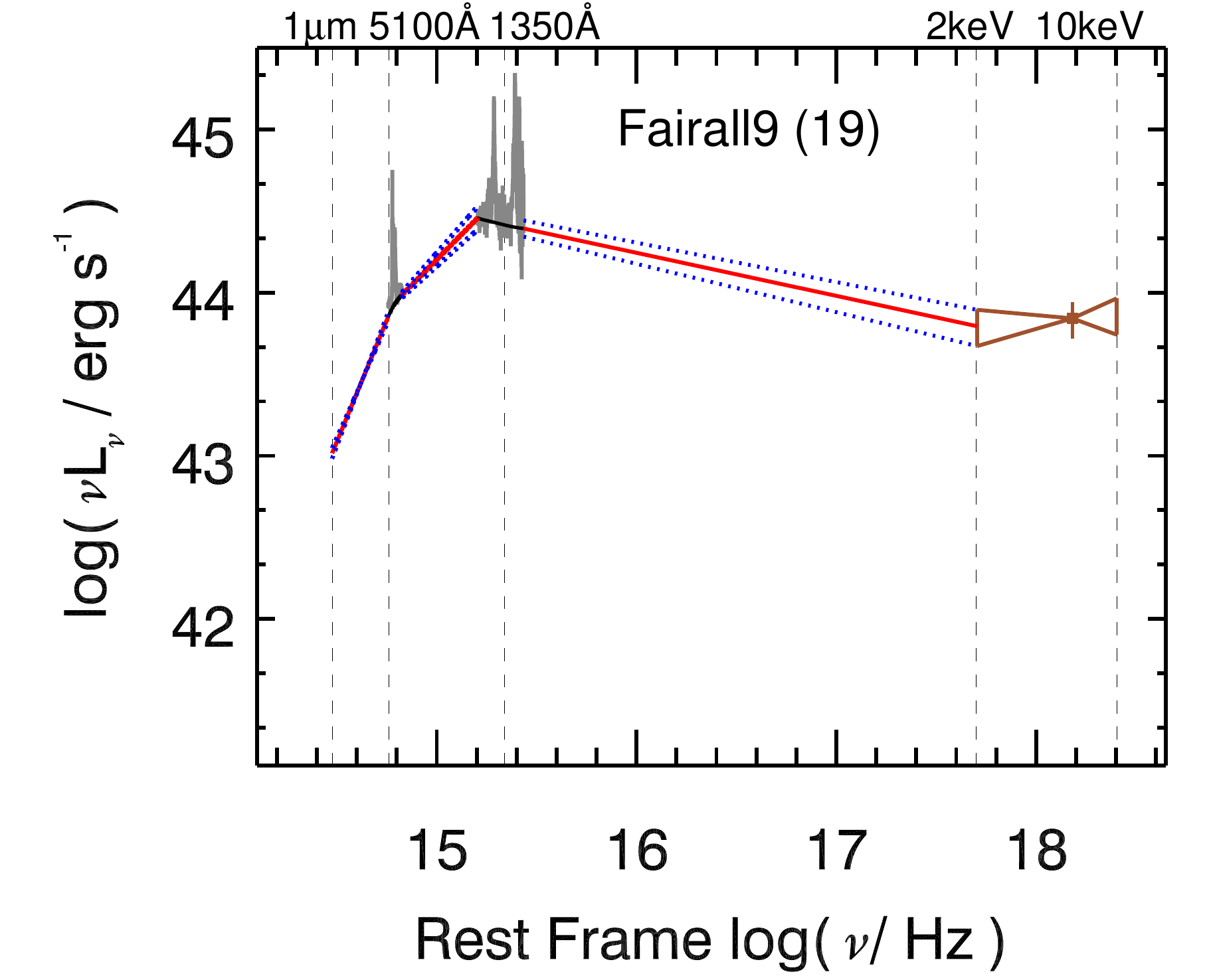}\\
\end{array}$
\end{center}
\contcaption{}
\end{figure*}

\begin{figure*}
\begin{center}$
\begin{array}{cc}
\includegraphics[width=0.5\linewidth,scale=1.5]{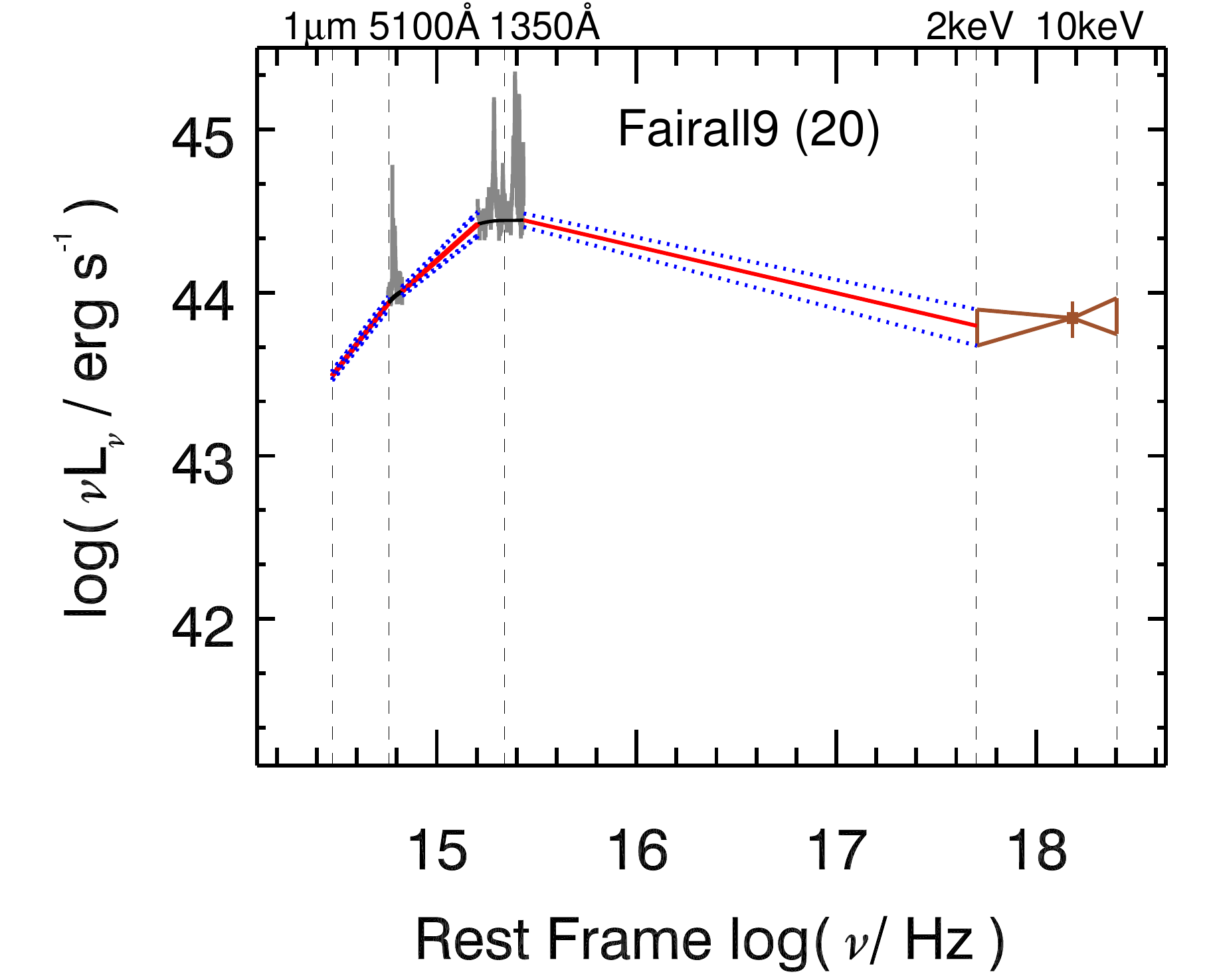} &
\includegraphics[width=0.5\linewidth,scale=1.5]{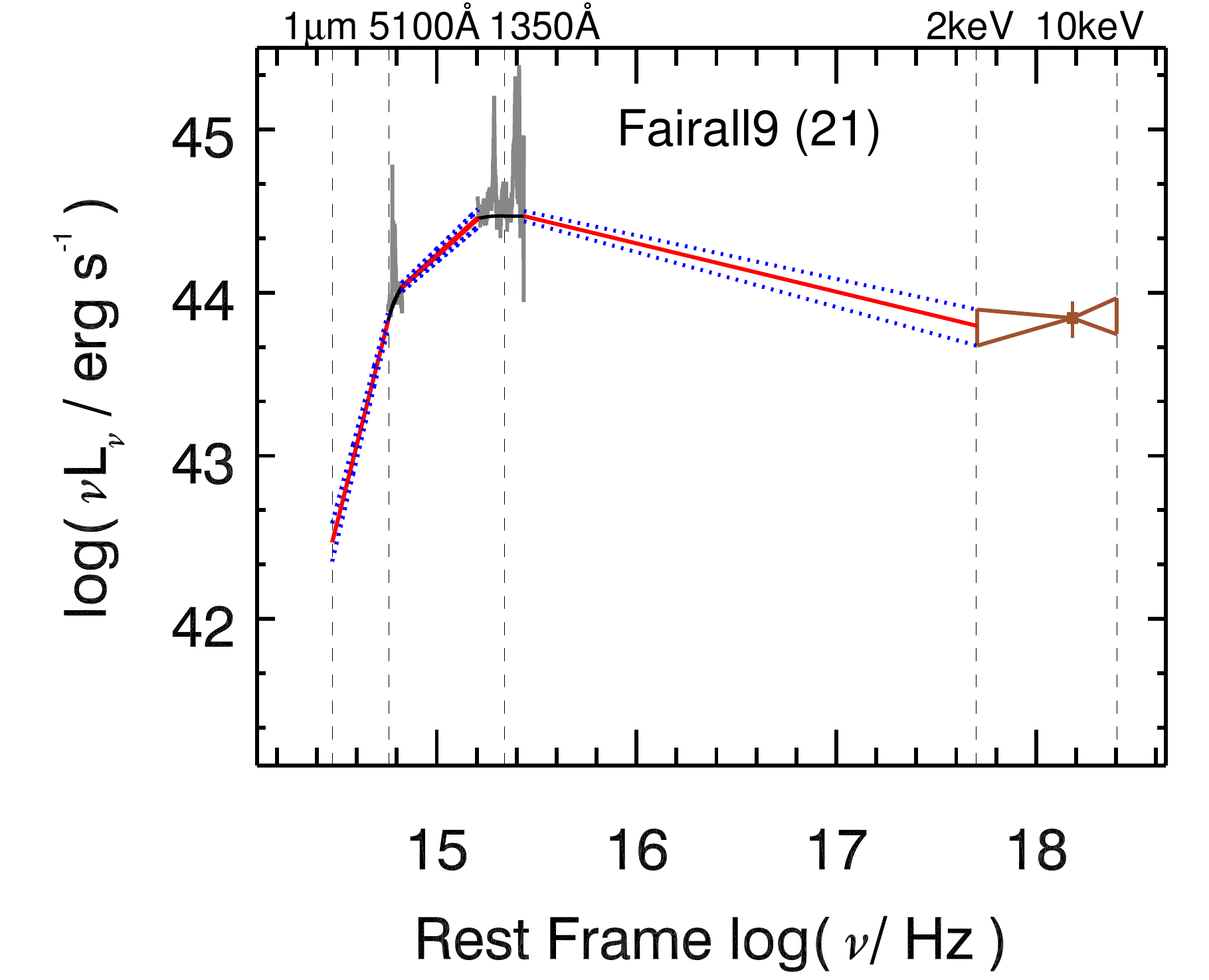}\\
\includegraphics[width=0.5\linewidth,scale=1.5]{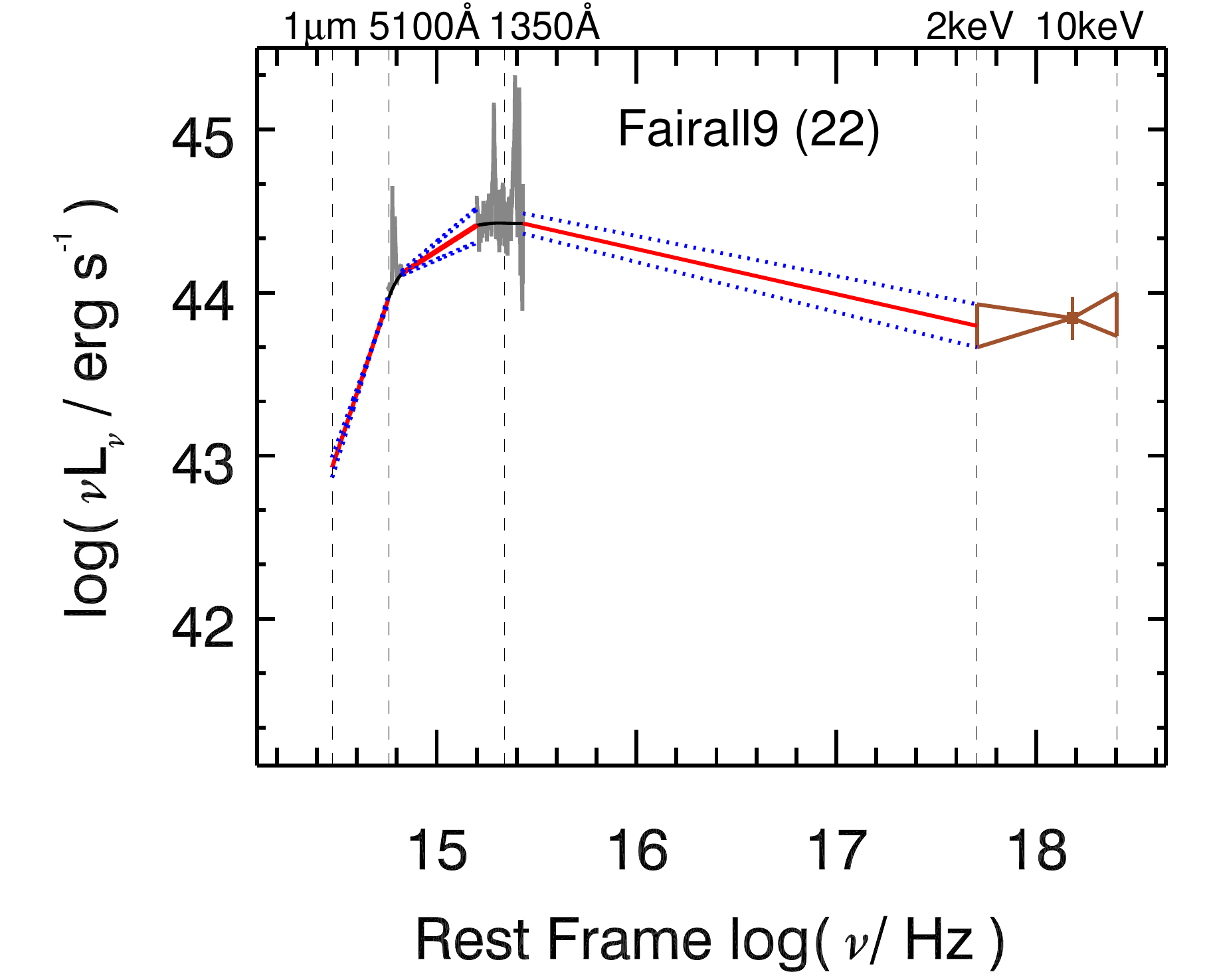} &
\includegraphics[width=0.5\linewidth,scale=1.5]{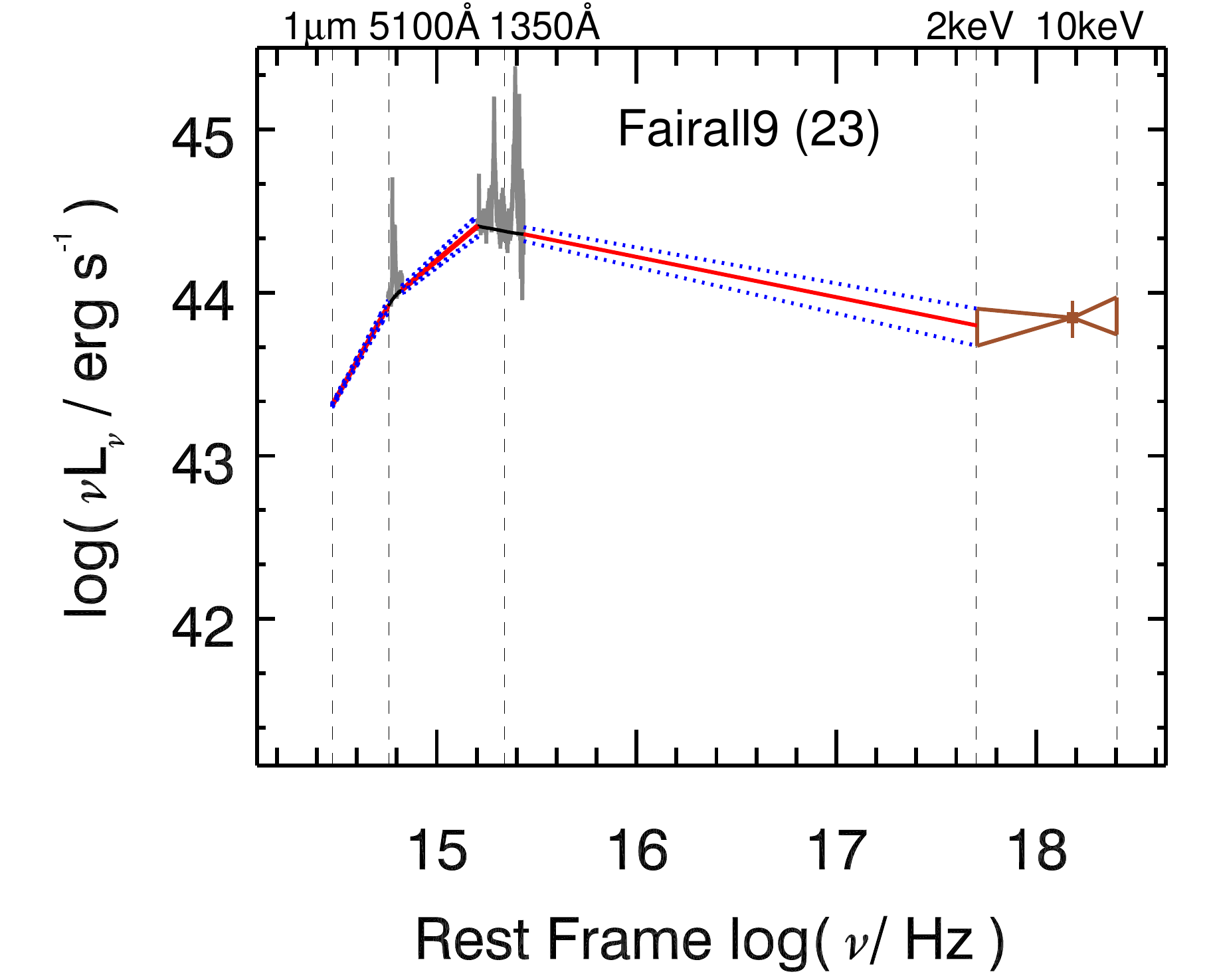}\\
\includegraphics[width=0.5\linewidth,scale=1.5]{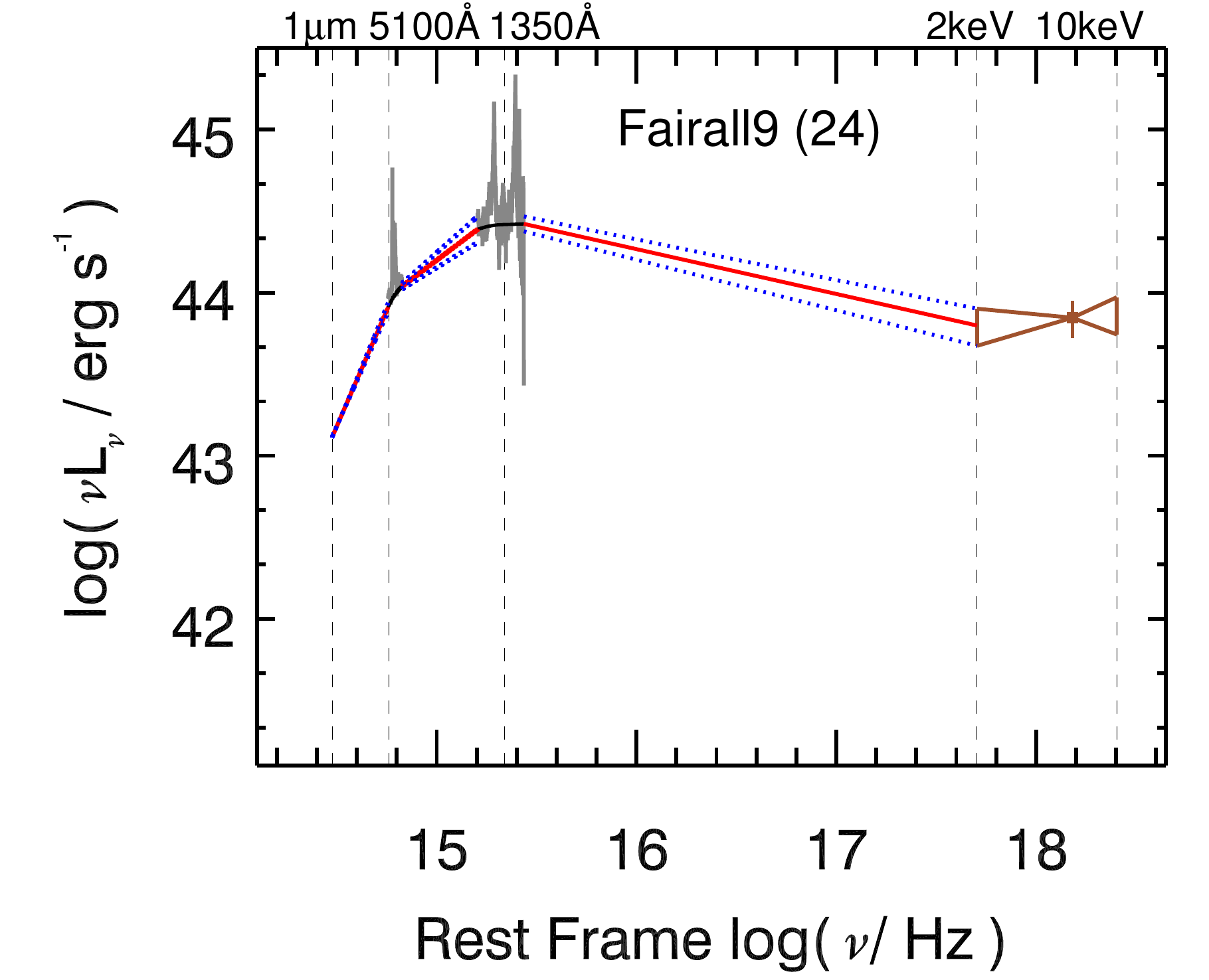} &
\includegraphics[width=0.5\linewidth,scale=1.5]{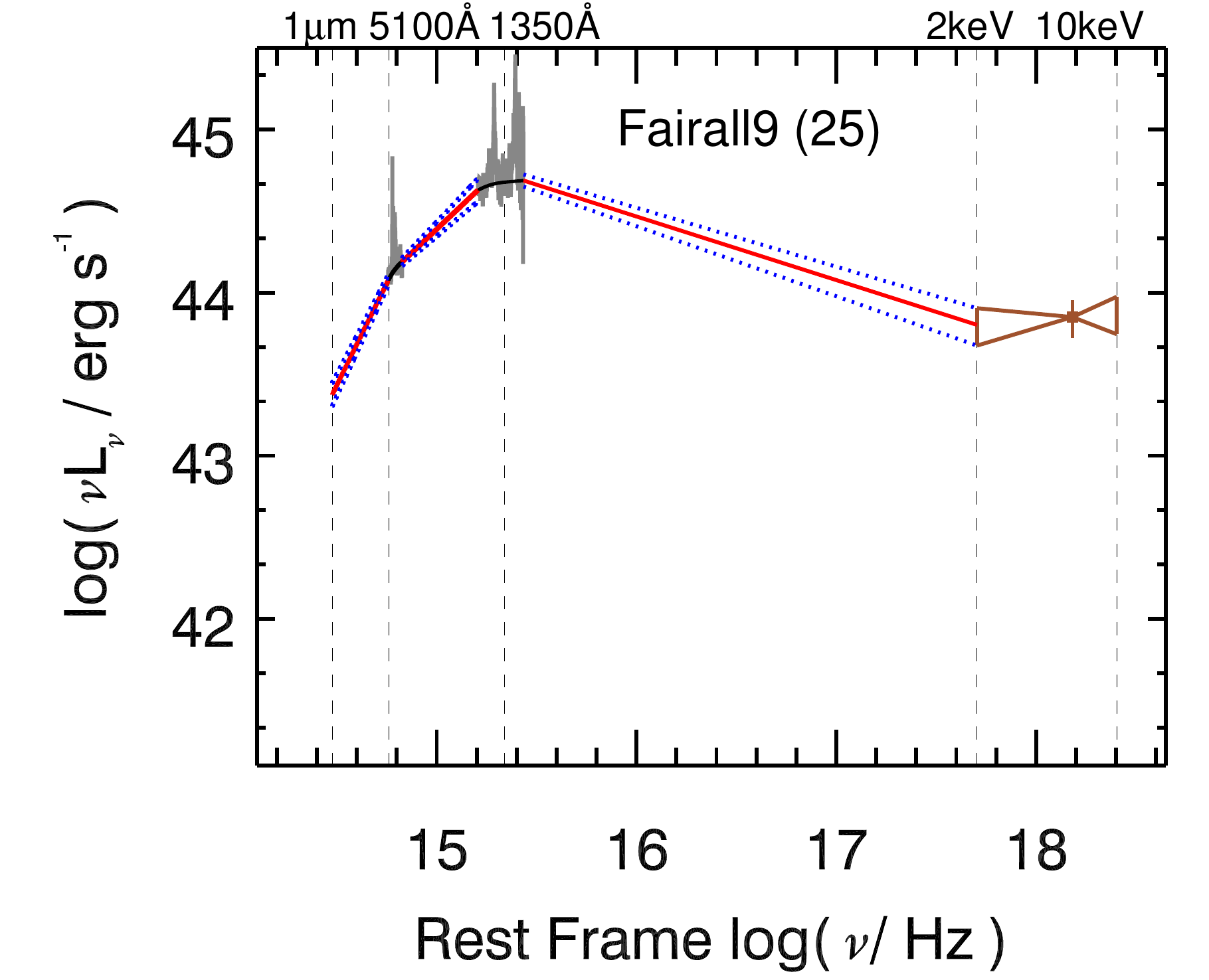}\\
\end{array}$
\end{center}
\contcaption{}
\end{figure*}

\begin{figure*}
\begin{center}$
\begin{array}{cc}
\includegraphics[width=0.5\linewidth,scale=1.5]{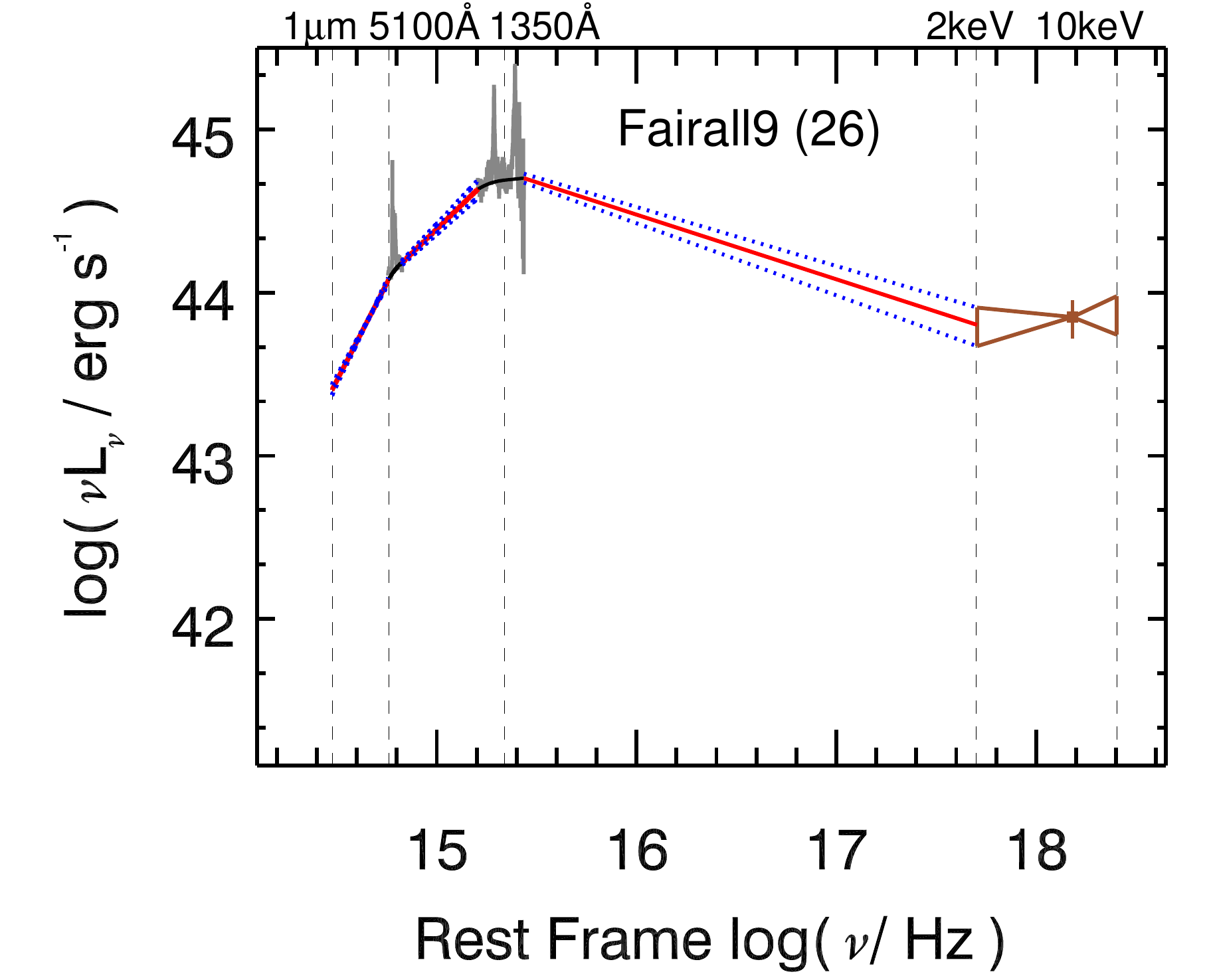} &
\includegraphics[width=0.5\linewidth,scale=1.5]{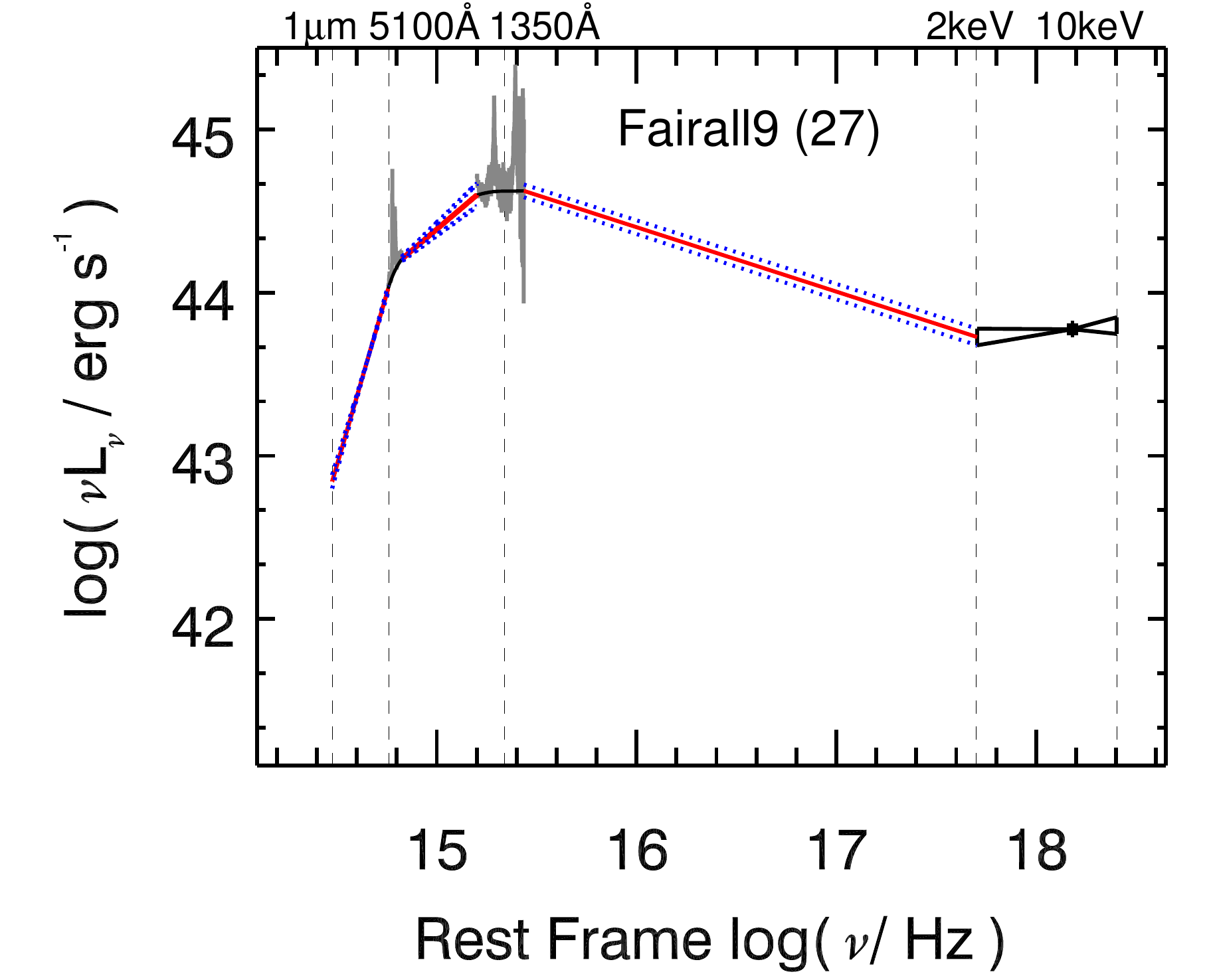}\\
\includegraphics[width=0.5\linewidth,scale=1.5]{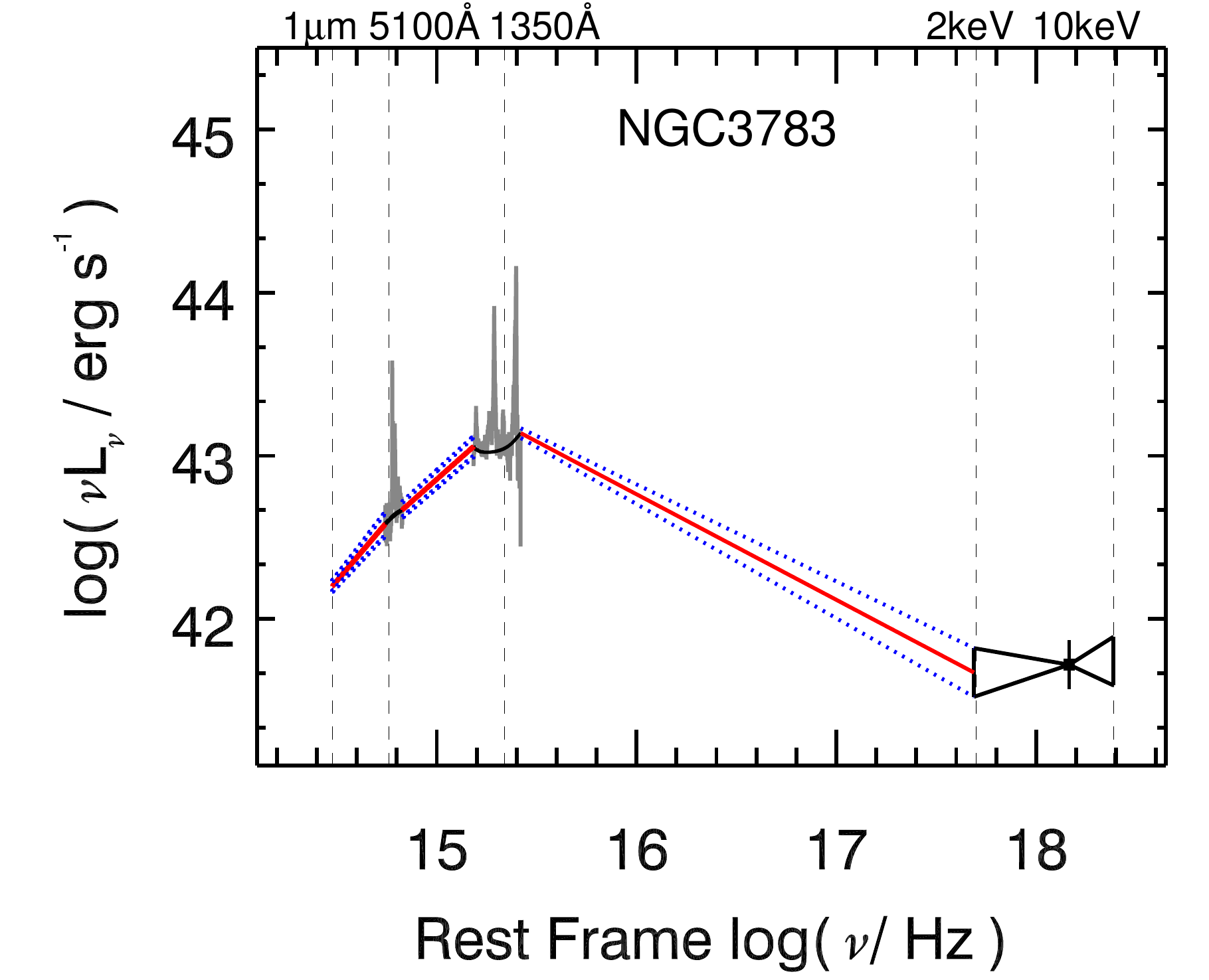} &
\includegraphics[width=0.5\linewidth,scale=1.5]{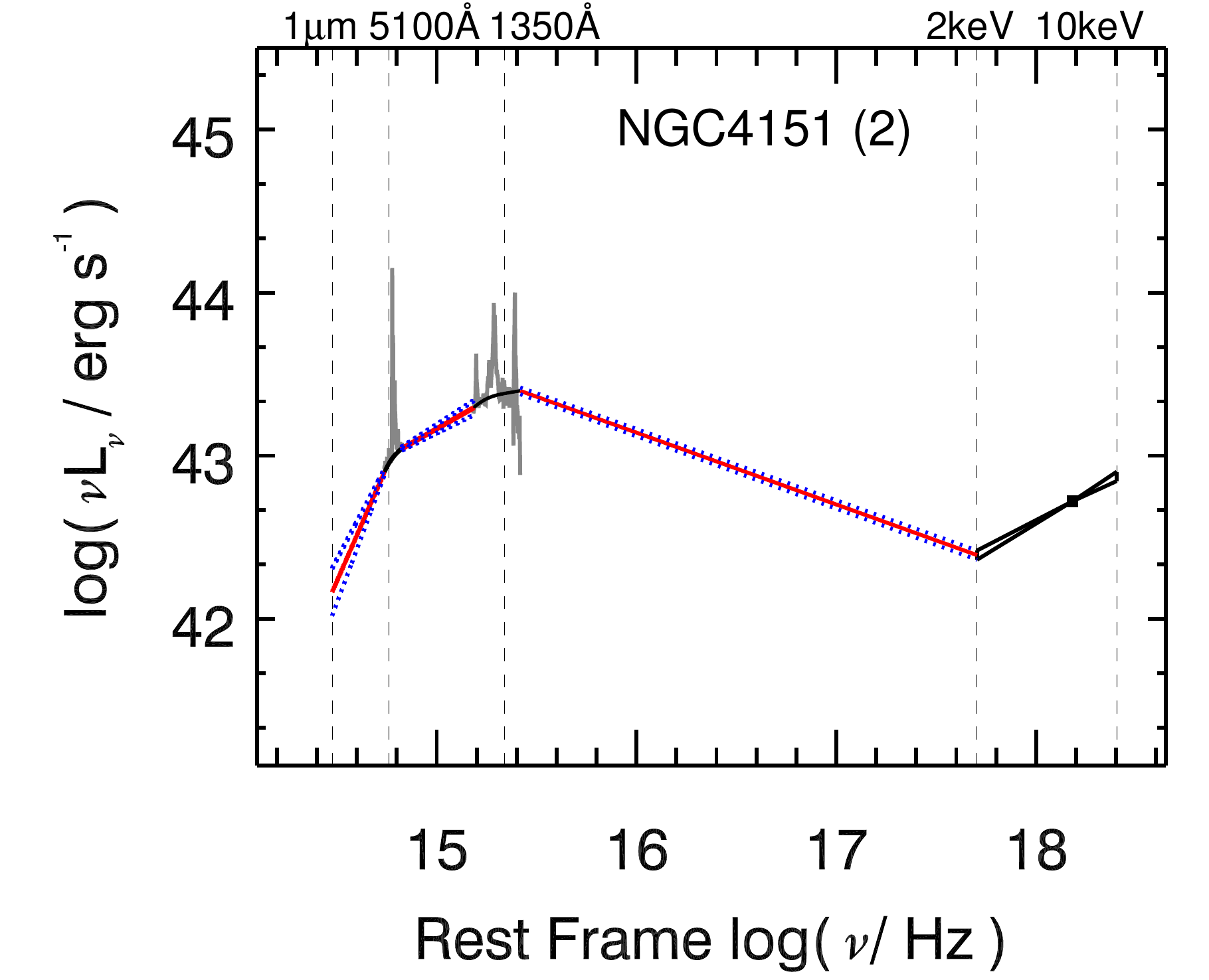}\\
\includegraphics[width=0.5\linewidth,scale=1.5]{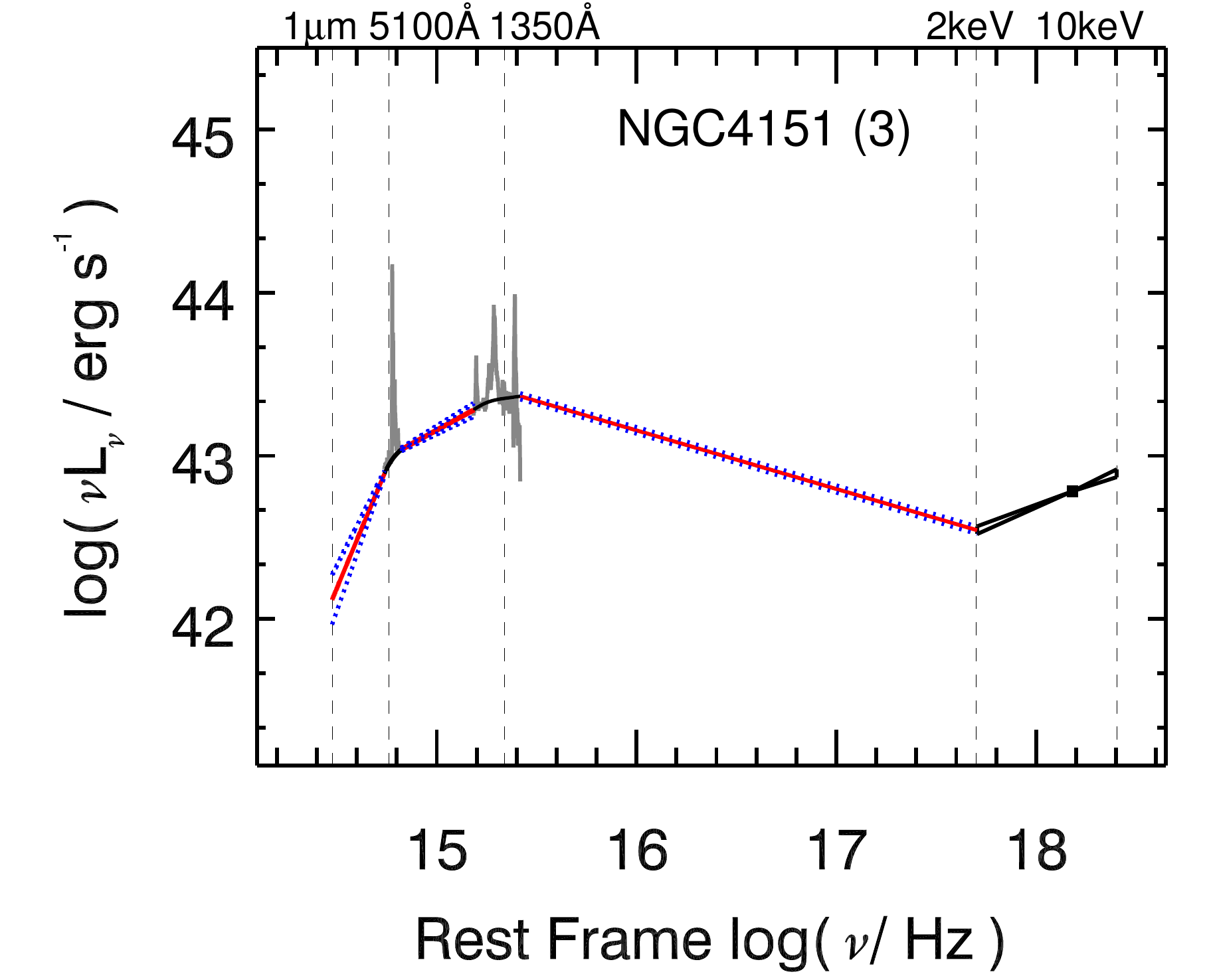} &
\includegraphics[width=0.5\linewidth,scale=1.5]{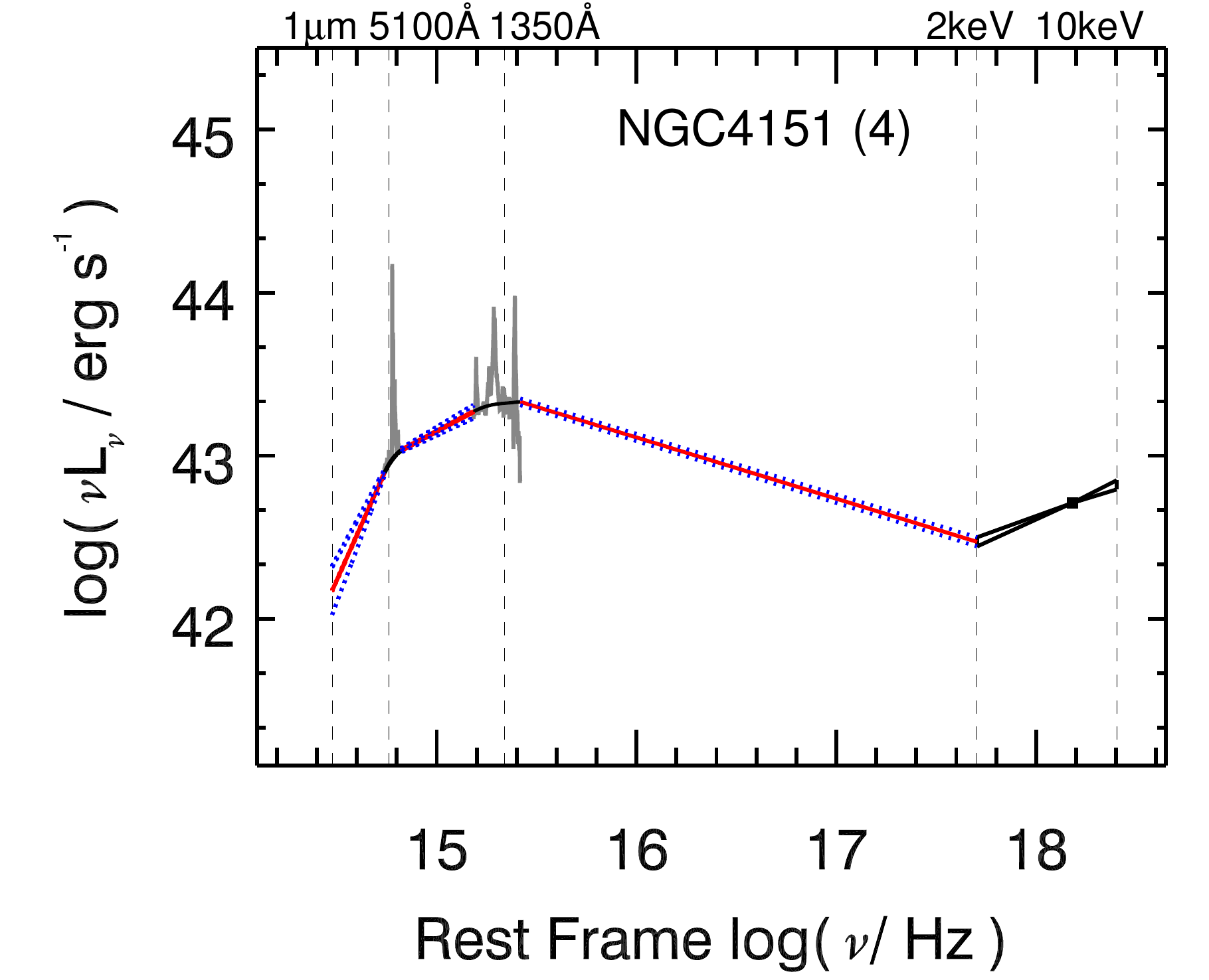}\\
\end{array}$
\end{center}
\contcaption{ }
\end{figure*}

\begin{figure*}
\begin{center}$
\begin{array}{cc}
\includegraphics[width=0.5\linewidth,scale=1.5]{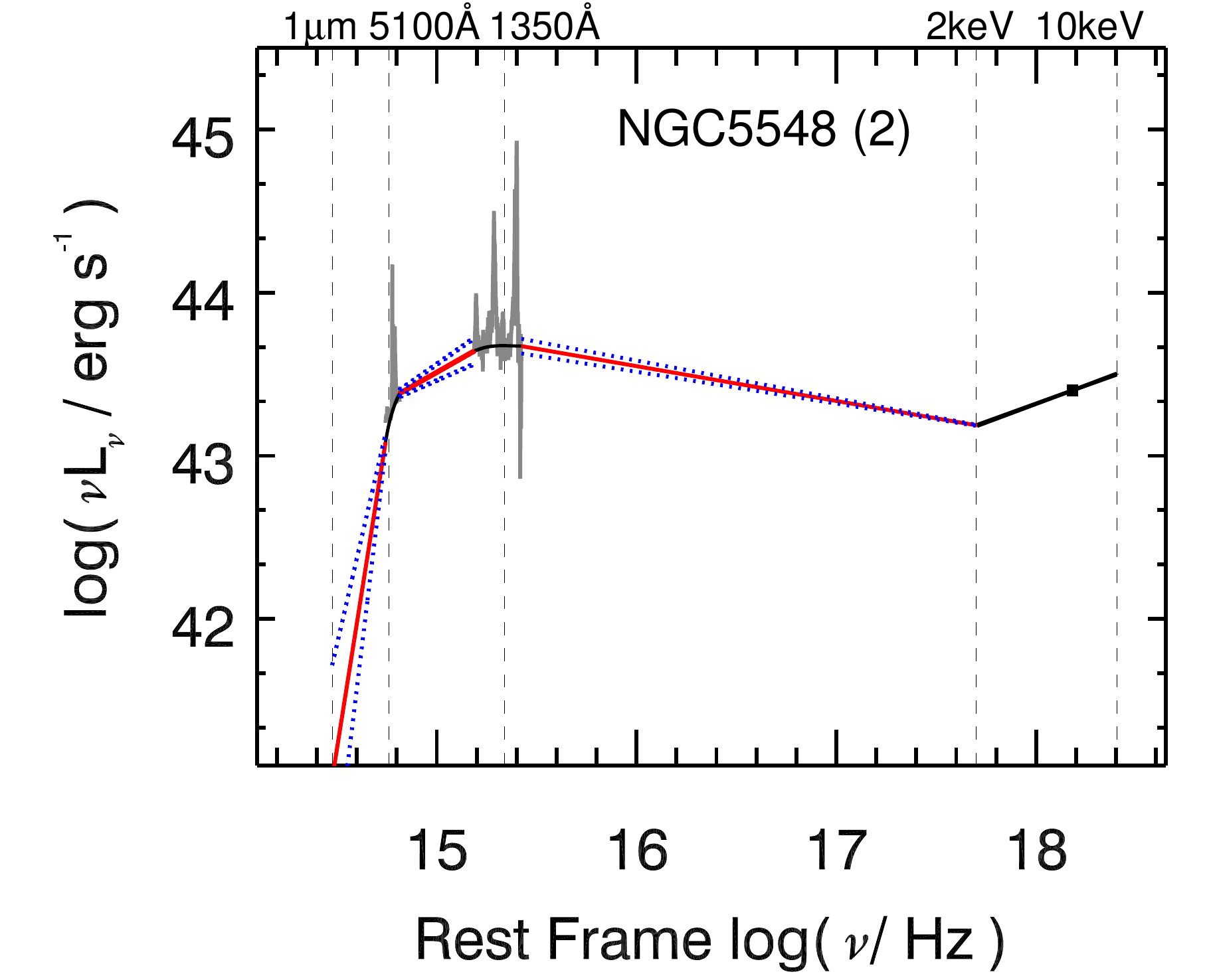} &
\includegraphics[width=0.5\linewidth,scale=1.5]{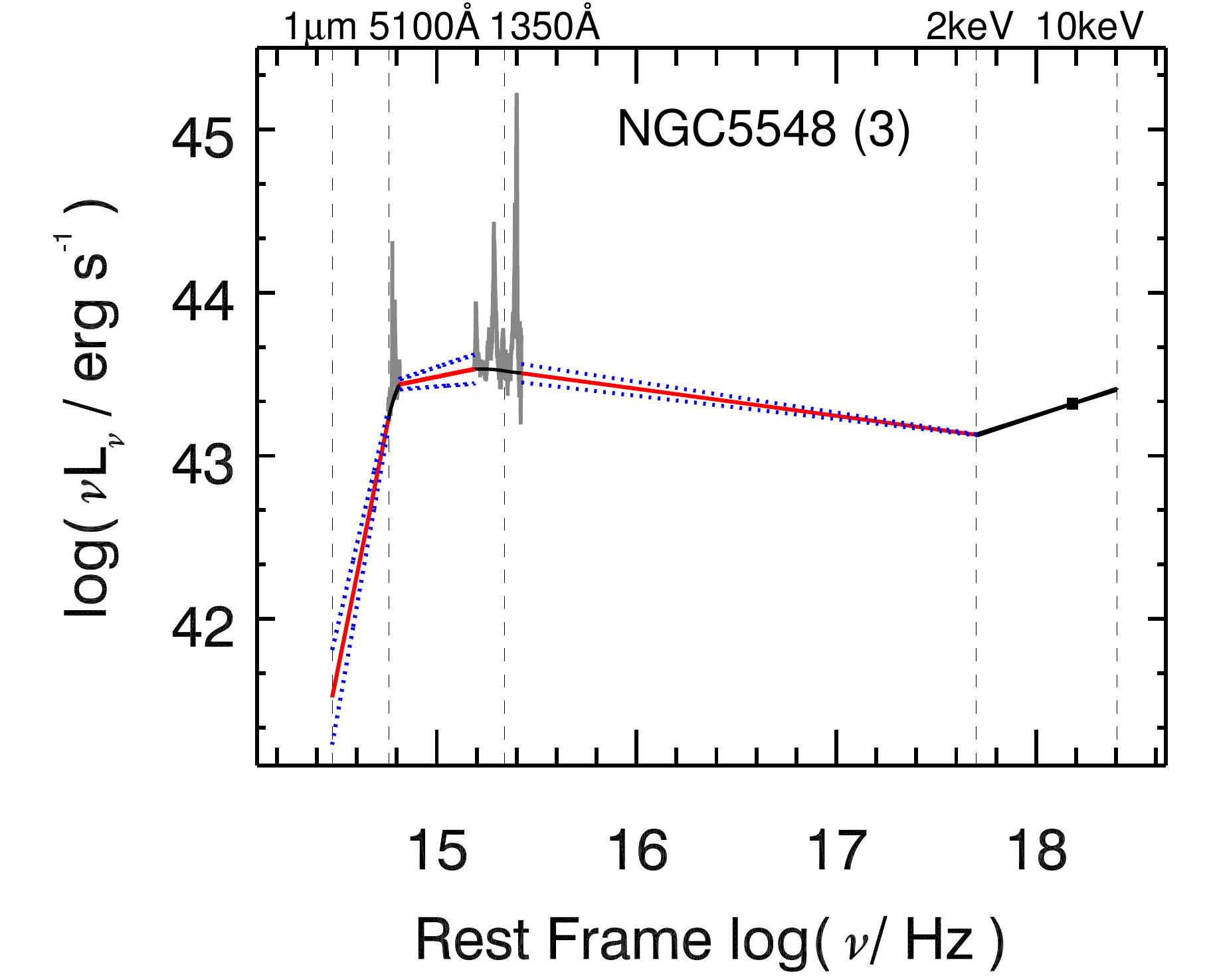}\\
\includegraphics[width=0.5\linewidth,scale=1.5]{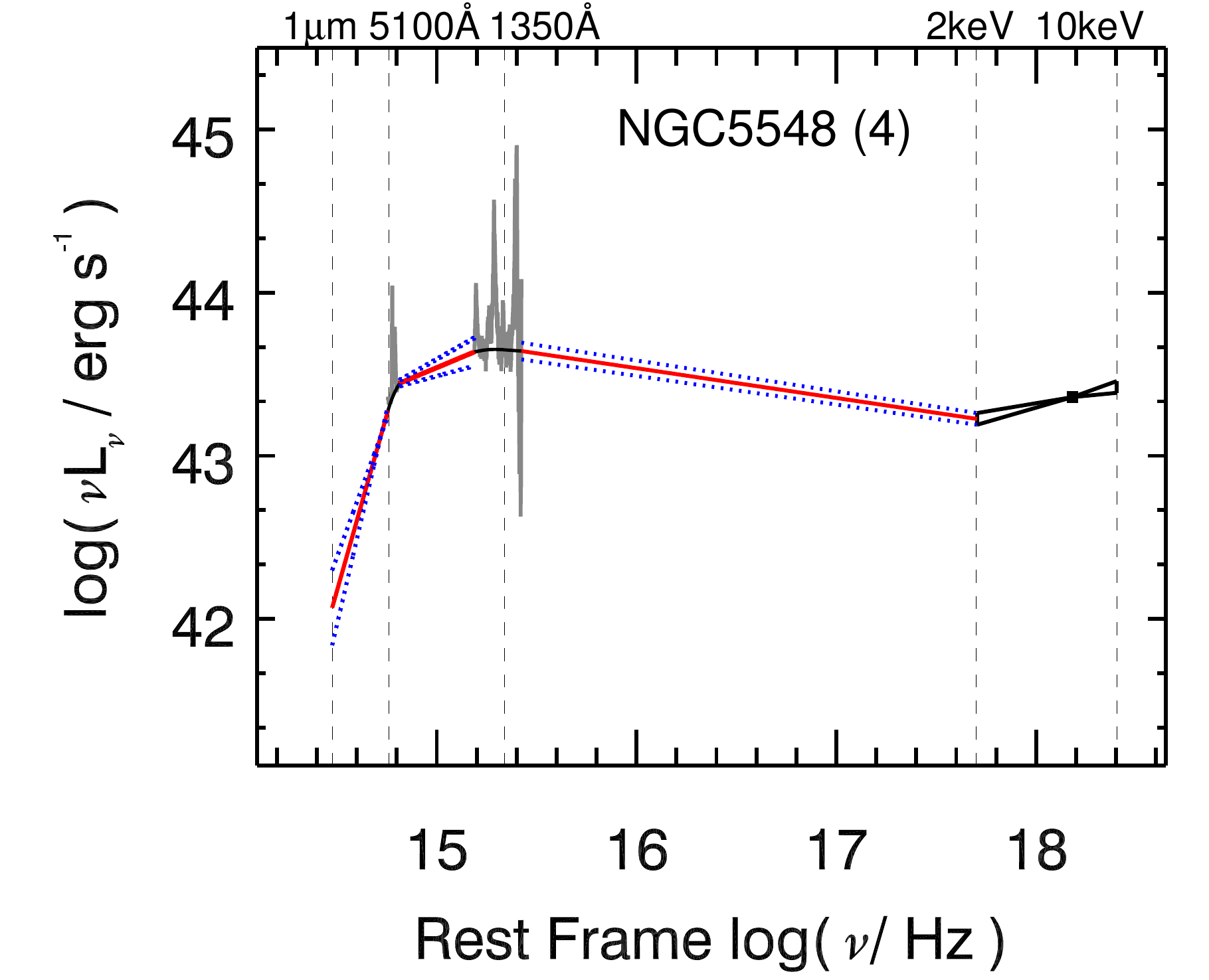} &
\includegraphics[width=0.5\linewidth,scale=1.5]{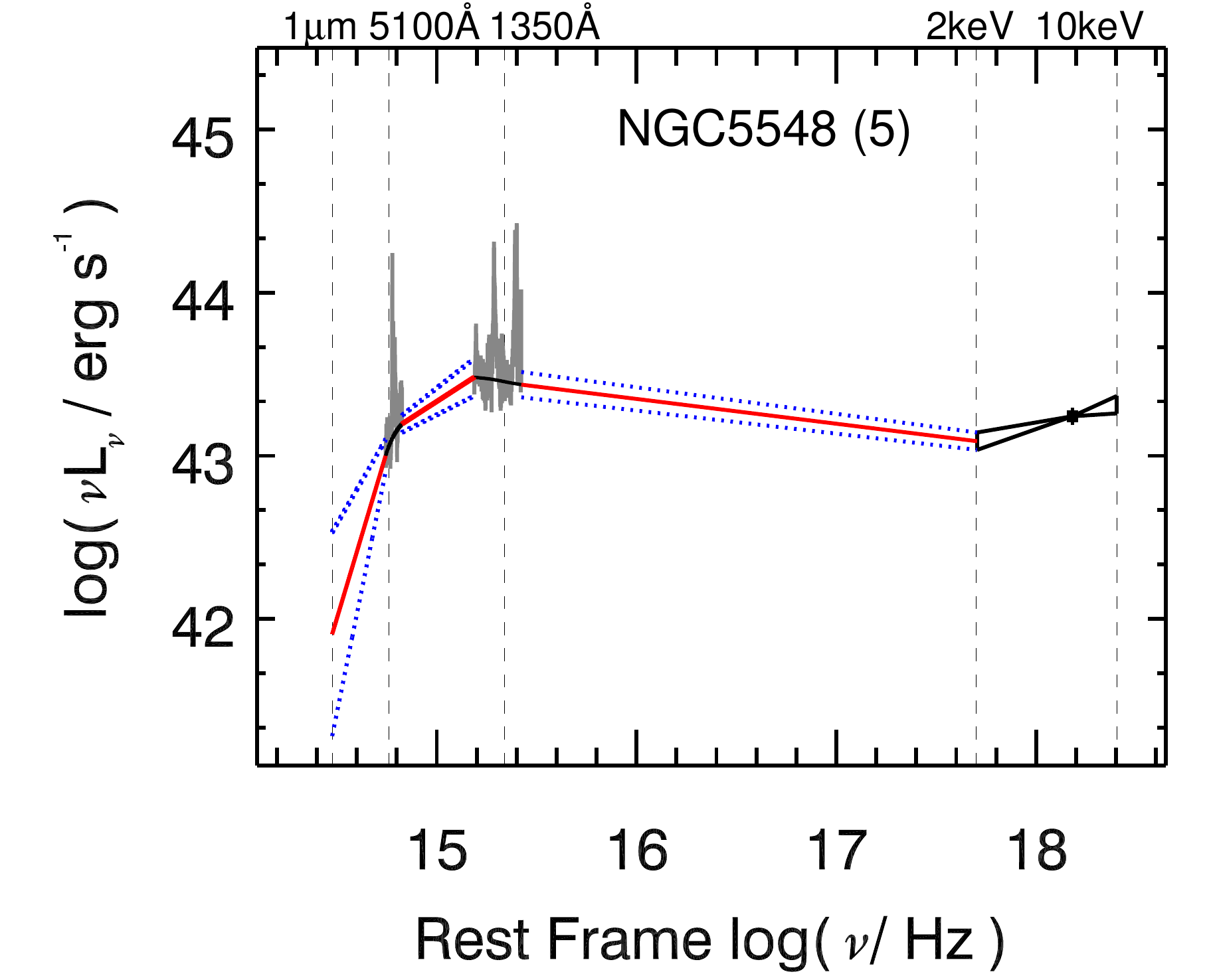}\\
\includegraphics[width=0.5\linewidth,scale=1.5]{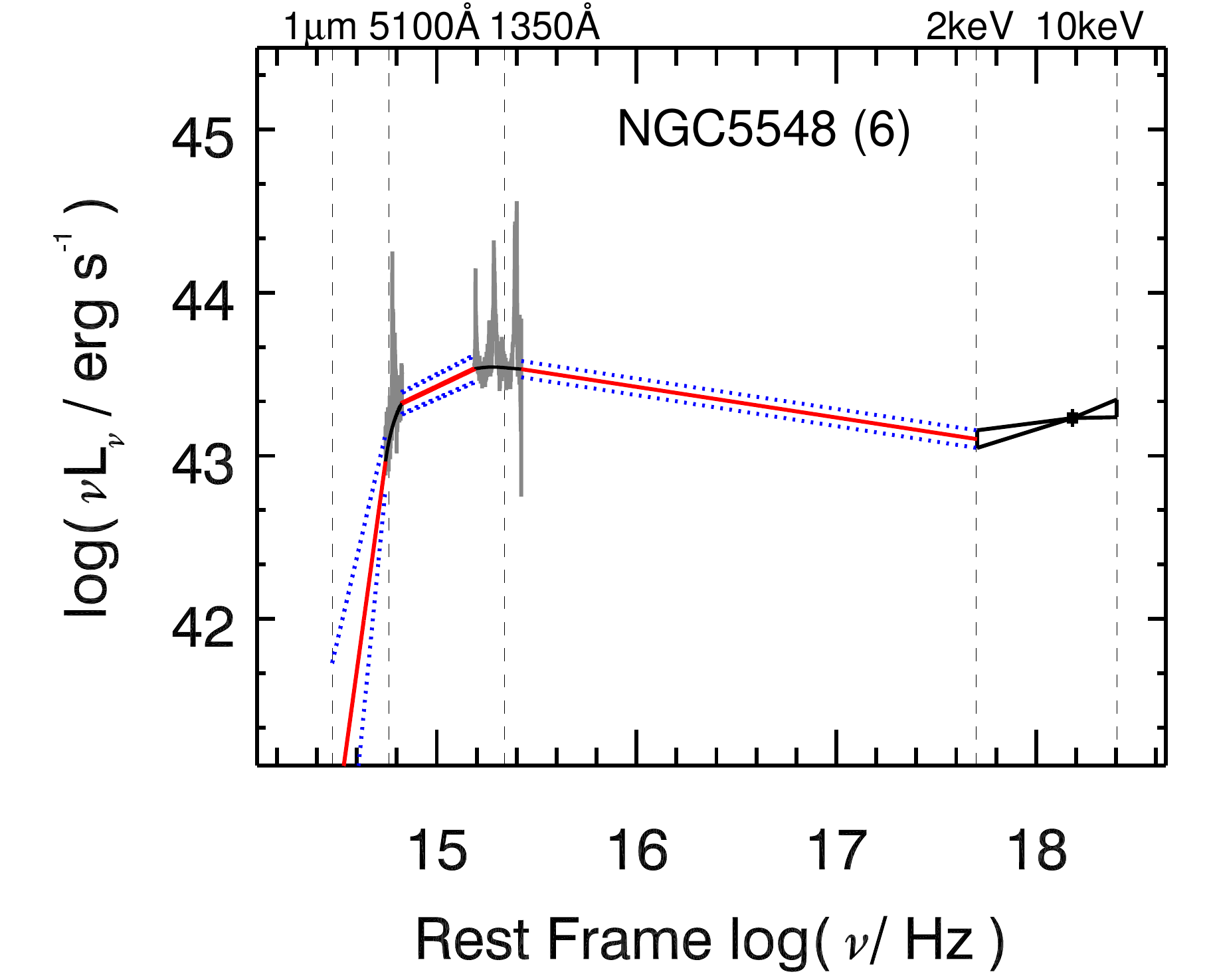} &
\includegraphics[width=0.5\linewidth,scale=1.5]{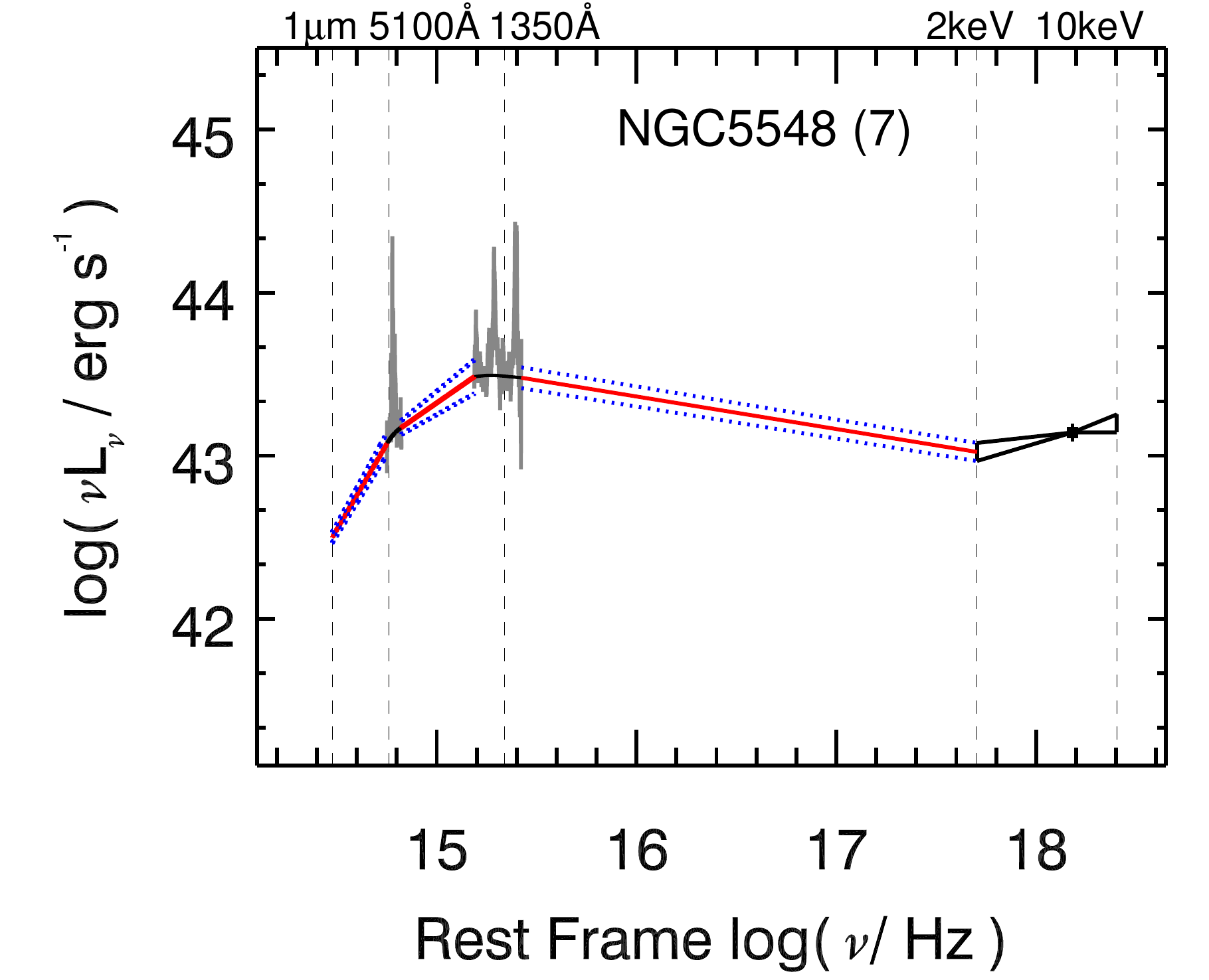}\\
\end{array}$
\end{center}
\contcaption{}
\end{figure*}

\begin{figure*}
\begin{center}$
\begin{array}{cc}
\includegraphics[width=0.5\linewidth,scale=1.5]{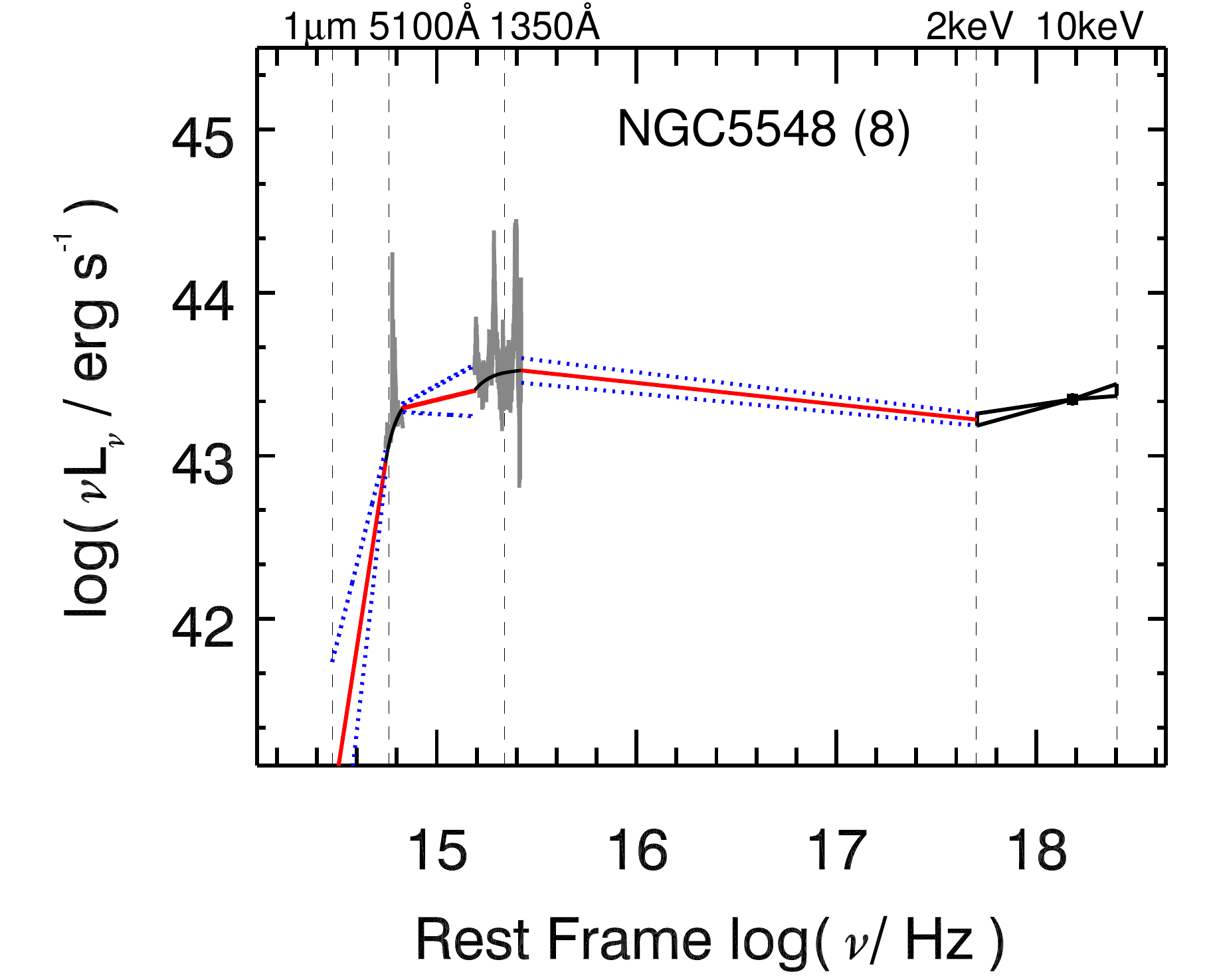} &
\includegraphics[width=0.5\linewidth,scale=1.5]{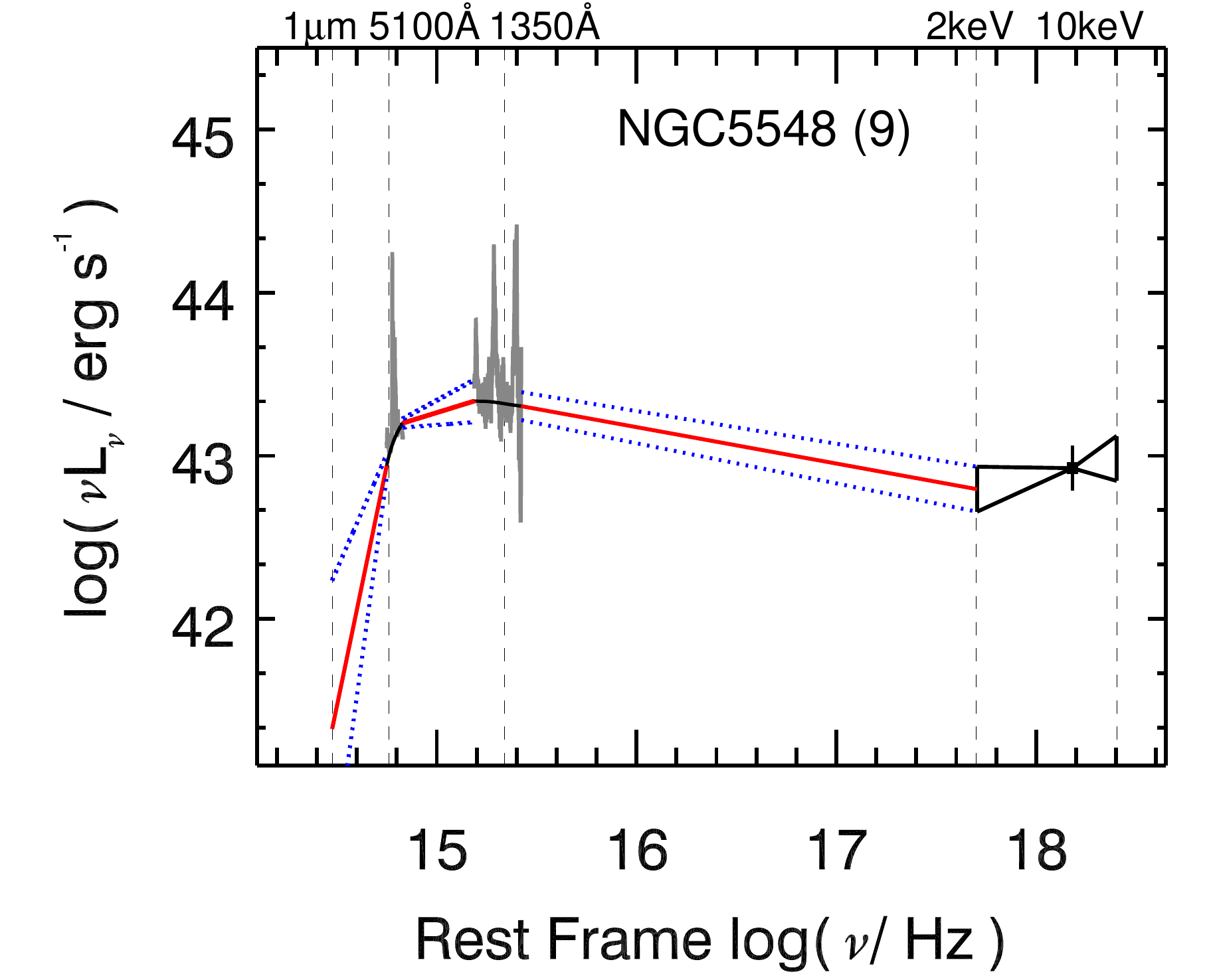}\\
\includegraphics[width=0.5\linewidth,scale=1.5]{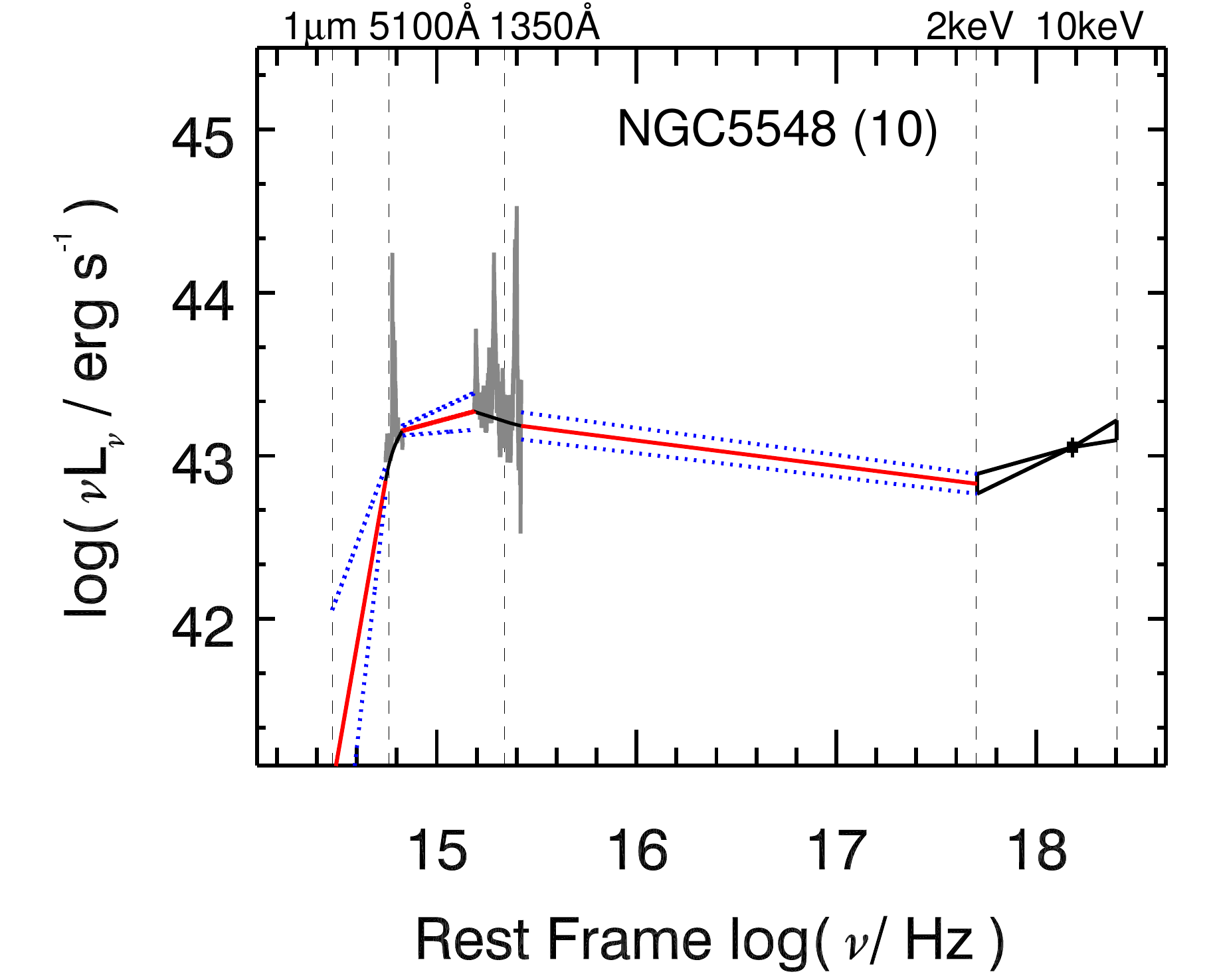} &
\includegraphics[width=0.5\linewidth,scale=1.5]{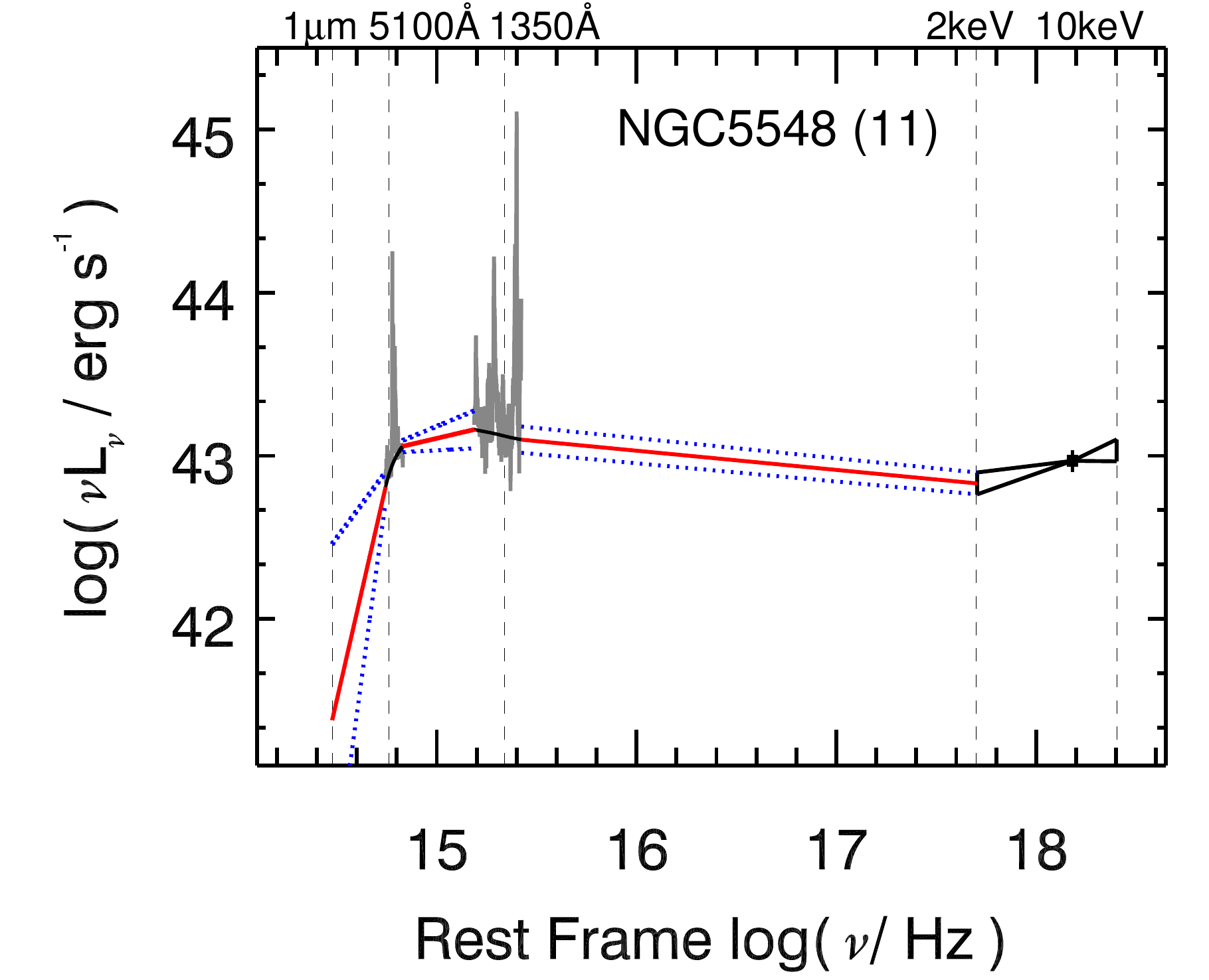}\\
\includegraphics[width=0.5\linewidth,scale=1.5]{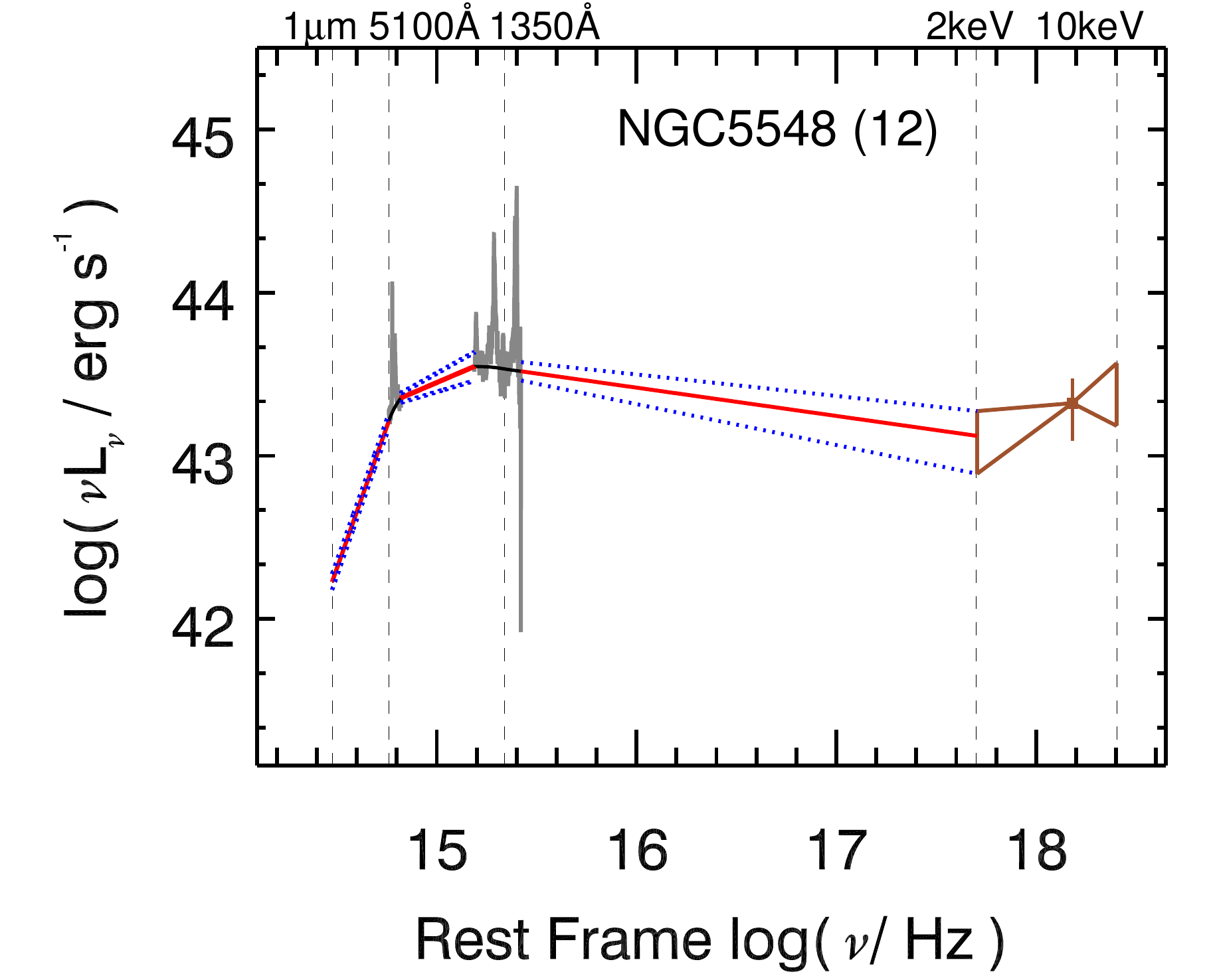} &
\includegraphics[width=0.5\linewidth,scale=1.5]{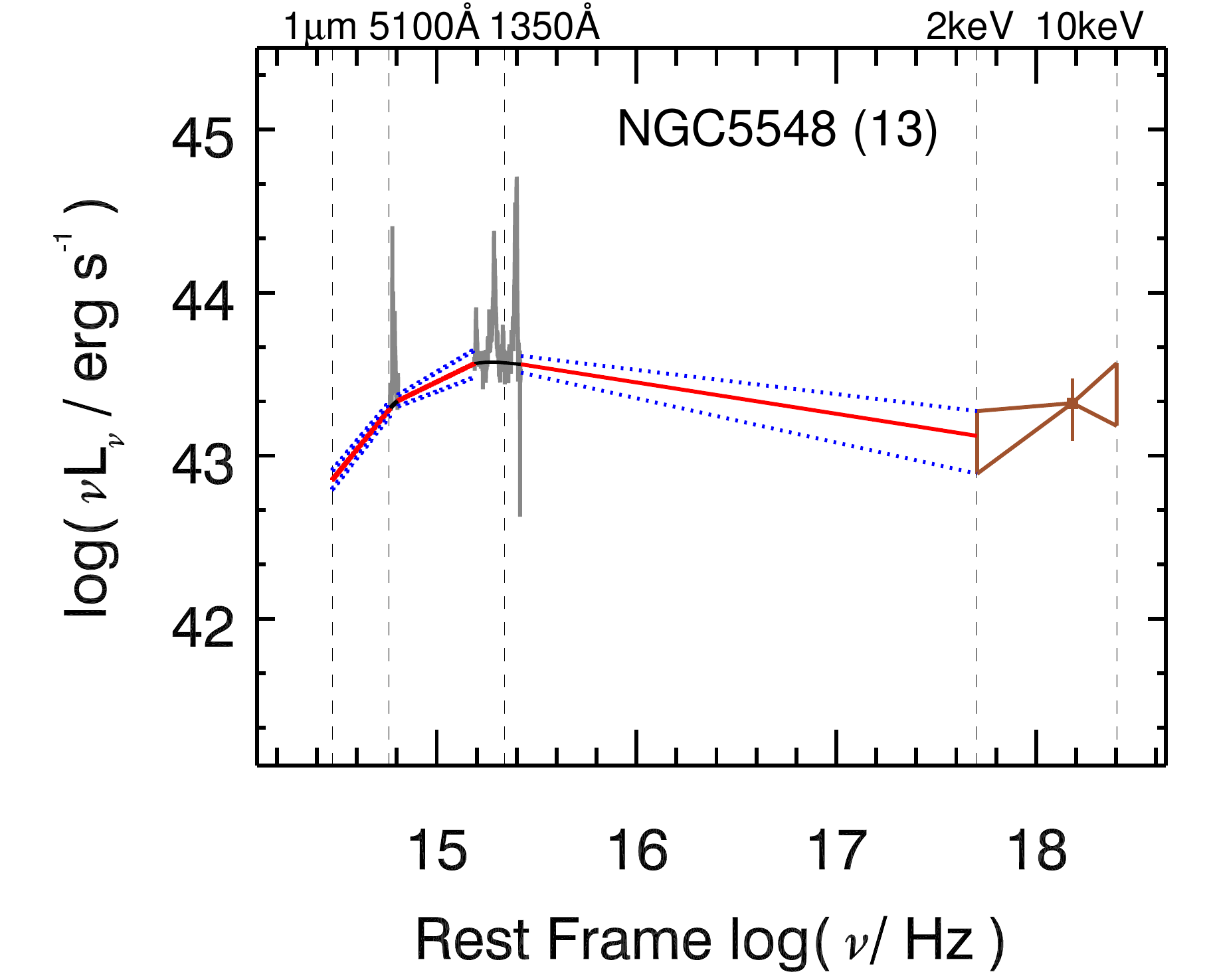}\\
\end{array}$
\end{center}
\contcaption{}
\end{figure*}

\begin{figure*}
\begin{center}$
\begin{array}{cc}
\includegraphics[width=0.5\linewidth,scale=1.5]{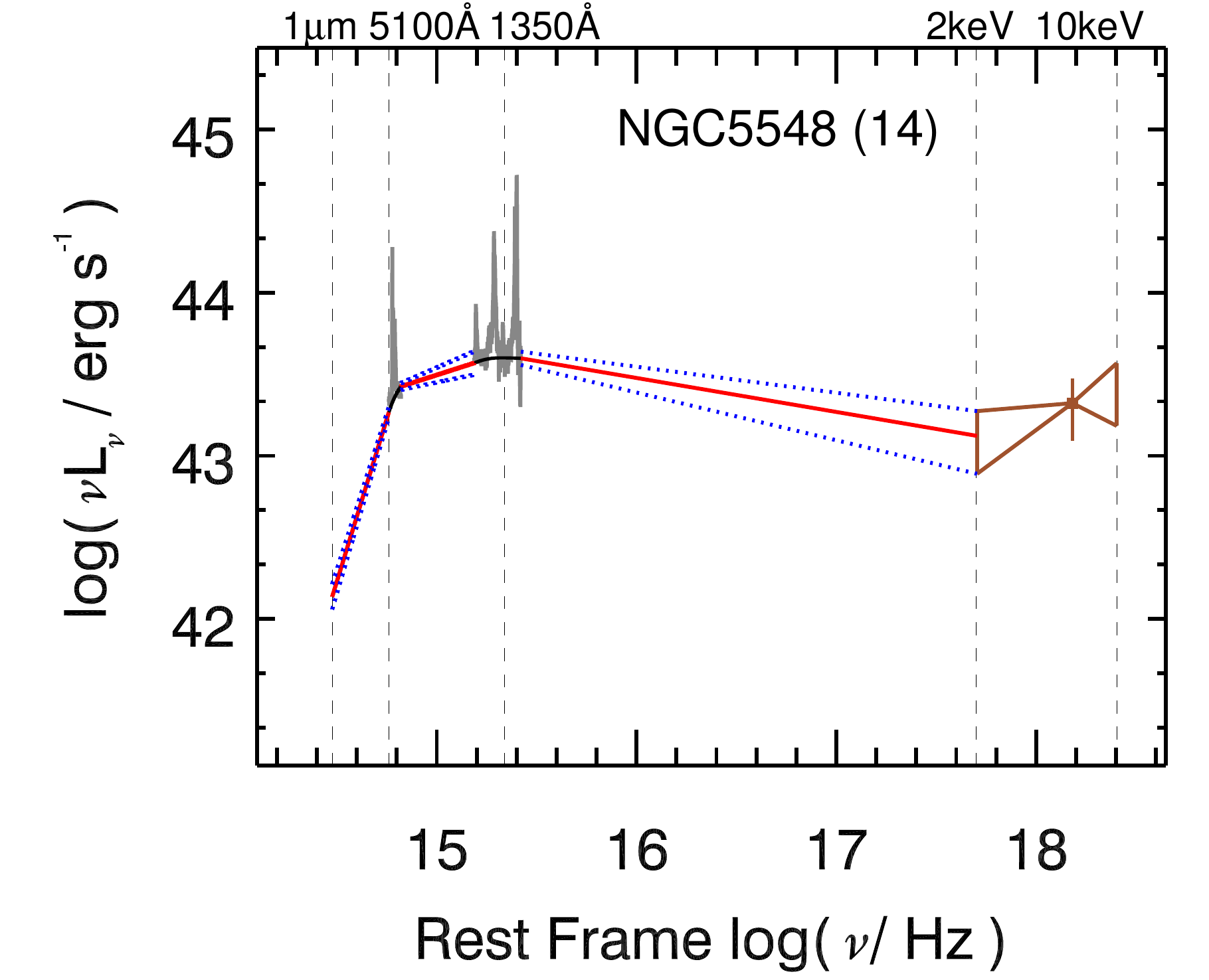} &
\includegraphics[width=0.5\linewidth,scale=1.5]{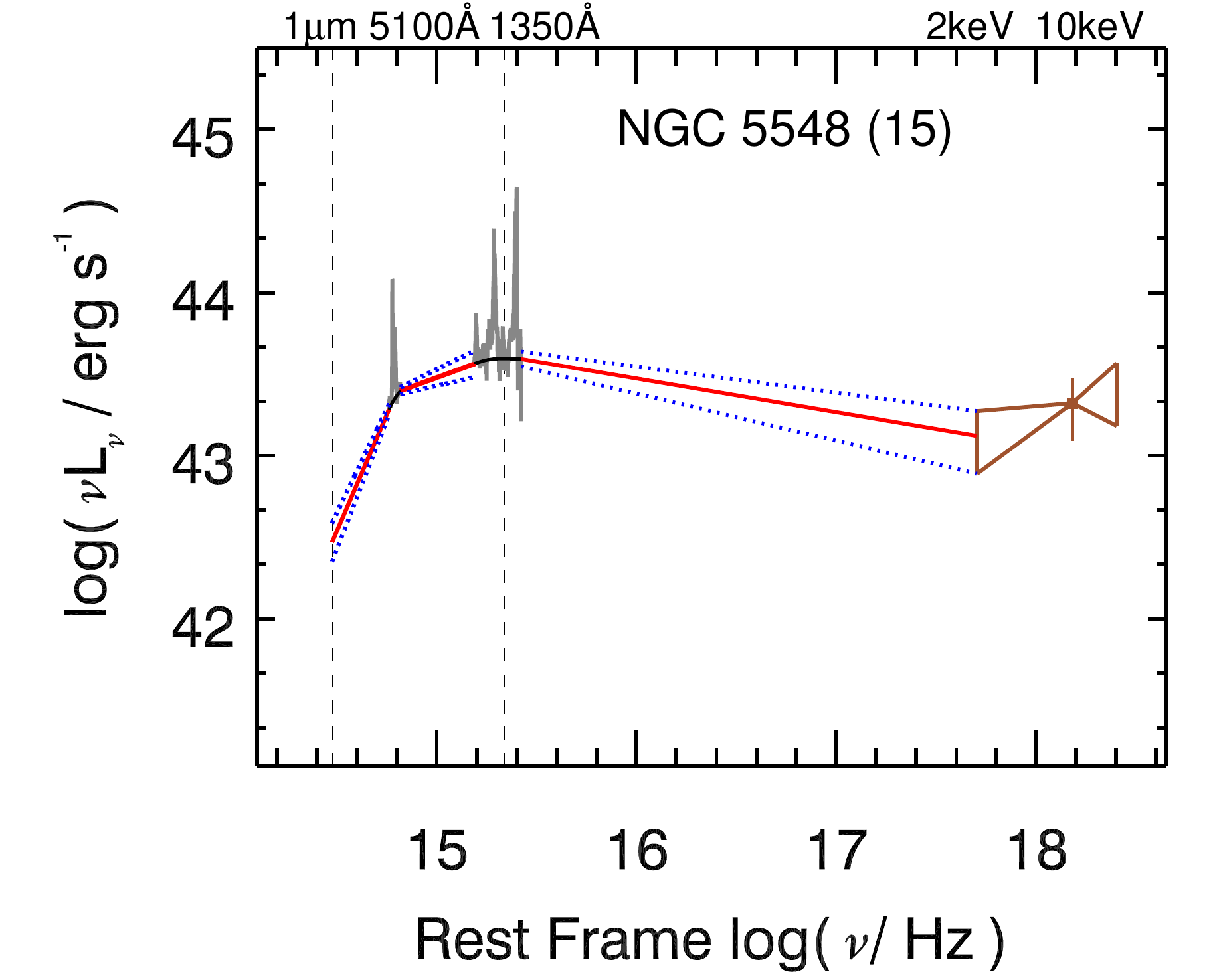}\\
\includegraphics[width=0.5\linewidth,scale=1.5]{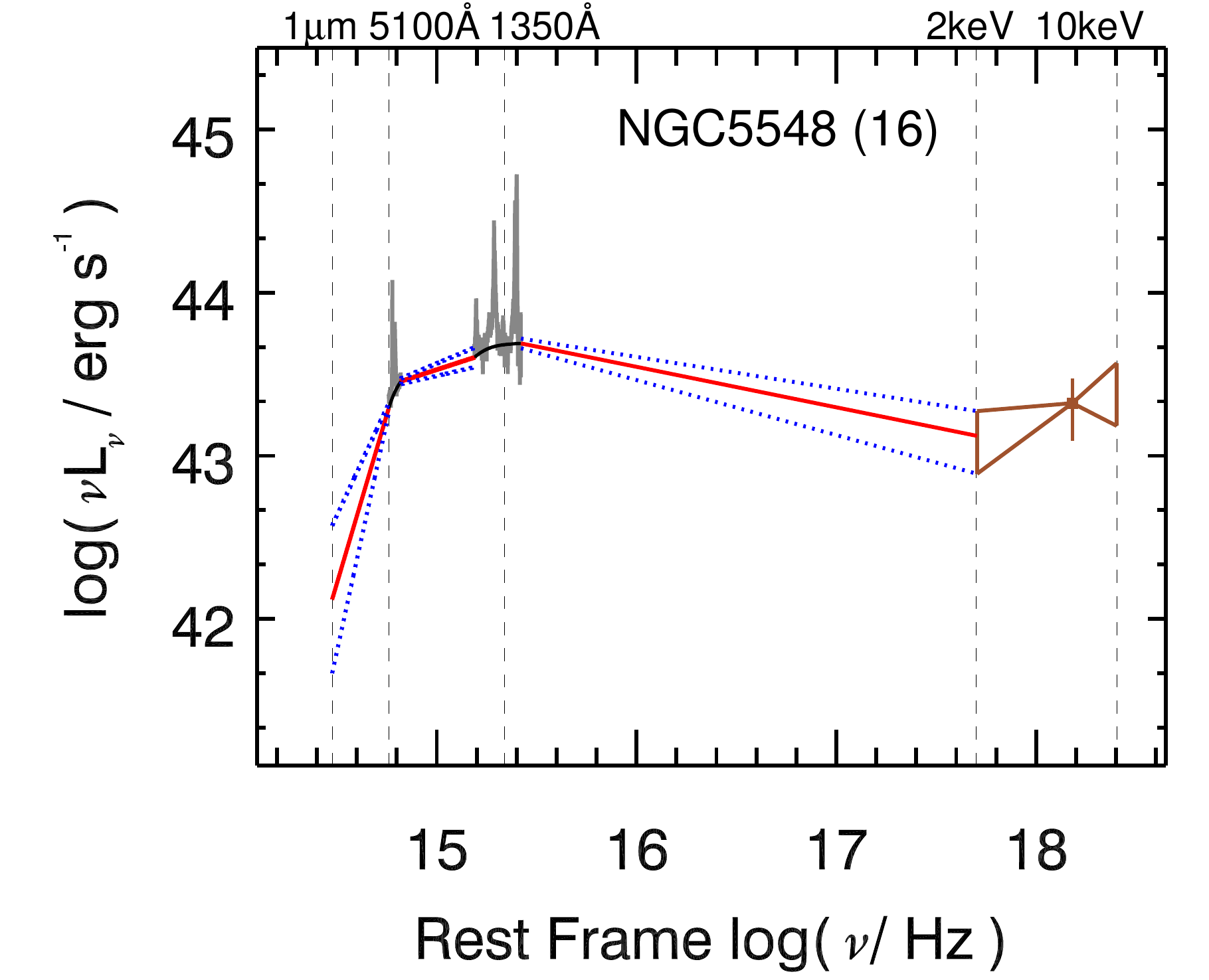} &
\includegraphics[width=0.5\linewidth,scale=1.5]{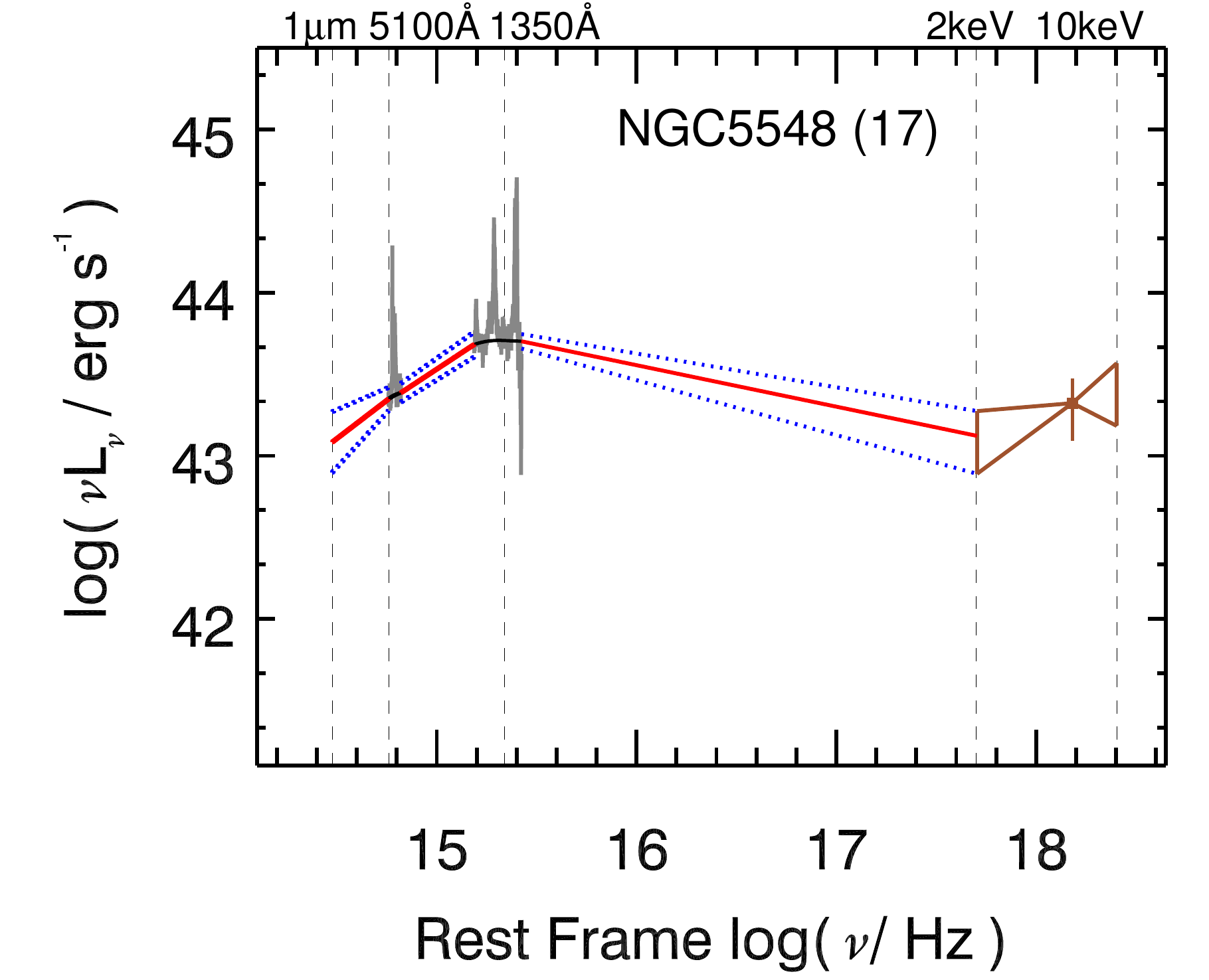}\\
\includegraphics[width=0.5\linewidth,scale=1.5]{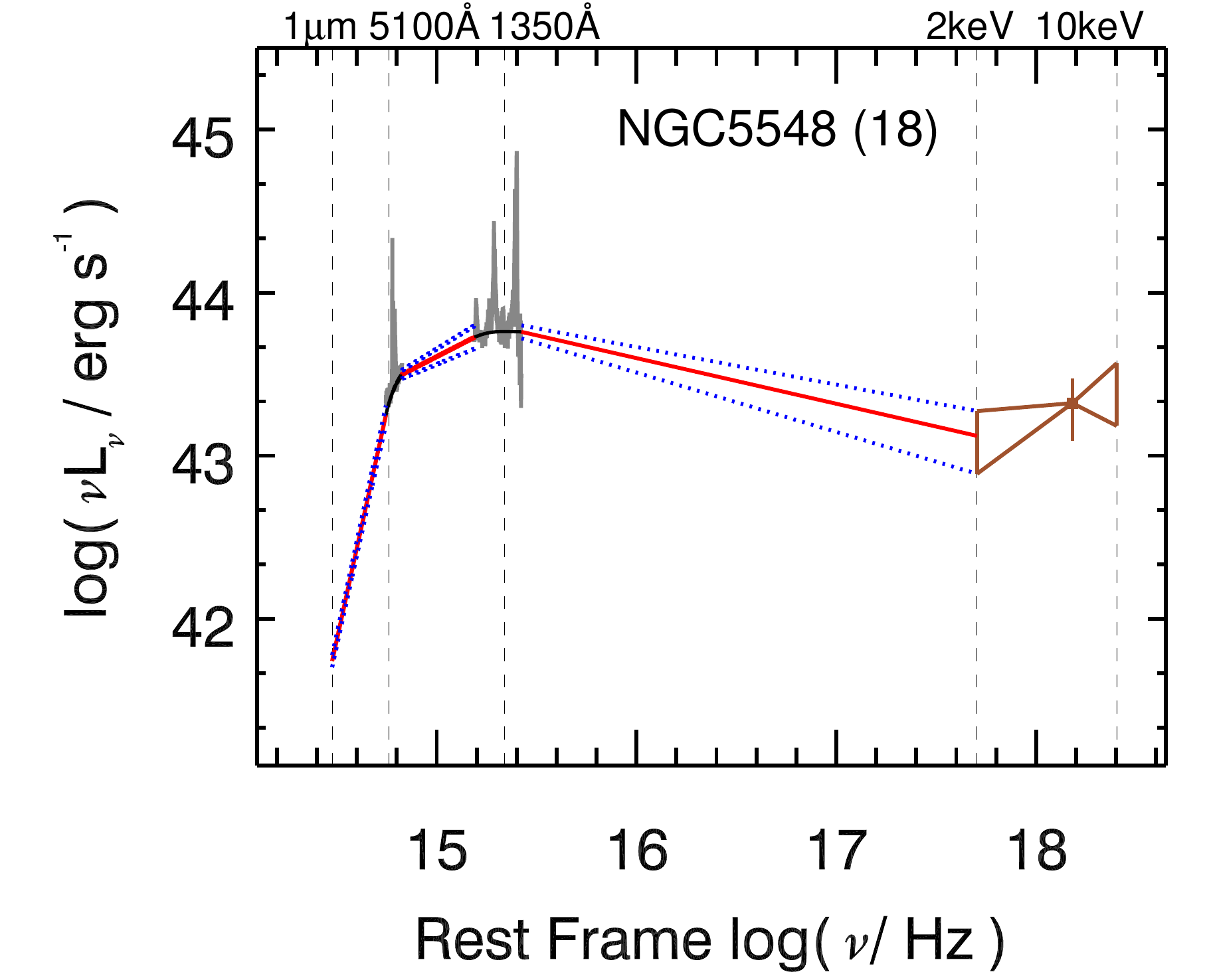} &
\includegraphics[width=0.5\linewidth,scale=1.5]{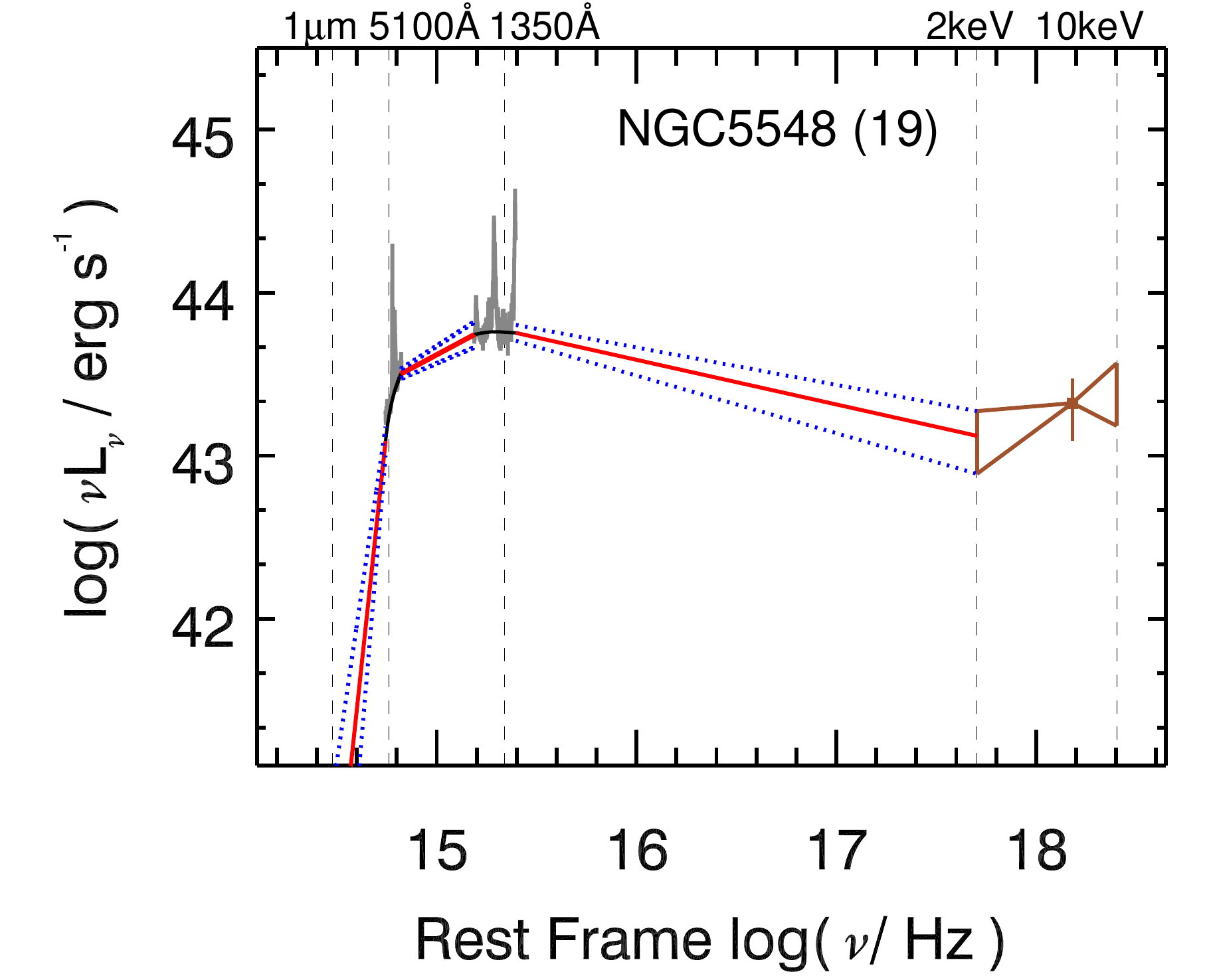}\\
\end{array}$
\end{center}
\contcaption{}
\end{figure*}

\begin{figure*}
\begin{center}$
\begin{array}{cc}
\includegraphics[width=0.5\linewidth,scale=1.5]{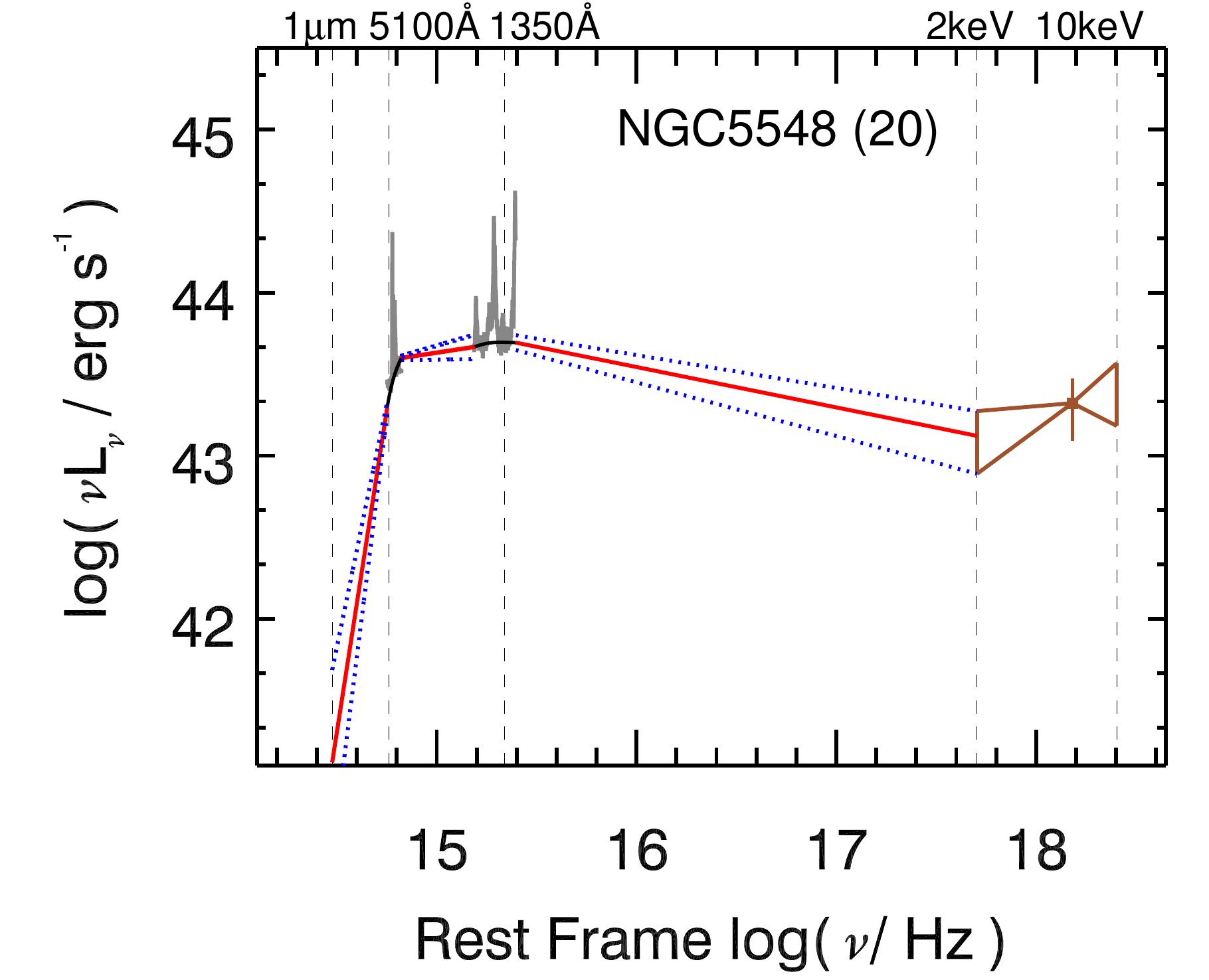} &
\includegraphics[width=0.5\linewidth,scale=1.5]{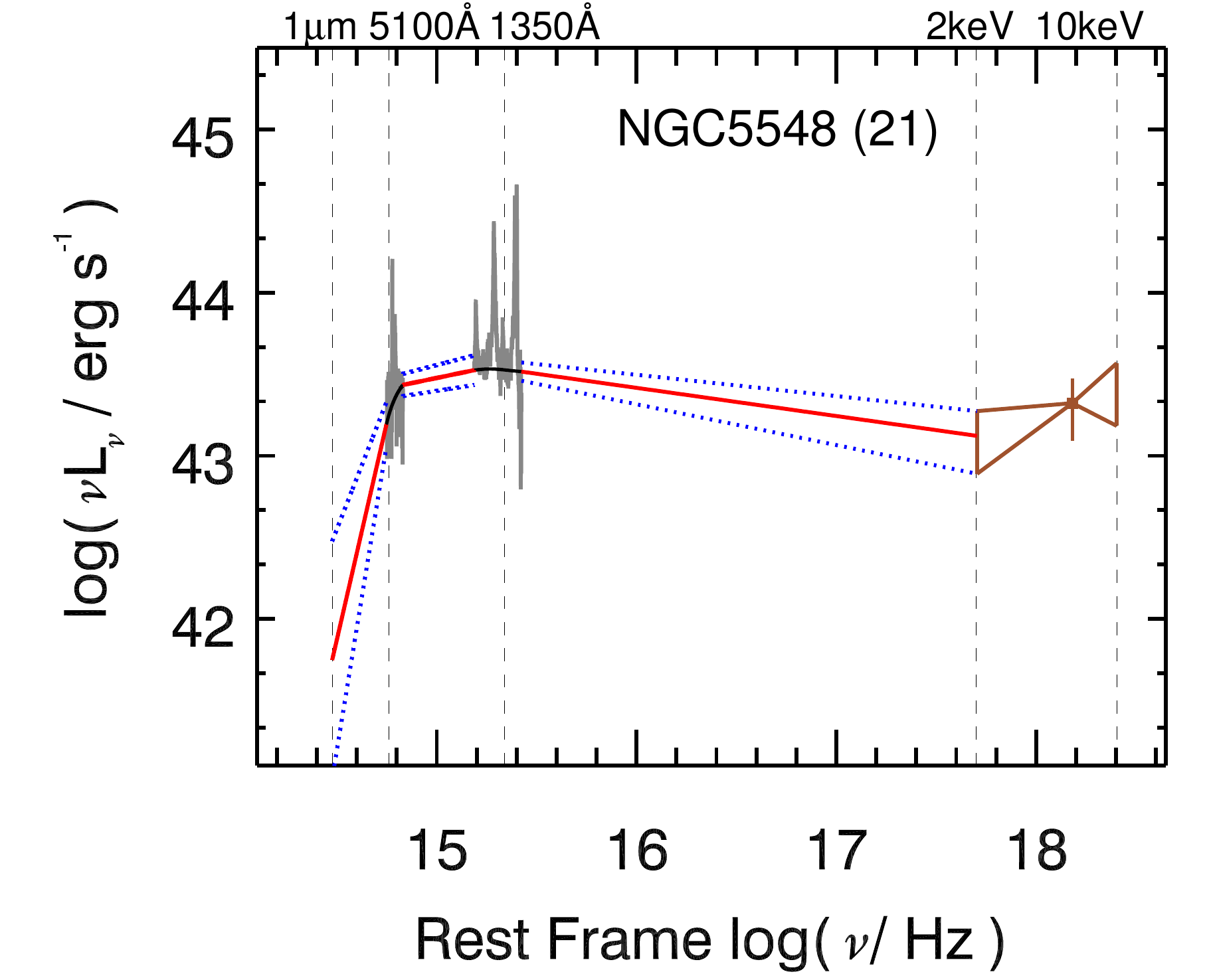}\\
\includegraphics[width=0.5\linewidth,scale=1.5]{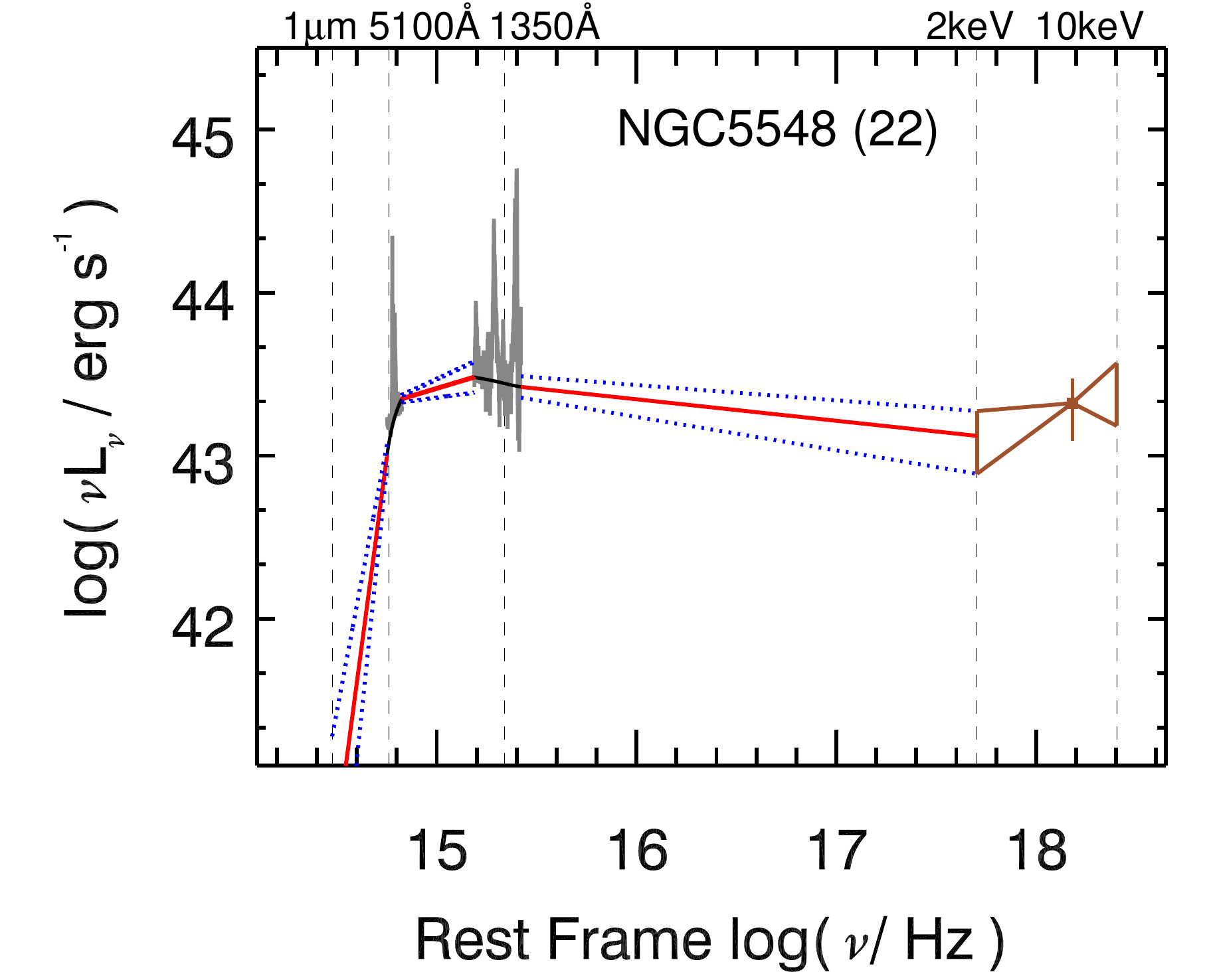} &
\includegraphics[width=0.5\linewidth,scale=1.5]{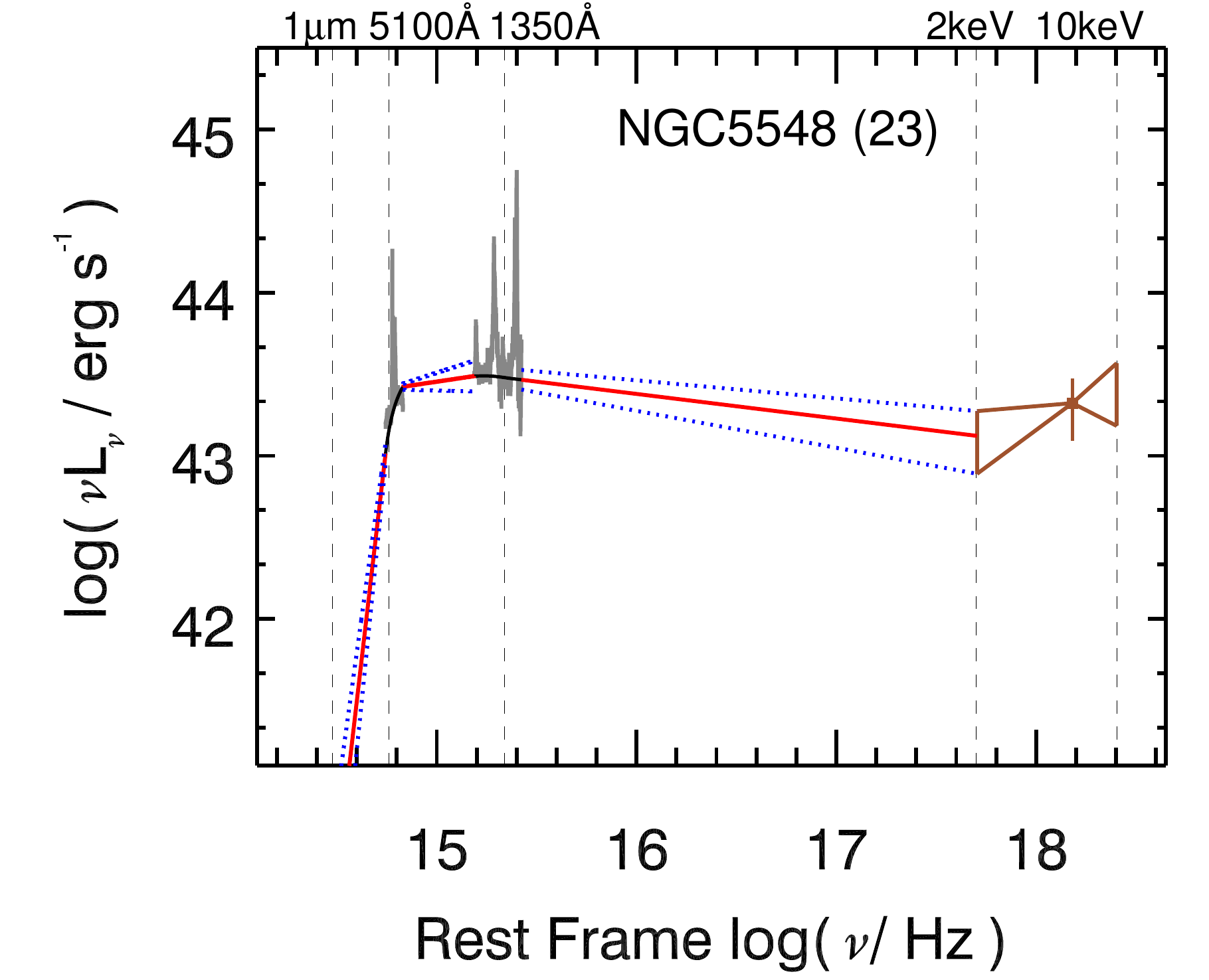}\\
\includegraphics[width=0.5\linewidth,scale=1.5]{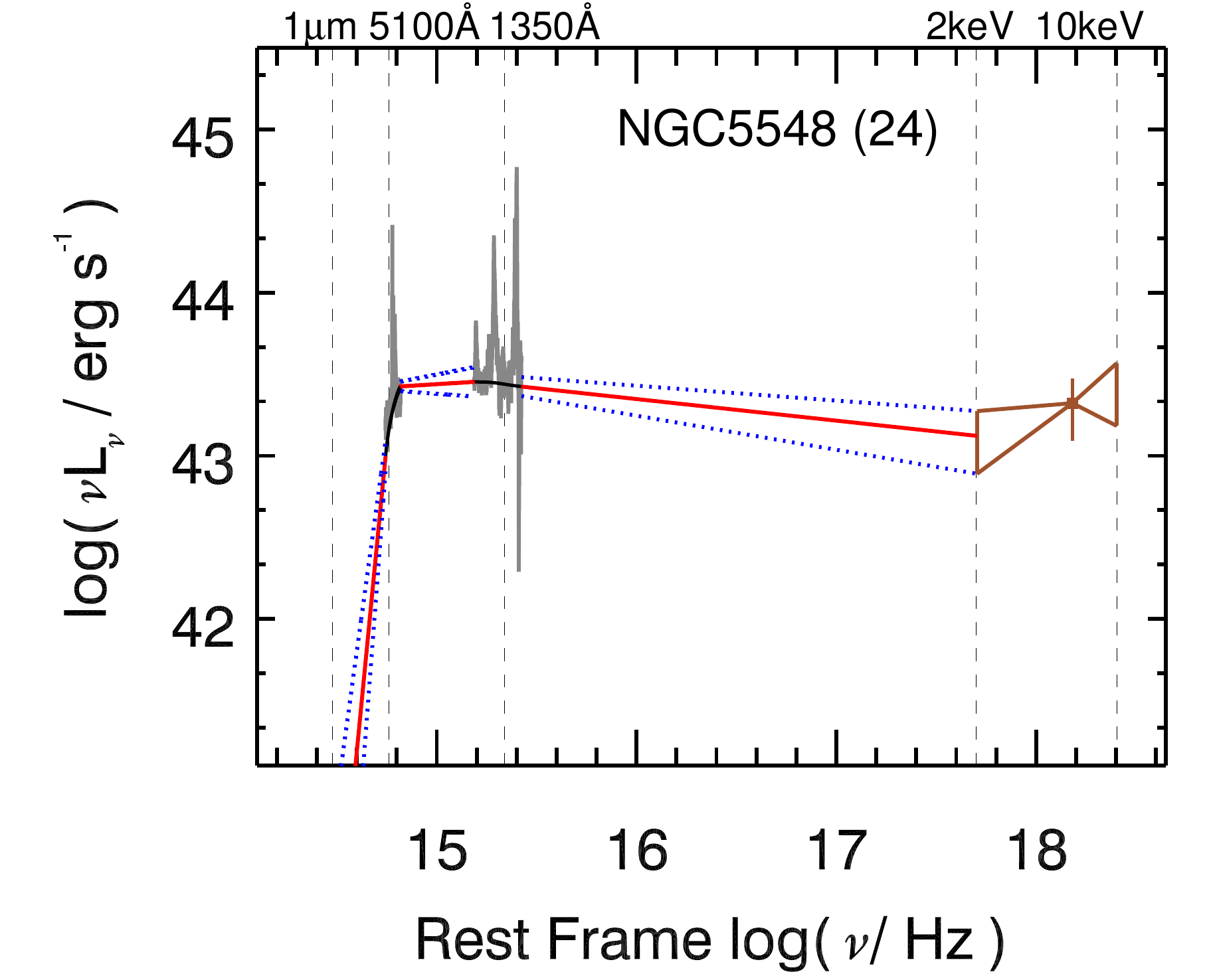} &
\includegraphics[width=0.5\linewidth,scale=1.5]{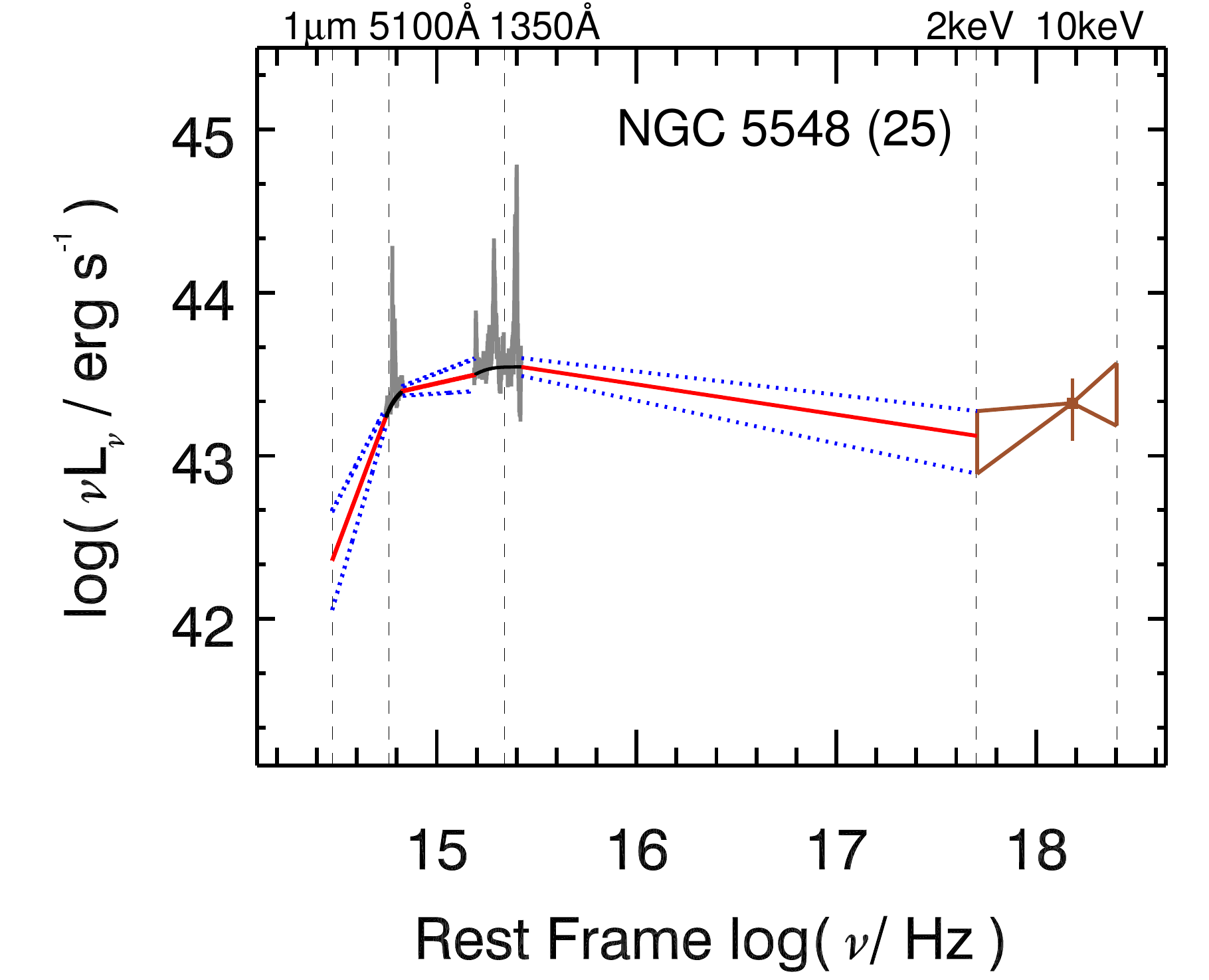}\\
\end{array}$
\end{center}
\contcaption{}
\end{figure*}

\begin{figure*}
\begin{center}$
\begin{array}{cc}
\includegraphics[width=0.5\linewidth,scale=1.5]{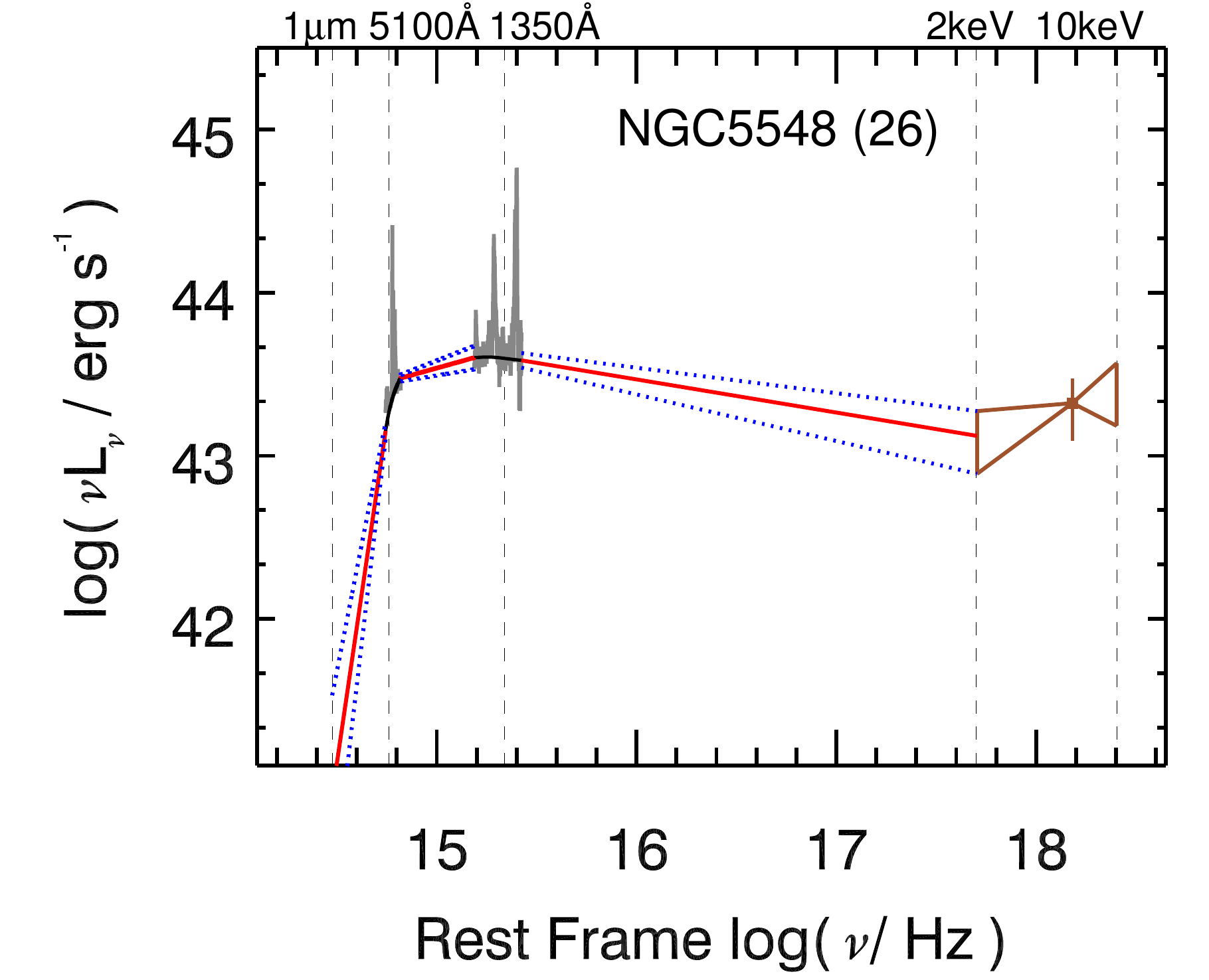} &
\includegraphics[width=0.5\linewidth,scale=1.5]{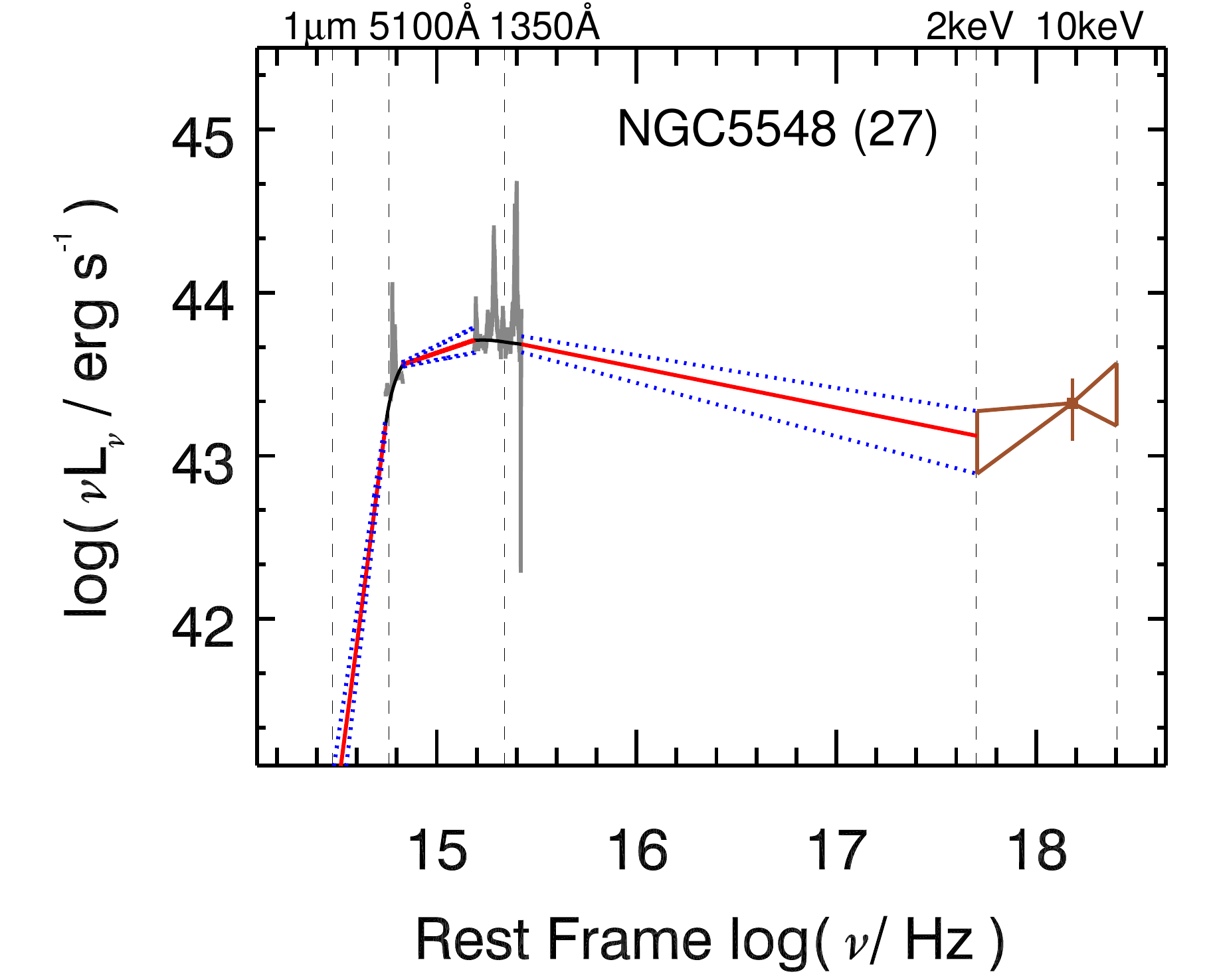}\\
\includegraphics[width=0.5\linewidth,scale=1.5]{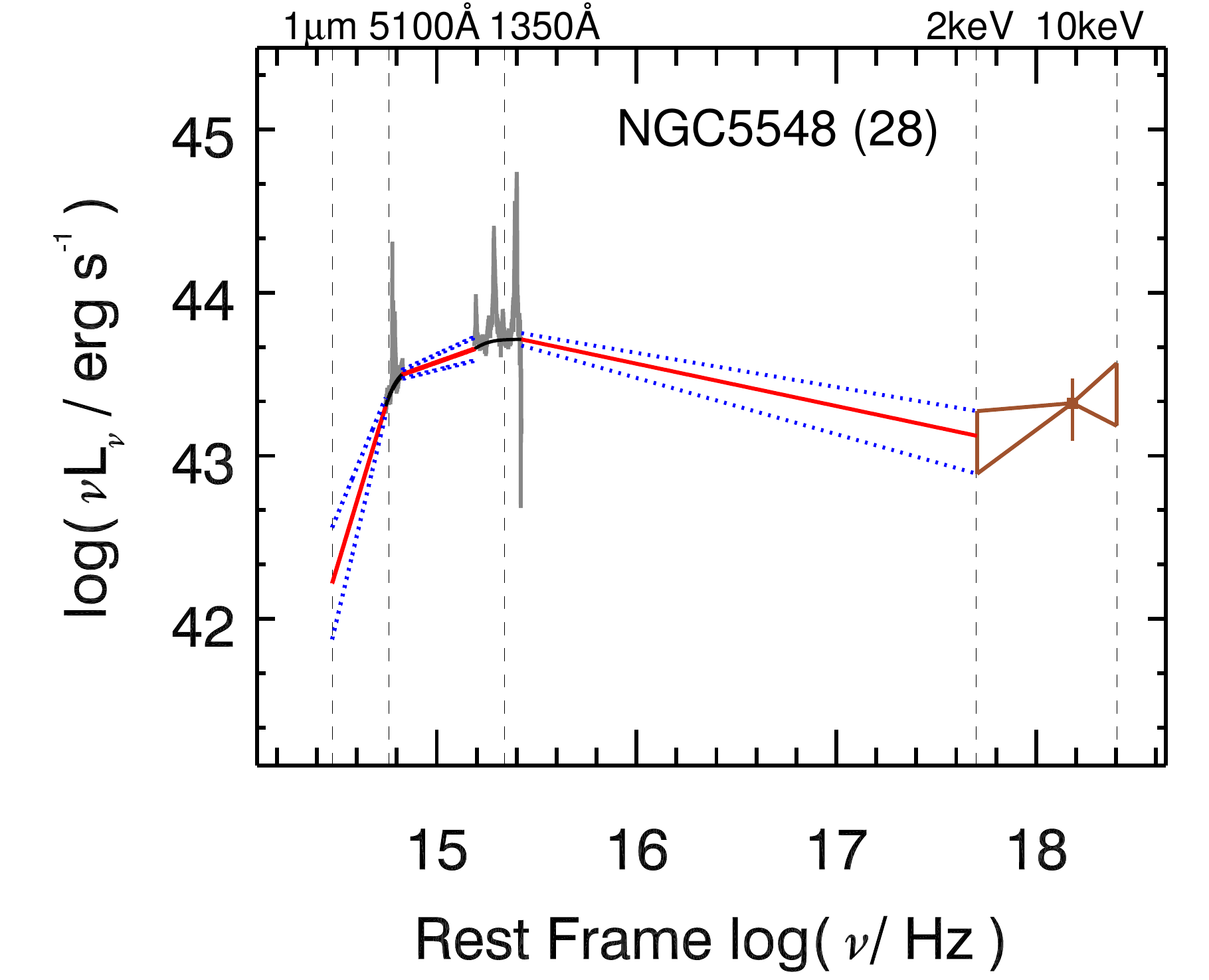} &
\includegraphics[width=0.5\linewidth,scale=1.5]{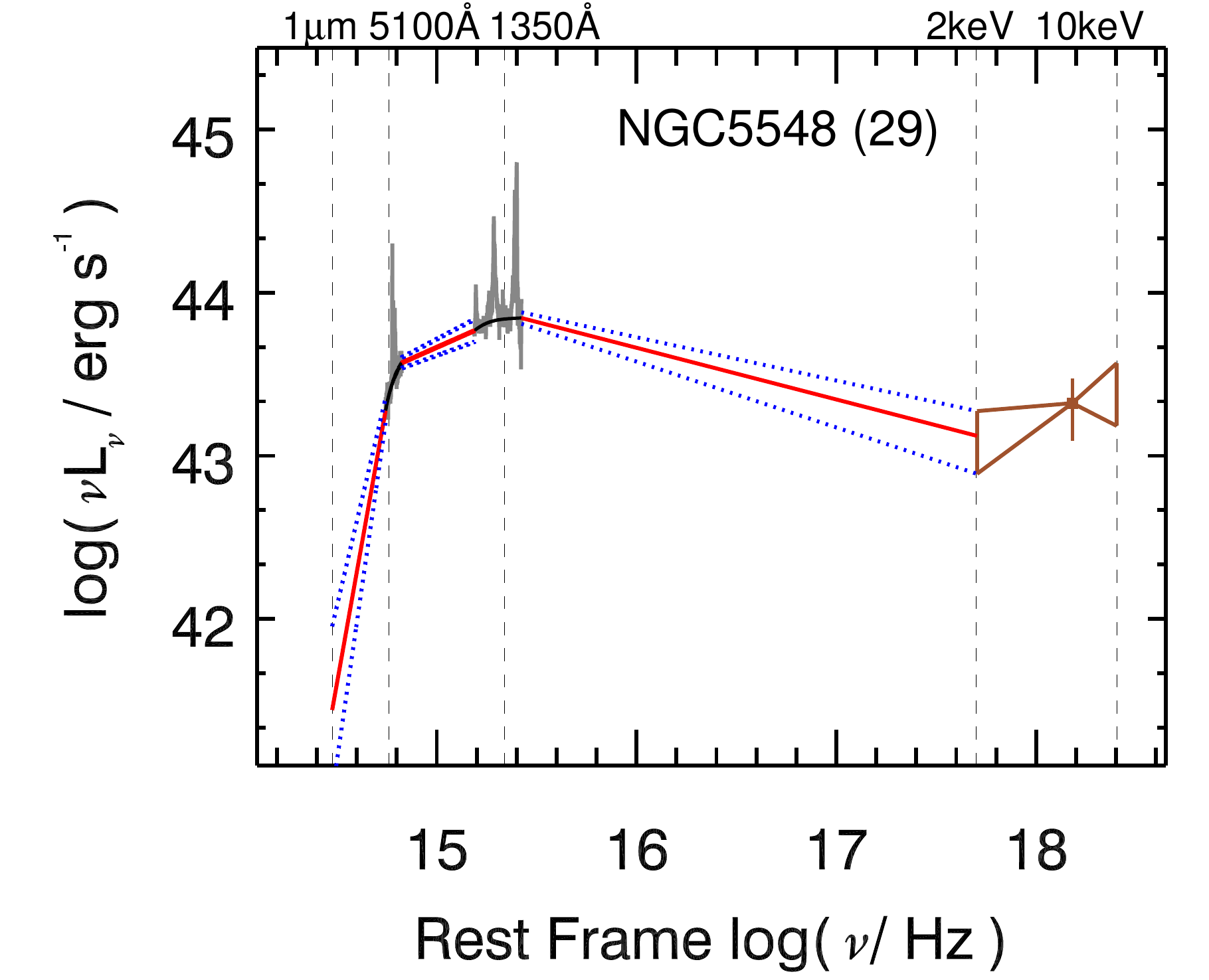}\\
\includegraphics[width=0.5\linewidth,scale=1.5]{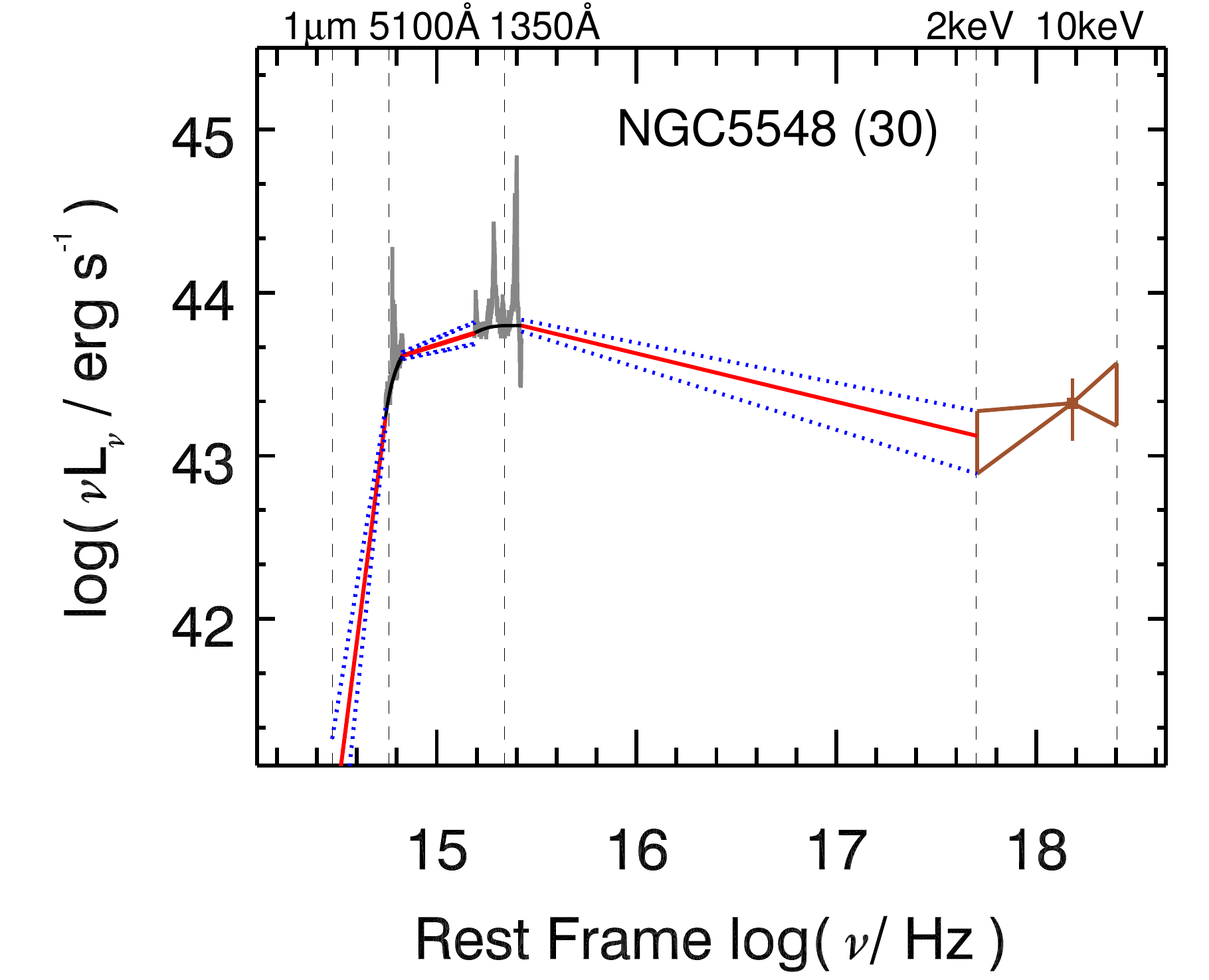} &
\includegraphics[width=0.5\linewidth,scale=1.5]{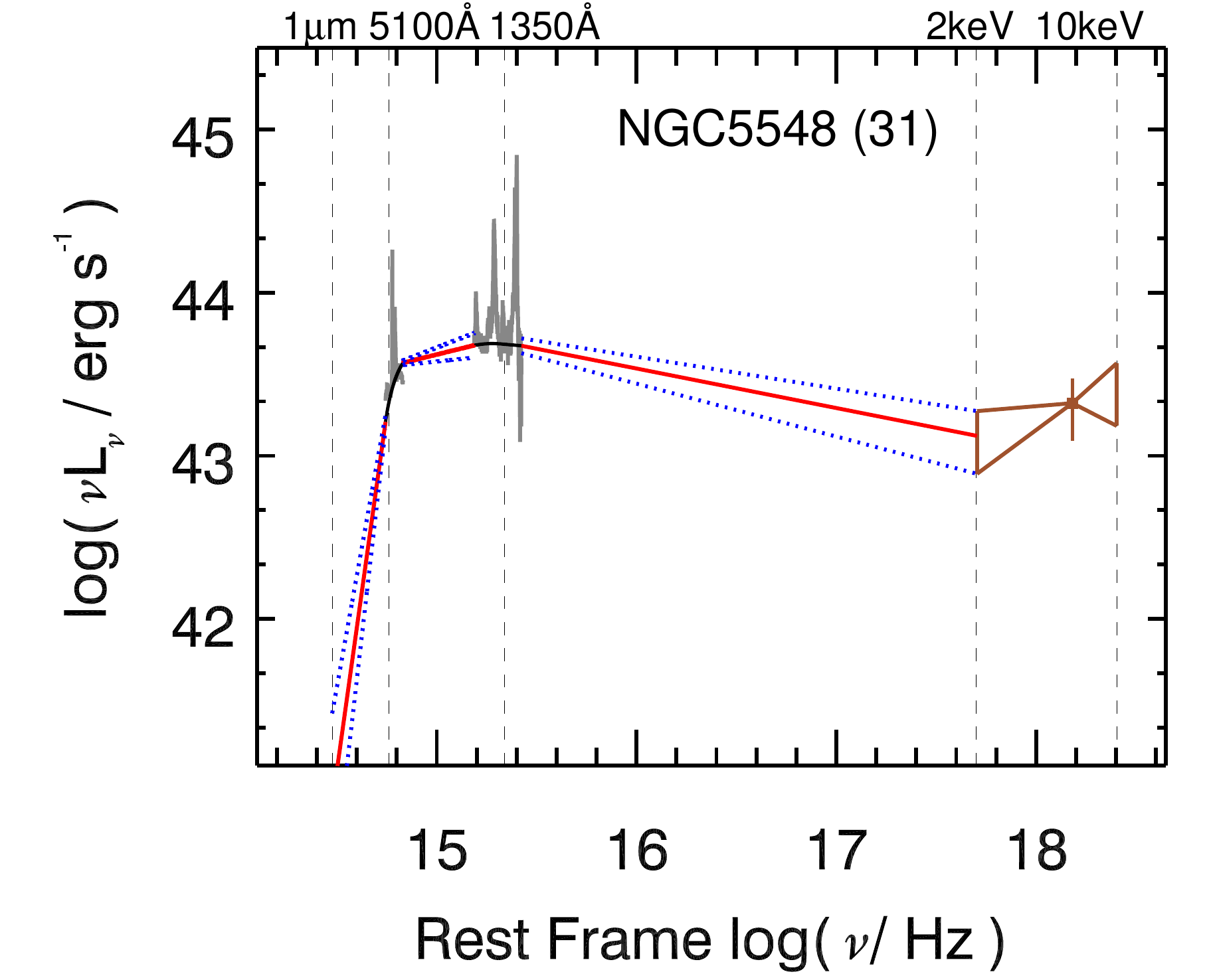}\\
\end{array}$
\end{center}
\contcaption{}
\end{figure*}

\begin{figure*}
\begin{center}$
\begin{array}{cc}
\includegraphics[width=0.5\linewidth,scale=1.5]{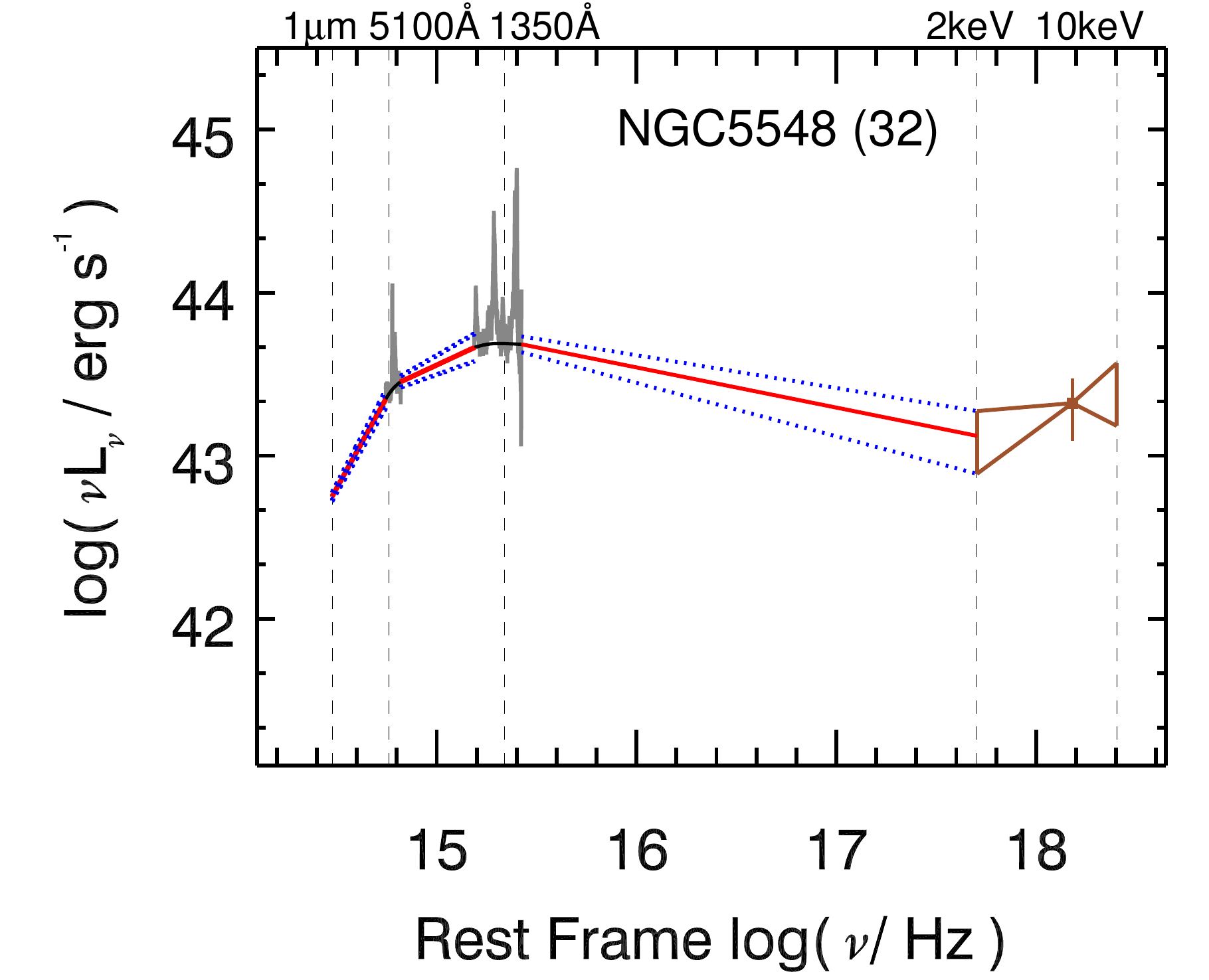} &
\includegraphics[width=0.5\linewidth,scale=1.5]{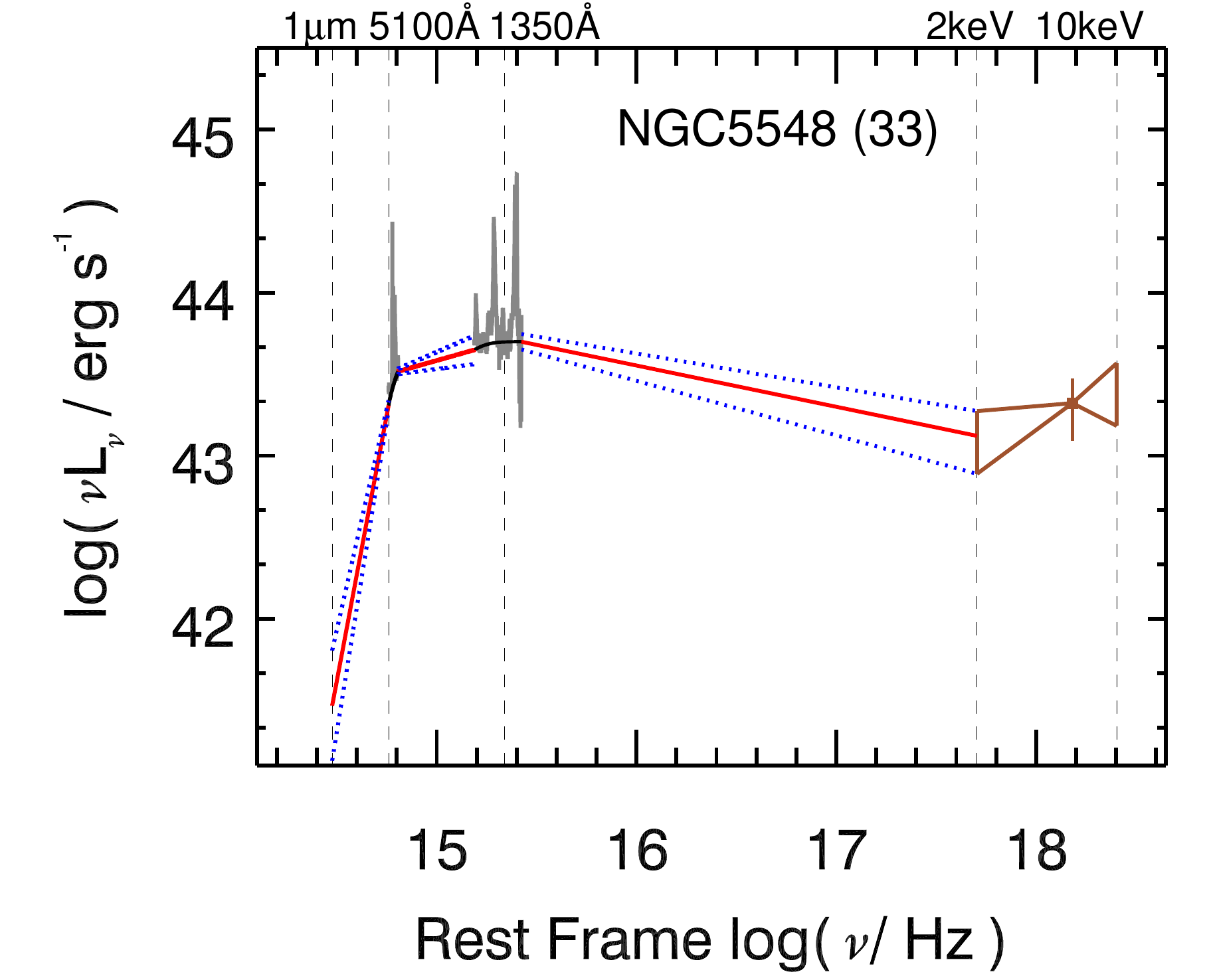}\\
\includegraphics[width=0.5\linewidth,scale=1.5]{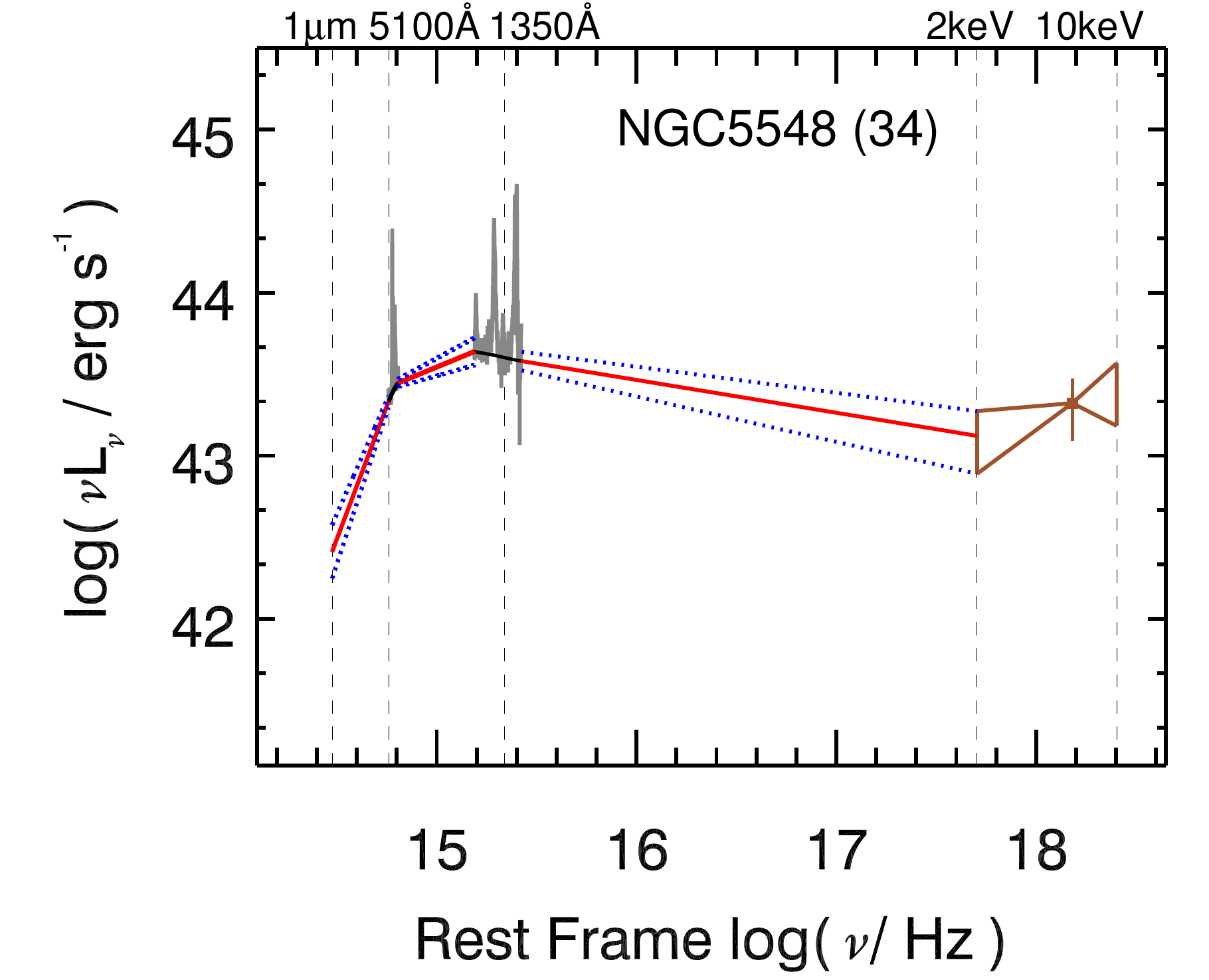} &
\includegraphics[width=0.5\linewidth,scale=1.5]{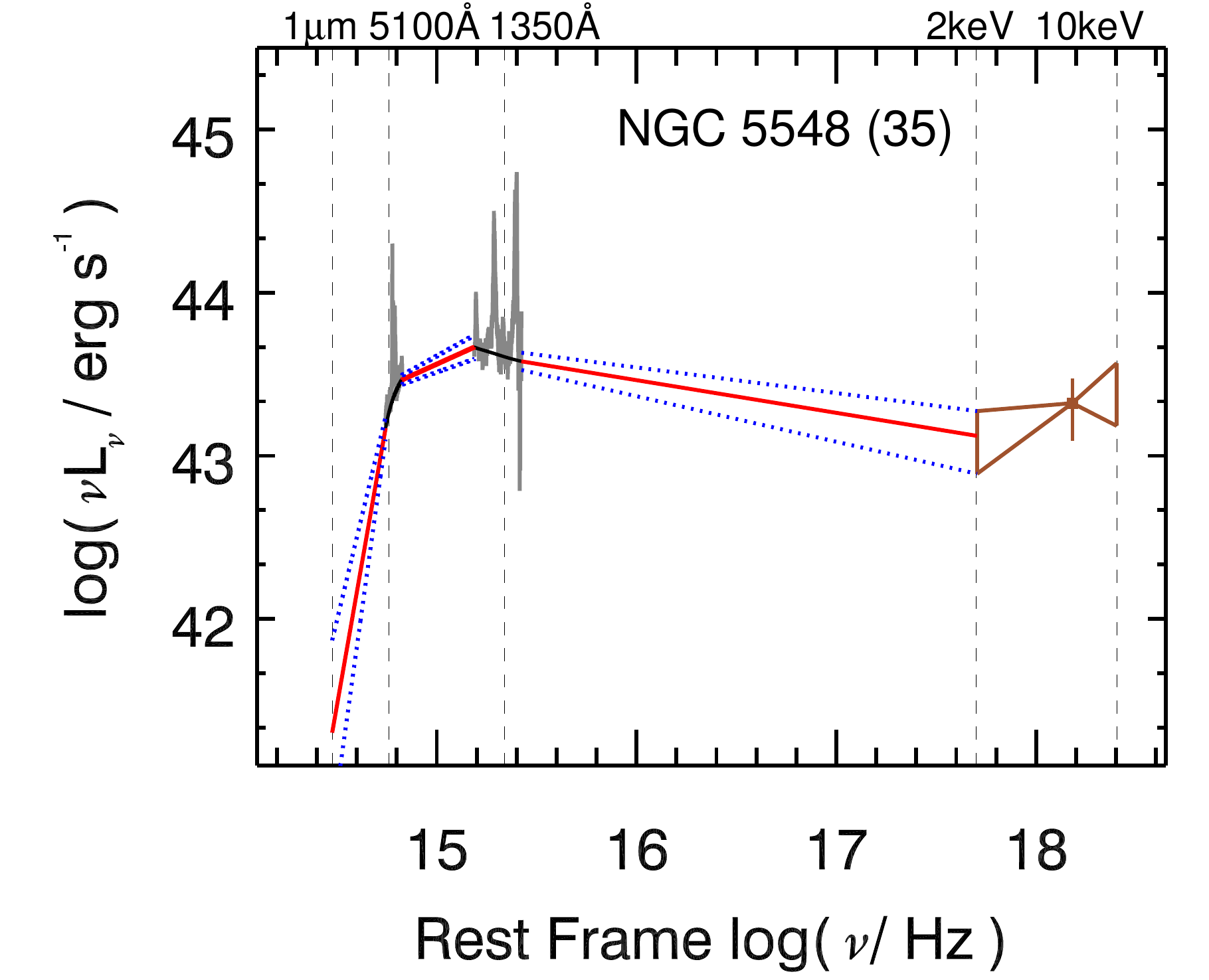}\\
\includegraphics[width=0.5\linewidth,scale=1.5]{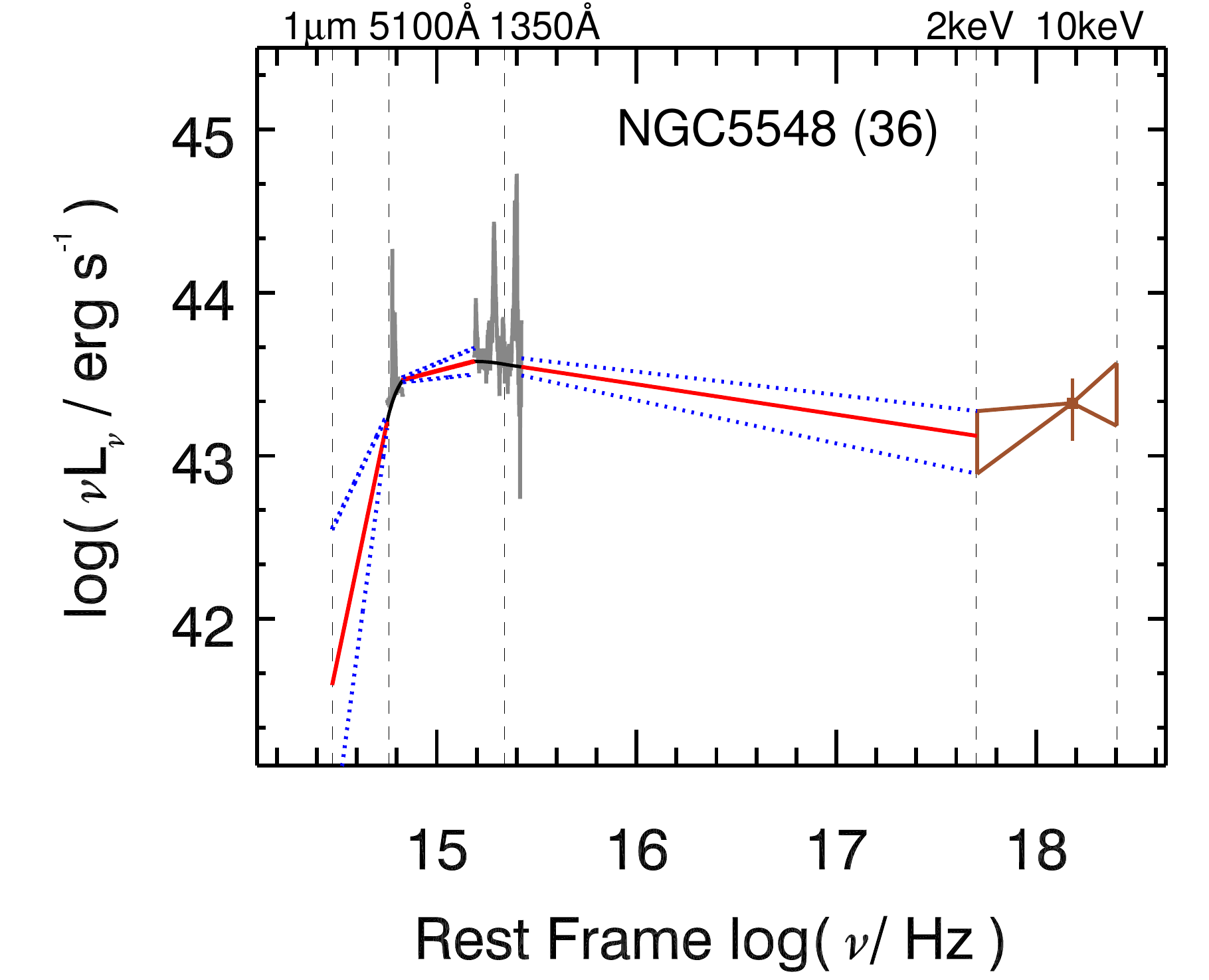} &
\includegraphics[width=0.5\linewidth,scale=1.5]{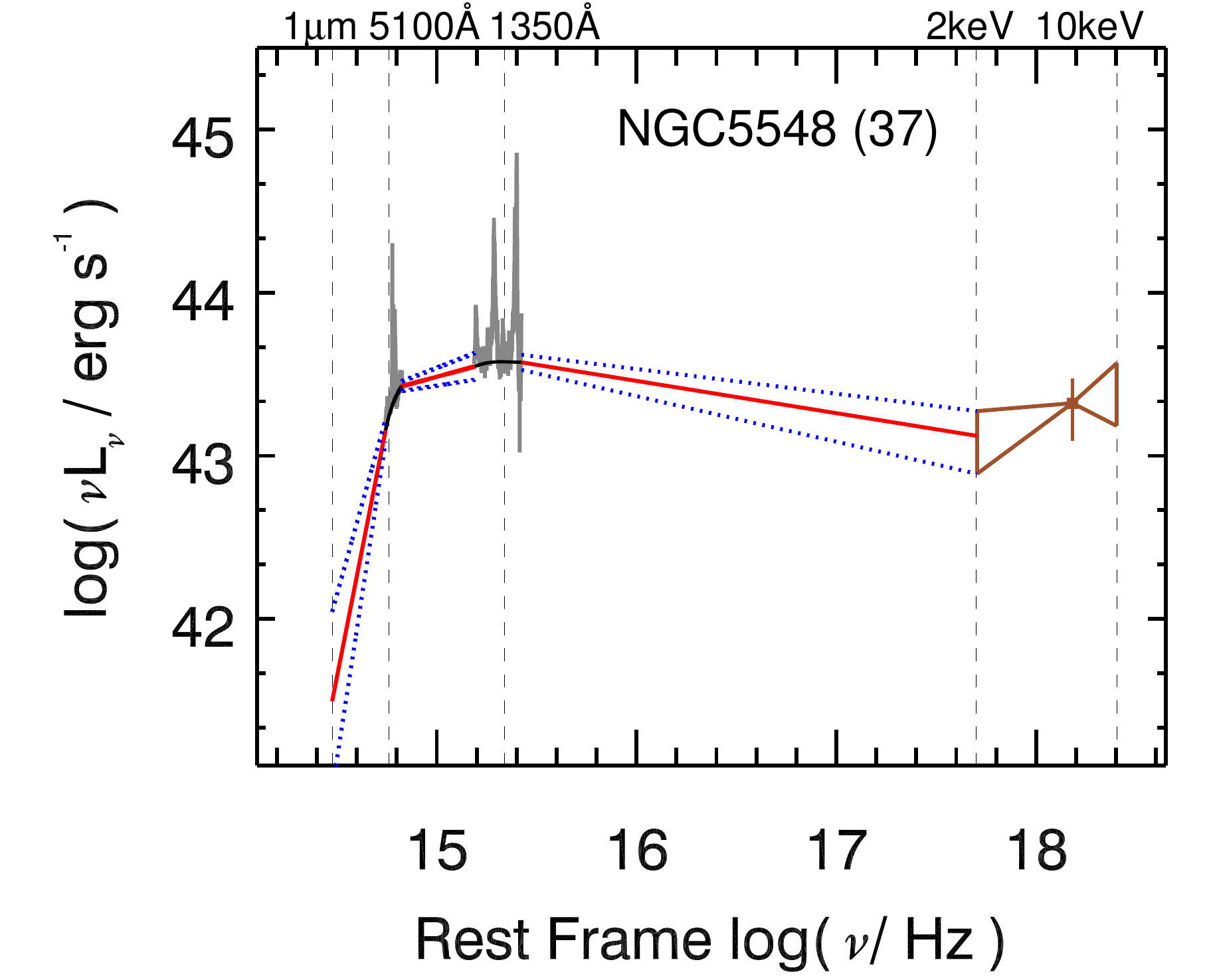}\\
\end{array}$
\end{center}
\contcaption{}
\end{figure*}

\begin{figure*}
\begin{center}$
\begin{array}{cc}
\includegraphics[width=0.5\linewidth,scale=1.5]{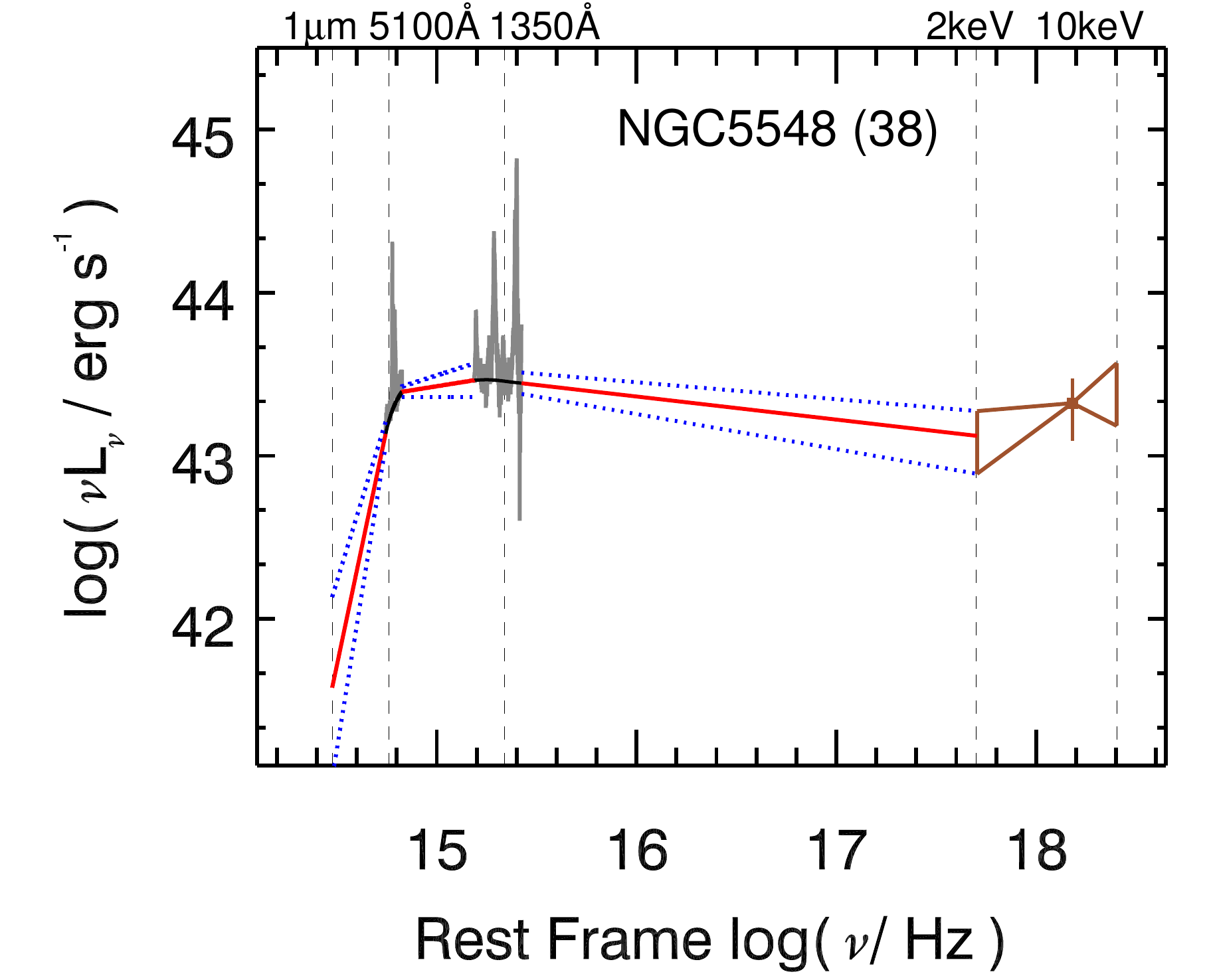} &
\includegraphics[width=0.5\linewidth,scale=1.5]{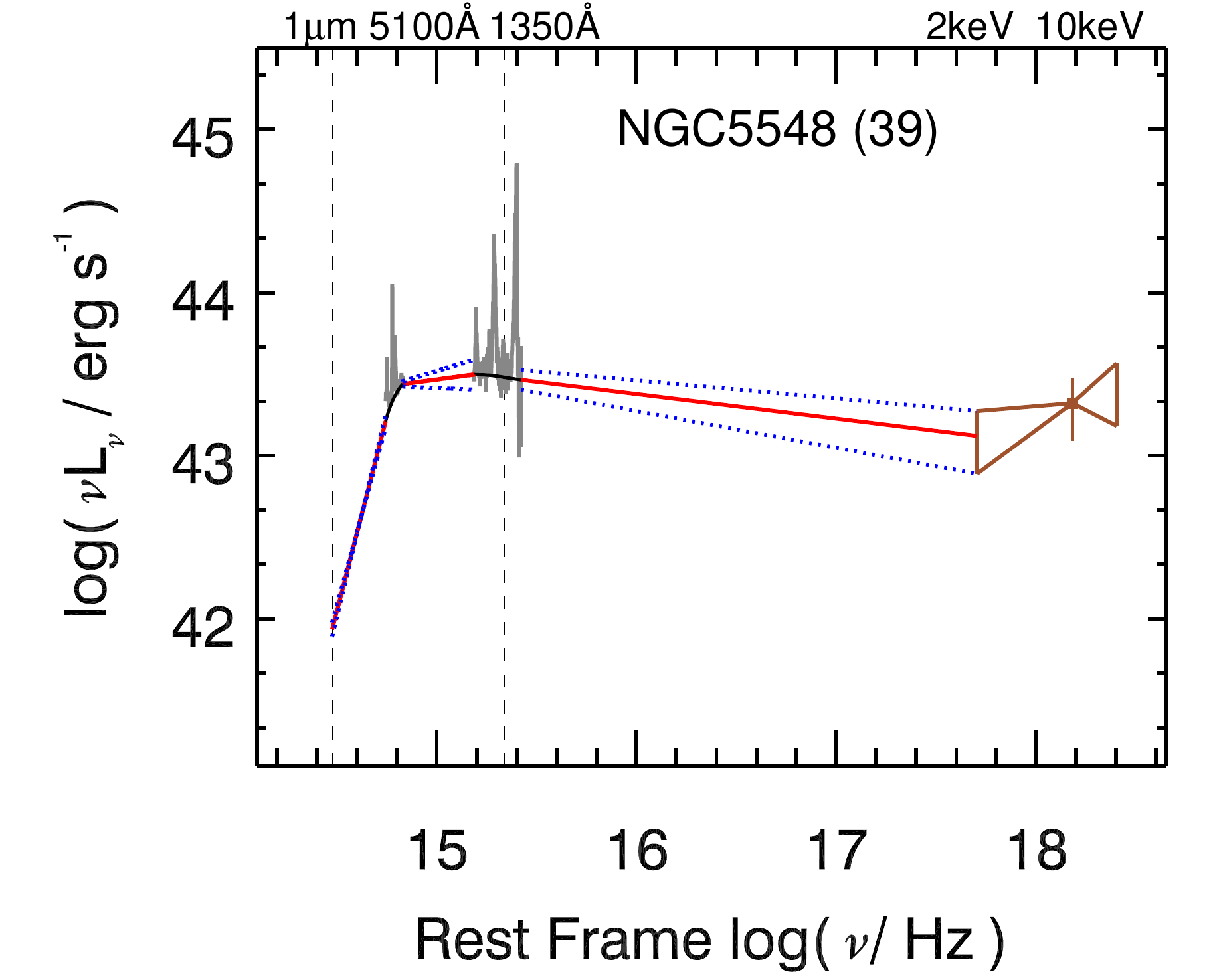}\\
\includegraphics[width=0.5\linewidth,scale=1.5]{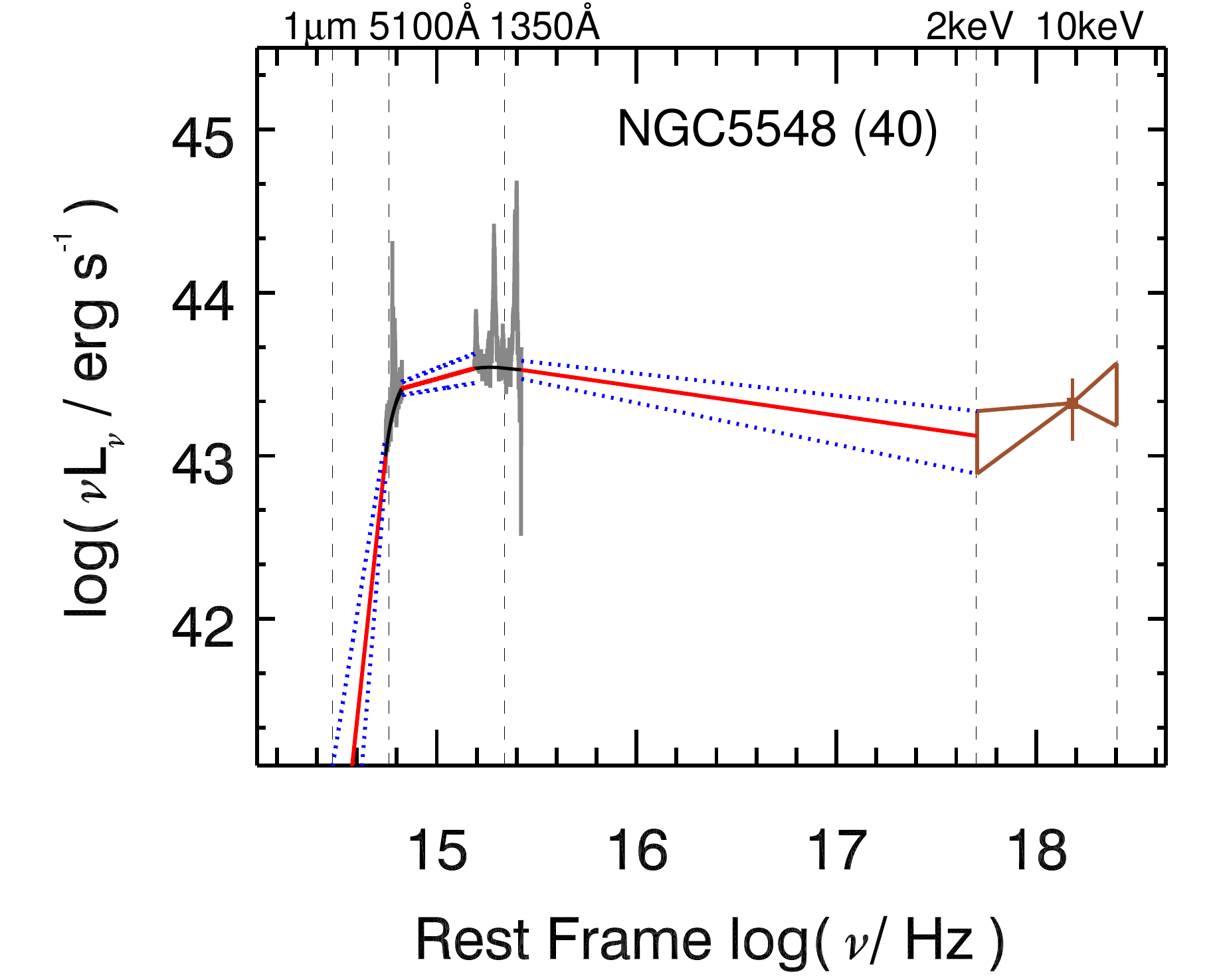} &
\includegraphics[width=0.5\linewidth,scale=1.5]{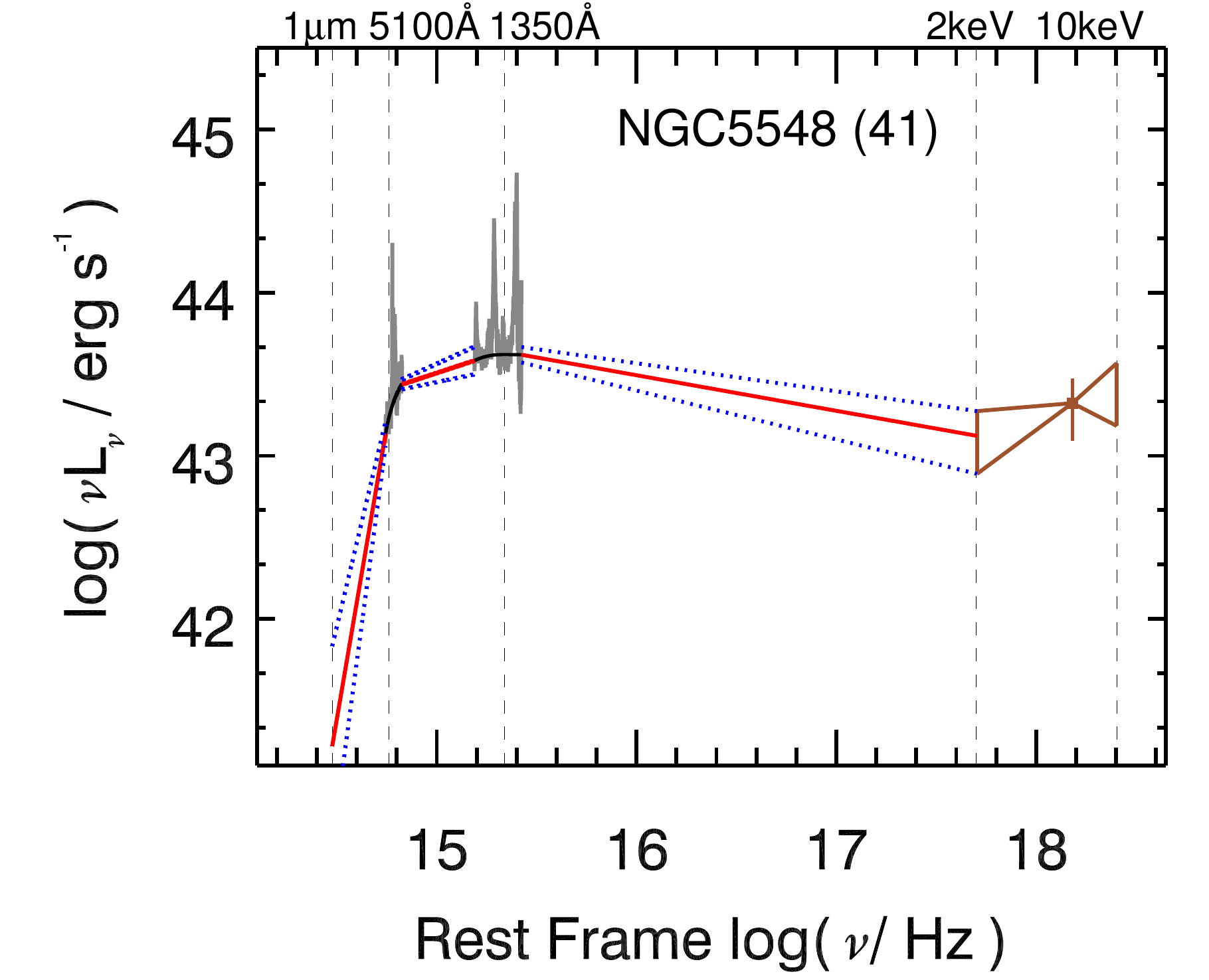}\\
\includegraphics[width=0.5\linewidth,scale=1.5]{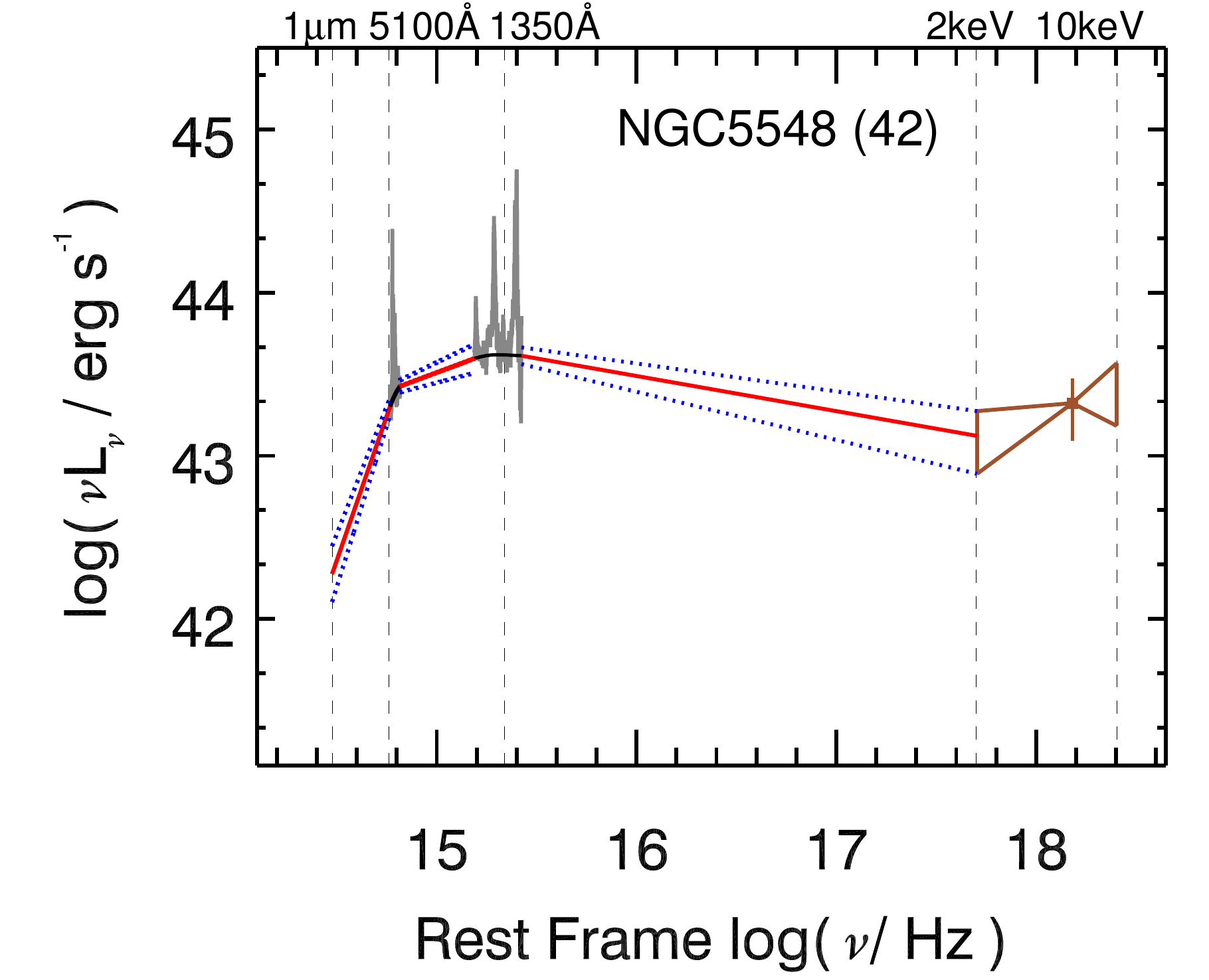} &
\includegraphics[width=0.5\linewidth,scale=1.5]{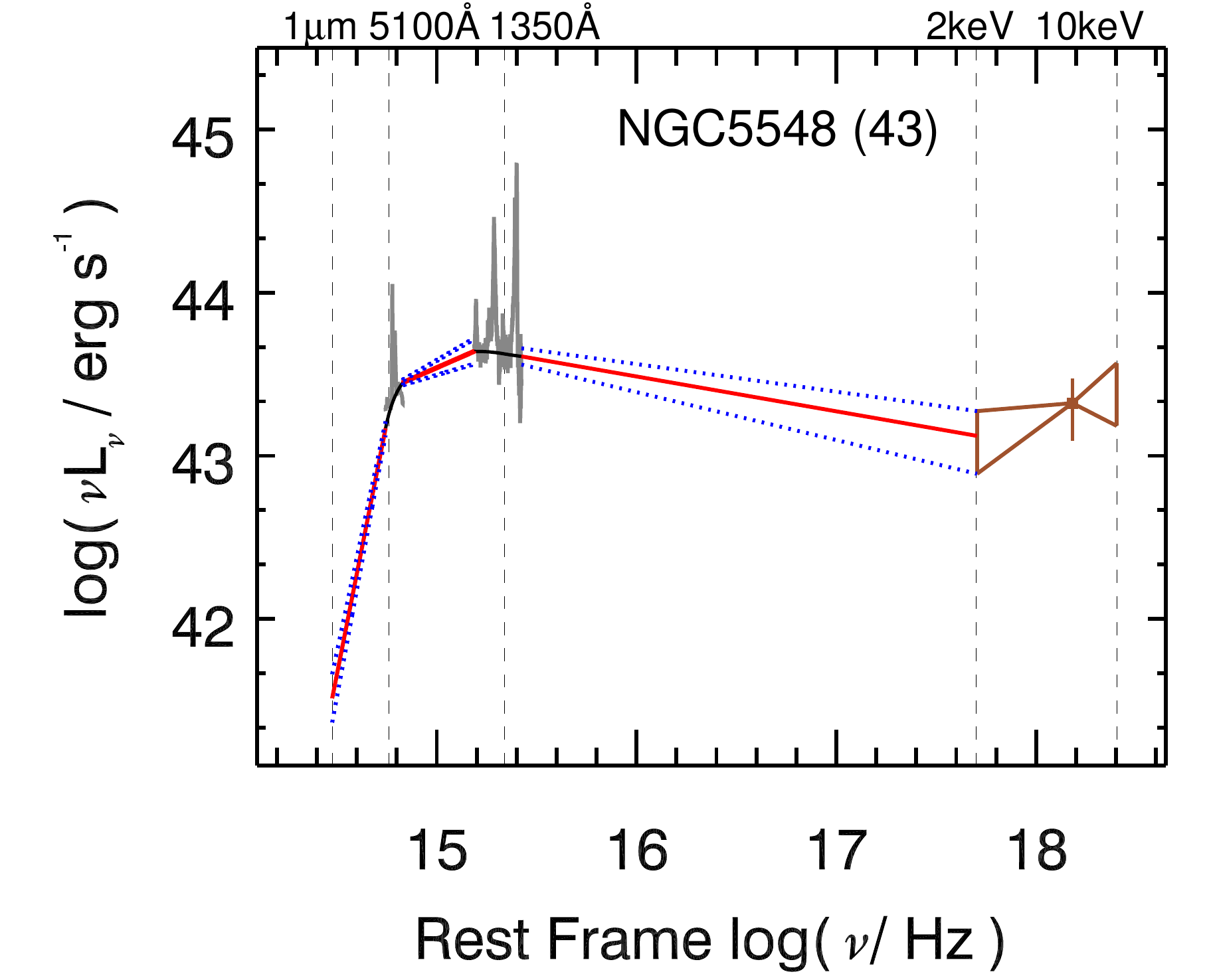}\\
\end{array}$
\end{center}
\contcaption{}
\end{figure*}

\begin{figure*}
\begin{center}$
\begin{array}{cc}
\includegraphics[width=0.5\linewidth,scale=1.5]{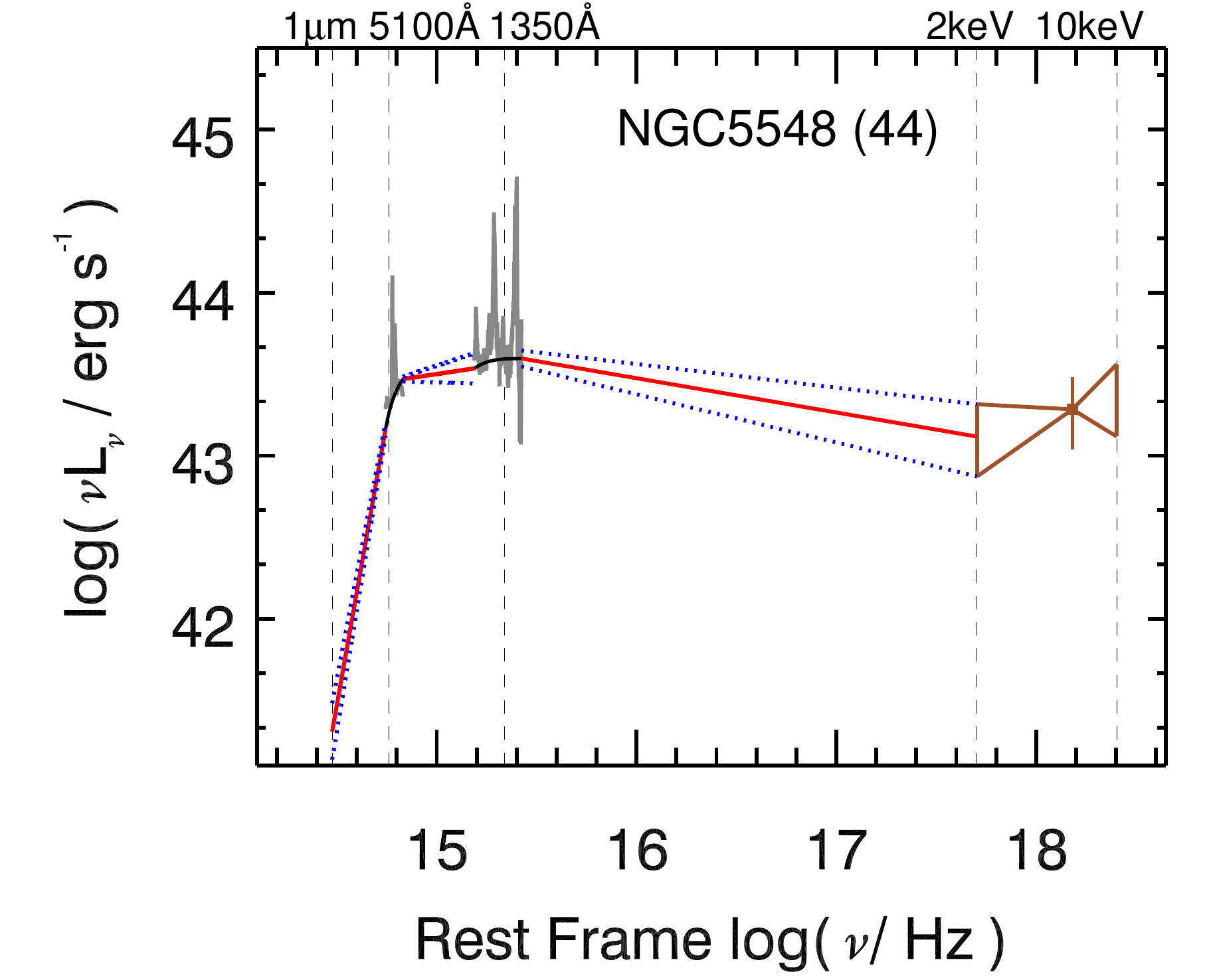} &
\includegraphics[width=0.5\linewidth,scale=1.5]{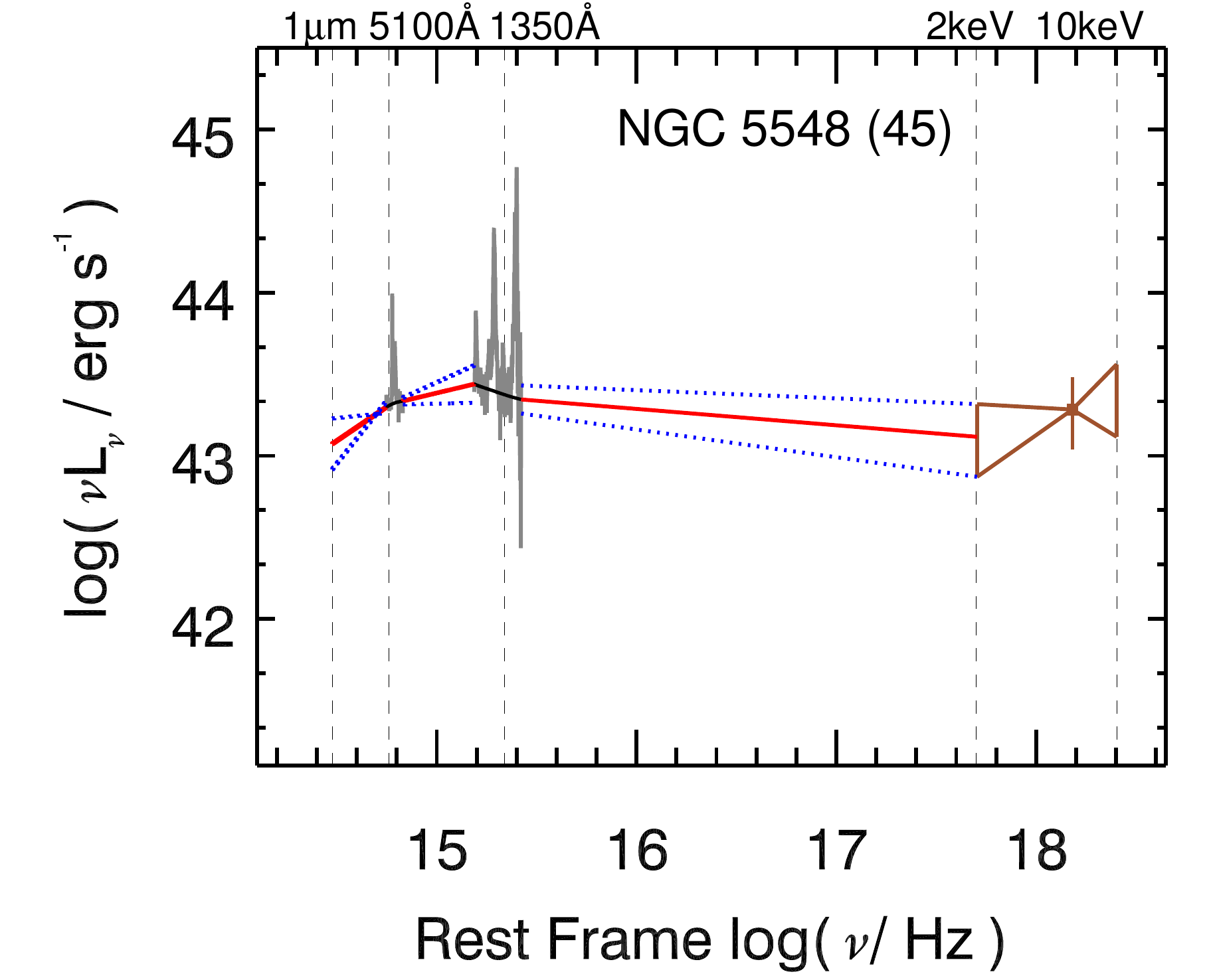}\\
\includegraphics[width=0.5\linewidth,scale=1.5]{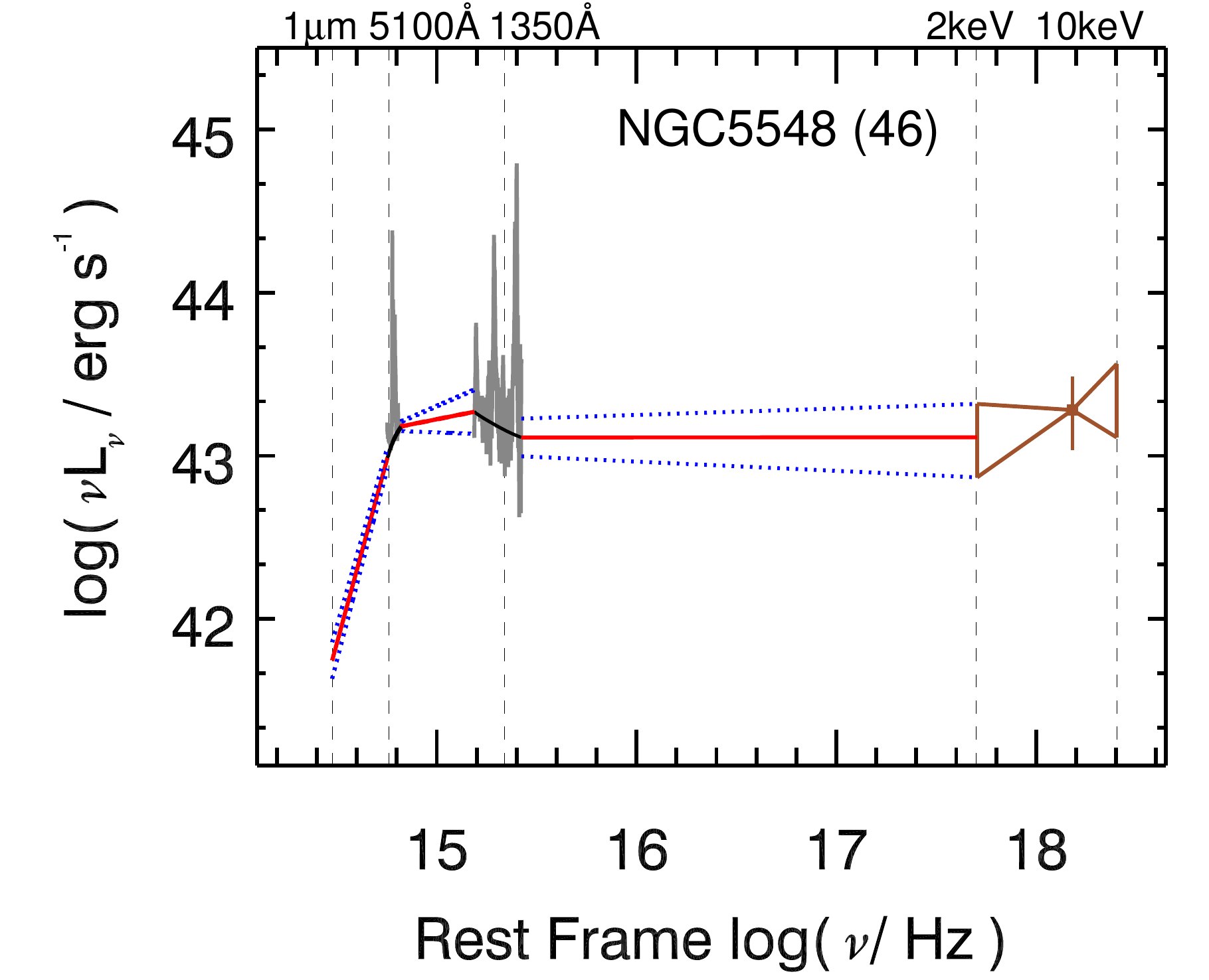} &
\includegraphics[width=0.5\linewidth,scale=1.5]{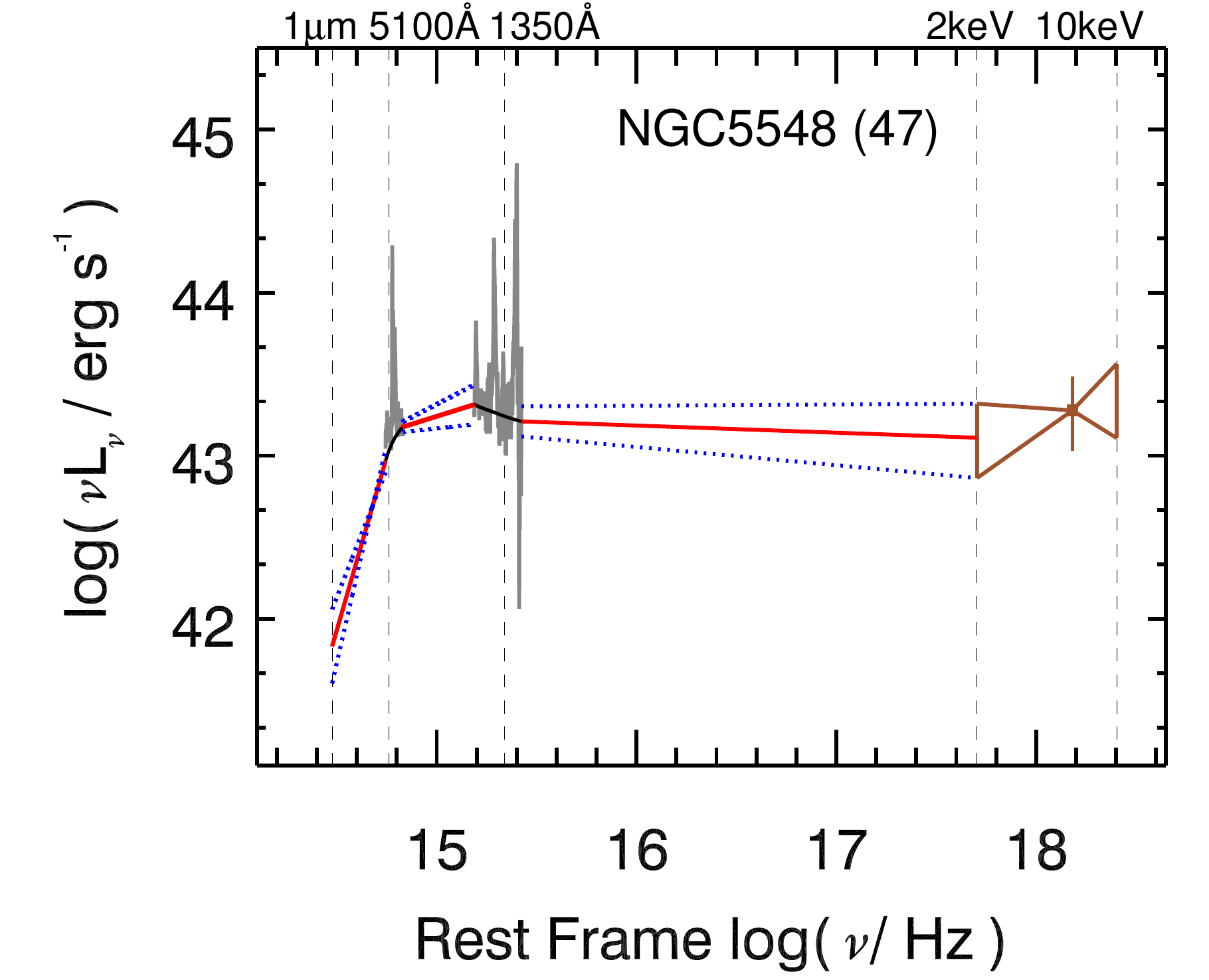}\\
\includegraphics[width=0.5\linewidth,scale=1.5]{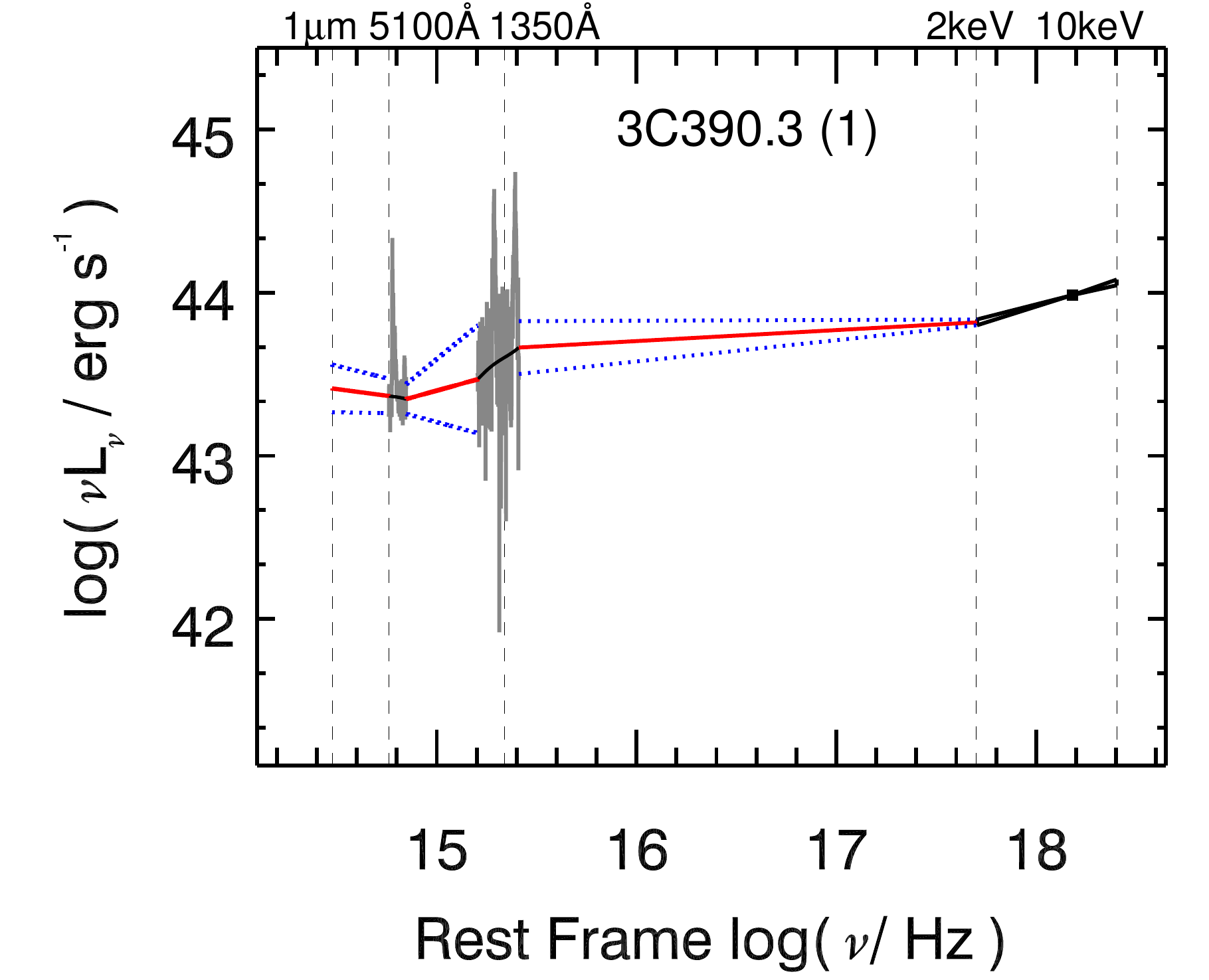} &
\includegraphics[width=0.5\linewidth,scale=1.5]{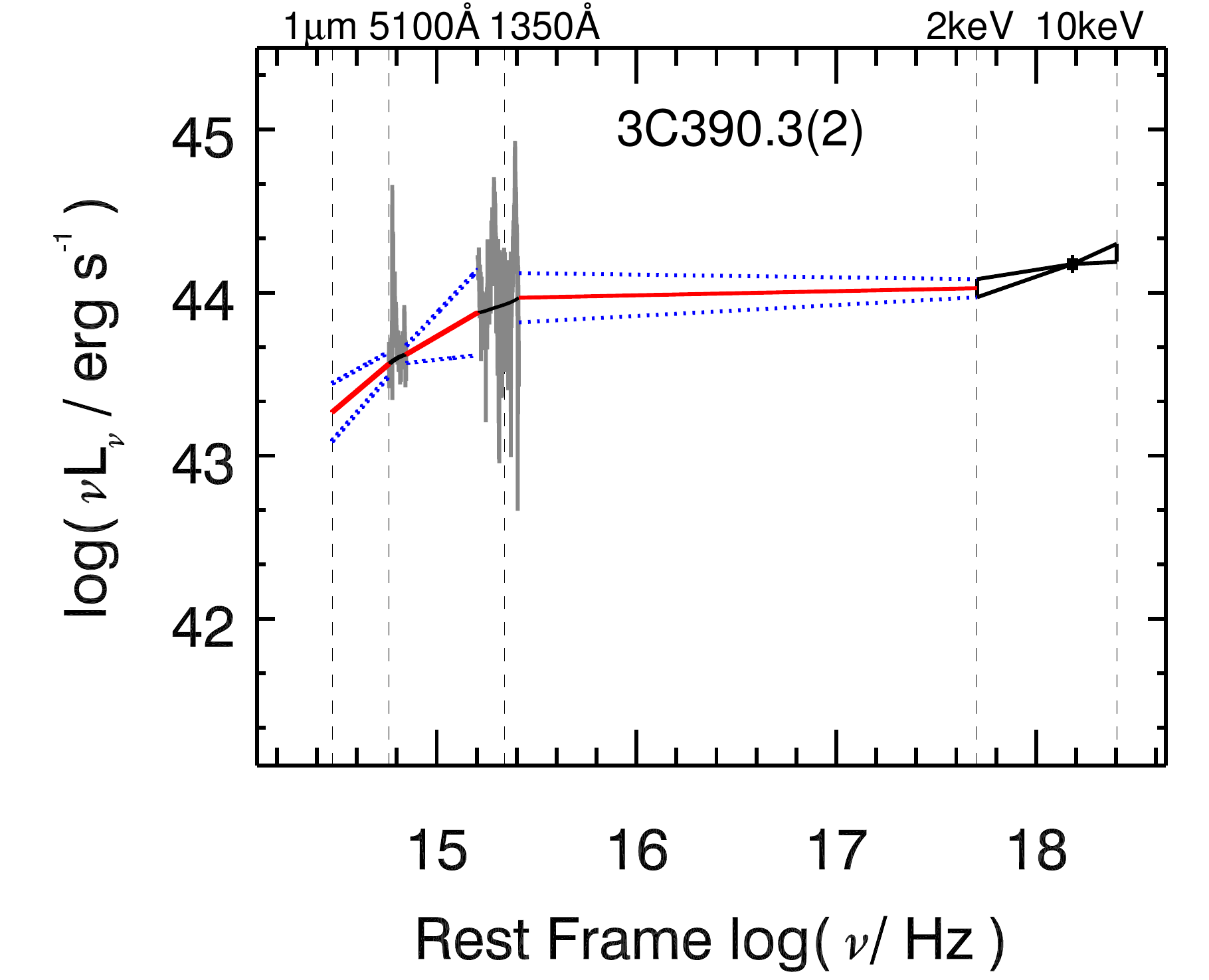}\\
\end{array}$
\end{center}
\contcaption{}
\end{figure*}

\begin{figure*}
\begin{center}$
\begin{array}{cc}
\includegraphics[width=0.5\linewidth,scale=1.5]{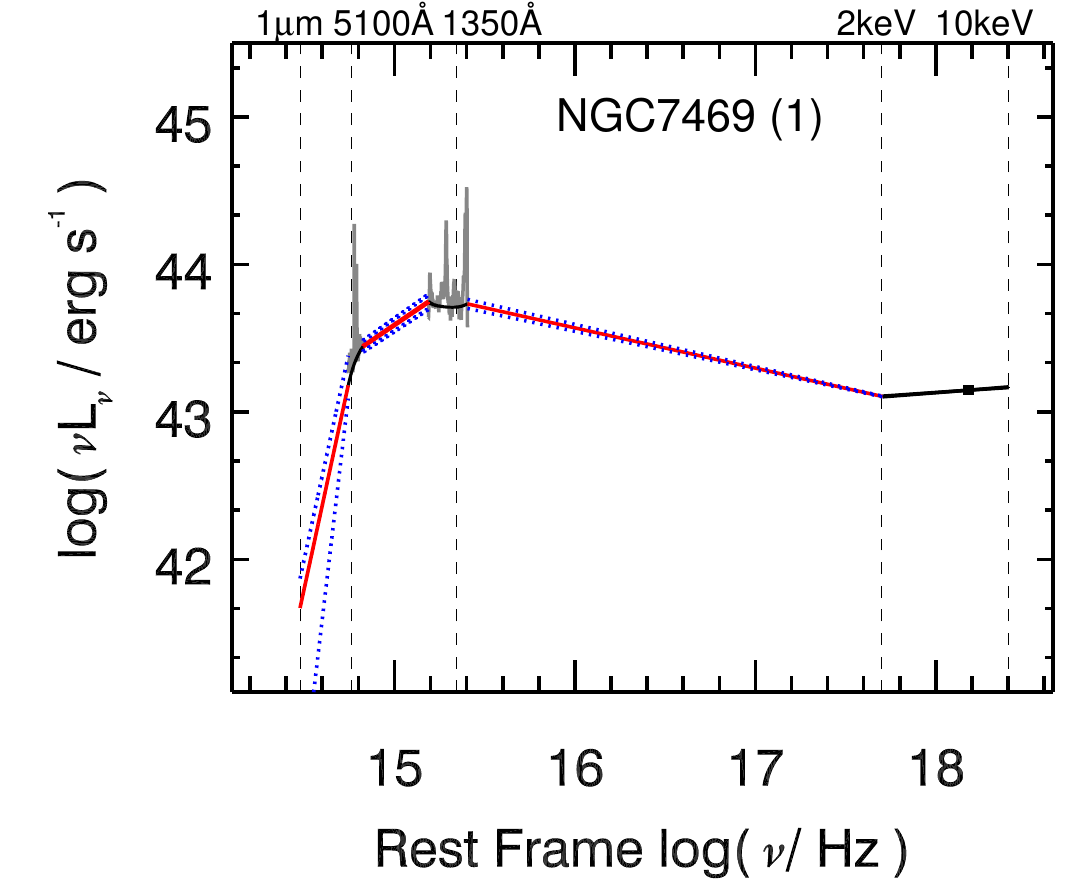} &
\includegraphics[width=0.5\linewidth,scale=1.5]{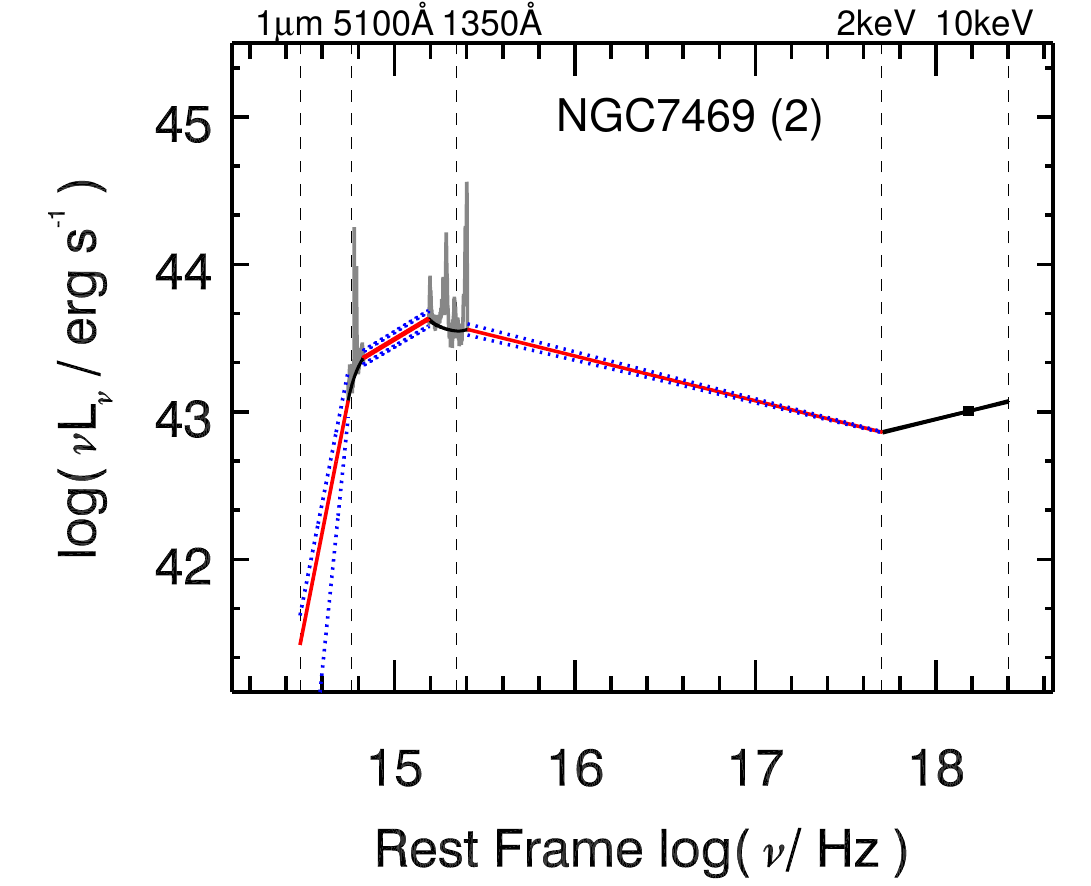}\\
\includegraphics[width=0.5\linewidth,scale=1.5]{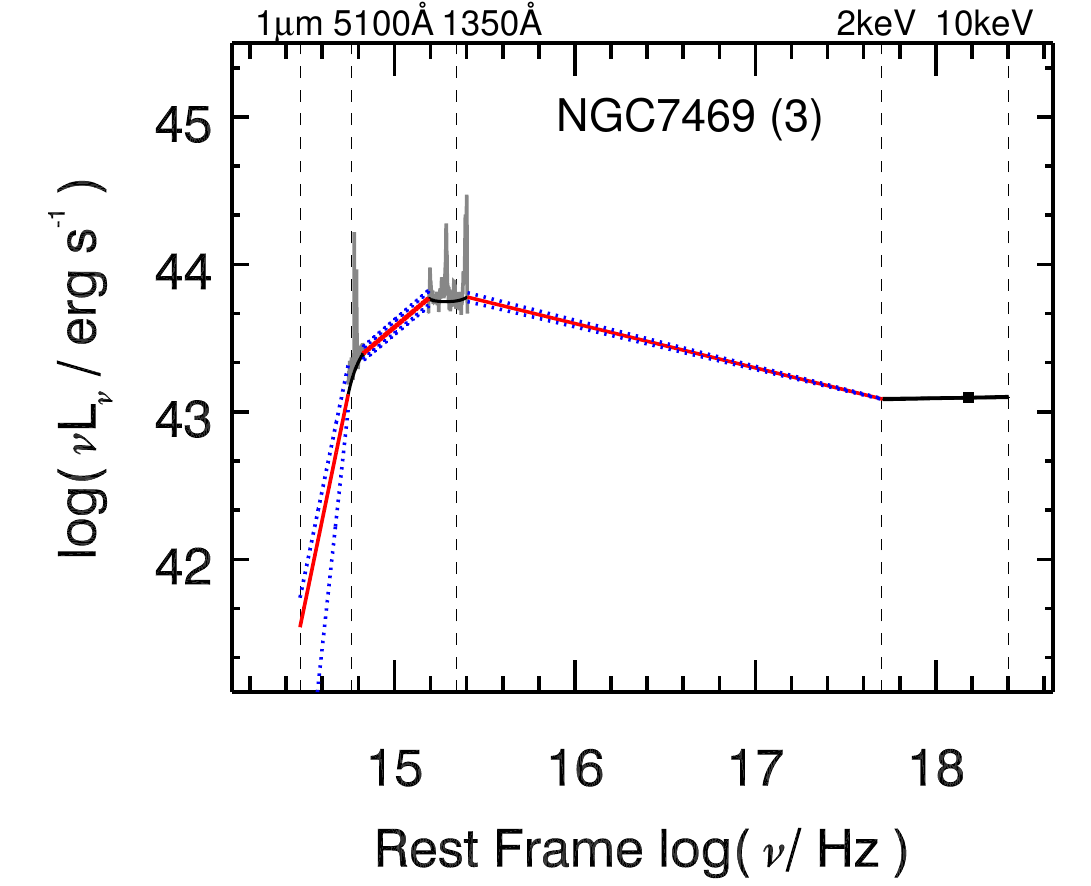} &
\includegraphics[width=0.5\linewidth,scale=1.5]{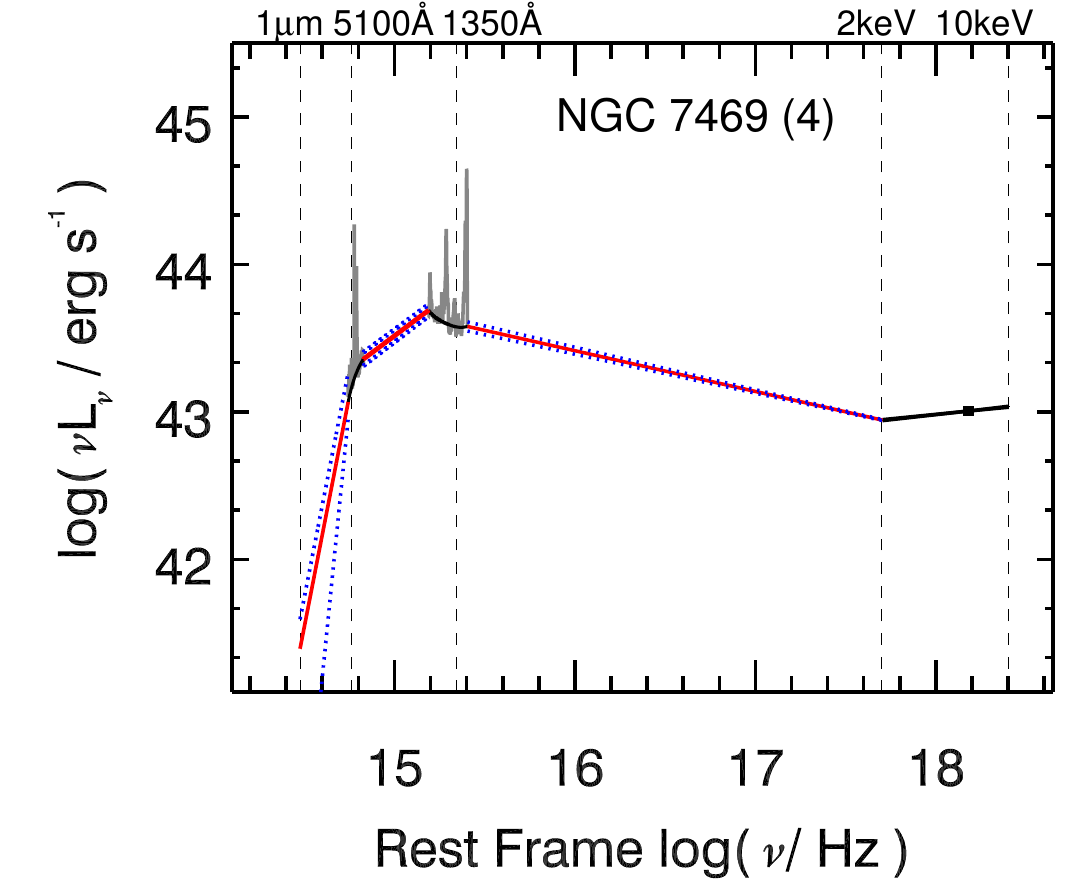}\\
\end{array}$
\end{center}
\contcaption{}
\end{figure*}

\section{ACCRETION LUMINOSITY
BASED ON THE OPTXAGNF MODEL} \label{S:optxangf} 

We attempt to fit one SED of NGC\,5548 [namely, epoch (1)] in order to test  the effectiveness of the \textit{optxagnf} model to represent our multiwavelength quasi-simultaneous SEDs. 
This model connects the UV and X-ray emission by the assumption of energy conservation 
such that the accretion disc generates the total energy output and the soft and hard X-ray emission is restricted to reprocess the radiation  produced by the disc. 
The \textit{optxagnf} model includes three different emission components to model the optical to X-ray SED. 
The optical-UV emission is modelled as blackbody radiation produced in the outer disc.  
The X-ray emission is produced in a two-phase medium located between the last stable orbit and the corona radius ($R_{cor}$), the transition radius between the blackbody and X-ray radiation components. 
The low temperature ($\sim$ 0.1--0.3 \,keV) plasma generates the soft excess X-ray emission (below $\sim$ 2 keV) by inverse-Compton scattered optical/UV photons. 
The hot ($\sim$ 100\,keV) plasma produces the hard X-ray power law between 2 and 10\,KeV. 

We perform the SED fitting with the XSPEC v12.8.1 \citep{Arnaud1996} software.  
In order to have a nearly equal number of data points in the optical/UV and X-ray regimes, we bin each optical and UV continuum into $\sim$10 bins.
The optical, UV and X-ray continuum fluxes are converted to the required format for XSPEC using the {\rm FLX2XSP} tool of HEASOFT. 

The \textit{optxagnf} model has 12 parameters: black hole mass, $M_{\rm BH}$; distance of the source, $D_{L}$; Eddington ratio, $\lambda_{Edd}$; spin of the black hole, $a_{\star}$; radius of the corona,$R_{cor}$, in units of gravitational radii; 
outer radius of the disc, $R_{out}$; electron temperature of the soft Comptonization component, $kT_{e}$; optical depth of the soft Comptonization region, $\tau$; slope of the hard X-ray power law, $\Gamma$; 
the fraction of the power below $R_{cor}$ which is emitted in the hard Comptonization component, $f_{pl}$; redshift of the source, $z$; and normalization.
 
\begin{figure}
\begin{center}$
\begin{array}{c}
\includegraphics[scale=0.45]{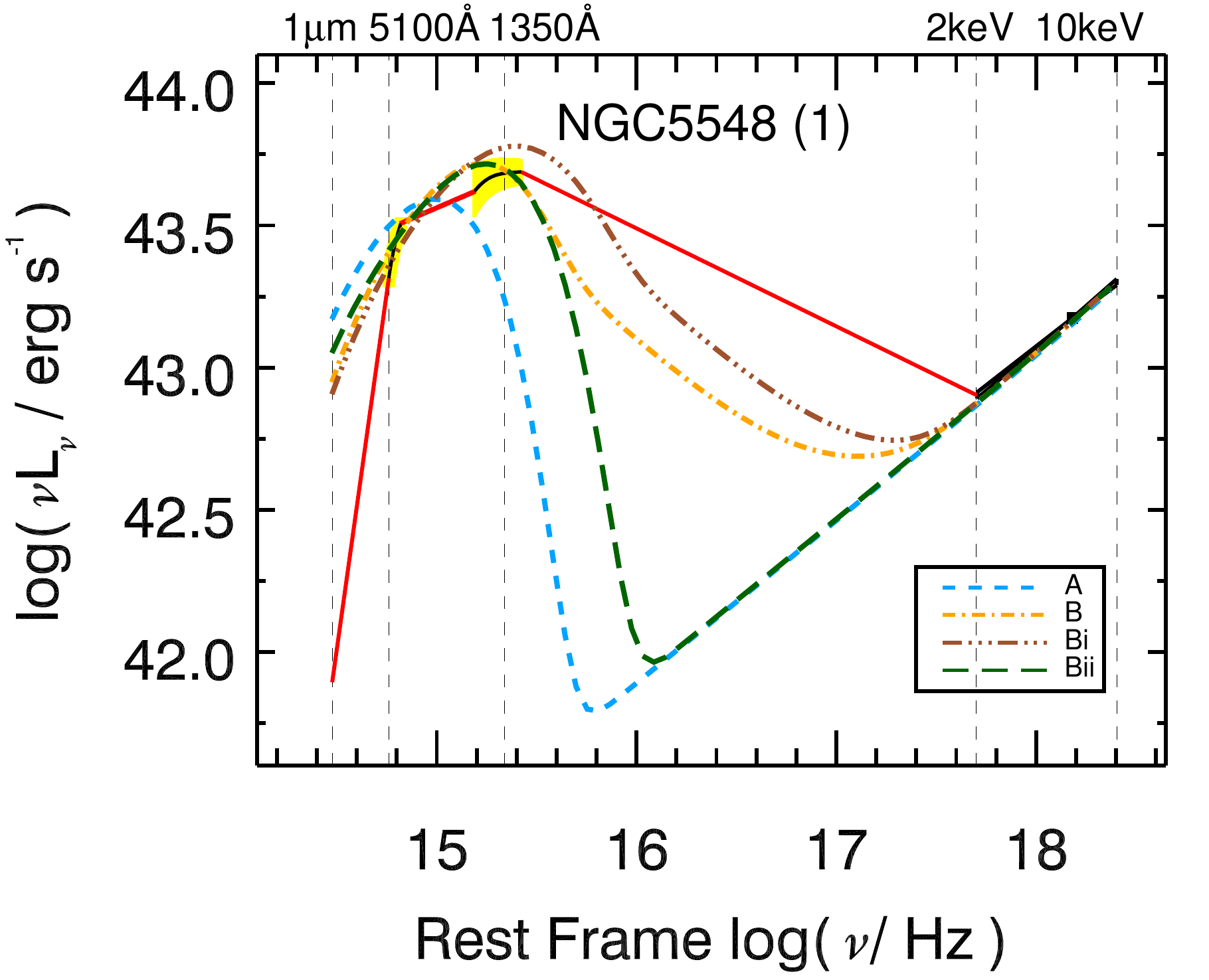}\\
\end{array}$
\end{center}
\caption{OPTXAGNF model fits for NGC\,5548 SED epoch (1). 
The dashed blue line (A), the dot--dashed orange curve (B), the three dot--dashed brown curve (Bi) and the long dashed green curve (Bii) 
show the best-fitting models for different parameter settings (see Appendix \ref{S:optxangf}). 
Model A has $M_{\rm BH}=6.71\times\,10^{6}$\,$M_{\sun}$, $a_{\star}$=0.0, $\lambda_{Edd}$=0.043; 
Model B has $M_{\rm BH}=2.97\times\,10^{7}$\,$M_{\sun}$, $a_{\star}$=0.0,  $\lambda_{Edd}$=0.094; 
Model Bi has $M_{\rm BH}=5.49\times\,10^{7}$\,$M_{\sun}$, $a_{\star}$=0.73,  $\lambda_{Edd}$=0.048; 
Model Bii has $M_{\rm BH}=2.97\times\,10^{7}$\,$M_{\sun}$, $a_{\star}$=0.5,  $\lambda_{Edd}$=0.10.  
The red solid lines show the linearly interpolated SED, the solid black curves represent the observed optical and UV continua and the yellow shades show the continuum uncertainties (\S \ref{S:sedssection}). }
\label{fig:figb1}
\end{figure}

Four of these parameters is held fixed in value during the modelling by default: $D_{L}$ and $z$ as given in Table \ref{tab:table1}, $\Gamma$ to the value adopted from the literature (references are listed in Table \ref{tab:table2}) and the normalization to 1.0. 
In our first test, we set $M_{\rm BH}$ to the value obtained by the RM analysis \citep{Peterson04} and assume a non-rotating Schwarzschild black hole ($a_{\star}$ = 0).  
Since our X-ray data do not cover the soft X-ray emission (\S~\ref{S:Xray_cont}), our data base cannot constrain the soft excess emission sufficiently. 
With no soft X-ray data at hand, we set $f_{pl}$ to 1, such that the soft excess Comptonization is not considered. 
We refer to this case as model `A', the resultant model parameters are tabulated in Table~\ref{tab:tabb1} and the model is shown in Fig. \ref{fig:figb1} (the light blue dashed lines). 
Model `A' significantly underestimates the UV emission and overestimates the optical continuum and therefore  it is not satisfactory. 

Next, we tested different parameter settings in order to obtain a good fit. 
In order to account for soft excess emission with no data to constrain it, we set the Comptonization components to the typical values given by \citet{Jin2012}, namely 
$kT_{e}$=0.2 and $\tau$=16 and let $f_{pl}$ be a freely varying parameter. 
We also set $M_{\rm BH}$ to be in the range between $2.4\times\,10^{7}$ and $10.2\times\,10^{7}$, consistent with the $M_{\rm BH}$  estimates of a few recent studies \citep{Peterson04,Bentz09b,Pancoast2014}. 
This model, labelled `B', is shown in Fig.~\ref{fig:figb1} as the orange dashed-dotted curve. 
For a lower $M_{\rm BH}$ value of $2.97\times\,10^{7}$, we obtain a somewhat better fit compared to model `A'. 

We then test the effect of a non-zero spin model `B'. 
First we set the $M_{\rm BH}$ value to the best-fitting result of model `B' and allow the spin to be a free parameter. 
In this case, we obtain a spin parameter that is consistent with zero. 
When we set the spin $a$=0.5 for $M_{\rm BH}=2.97\times\,10^{7}$, and the resultant model overestimates the UV continuum (model `B(i)' in Fig. \ref{fig:figb1}; brown dash-dots). 
Then we set $M_{\rm BH}$ and spin as a free parameters ($f_{pl}$ is also a free parameter). 
We obtain $a_{\star}=0.73$ and $M_{\rm BH}=5.49\times\,10^{7}$ (model `B(ii)' in Fig.~\ref{fig:figb1}). 
The latter comparison shows that 
$M_{\rm BH}$ and the spin are degenerate parameters, and it is not possible to measure these parameters robustly with the current data.

Although we have RM based $M_{\rm BH}$ estimates and simultaneous optical-UV and X-ray continuum data, 
the results of the obtained fits are clearly not robust. 
Therefore we consider the accretion luminosity obtained from model `B' a lower limit. It is 23 per cent\ lower compared to $L_{BOL}(acc,ip)$. 
The Eddington ratio constrained from model `B' is a factor of $\sim$3 higher than the one computed from the interpolated SED, because model `B' has a lower $M_{\rm BH}$. 
If we also manipulate parameters as in model `Bi' we may obtain a higher $L_{BOL}(acc)$ that is only 5 per cent\ lower than $L_{BOL}(acc,ip)$, but of course these parameters are highly uncertain and the process is ad hoc.  
 Since the results of the \textit{optxagnf} model are not sufficiently robust for the data available to us, we do not attempt to model any other SED with this model.  
Therefore, we conclude that it is reasonable for our purposes to adopt the model independent approach of \S~\ref{S:bomeasurements}. 

\begin{table*}
           \centering
           \caption{
           OPTXAGNF fit results. Columns: (1) Object name. (2) SED epoch. (3) Black hole mass in $10^{7}$ $M_{\sun}$. (4) Spin of the black hole. (5) Eddington ratio. (6) Radius of the corona in units of gravitational radii. (7) Log of outer radius of the disc in units of gravitational radii. (8) Electron temperature of the soft Comptonization component. (9) Optical depth of the soft Comptonization region. (10) The fraction of the power below $R_{cor}$. (11) Log of accretion luminosity in units of erg s$^{-1}$. 
	}
           \label{tab:tabb1}
           \begin{tabular}{ccccccccccc}
           \hline
           Object & Epoch & $M_{\rm BH}$ & $a_{\star}$ & $\lambda_{Edd}$  & $R_{cor}$ & $\log R_{out}$& $kT_{e}$ & $\tau$  & $f_{pl}$ & $\log(L_{BOL}(acc))$ \\
                       &            & ($10^{7} M_{\sun}$) &          &                               & ($R_{g}$) & ($R_{g}$)       & (keV) &                    &              &                                    \\
           (1) & (2) & (3) & (4) & (5) & (6) & (7) & (8) & (9) & (10) & (11) \\
           \hline
\multicolumn{9}{c}{\boldmath{A}}\\
\hline
NGC\,5548 &1& 6.71 &0.0 & 0.043 &  89.02 & 3.0 & 0.20 & 16.00 & 1.00 &  43.99 \\
\hline 
\multicolumn{9}{c}{\boldmath{B}}\\ 
\hline
NGC\,5548 &1& 2.97 &0.0 & 0.094 &  99.56 & 3.0 & 0.20 & 16.00 & 0.83 &  44.19 \\
\hline 
\multicolumn{9}{c}{\boldmath{Bi}}\\ 
\hline
NGC\,5548 &1& 2.97 & 0.5 & 0.10 & 50.70 & 3.0 & 0.20 & 16.00 & 0.83 &  44.12 \\
\hline
\multicolumn{9}{c}{\boldmath{Bii}}\\ 
\hline
NGC\,5548 &1& 5.49 &0.73 & 0.048 & 32.90 & 3.0 & 0.20 & 16.00 & 1.00 &  44.28 \\
		\hline
	\end{tabular}
\end{table*}


\bsp	
\label{lastpage}
\end{document}